%% file: main.tex
\documentclass[
11pt, 
english, 
singlespacing, 
headsepline, 
]{MastersDoctoralThesis} 

\usepackage[utf8]{inputenc} 
\usepackage[T1]{fontenc} 

\usepackage{mathpazo} 
\usepackage{amssymb, amsmath,graphicx,color,enumitem,hyperref}
\usepackage{xcolor}
\usepackage{ulem}
\usepackage{overpic}
\usepackage[FIGTOPCAP,raggedright,nooneline,bf,footnotesize]{subfigure}

\usepackage[autostyle=true]{csquotes} 

\newcommand{\indep}{\perp \!\!\! \perp}
\newcommand{\figref}[1]{Fig.~\ref{#1}}
\newcommand{\eref}[1]{Eq.~\ref{#1}}

\newcommand{\Tau}{\mathrm{T}}

\DeclareMathOperator{\sign}{sign}
\DeclareMathOperator{\sinc}{sinc}

\thesistitle{Light guided in tailored environments: from basic aspects to applications} 
\supervisor{Dr hab. Piotr  \textsc{Kolenderski}} 
\examiner{} 
\degree{Doctor of Philosophy} 
\author{Andrzej \textsc{Gajewski}} 
\addresses{} 

\subject{Physics} 
\university{\href{http://www.umk.pl}{Nicolaus Copernicus University}} 
\department{\href{https://www.fizyka.umk.pl/en/}{Faculty of Physics, Astronomy and Informatics}} 
\group{\href{http://spa.fizyka.umk.pl/}{Single Photon Applications Laboratory}} 
\faculty{\href{https://www.ifiz.umk.pl/en/}{Institute of Physics}} 

\AtBeginDocument{
\hypersetup{pdftitle=\ttitle} 
\hypersetup{pdfauthor=\authorname} 
\hypersetup{pdfkeywords=\keywordnames} 
}

\begin{document}

\frontmatter

\pagestyle{plain}

\begin{titlepage}
\begin{center}

\vspace*{.06\textheight}
{\scshape\LARGE \univname\par}\vspace{1.5cm}
\textsc{\Large Doctoral Thesis (revised after submission)}\\[0.5cm]

\HRule \\[0.4cm] 
{\huge \bfseries \ttitle\par}\vspace{0.4cm} 
\HRule \\[1.5cm] 
 
\begin{minipage}[t]{0.4\textwidth}
\begin{flushleft} \large
\emph{Author:}\\
{\authorname}
\end{flushleft}
\end{minipage}
\begin{minipage}[t]{0.4\textwidth}
\begin{flushright} \large
\emph{Supervisors:} \\
{\supname} \\
{Dr hab. Karolina \textsc{Słowik}} 
\end{flushright}
\end{minipage}\\[3cm]
 
\vfill

\large \textit{A thesis submitted in fulfillment of the requirements\\ for the degree of \degreename}\\[0.3cm] 
\textit{in the}\\[0.4cm]
\groupname\\\deptname\\[2cm]
 
\vfill

{\large May 9, 2022}\\[4cm] 
 
\vfill
\end{center}
\end{titlepage}

\begin{acknowledgements}
\addchaptertocentry{\acknowledgementname} 

\setlength\parindent{24pt} The author would like to acknowledge and thank his supervisor ~\textbf{Dr Piotr Kolenderski} for the support and collaboration in research through the past 8 years. 

\vspace{\baselineskip}

Special thanks go to \textbf{ Dr Karolina Słowik} not only for the joint research, but also for the fact that she always had time for the author. The author would like to thank and apologize at the same time for all the weekends and holidays that \textbf{Dr Słowik }spent reading this doctoral dissertation.

\vspace{5\baselineskip}

\textit{The research presented within this thesis was supported and partially financed by Project HEIMaT No. Homing/2016-1/8.}
\end{acknowledgements}

\tableofcontents

\dedicatory{To my late father, who unfortunately didn't live long enough to see his son become a doctor.}

\mainmatter 

\pagestyle{thesis} 

\include{Chapters/introduction}

\include{Chapters/Chapter2} 
\include{Chapters/Chapter3}

\include{Chapters/Chapter5}
\include{Chapters/Chapter4}
\include{Chapters/summary}
\include{Appendices/AppendixA}

\appendix 
\bibliographystyle{unsrt}
\bibliography{FamoLab3}

\end{document}

%% file: Chapters/introduction.tex
\chapter{Introduction} 

\label{intro}

This dissertation is dedicated to the investigation of quantum optical effects involving single photons in tailored structures. All the research results presented in this thesis were the outcome of my work in the KL FAMO (Krajowe Laboratorium Fizyki Atomowej i Optycznej) laboratory at the Institute of Physics of the Nicolaus Copernicus University in Toruń, in the Single Photon Applications Laboratory (SPA) laboratory group led by Dr Piotr Kolenderski. Research in the SPA laboratory combines the use of single-photon sources based on the spontaneous parametric down-conversion (SPDC) phenomenon. This research is both experimental, theoretical and numerical in nature.

Spontaneous parametric down-conversion in a nonlinear crystal was first demonstrated in 1967 \cite{Harris1967}. Although that way of producing nonclassical light is relatively old, it is still one of the most efficient ways of generating pairs of photons including highly entangled photon pairs. The SPDC process plays a crucial role in experiments involving single photons \cite{Shalm2015, Giustina2015, Aharon2008, Kolenderski2011} but also it is very important in quantum communication applications, especially in quantum key distribution \cite{Khan2018,Liao2017, Pugh2017}.

The reasons why SPDC is widely used are its relative simplicity of implementation, the robustness of setup and the possibility of creating single photons on demand. There are other ways of creating single photons i.e using quantum dots \cite{kako2006gallium, shields2007semiconductor} or NV-centers \cite{yonezu2017efficient}. These types of sources have their advantages, for example, quantum dots might be integrated with photonic circuits. SPDC sources, on the other hand, would be very hard to integrate with microelectronics.  

My research focused on the theoretical and numerical aspects of the research work. A lot of my theoretical research and numerical simulations support the experiments performed in this laboratory. Others lead to suggestions for experiments that may be performed in the future. The scope of my research on issues related to quantum optics is wide: it includes the applied aspect, the development of the theoretical model and basic research. The applied aspect is the development of a methodological framework for the determination of Sellmeier coefficients \cite{Marcuse1980a} based on measurements of photon properties from SPDC sources.

Sellmeier equations together with the parameterization coefficients of these equations are a way to describe dispersion phenomena in crystals, also those used for SPDC \cite{Saleh2007}.
The precise determination of the Sellmeier coefficients is important both in the design of experiments based on SPDC and in non-linear optics \cite{Kato2017, Miyata2017}. The growing interest in quantum key distribution (QKD) based on single photon-pair sources \cite{Yin2017} also motivates research into the determination of Sellmeier coefficients. The process of forming crystals enabling SPDC often suffers from uncertainties originating in fabrication process, leading to different dispersion characteristics of the crystal than intended. Therefore, it is important to find an uncomplicated and precise method of their determination. I will describe the research on this issue in Chapter 3, and the results will be used in the next chapter.

The fourth chapter is devoted to the theory of the parametric frequency conversion phenomenon. I will develop a model describing the joint probability distribution of the frequencies of photon pairs generated in the process, called in the literature on the subject "biphoton wave function". My contribution here is a generalization of the pre-existing model limited to the case of non-periodic crystals. I also developed a numerical implementation of the analytical model, the performance of which was presented in the articles \cite{Lutz2013, Lutz2014, Gajewski2016}. The model presented in the fourth chapter calculates the "biphoton wave function" not only for uniaxial but also for biaxial crystals.
After a short introduction, I will conduct an extensive discussion to validate the model against two experiments performed in the SPA laboratory.

I will devote an extensive fifth chapter to modelling the modes of the bent waveguide, combining analytical considerations and numerical modelling. In this chapter, I find solutions to the Helmholtz equation \cite{Saleh2007} defining the modes of an optical waveguide, going beyond the approximations commonly used in the literature. Thanks to this, I will be able to correctly predict the number of modes in a bent waveguide and their spatial distribution. The correctness of my results is verified with numerical simulations obtained in COMSOL. 
In this chapter I emphasize the context of basic research: I show that the bent waveguide is described by equations analogous to the equations of the dynamics of a quantum particle in a space with axial symmetry. 
The motion of such a particle is described using an effective image based on a fictitious quantum potential related to quantum centrifugal force \cite{Wheeler2001,Cirone2001, Rembielinski2002, Birula2007, Dandoloff2011, Dandoloff2015}. According to Feynmann's famous remark that "the same equations have the same solutions" \cite{leighton1965feynman}, due to the analogical equations, a bent waveguide can provide an experimental platform for studying fictitious quantum forces. The theoretical considerations described above use the mathematical apparatus described in detail in the introductory Chapter 2. There, I will introduce basic information about the statistical methods used in my work and about the properties of non-classical light. In addition, the appendices devoted to the form of the differential operator in cylindrical coordinates, the analysis of the influence of photon propagation in the optical fibre on the photon wave function and the Schrödinger equation for a particle in two-dimensional space with axial symmetry will be helpful to understand the content.

%% file: Chapters/Chapter2.tex
\chapter{Theoretical background} 

\label{Chapter1} 
This chapter introduces basic mathematical tools to be used throughout the thesis. First, we shortly review Maxwell and Helmholtz equations. We will need both in the last chapter of this thesis. Next, we introduce quantum coherence functions and with their help, we describe the most important quantum states. We move on to a description of the spontaneous parametric down-conversion process, crystal anisotropy and phase matching conditions. We use this description to introduce the formalism of biphoton wavefunction in the next chapter.

\section{Classical light}
Classical electromagnetic radiation is fully described by a set of Maxwell equations. Maxwell equations in a linear dielectric, when no sources or charges are present have the following form \cite{Jackson1999}:
\begin{align}
    \nabla \cdot \boldsymbol E &=  0 \label{eq:Ediv} \\
    \nabla \cdot \boldsymbol B &= 0 \label{eq:Bdiv} \\
    \nabla \times \boldsymbol E  &= -\frac{\partial \boldsymbol B}{\partial t}\label{eq:Erot}\\
    \nabla \times \boldsymbol B  &= \frac{\mu_r \epsilon_r}{c^2} \frac{\partial \boldsymbol E}{\partial t}. \label{eq:Brot}
\end{align}
Here, $\mu_r$ is magnetic permeability and $\epsilon_r$ is relative electric permittivity. As usual, $c$ stands for light velocity in vacuum. Bold symbols are used to denote vectors and $\boldsymbol E$ and $\boldsymbol B$ represent electric and magnetic vector fields, respectively. The first two equations describe the divergence of fields, and the last two describe the curl of the fields. Both expressions involve the nabla symbol $\nabla$. The dot product is denoted with $\cdot$ and the vector product with $\times$. From the above equations, one can easily obtain the wave equation. Wave equations in vacuum read:
\begin{align}
    \frac{\partial^2 \boldsymbol E}{\partial t^2} - c^2 \nabla^2 \boldsymbol E &= 0 \\
    \frac{\partial^2 \boldsymbol B}{\partial t^2} - c^2 \nabla^2 \boldsymbol B &= 0
\end{align}
Both the electric and magnetic fields are position and time dependent $\boldsymbol E = \boldsymbol E(\boldsymbol r, t)$, $\boldsymbol B = \boldsymbol B(\boldsymbol r, t)$. If we assume that separation of variables is justified in our situation, we can simplify the wave equation to the Helmholtz equation. Let us assume that linearly polarized electric field can be separated into two functions $\boldsymbol E(\boldsymbol r, t) = R(\boldsymbol r) T(t) \boldsymbol e$, with $\boldsymbol e$ being the polarization vector, then \cite{Jackson1999}:
\begin{align}
    \frac{\nabla^2 R(\boldsymbol r)}{R(\boldsymbol r)} &= -k^2 \label{eq:Helmholtz} \\
    \frac{1}{c^2 T(t)} \frac{d^2 T(t)}{dt^2} &= - k^2.
\end{align}
The above equation is the Helmholtz equation and $k = |\boldsymbol k|$, where $\boldsymbol k$ stand for wavevector in vacuum.

\section{Non-classical light}

Electromagnetic radiation is very well described by Maxwell equations as long as intensity of radiation is significant. When intensity is low, the corpuscular nature of radiation becomes apparent. At that point, Maxwell equations are not sufficient for description of physical phenomena caused by low intensity radiation. The reason why that happens is that Maxwell equations do not impose a limit of how low intensity the radiation can get. This result is in contradiction with observations made in experiment i.e Hanbury Brown and Twiss. We know that the smallest possible portions of radiation energy is carried by photons. 
Another difference between the classical and quantum description of electromagnetic fields comes from non determinism of quantum theory. 
In quantum optics, solutions of Maxwell's equations can be interpreted as probability amplitude distributions for photons. We will make use of such interpretation in the spectral domain when considering biphoton wavefunctions or in the spatial domain discussing waveguide modes.This interpretation is in agreement with observed statistical nature of very low intensity radiation. 

To describe physical phenomena which involve very low intensity radiation we need a different approach in experiments and different mathematical apparatus. In this chapter I will address both problems. I will start with short introduction to stochastic processes. 

First, I will introduce the notion of the Poisson process, which is an important stochastic process in the theory of quantum optics. Properties of that process will be used for proving the non-classical behaviour of some light states. Then, I will describe classical correlation functions and their quantum analogues: quantum coherence function. For completeness of description, discussion of the first and second-order correlation function will be supplemented with a description of the Young interference experiment and Hanbury Brown and Twiss experiment. At this point, all the necessary tools will have been introduced and I will move on to a description of quantum states of light. At the end of the chapter, we explain the notion of a squeezed state which will be important in the following chapter.

\subsection{Statistical approach to light states}
Let us imagine an idealized experimental scenario where we have a source of photons and a perfect detector capable of detecting every photon with $100 \%$ efficiency and without any noise. In that setup, there are no other sources of photons. Such a detector would be connected to a simple counter, which would show how many photons have already hit the detector since the beginning of the experiment. Such a setup would provide us with statistics about the arrival times of photons at the detector location. Such a process in the theory of stochastic process is called a counting process \cite{gallager2013stochastic}.

Counting processes are characterized by a set of arrival times $T_n$ given by
\begin{equation}
T_n = T_{n-1} + \xi_n,
\end{equation}
where $ n \in \mathrm{N}$ , $T_0 = 0$ and $\xi_1, \xi_2, \dots, \xi_n$ are positive random variables. We denote stochastic processes in curly brackets $\{ \cdot \}$:
\begin{equation}
\{X\}  : T \times \Omega \rightarrow S,
\end{equation}
where $T \in R^+$ stands for time, $\times$ is a Cartesian product, $\Omega$ is a sample space and $S$ is a state space. It is important to note that $X(t)$ is a random variable. In the case of the counting process, the sample space $\Omega$ is a set of non-negative integers (set of possible outcomes).

Values of the counting process $\{N(t) \}$ are non-negative and non decreasing. In the case of our idealized experimental scenario, $N(t)$ is a number of photons that hit detector before time $t$, and $\xi_n$ is a difference between the arrival of $n$-th photon and $(n-1)$th photon. In that sense, process $\{N(t) \}$ characterizes the photon source.

Renewal processes are a subclass of the counting processes, where  $\xi_1, \xi_2, \dots, \xi_n$ are not only positive but also independent and identically distributed (IID) - after each photon  emission the source 'resets' itself to its default state.  One way to understand that statement is to imagine a source where photons are not independent (e.g bunched or anti-bunched, not randomly distributed in time) or not identically distributed (e.g laser with decreasing power: the source is getting weaker with each emission, and arrival times are getting larger with time). These are examples of processes that do not belong to the class of renewal processes. Renewal process $N_t$ is modelled by equation:
\begin{equation}
N(t) = \sup \{k: T_k < t\} = \sum_{k = 1}^{\infty} \mathrm{I} (T_k < t), \label{ref:Poisson}
\end{equation}
Where $\mathrm{I \{\cdot \}}$ is an indicator function, which is $1$ anytime a statement is true, and $0$ otherwise. The function $N(t)$ defined in \eqref{ref:Poisson} describes number of photons which arrived at the detector before time $t$.

The Poisson process is a renewal process where the probability $Pr\{\cdot \}$ of different time jumps $\xi_n$ is given by the exponential distribution:
\begin{equation}
\forall_{\tau \geq 0} Pr \{T_{n} - T_{n - 1} < \tau \} =Pr \{ \xi_n < \tau \} = 1 - \exp(-\lambda \tau), 
\end{equation}
where $\lambda$ is called the intensity parameter of the Poisson process. The above equation can be reformulated to:
\begin{equation}
\forall_{\tau \geq 0, t \geq 0} Pr \{N_{t+\tau} - N_{t} \geq 1 \} = 1- Pr \{N_{t+\tau} - N_{t} = 0 \}  = 1 - \exp(-\lambda \tau). 
\end{equation}
The exponential distribution is called 'memoryless', since the probability of an event happening (photon arrival at the detector) in some time interval $\tau$, given that amount of time $t$ passed, is independent of $t$:
\begin{equation}
\forall_{Pr \{X > t \} \neq 0}Pr\{X > t + \tau | X > t \} = Pr\{X > \tau \},
\end{equation}
where the probability distribution of random variable $X$ is given by exponential distribution. The well known Poisson process formula for probability of exactly $k$ events happening during period $t$ is \cite{gallager2013stochastic}:
\begin{equation}
Pr\{N_t = k\} = \exp( - \lambda t) \frac{(\lambda t) ^k}{k!}. \label{eq:poissondist}
\end{equation}
The Poisson process has an expected value equal to its variance:
\begin{equation}
\langle N_t \rangle =(\Delta N_t)^2 = \lambda t, \label{eq:poissonmean}
\end{equation}
where $\langle N_t \rangle$ is an ensemble average, $(\Delta N_t)^2$ is an ensemble variance.  
Another important property of the Poisson process is called branching.

If each event, as it is happening, we assign randomly with probability of $p$ and $q = 1-p$ to processes $\{ N^p_t \}$ and $\{ N^q_t \}$, so that at any point of time $N_t = N^p_t + N^q_t$, where $\{ N_t \}$ is a Poisson process, then $\{ N^p_t \}$ and $\{ N^q_t \}$ are Poisson process too. The intensity parameters of those processes are $\lambda^p = p \lambda$ and $\lambda^q = (1-p) \lambda$.

Proof of that can be found in \cite{gallager2013stochastic}. 
The Poisson process is therefore a very useful tool for the description of some states of light, as measured by single-photon detectors. Such detectors count photons, zero or one per detection time interval. Such a process is obviously a counting process. If the time arrival of photons is IID (independent, identically distributed) then one might expect that counts will be well modelled by the Poisson process. Even if the detector does not work with perfect efficiency, because of branching property, counts still should be well described by the Poisson process. It can be understood as follows: photons are randomly divided into a group of detected photons and a group of undetected photons. Since division is random but with constant probability given by detector efficiency, then (for a set time) probability of finding a given number of photons in any of these group is given by the Poisson distribution.

To sum up, if events happen in independent time intervals given by memoryless distribution then the process is a Poissonian. Equivalently, if counting process is not described by the Poisson distribution then either the system is not memoryless or counting events are not independent. 

Of course, photons can be correlated and the source which generates photons can have memory. In that case, the statistic will deviate from the Poisson distribution. If the variance in such a process is larger than the corresponding expectation value then a process is called super-Poissonian. Similarly, if the variance is smaller than the expected value, the process is called sub-Poissonian.

\subsection{Correlation functions}

In statistics, there is a well-known coefficient describing linear dependence between two variables. It is known as the Pearson correlation coefficient \cite{gallager2013stochastic}:
\begin{equation}
\rho_{XY} = \frac{Cov(X,Y)}{\sigma_X \sigma_Y}, \label{eq:rho}
\end{equation}
where $Cov(\cdot{, \cdot})$ is a covariance function. In theory of stochastic processes there is a similar quantity describing linear dependence between stochastic processes:
\begin{equation}
\rho_{XY}(t_1, t_2) = \frac{K_{XY}(t_1,t_2)}{\sqrt{\sigma^2_{X}(t_1) \sigma^2_{Y} (t_2)}},
\label{eq:time_dependent_rho}
\end{equation}
where $K_{XY}(\cdot,\cdot)$ is a cross-covariance function. $\rho_{XY}(t_1, t_2)$ is called a time-dependent correlation coefficient. Covariance function is defined as: 
\begin{equation}
K_{XY}(t_1,t_2) = Cov(X(t_1),Y(t_2)) = \langle X(t_1) Y(t_2) \rangle - \langle X(t_1) \rangle \langle Y(t_2) \rangle, 
\end{equation}
for a real-valued process (sample space $\Omega$ is a set of real values). For complex valued processes, which are what we are interested in, definition is:
\begin{equation}
K_{XY}(t_1,t_2) = Cov(X(t_1),Y(t_2)) = \langle (X(t_1) - \mu_{X}(t_1)) (\overline{Y(t_2) - \mu_{Y}(t_2)}) \rangle, 
\end{equation}
where the over-line denotes complex conjugate and $\mu_{Z}(t) = \langle Z(t) \rangle$. The average value is generally dependent on $t$. Note that $\langle \cdot \rangle$ denotes ensemble average, not the time average. It is important to notice that even $K_{XX}(t_1, t_2)$ in general might be dependent on $t_1$ and $t_2$. The function $K_{XX}(t_1, t_2) = 0$ only for a process $\{ X(t) \}$ where $\forall_{t , s \geq 0, t \neq s} X(t) \indep X(s)$, which means that the process is a set of random variables independent of each other. The symbol $\indep$ denotes independence of two random variables. The theory of processes with that property (i.e. with outcomes independent of each other, e.g., throwing dice) is part of probability theory, but (in general) independence of outcomes for two different times is not true for stochastic processes. That is why quantity $K_{X X}(t_1, t_2)$, called auto-covariance, is very important in studying properties of stochastic processes.

In signal theory, cross-covariance (and auto-covariance) function is not used. Instead, the subject of analysis are cross-correlation and auto-correlation functions which have the form :
\begin{align}
G_{XY}(t_1, t_2) &= \langle X(t_1) \overline{Y(t_2)} \rangle \label{eq:crosscorrelation} \\
G_{XX}(t_1, t_2) &= \langle X(t_1) \overline{X(t_2)} \rangle \label{eq:autocorrelation}
\end{align}
For a very important class of processes the difference between auto-correlation and auto-covariance functions comes down to a constant. These processes are called weakly stationary (also: wide sense stationary, or WSS for short).
If a process is WSS, the auto-covariance function depends only on time difference:
\begin{equation}
K_{XX}(t_1,t_2) = \gamma(|t_1 - t_2|).
\end{equation} 
The above equation is one of the conditions for a process to be categorized as a WSS process. The other conditions state that the expected value has to be time-independent and $\langle |X_t|^2 \rangle < \infty$. In the case of photons, it can be understood as follows: the process of generating photons has to be stable so that the expected number of photons arriving in some time window does not depend on time. The restriction that $\langle |X_t|^2 \rangle < \infty$ means that a process has to have a defined variance and that variance has to be finite.  Therefore, functions can be reformulated as follows:
\begin{align}
K_{XX}(\tau) &= \langle X(t) \overline{X(t + \tau)} \rangle - \mu_X \overline{\mu_X}, \\
G_{XX}(\tau) &= \langle X(t) \overline{X(t + \tau)} \rangle = K_{XX}(\tau) + \mu_X \overline{\mu_X},
\end{align}
for $\tau \geq 0$.

\subsection{First order coherence functions: Young experiment}
Let us consider the electromagnetic wave of the form:
\begin{equation}
\boldsymbol{E} (\boldsymbol{r}, t) \equiv \boldsymbol{e}(E^+ (\boldsymbol{r}, t) + E^- (\boldsymbol{r}, t)),
\end{equation} 
where $\boldsymbol e$ is a unitary vector. $\boldsymbol{E^+}$ and $(\boldsymbol{E}^-)$ are called, respectively, a positive and negative frequency components of an electromagnetic wave and in general, are complex. These field components are complex conjugate of each other: $E^- (\boldsymbol r, t) \equiv \overline{E^{+} (\boldsymbol r, t)}$. The first-order correlation function is defined as \cite{loudon1973}:
\begin{equation}
G^{(1)}(\boldsymbol r_1, t_1 ;\boldsymbol r_2, t_2) = \langle E^- (\boldsymbol r_1, t_1) E^+ (\boldsymbol r_2, t_2) \rangle. \label{eq:G1}
\end{equation} 
Note the similarity between the first-order correlation function and cross-correlation function \eref{eq:crosscorrelation}.

\begin{figure}[tbh]
	\centering
	\includegraphics[width=0.5\columnwidth,keepaspectratio]{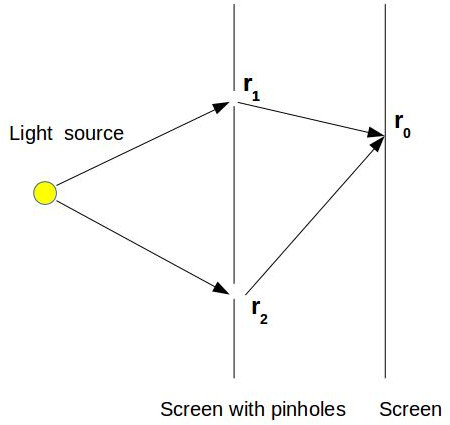}
	\caption{Simplified setup of The Young interference experiment.}
	\label{fig:Young}
\end{figure}

The first-order correlation function appears in the context of Young's interference experiment, where spatial light intensity distribution is measured after passing through a screen with two pinholes (\figref{fig:Young}). The intensity of light averaged over a cycle of oscillation at position $\boldsymbol r_0$ is given by a weighted sum of the intensity of light going through two pinholes at positions $\boldsymbol r_1$ and $\boldsymbol r_2$:
\begin{align}
I(\boldsymbol r_0, t) &= \frac{\epsilon_0}{2} c \{|a|^2 |E(\boldsymbol r_1, t_1)|^2 + |b|^2 |E(\boldsymbol r_2, t_2)|^2 + 2 |a b| \mathrm{Re}(\overline E(\boldsymbol r_1, t_1)E(\boldsymbol r_2, t_2)) \},  \\
t_{1,2} &= t - \frac{|\boldsymbol r_{1,2} - \boldsymbol r_0|}{c},
\end{align}  
where $a$ and $b$ depend on the geometry of the setup and are purely imaginary \cite{loudon1973}. With the above equation, we describe a superposition of fields as measured at time $t$. These fields have been emitted from their respective sources at times $t_1$ and $t_2$.  Now if we take an ensemble average and use \eref{eq:G1} we get: 
\begin{equation}
\langle I(\boldsymbol r_0, t) \rangle = \frac{\epsilon_0}{2} c \{|a|^2 \langle |E(\boldsymbol r_1, t_1)|^2 \rangle + |b|^2 \langle |E(\boldsymbol r_2, t_2)|^2 \rangle + 2 |a b| \mathrm{Re}( G^{(1)}(\boldsymbol r_1, t_1 ;\boldsymbol r_2, t_2) ) \}.\label{eq:non_stat_inten}
\end{equation}  
This result shows that first-order correlation function arises naturally in the context of light interference. Let us take a closer look at this result. First, note that the stationary light correlation function  is, by ergodic theorem,  a time average \cite{loudon1973}:
\begin{align}
G^{(1)}(\boldsymbol r_1, t_1 ;\boldsymbol r_2, t_2) &= \lim_{T \rightarrow \infty} \frac{1}{T} \int_0^T \overline E(\boldsymbol r_1, t_1)  E(\boldsymbol r_2, t_1 + \tau) dt_1 = G^{(1)}(\boldsymbol r_1;\boldsymbol r_2; \tau) ,
\end{align}
where $\tau$ again stands for a time difference $\tau = t_2 - t_1$. 
If a light source is stationary, then equation \eref{eq:non_stat_inten} can be further simplified to:
\begin{equation}
I(\boldsymbol r) = \frac{\epsilon_0}{2} c \{|a|^2 |E(\boldsymbol r_1)|^2  + |b|^2  |E(\boldsymbol r_2)|^2  + 2 |a b| \mathrm{Re}( G^{(1)}(\boldsymbol r_1;\boldsymbol r_2; \tau)) \}.
\end{equation}
Although the $G^{(1)}$ function arises naturally in many experiments, its form is not the most convenient one for discussing the level of coherence. Analogically to correlation coefficient \eref{eq:time_dependent_rho}, it can be normalized
\begin{align}
g^{(1)}(\boldsymbol r_1, t_1 ;\boldsymbol r_2, t_2) &= \frac{G^{(1)}(\boldsymbol r_1, t_1 ;\boldsymbol r_2, t_2)}{(G^{(1)}(\boldsymbol r_1, t_1 ;\boldsymbol r_1, t_1) G^{(1)}(\boldsymbol r_2, t_2 ;\boldsymbol r_2, t_2))^{1/2}} = \dots \nonumber\\ 
\dots &= \frac{\langle E^- (\boldsymbol r_1, t_1) E^+ (\boldsymbol r_2, t_2) \rangle}{(\langle E^- (\boldsymbol r_1, t_1) E^+ (\boldsymbol r_1, t_1) \rangle \langle E^- (\boldsymbol r_2, t_2) E^+ (\boldsymbol r_2, t_2) \rangle )^{1/2}},
\end{align} 
where $g^{(1)}$ is normalized first-order correlation function and \mbox{$|g^{(1)}| \leq 1$}. When $|g^{(1)}| = 1$ then we say that a field is first-order coherent.
Similarly, we can define higher-order correlation functions:
\begin{align}
G^{(n)}(\boldsymbol r_1, t_1 ;\dots ; \boldsymbol r_n, t_n;\boldsymbol r_{n+1}, t_{n+1} ;\dots ; \boldsymbol r_{2n}, t_{2n}) &=  \langle E^- (\boldsymbol r_1, t_1) \dots E^+ (\boldsymbol r_{2n}, t_{2n}) \rangle \label{eq:Gfun},
\end{align}
and higher-order normalized correlation functions:
\begin{equation}
g^{(n)}(\boldsymbol r_1, t_1 ;\dots ; \boldsymbol r_n, t_n;\boldsymbol r_{n+1}, t_{n+1} ;\dots ; \boldsymbol r_{2n}, t_{2n}) = \frac{G^{(n)}(\boldsymbol r_1, t_1 ;\dots ; \boldsymbol r_n, t_n;\boldsymbol r_{n+1}, t_{n+1} ;\dots ; \boldsymbol r_{2n}, t_{2n};)}{[\Pi^{2n}_{j=1} G^{(1)}(\boldsymbol r_j, t_j ;\boldsymbol r_j, t_j)]^{1/2}}.
\end{equation}
Field is coherent to the $N$-th order when $\forall_{j < N} |g^{(j)}| = 1$. In that sense, $N$-th order coherent light stands for electromagnetic field whose $g^{(n)}$- functions are equal to $1$ up to $g^{(N)}$ function.

\subsection{Second-order coherence functions: Hanbury Brown and Twiss experiment}

In the Young experiment we deal with a long time average intensity, but in Hanbury Brown and Twiss (HBT) experiment what we measure is the covariance function of field intensity. In the original experiment, researchers were interested in estimating the angular size of celestial radio sources \cite{Brown1956}. 
Measuring the distance between a pair of antennas, they could measure the interference effect. That effect is directly connected to the angular size of the source. In our case, we discuss a simplified experimental setup. Instead of two antennas, we will split the light coming from the source with the help of a beam splitter. 

\begin{figure}[tbh]
	\centering
	\includegraphics[width=0.7\columnwidth,keepaspectratio]{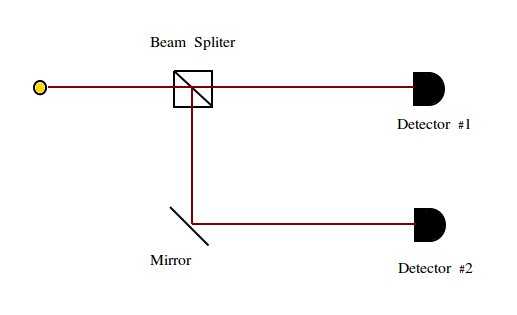}
	\caption{A simplified setup of the Hanbury Brown and Twiss experiment.}
	\label{fig:HBT}
\end{figure}

In the HBT experiment, the light source is stationary. In the following discussion, we will also assume that light is linearly polarized.  Covariance of intensity, evenly split by the beam splitter and measured at two points of spacetime, is given by:
\begin{equation}
Cov(\boldsymbol r_1, t_1; \boldsymbol r_2, t_2) = \langle (I(\boldsymbol r_1, t_1) - I )(I(\boldsymbol r_2, t_2) - I )\rangle = \langle I(\boldsymbol r_1, t_1)I(\boldsymbol r_2, t_2) \rangle - I^2 \label{eq:cov},
\end{equation} 
where $I$ denotes a half of the long-time average intensity and $I(\boldsymbol r_{1,2}, t_{1,2})$ denotes cycle average intensity of light source.
An example setup for such measurement is shown in \figref{fig:HBT}. 
We can replace $\boldsymbol r_{1,2}$ by $x_{1,2}$ which will denote the distance from the beam splitter measured to the detection device on respective optical paths. The detectors measure the variation of the field intensity and their output is multiplied. Fields that contribute to the correlation function may be measured at different times denoted by $t_{1,2}$. The long time average of product of that multiplication is given by the covariance function $Cov(x_1, t_1; x_2, t_2)$. The covariance function in \eref{eq:cov} has an interesting component representing correlation in intensities. That component might be recast in terms of the electric field:
\begin{equation}
\langle I(\boldsymbol x_1, t_1)I(\boldsymbol x_2, t_2) \rangle = (\frac{1}{2} \epsilon_0 c)^2 \langle E^-(x_1, t_1) E^-(x_2, t_2) E^+(x_2, t_2)E^+(x_1, t_1) \rangle.
\end{equation}
According to equation \eref{eq:Gfun}, the above equation describes the $G^{(2)}(x_1, t_1;x_2, t_2)$ function. For our stationary light source we can simplify the above formula to:
\begin{equation}
G^{(2)}(x_1;x_2; \tau) = (\frac{1}{2} \epsilon_0 c)^2 \langle E^-(x_1, 0) E^-(x_2, \tau) E^+(x_2, \tau)E^+(x_1, 0) \rangle,
\end{equation}
which means that the corresponding $g^{(2)}$ function has the form:
\begin{equation}
g^{(2)}(x_1;x_2; \tau) =  \frac{G^{(2)}(x_1;x_2; \tau)}{|G^{(1)}(x_1;x_2; \tau)|^2}.
\end{equation}

If measurements are taken in equal distances from the source, arguments $x_{1,2}$ can be dropped which results in the most commonly used form of the $g^{(2)}$ function which only depends on the time difference between the measurements. This expression is only valid in case where electromagnetic field in both arms has the same spatial dependence on wavevector.

Let us take a closer look at the $g^{(2)}(\tau)$ function which is very important for establishing the quantum behaviour of electromagnetic fields. For the following discussion, we will assume measuring light at points in equal distances from source ($x_1 = x_2$). For such a setup, the  $g^{(2)}$ function has the form:
\begin{equation}
g^{(2)}(\tau) = \frac{G^{(2)}(\tau)}{|G^{(1)}(0)|^2},
\end{equation} 
where $\tau$ stands for the time difference $t_2 - t_1$ and $G^{(2)}(\tau) = \langle E^- (0)  E^- (\tau) E^+ (\tau)  E^+ (0) \rangle$. For $\tau = 0$ we get:
\begin{equation}
g^{(2)}(0) = \frac{G^{(2)}(0)}{|G^{(1)}(0)|^2},
\end{equation} 
which can be rewritten in terms of intensities $I = |E(0)|^2 = E^- (0) E^+ (0)  $:
\begin{equation}
g^{(2)}(0) = \frac{\langle I^2 \rangle}{ \langle I \rangle^2} = 1 + \frac{\langle I^2 \rangle - \langle I \rangle^2}{ \langle I \rangle^2}.
\end{equation} 
The expression $\langle I^2 \rangle - \langle I \rangle^2$ is just a variance of the intensity and can be rewritten as:
\begin{equation}
(\Delta I)^2 = \langle I^2 \rangle - \langle I \rangle^2 = \int_0^\infty \Pr \{ I \} (I - \langle I \rangle)^2 dI,\label{eq:quasiprob}
\end{equation}
where $Pr \{ I \} $ is the probability density function of intensity for a fixed position and time, as in the discussion above. Classically, the above equation for variance cannot result in negative values since the probability and intensity are non-negative. This results in $g^{(2)}(0) \geq 1$ for classical fields. For fields displaying quantum behaviour, it is possible for $g^{(2)}(0)$ to be smaller than $1$. Then, the probability distribution has to be replaced by a quasi-probability distribution which allows for negative values.

\subsection{Quantum coherence functions}

To rewrite the correlation function so that we can use them for nonclassical fields we make two observations. The first observation is a need to replace electric and magnetic vectors ($\boldsymbol E$, $\boldsymbol B$) which describe electromagnetic field by field operators ($\hat{\boldsymbol E}$, $\hat{\boldsymbol B}$) in the formalism of second quantisation. In that formalism, the electric field operator associated with an electromagnetic field is given by: 
\begin{equation}
	\hat{\boldsymbol E}(\textbf r, t) =  \hat{\boldsymbol E}^+(\textbf r, t) + \hat{\boldsymbol E}^ -(\textbf r, t), \label{eq:hatE}
\end{equation}
where $\hat{\boldsymbol E}^+$ is called positive and $\hat{\boldsymbol E}^-$ is called negative frequency part of the electric field operator. The positive and negative frequency parts of the field operator are Hermitian conjugates of each other $\hat{\boldsymbol E}^{- \dagger} (\boldsymbol r, t) \equiv {\hat{\boldsymbol E}^{+} (\boldsymbol r, t)}$. 
With that, we can rewrite the correlation function as:
\begin{align}
G^{(1)}(\boldsymbol r_1, t_1; \boldsymbol r_2, t_2) &= \langle \hat E^- (\boldsymbol r_1, t_1) \hat E^+ (\boldsymbol r_2, t_2) \rangle,  \label{eq:quantum_G1}  \\
G^{(2)}(\boldsymbol r_1, t_1; \boldsymbol r_2, t_2;\boldsymbol r_3, t_3; \boldsymbol r_4, t_4) &= \langle \hat E^- (\boldsymbol r_1, t_1) \hat E^- (\boldsymbol r_2, t_2) \hat E^+ (\boldsymbol r_3, t_3) \hat E^+ (\boldsymbol r_4, t_4) \rangle. \label{eq:quantum_G2}
\end{align}
 From \eref{eq:quantum_G1} and  \eref{eq:quantum_G2} we can see that similarly to its classical analogue, $G^{(n)}$ function is non-negative for $\{ \textbf r_1, t_1; \dots ; \textbf r_{n}, t_{n} \}  = \{ \textbf r_{n+1}, t_{n+1}; \dots; \textbf r_{2n}, t_{2n} \}$.

In the second quantisation, the positive and negative frequency parts for the linearly polarized electric field operator for transverse electromagnetic waves in the vacuum can be decomposed in terms of plane waves as follows \cite{Knight2005}:
\begin{align}
\hat{\textbf{E}}^+ (\textbf r, t) &= i  \sum_{\lambda \in \{1, 2\} } \sum_\mathbf{k} \textbf{e}_\lambda \frac{C_{\mathbf{k}}}{\sqrt{V}}  \hat a_{\textbf k, \lambda} \exp(- i \omega_{\textbf k} t + i \textbf k \cdot \textbf r) \label{eq:Eplus}\\
\hat{\textbf{E}}^- (\textbf r, t) &= -i  \sum_{\lambda \in \{1, 2\} } \sum_\mathbf{k} \overline{\textbf{e}}_\lambda  \frac{C_{\mathbf{k}}}{\sqrt{V}}  \hat a^\dagger_{\textbf k, \lambda} \exp(i \omega_{\textbf k} t - i \textbf k \cdot \textbf r)\label{eq:Eminus},
\end{align}
were $V$ is volume of quantization, $\omega_{\textbf k}$ is photon angular momentum and $\hat a_{\textbf k, \textbf e}$ and $\hat a_{\textbf k, \textbf e}^\dagger$  are (respectively) annihilation and creation operators of a photon with momentum $\textbf{k}$ and polarization $\textbf{e}_\lambda$ where $\textbf{k} \perp \textbf{e}_\lambda$.  The quantity $C_{\mathbf{k}}/\sqrt{V}$ might be regarded as magnitude of electric field "per photon". The constant $C_{\mathbf{k}}$ is given by:
\begin{equation}
C_\mathbf{k} = \sqrt{\frac{\hbar \omega_{\textbf k}}{2 \epsilon_0}}, \label{eq:quantConst}   
\end{equation}
where $\epsilon_0$ is electric constant, $\hbar$ is reduced Planck's constant.

The second observation is that ensemble average over any operator $\langle \hat \Phi(x) \rangle$ has to be replaced by the expectation value of that operator for a system in the state given by a density matrix $\rho$:
\begin{equation}
    \langle \hat \Phi(x) \rangle \equiv \mathrm{Tr} [\rho \hat \Phi(x)].     
\end{equation}

For stationary light and two measurement devices, the above equation simplifies to \cite{Glauber1963a}:
\begin{equation}
g^{(2)}(\boldsymbol r_1; \boldsymbol r_2; \tau) = \frac{\mathrm{Tr}[\rho \hat E^- (\boldsymbol r_1, t) \hat E^- (\boldsymbol r_2, t+\tau) \hat E^+ (\boldsymbol r_2, t+\tau) \hat E^+ (\boldsymbol r_1, t)]}{G^{(1)}(\boldsymbol r_1, t; \boldsymbol r_1, t) G^{(1)}(\boldsymbol r_2, t +\tau; \boldsymbol r_2, t +\tau)}. \label{eq:statg2}
\end{equation}
If we calculate the value of the normalized auto-correlation $g^{(2)}(\cdot)$ function \eref{eq:statg2} for stationary light measured at the same point and at the time ($\tau = 0$), we get:
\begin{equation}
g^{(2)}(\boldsymbol r; \boldsymbol r;0) = \frac{\langle \hat E^- (\boldsymbol r, t) \hat E^- (\boldsymbol r, t) \hat E^+ (\boldsymbol r, t) \hat E^+ (\boldsymbol r, t)\rangle }{G^{(1)}(\boldsymbol r, t; \boldsymbol r, t) G^{(1)}(\boldsymbol r, t ; \boldsymbol r, t)}.
\end{equation}
For a single-mode light, that equation simplifies to:
\begin{equation}
g^{(2)}(0) = \frac{\langle a^\dagger a^\dagger  a a  \rangle}{\langle a^\dagger a \rangle \langle a^\dagger a \rangle} = \frac{\langle a^\dagger (a a^\dagger  - 1) a  \rangle}{\langle a^\dagger a \rangle ^2} = \frac{\langle a^\dagger a a^\dagger a  \rangle -\langle a^\dagger a \rangle }{\langle a^\dagger a \rangle ^2}, \label{eq:g2singlemode}
\end{equation}
where we used equations \eref{eq:quantum_G1}, \eref{eq:Eplus} and \eref{eq:Eminus}. We now introduce the single-mode photon number operator $\hat n$. Annihilation and creation operator are non-Hermitian but a photon number operator is:
\begin{equation}
\hat n \equiv \hat a^\dagger \hat a.
\end{equation}
because $\langle \hat a^\dagger \hat a \rangle = \langle n \rangle$, then our function takes a form:
\begin{equation}
g^{(2)}(0) = \frac{\langle \hat n^2  \rangle - \langle \hat n \rangle}{\langle \hat n \rangle^2}.
\end{equation}
If we now introduce variance of photon number $(\Delta \hat n)^2 = \langle \hat n^2  \rangle - \langle \hat n \rangle^2$, then we can rewrite above equation:
\begin{equation}
g^{(2)}(0) = \frac{\langle \hat n \rangle^2 + (\Delta \hat n)^2 - \langle \hat n \rangle}{\langle \hat n \rangle^2} = 1 + \frac{ (\Delta \hat n)^2 - \langle \hat n \rangle}{ \langle \hat n \rangle^2}. \label{eq:g2}
\end{equation}
We will use this function to characterize the statistical properties of different types of light.

\subsection{Sub-Poissonian, Poissonian and super-Poissonian light}
As we mentioned before, the Poisson process helps to characterize the nature of light. 
We will refer to states generated in Poissonian, sub-Poissonian and super-Poissonian processes respectively as Poissonian, sub-Poissonian and super-Poissonian states.
In this paragraph, we will show how. 
As mentioned before, a  Poisson process has its mean equal to variance. Consequently, the auto-correlation function of light in a Poissonian state simplifies to:
\begin{equation}
g_{\mathrm{Poiss}}^{(2)}(0) = 1. 
\end{equation}
For a sub-Poissonian state we have:
\begin{equation}
(\Delta \hat n)^2 < \langle \hat n \rangle \Leftrightarrow g_{\mathrm{sub}}^{(2)}(0) < 1.
\end{equation}
As mentioned before, this result cannot be explained with the classical probability theory since from discussion of \eref{eq:quasiprob} we know that it leads to probability distribution with negative probability values. Similarly to the sub-Poisson process, we can define the super-Poisson process via:
\begin{equation}
(\Delta n)^2 > \langle \hat n \rangle \Leftrightarrow g_{\mathrm{Super}}^{(2)}(0) > 1.
\end{equation}

\subsection{Thermal state} 
The thermal state can be fully described on the grounds of the classical theory without a need to use quantum mechanical results. Since the quantum mechanical calculation would lead us to the same properties of the thermal state as classical calculations, they are not necessary.

In a single-mode thermal state, the probability of the system to be found in the energy state  $\hbar n \omega$, $\omega$ being the frequency of the mode, is
\begin{equation}
P(n) = e^{- \beta n \hbar \omega}(1 - e^{- \beta \hbar \omega}),
\end{equation}
and $\beta = (k_b T)^{-1}$, where $k_b$ stands for the the Boltzman constant and $T$ stands for temperature. The expectation value of the photon number is given by
\begin{equation}
\langle n \rangle = \sum_{n = 0}^{\infty} n P(n) = \sum_{n = 0}^{\infty} n e^{- \beta n \hbar \omega}(1 - e^{- \beta \hbar \omega}),
\end{equation}
where the last term can be easily calculated by noting that
\begin{equation}
\sum_{n = 0}^{\infty} n e^{- \beta n \hbar \omega}(1 - e^{- \beta \hbar \omega}) = - (1 - e^{- \beta \hbar \omega})\frac{\partial}{\partial \beta \hbar \omega } \sum_{n = 0}^{\infty}  e^{- \beta n \hbar \omega},
\end{equation}
hence the expectation value of photon number is equal to 
\begin{equation}
\langle n \rangle = \frac{1}{e^{\beta \hbar \omega } - 1}.
\end{equation}
In a similar fashion $\langle n^2 \rangle$ can be calculated. If we follow the procedure above and use the second derivative in our calculation we end up with:
\begin{equation}
\langle n^2 \rangle = \frac{e^{\beta \hbar \omega }+ 1}{(e^{\beta \hbar \omega}- 1)^2}.
\end{equation}
With $\langle n \rangle$ and $\langle n^2 \rangle$ we can calculate $\Delta n$,
\begin{equation}
\Delta n^2 = \langle n^2 \rangle - \langle n \rangle^2 = \frac{e^{\beta \hbar \omega }}{(e^{\beta \hbar \omega}- 1)^2} =  \langle n \rangle^2 + \langle n \rangle.
\end{equation}
If we now plug this result into equation \eref{eq:g2} we get
\begin{equation}
g^{(2)}(0) = 2.
\end{equation}
This result shows that thermal light has stronger fluctuations of intensity then second-order coherent light (e.g monochromatic light).

\subsection{Fock states of light} 

As mentioned before, any state for which the corresponding  $g^{2}(0)$ is less than $1$ exhibits behaviour that cannot be fully explained on the ground of classical electrodynamics. 
One class of such states are Fock or number states. Focks states form a basis of single mode Hilbert space of light states. Fock states are eigenstates of single-mode photon number operator $\hat n$:
\begin{equation}
 \hat n |m \rangle = m |m \rangle,
\end{equation}
where the notation $|m \rangle$ means that there are $m$ photons in the mode, where $m$ is a non-negative integer. Such state has a very well defined energy which is $E = m \hbar \omega$ and a completely random phase \cite{loudon1973}. States for which $m =0 $ are called vacuum states. In principle, all number states have no variance associated with the number of photons.
\begin{figure}[tbh]
	\centering
	\includegraphics[width=0.75\columnwidth,keepaspectratio]{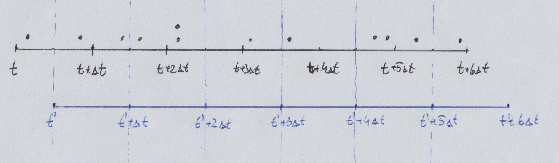}
	\caption{Photon counting measurement. Dots represent photon distribution in time relative to two clocks - one starting at time $t$ (black line) and the other at time $t'$ (blue line). It can be seen that in time bins of duration $\Delta t$ there are always two and only two photons if measurement starts in $t$. If the measurement starts in $t'$ we can see that there is a variation of photon number.}
	\label{fig:counts}
\end{figure}
Let us take a look at possible values that $g^{(2)} (0)$ can take.  
Let us assume we have a stationary source that emits $m$ and only $m$ single-mode photons during some constant period $\Delta t$. Then  the expectation value of the number of photons  $\langle N \rangle$ emitted in time much larger then the period $\Delta t$ is equal to the number of periods $s$ multiplied by expectation value $\langle n \rangle$ of photons emitted during one time period $\Delta t$ which is $m$:
\begin{equation}
    \langle N \rangle_{s \Delta t} = s \langle n \rangle_{\Delta t} \approx sm.
\end{equation}
Note that with this formulation we cannot claim that the standard deviation of the number of photons found in the time bins of duration $\Delta t$ is equal to $0$ since we have not said anything about the distribution of these $m$ photons during that time. Also, we have not mentioned when we start our measurement (when the clock starts).  But we can claim two things: that it is possible to start the measurement at such a time that per each time bin we would get exactly $m$ photons and that moment can be found retrospectively after a time much longer than $\Delta t$. In that time frame the standard deviation $\Delta n = 0$.  This situation is illustrated in Fig.~\ref{fig:counts} for $m=2$. We can see that in time bins of duration $\Delta t$ there are always two photons if a clock of the counting device is synchronised with the source and counting starts at time $t$. If we start at different time $t'$ (Fig. \ref{fig:counts}), then although the source still emits two photons in time bins of duration $\Delta t$, our counting method has introduced a variation to the system.

This discussion may seem unnecessarily confusing it is rather important since all measurement results are strongly influenced by how the measurement is performed. 
Therefore, for a state with a set number of $m$ photons in a measurement time bin:
\begin{equation}
g^{(2)} (0) = 1 - \frac{1}{m}, \label{eq:subpoi}
\end{equation}
where the smallest possible value is $g^{(2)} (0) = 0$ for $m = 1$. 

\subsection{Coherent states of light}
In limits of large energy, coherent states resemble classical electromagnetic waves. Coherent states describe very well laser light, especially very-low intensity light, where the corpuscular nature of light makes classical theory unfit for a description of the physical process observed. Coherent states are traditionally denoted by $| \alpha \rangle$ and are defined as eigenstates of the annihilation operator:
\begin{equation}
\hat a |\alpha \rangle = \alpha |\alpha \rangle,
\end{equation} 
where $\alpha$ is any complex number. From this definition it follows that the representation of an arbitrary coherent state in Fock state basis has a form:
\begin{equation}
|\alpha \rangle = e^{-\frac{1}{2} |\alpha|^2} \sum_{n=0}^{\infty} \frac{\alpha^n}{(n!)^{\frac{1}{2}}} |n \rangle. 
\end{equation} 
An important property of coherent states is that they do not form an orthogonal set of states.  In terms of coherence functions, these states lay on the border of classical and quantum realms. Their $g^{(2)}(0)$ function is equal to $1$ which is a consequence of the Poissonian nature of those states - the probability of finding $n$ photons in a coherent state $|\alpha\rangle$ is:
\begin{equation}
	|\langle n| \alpha \rangle|^2 = \exp( - |\alpha|^2) \frac{|\alpha|^{2n}}{n!}.
\end{equation}
The above equation describes the Poisson distribution (see \eref{eq:poissondist}) with an expectation value $|\alpha|^2$, meaning that:
\begin{equation}
|\langle \alpha | \hat n | \alpha \rangle|^2 = |\alpha|^2.
\end{equation}
The Poisson distribution has a property that the larger the mean value is the smaller the standard deviation becomes with respect to that value (see \eref{eq:poissonmean}):
\begin{equation}
	\frac{\Delta n}{\langle n \rangle} = \frac{1}{\sqrt{\langle n \rangle}}.
\end{equation}
Additionally, one can show that the phase is well defined for the expectation value of the field operator in a coherent state. In this sense, in the limit of large energies, the coherent state has a well-defined amplitude and phase.

Single mode coherent state is coherent to all orders of coherence function - function $g^{(n)}(\tau) = 1$ for any $n$. 

\subsection{Quadrature squeezed coherent states}

Another important class of quantum states are quadrature squeezed coherent states. These states are eigenstates of the operator $\hat A_Z$ which is defined as \cite{Mandel1995}:
\begin{align}
    \hat A_Z &= \hat a \cosh R + e^{i \theta} \hat a^\dagger \sinh R ,
\end{align}
where $Z = R e^{i \theta}$ and $R$ and $\theta$ are non-negative real numbers. Eigenstates of this operator, also called two-photon coherent states, are given by the expression:
\begin{align}
    \hat A_Z | \alpha, Z \rangle &= \alpha | \alpha, Z \rangle. \label{eq:squeezedstate}
\end{align}
Note that when $Z =0$ then $\hat A_0 = \hat a$ which implies that state $|\alpha, 0 \rangle$ is a coherent state:
\begin{align}
    |\alpha, 0 \rangle & = |\alpha \rangle .    
\end{align}
Relation between a coherent state $|\alpha \rangle$ and the corresponding squeezed coherent state $| \alpha, Z \rangle$ is given by the equation:
\begin{equation}
    | \alpha, Z \rangle = \hat S(Z) |\alpha \rangle,
\end{equation}
where $\hat S(Z)$ is the squeezing operator defined as:
\begin{equation}
    \hat S(Z) \equiv \exp[\frac{1}{2}(\overline{Z} \hat a^2 - Z ( \hat a^\dagger)^2)].
\end{equation}
Relation of operators $\hat A_Z$ and $\hat S(Z)$ is given by expression:
\begin{equation}
    \hat A_Z = \hat S(Z) \hat a \hat S^\dagger(Z).
\end{equation}
The importance of two-photon coherent states becomes apparent when we analyze the variance of the electric field. First, we need to define two operators \cite{Mandel1995}:
\begin{align}
    \hat Q_{\beta} &= \hat a^{\dagger} e^{i \beta} + \hat a e^{- i \beta} \\
    \hat P_{\beta} &= \hat a^{\dagger}e^{i (\beta +\pi/2)} + \hat a e^{-i (\beta + \pi/2)}.
\end{align}
For $\beta=0$, we will refer to these operators as quadrature operators $\hat Q_0$ and $\hat P_0$. Quadrature operators represent the real and imaginary parts of the complex amplitude of the electric field. The commutation relation is:
\begin{equation}
    [\hat Q_0, \hat P_0] = 2i.
\end{equation}
This commutation relation leads to uncertainty relation of a form:
\begin{equation}
    \langle (\Delta \hat Q_0)^2 \rangle \langle (\Delta \hat P_0)^2 \rangle \geq 1,
\end{equation}
which is true for any state of an electric field. For coherent states, the above relation transforms into equality \cite{Mandel1995}. For that reason, coherent states are called minimum uncertainty states.

The standard deviations  of operators $\hat Q_\beta$ and $\hat P_\beta$ are related to deviations of quadratic operators in the following manner \cite{Mandel1995}: 
\begin{align}
    \Delta \hat Q_\beta &= \Delta \hat Q_0 \cos \beta + \Delta \hat P_0 \sin \beta \\
    \Delta \hat P_\beta &= -\Delta \hat Q_0 \sin \beta + \Delta \hat P_0 \cos \beta,
    \end{align}
these equations might be regarded as a rotation in a plane spanned by the quadrature operators. With the above we can state the definition of the quadrature squeezed state, which is:
\begin{equation}
    \exists_{\beta \in [0, 2\pi)} \langle (\Delta \hat Q_\beta)^2 \rangle < 1. \label{eq:squeezeddefinition}
\end{equation}
Let us take a look at the variance of operator $\hat Q_\beta$ for the squeezed coherent state defined in \eref{eq:squeezedstate} \cite{Mandel1995}:
\begin{equation}
     \langle \alpha, Z |(\Delta \hat Q_\beta)^2 | \alpha, Z \rangle = \cosh 2R - \cos(\theta - 2 \beta)\sinh 2R. 
\end{equation}
That expectation value is smallest for $\beta = \theta /2$:
\begin{equation}
     \langle \alpha, Z |(\Delta \hat Q_{\theta /2})^2 | \alpha, Z \rangle = e^{-2R},
\end{equation}
which for all $R$ except $R = 0$ leads to variance smaller than $1$ and by definition \eref{eq:squeezeddefinition} states given by \eref{eq:squeezedstate} are quadrature squeezed states.

\subsection{Squeezed vacuum states}

Vacuum states denoted as $|0 \rangle$ are coherent states of the electromagnetic field. They describe the ground state of an electromagnetic field. One can define quadrature squeezed vacuum states (also called squeezed vacuum states) as:
\begin{equation}
    |0,Z \rangle = \hat S(Z) | 0 \rangle. \label{eq:squeezed0}
\end{equation}
In the limit of weak squeezing $|Z| \ll 1$, we can expand the squeezing operator:
\begin{align}
    \hat S(Z) &= 1 + \frac{1}{2} (\overline Z a^2 - Z {a^\dagger}^2)+ \frac{1}{8} (\overline Z a^2 - Z {a^\dagger}^2)^2 + + \frac{1}{48} (\overline Z a^2 - Z {a^\dagger}^2)^3 + \dots.
\end{align}
If we now use the above form to decompose the squeezed vacuum state in the basis of Fock states we get the following expression:
\begin{align}
    |0,Z \rangle &= (1 - \frac{1}{4}|Z|^2 + \dots)|0 \rangle + (- \frac{\sqrt{2}}{2} Z + \frac{7 \sqrt{2}}{24} Z |Z|^2 + \dots) |2 \rangle + \dots  \nonumber \\
    & \dots + (\frac{\sqrt{6}}{4}Z^2 + \dots) |4 \rangle + (- \frac{\sqrt{5}}{4}Z^3+ \dots)|6 \rangle + \dots     
\end{align}
For $|z|<0$, higher-order terms have quickly decreasing amplitudes, which justifies the above expansion. It is clear that the squeezed vacuum state is composed of even number Fock states - photons in squeezed vacuum always come in pairs.
The squeezed vacuum is an important state in the theory of the parametric down-conversion process.

\subsection{Two-mode squeezed vacuum states}

In all the above paragraphs we considered only single-mode states and operators, taking into account only one term from the sum in \eref{eq:Eminus} and \eref{eq:Eplus}. 
The presented formalism can be extended to accommodate two-mode states.
For such a state the squeezing operator has the following form:
\begin{equation}
    \hat S_2(Z) = \exp(\overline Z \hat a_{\boldsymbol k_1, \boldsymbol e_1} \hat a_{\boldsymbol k_2, \boldsymbol e_2} - Z \hat a^\dagger_{\boldsymbol k_1, \boldsymbol e_1} \hat a^\dagger_{\boldsymbol k_2, \boldsymbol e_2}), \label{eq:2modesqueezedoperator}
\end{equation}
where the squeezing operator subscript $2$ stands for the number of modes, $\boldsymbol{k}_{s}$ is the wavevector in mode $s$ which, together with the polarization vector $\boldsymbol{e}_{s}$, describes a single mode of the electromagnetic field. In analogy to \eref{eq:squeezed0} we define a two-mode squeezed vacuum state as:
\begin{equation}
    |0,Z \rangle_2 \equiv \hat S_2(Z) |0 \rangle_{\boldsymbol k_1, \boldsymbol e_1} |0 \rangle_{\boldsymbol k_2, \boldsymbol e_2}. 
\end{equation}
If $Z$ is real then quadrature operators $\hat Q_0$ and $\hat P_0$ for which squeezing occurs are defined as \cite{Knight2005}:
\begin{align}
    \hat Q_0 &= \sqrt{\frac{1}{2}} (\hat a_{\boldsymbol k_1, \boldsymbol e_1} + \hat a_{\boldsymbol k_2, \boldsymbol e_2} + \hat a^\dagger_{\boldsymbol k_1, \boldsymbol e_1} + \hat a^\dagger_{\boldsymbol k_2, \boldsymbol e_2}) \\
    \hat P_0 &= -i \sqrt{\frac{1}{2}} (\hat a_{\boldsymbol k_1, \boldsymbol e_1} + \hat a_{\boldsymbol k_2, \boldsymbol e_2} - \hat a^\dagger_{\boldsymbol k_1, \boldsymbol e_1} - \hat a^\dagger_{\boldsymbol k_2, \boldsymbol e_2}).    
\end{align}
It can be shown \cite{Knight2005} that second moments of these operators are (for real $Z$):
\begin{align}
    \langle (\hat Q_0)^2 \rangle &= e^{- 2 Z} \\
    \langle (\hat P_0)^2 \rangle &= e^{2 Z}.
\end{align}
State $|0, Z \rangle_2$ can be decomposed into Fock states as follows \cite{Knight2005}:
\begin{equation}
    |0, Z \rangle_2 = \frac{1}{\cosh R} \sum_{n=0}^\infty (-1)^n e^{i n \theta} (\tanh R)^n |n,n \rangle,    \label{eq:2modesqueezedstate}
\end{equation}
where $Z = R e^{i \theta}$. Let us take a look at the statistical properties of this state. First, let us define the photon number operators for each mode as:
\begin{align}
    \hat n_{\boldsymbol k_1, \boldsymbol e_1} &= \hat a^\dagger_{\boldsymbol k_1, \boldsymbol e_1} \hat a_{\boldsymbol k_1, \boldsymbol e_1} \\
    \hat n_{\boldsymbol k_2, \boldsymbol e_2} &= \hat a^\dagger_{\boldsymbol k_2, \boldsymbol e_2} \hat a_{\boldsymbol k_2, \boldsymbol e_2}
\end{align}
and let us simplify the notation with $\hat n_1 = \hat n_{\boldsymbol k_1, \boldsymbol e_1}$, $\hat n_2 = \hat n_{\boldsymbol k_2, \boldsymbol e_2}$
For two-mode squeezed vacuum states the expected number of photons is the same for each mode and is equal to \cite{Knight2005}:
\begin{equation}
    \langle \hat n_1 \rangle =\langle \hat n_2 \rangle =\sinh^2 R.
\end{equation}
Variances of photon numbers in each state are also equal \cite{Knight2005} :
\begin{equation}
    \langle (\Delta \hat n_1)^2 \rangle =\langle ( \Delta \hat n_2 )^2 \rangle = \frac{1}{4} \sinh^2 2R
\end{equation}
It is easy to see that $ \langle (\Delta \hat n_{1,2})^2 \rangle >  \langle \hat n_{1,2} \rangle$ which means that photon statistics for each mode is super-Poissonian and does not exhibit non-classical traits. This is the case when a measurement is conducted with regard to one mode only. If we look at two-mode properties we will note that the expected photon number of each mode is always the same 
and that there is no variance for the operator of photon number difference between the modes:
\begin{equation}
   \langle [ \Delta (\hat n_1 - \hat n_2)]^2 \rangle  = 0.
\end{equation}
In other words, photons are strongly correlated. If we use the Pearson correlation coefficient (see \eref{eq:rho}) to quantify the strength of correlation between the numbers of photons we get the maximal possible value of correlation \cite{Knight2005}:
\begin{equation}
    \rho_{\hat n_1, \hat n_2} = \frac{\mathrm{Cov} (\hat n_1, \hat n_2)}{\sqrt{\langle (\Delta \hat n_1)^2 \rangle \langle ( \Delta \hat n_2 )^2 \rangle}} = 1.
\end{equation}

\section{Spontaneous parametric down-conversion}

Many quantum optical experiments require the use of light in a Fock state. While low-intensity coherent and thermal states can be easily obtained through filtering down radiation field generated by a laser (which approximates very well coherent light) or thermal radiation, generation of Fock states is challenging.  A single-photon state is of special interest because it allows for conducting experiments involving just one photon. There are many ways of generating single-photon states e.g from emission from a quantum dot or a nitrogen-vacancy centre \cite{beveratos2002room}. Arguably, one of the most widely used ways of preparing single-photon states is by means of spontaneous parametric down-conversion (SPDC). In that process, photons are generated in pairs but can be easily spatially separated. Since both photons are created in the same act of down conversion they are strongly correlated in time. Exploiting that correlation allows for using one of the photons to indicate the time of arrival of the other, therefore assuring the single-photon nature of the state to come. In other words, we use one of the photons to prepare the detector for detection of the other photon. Then, we can claim with a very high degree of confidence that we are dealing with single photons. That technique in which we use one of the photons from the pair to force the other into a single-photon Fock state is called 'heralding'. What makes that process popular is the relative ease of implementation, the possibility of manipulating pair characteristics and the fact that photons are generated at the same instance and their positions in space are strongly correlated. It will be shown later in this chapter that pairs of photons generated in SPDC are well approximated by two-mode squeezed vacuum states.

\begin{figure}[tbh]
	\centering
	\includegraphics[width=0.25\columnwidth,keepaspectratio]{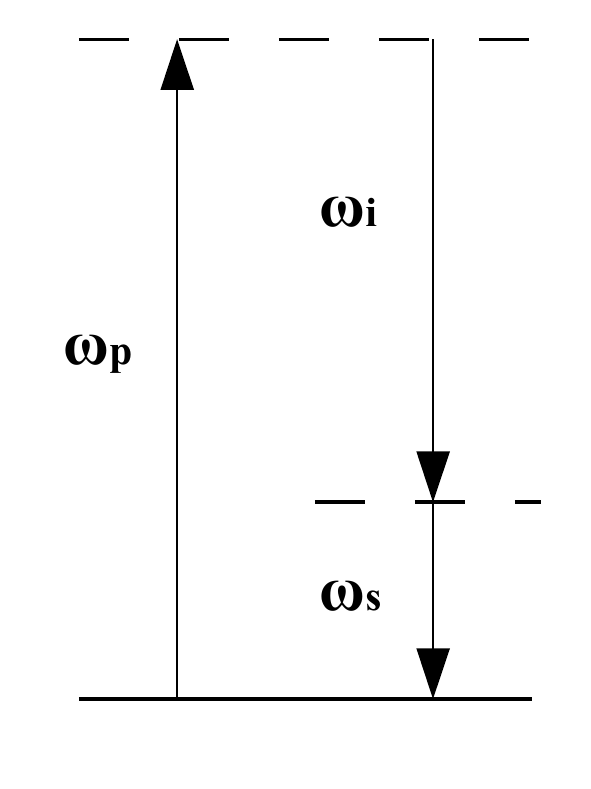}
	\caption{Simplified energy diagram for SPDC and DFG process. Pumping beam frequency must be a sum of idler and signal frequencies.}
	\label{fig:downconv}
\end{figure}

\begin{figure}[tbh]
	\centering
	\includegraphics[width=0.75\columnwidth,keepaspectratio]{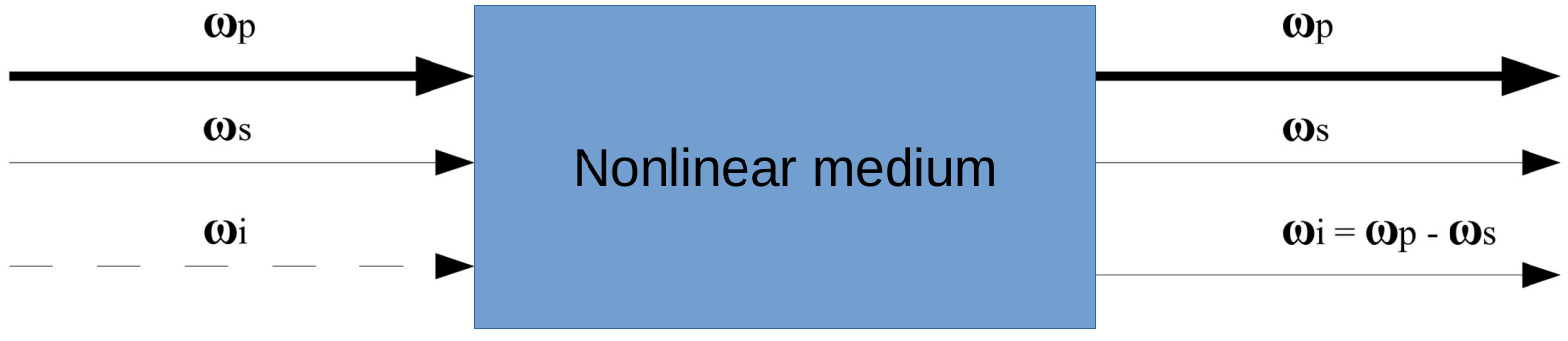}
	\caption{Schematic representation of the DFG process. A nonlinear crystal is illuminated by a strong pumping laser beam of frequency $\omega_p$. There are other two laser beams incident on the crystal surface: signal beam $\omega_s$ and idler beam $\omega_i$. If phase-matching conditions are fulfilled both idler and signal beam intensities are amplified. For DFG to take place the idler beam does not need to be present.}
	\label{fig:DFG}
\end{figure}

Spontaneous parametric down-conversion is an optical process similar to difference frequency generation (DFG)- its classical analogue. The DFG is a three-wave mixing process. It involves three beams of light called the pumping beam, signal beam and idler beam. The nonlinearity of the medium provides a mechanism of coupling all three light beams so that energy transfer between them is possible. The energy conservation requires that the frequencies of beams in the DFG process must be such that the idler and signal beam frequencies sum up to the frequency of the pumping beam. A simplified energy diagram is depicted in Fig. \ref{fig:downconv}. If such beams overlap in a nonlinear medium, and if their wavevectors fulfil the so-called phase-matching conditions then DFG (and SPDC) can take place. A schematic representation of the DFG process is presented in Fig. \ref{fig:DFG}. 
In order for SPDC to occur, the signal and idler beams are not necessary as input and can therefore be set in the vacuum state. All the other conditions for DFG and SPDC are the same: the medium needs to support phase matching and (in principle) has to allow for efficient three-wave mixing. Both conditions are described below.

\subsection{Three wave mixing Hamiltonian}
\begin{figure}[tbh]
	\centering
	\includegraphics[width=0.9\columnwidth,keepaspectratio]{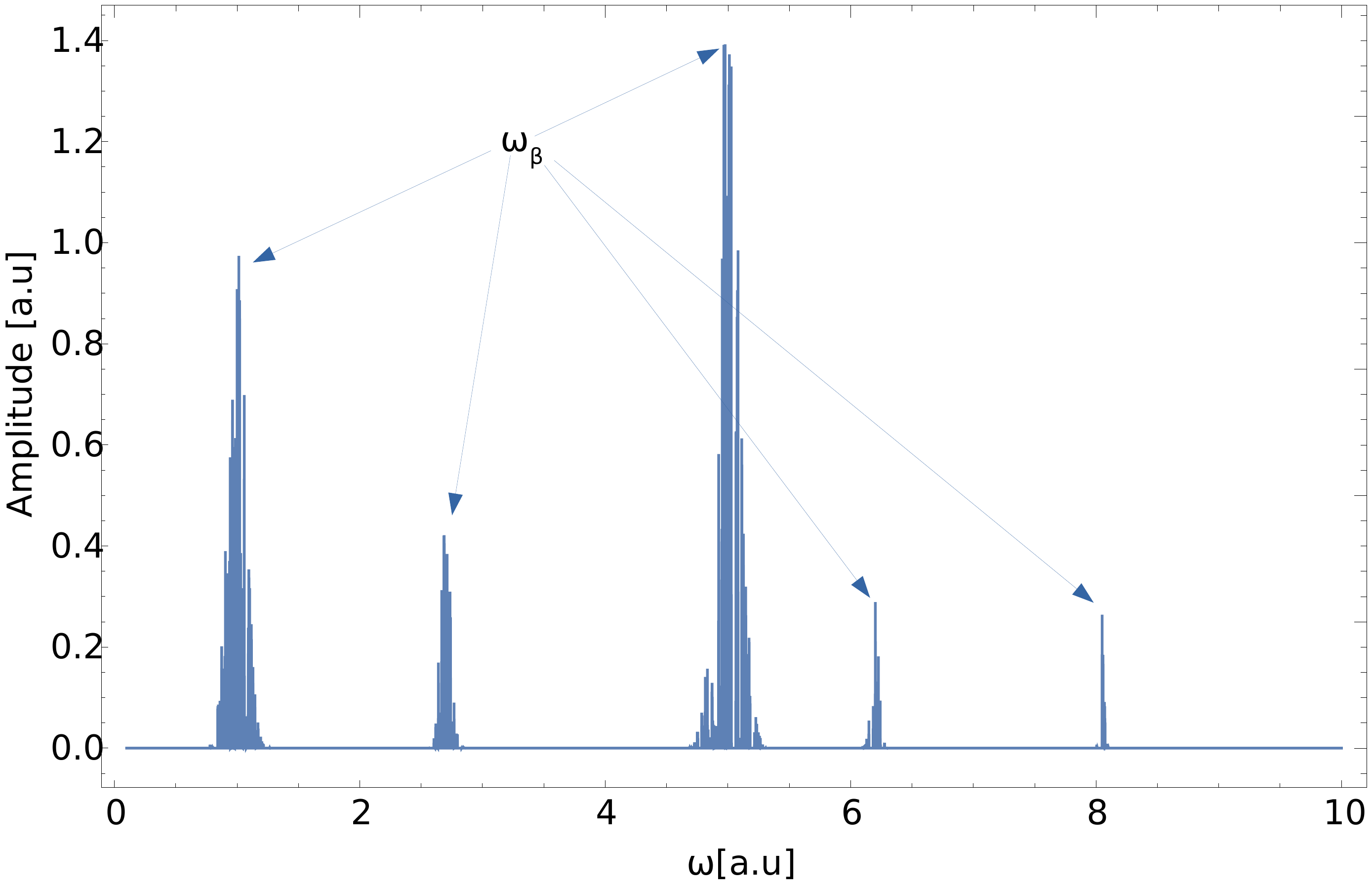}
	\caption{Schematic representation of the electromagnetic spectrum. Sharp peaks correspond to detunings frequencies and are grouped around carrier frequencies. Spacing between carrier frequencies should be much larger than the spacing between any given carrier frequency and detuning according to \eref{eq:spacing} (for visualization we decided to make them broader on this picture).}
	\label{fig:grzebien}
\end{figure}

To show that two-mode squeezed vacuum states approximate well the states generated in SPDC and to talk about SPDC itself, first we need to define a Hamiltonian governing the optical process occurring in the crystal. For a non-magnetic medium the most general form of the Hamiltonian is \cite{Mandel1995}:

\begin{equation}
    H = \int \frac{1}{2 \mu_0} \boldsymbol{B}^2 (\boldsymbol r, t) d^3 r + \int d^3 r \int_{\boldsymbol{0}}^{\boldsymbol{D} (\boldsymbol r, t)} \boldsymbol{E} (\boldsymbol r, t) \cdot d \boldsymbol{D} \boldsymbol (\boldsymbol{r}, t), \label{eq:emHamiltonian}
\end{equation}
where $\boldsymbol{D} (\boldsymbol{r}, t)$ is electric displacement vector:
\begin{equation}
    \boldsymbol{D} (\boldsymbol{r}, t) = \epsilon_0 \boldsymbol{E}  (\boldsymbol{r}, t) + \boldsymbol{P} (\boldsymbol{r}, t),
\end{equation}
and $\boldsymbol{P} (\boldsymbol{r}, t)$ is a vector of polarization. In a linear medium, its relation with the electric field is simple but in a nonlinear medium it has a complicated form \cite{Garrison2008}: 
\begin{align}
    \frac{P_i (\boldsymbol r, t)}{\epsilon_0} = &\int \sum_j \chi_{ij}^{(1)}(t - t_1) E_j(\boldsymbol r, t_1) dt_1 + \iint \sum_{jk} \chi^{(2)}_{ijk}(t-t_1,t-t_2) E_j(\boldsymbol r, t_1) E_k(\boldsymbol r, t_2) dt_1 dt_2 + \nonumber \\ +& \iiint \sum_{jkl} \chi_{ijkl}^{(3)} (t-t_1, t - t_2, t - t_3) E_j(\boldsymbol r, t_1) E_k(\boldsymbol r, t_2) E_l(\boldsymbol r, t_3)dt_1 dt_2dt_3  + \dots
\end{align}
where $\chi^{(n)}$ is susceptibility tensor of rank $n+1$ and sums go over all three components of the electric field. The above equation is true only for weakly dispersive media. A more convenient form of that equation is obtained through the Fourier transform:
\begin{align}
       \frac{P_i (\boldsymbol r, \omega_1)}{\epsilon_0}&= \sum_j \chi_{ij}^{(1)} (\omega_1; \omega_1) E_j(\boldsymbol r,\omega_1) + \int \sum_{jk} \chi^{(2)}_{ijk}(\omega_1; \omega_1 - \omega_2, \omega_2) E_j(\boldsymbol r,\omega_1 - \omega_2) E_k(\boldsymbol r,\omega_2) d\omega_2  \nonumber \\
       + \iint \sum_{jkl} & \chi_{ijkl}^{(3)}(\omega_1; \omega_1 - \omega_2 - \omega_3, \omega_2, \omega_3) E_j(\boldsymbol r, \omega_1 - \omega_2 - \omega_3) E_k(\boldsymbol r, \omega_2) E_l(\boldsymbol r, \omega_3) d\omega_2 d\omega_3+ \dots \label{eq:nonlinearP}
\end{align}
We can divide the above equation with regard to the power of the fields: 
\begin{align}
    [\boldsymbol P^L (\boldsymbol r, \omega)]_i  &= \epsilon_0 \sum_{j} \int \delta(\omega - \omega ') \chi_{ij}^{(1)}(\omega; \omega') E_j (\boldsymbol r, \omega') d \omega' \\
    [\boldsymbol P_1^{NL}(\boldsymbol r, \omega)]_i &= \epsilon_0 \sum_{jk} \int \int \delta(\omega - \omega ' - \omega'') \chi_{ijk}^{(2)}(\omega;\omega', \omega'') E_j (\boldsymbol r, \omega') E_k (\boldsymbol r, \omega'')d \omega' d \omega'' \\
     \dots \text{ and so on.}& \nonumber
\end{align}
The term $[\boldsymbol P^L (\boldsymbol r, \omega)]_i$ describes a Cartesian component of the linear response of the medium to an electric field, where the superscript $L$ stands for 'linear'. Following that convention, we can express the polarization vector as a sum of its linear and nonlinear parts:
\begin{equation}
    \boldsymbol P (\boldsymbol r, \omega_1) =\boldsymbol P^L (\boldsymbol r, \omega_1) + \boldsymbol P^{NL}_1 (\boldsymbol r, \omega_1) + \boldsymbol P^{NL}_2 (\boldsymbol r, \omega_1) + \dots. 
\end{equation}
We can rewrite Hamiltonian from \eref{eq:emHamiltonian} as:
\begin{equation}
    H(t) = \int [\frac{1}{2 \mu_0} \boldsymbol{B}^2 (\boldsymbol r, t) + \frac{1}{2 \epsilon_0} \boldsymbol{E}^2 (\boldsymbol r, t) + \mathcal{E}^L (\boldsymbol r, t) + \mathcal{E}^{NL}_1(\boldsymbol r, t) + \mathcal{E}^{NL}_2(\boldsymbol r, t) + \dots] d^3r,
\end{equation}
where the $\mathcal{E}$ terms denote the linear and nonlinear parts of the energy density of interaction of the field with the medium. Terms $\mathcal{E}^L$ and $\mathcal{E}^{NL}_n$ are defined as:

\begin{align}
    \mathcal{E}^L (\boldsymbol r, t) &= \frac{1}{2} \langle \boldsymbol{P}^L(\boldsymbol r, t) \cdot \boldsymbol{E} (\boldsymbol r, t) \rangle  \\
    \mathcal{E}^{NL}_n (\boldsymbol r,t) &= \frac{1}{n + 2} \langle \boldsymbol{P}^{NL}_n(\boldsymbol r, t) \cdot \boldsymbol{E}  (\boldsymbol r, t) \rangle.
\end{align}
The average is taken over a time much longer then period of carrier frequency, but short compared to time period over which envelope function changes. This formulation is necessary since in a dielectric polarization is in general dependent on its previous values \cite{Garrison2008,Jackson1999}.  
Terms of order $n > 1$ will not contribute to Hamiltonians of the systems which we will discuss throughout this thesis. This rejection of higher-order terms is justified under the assumption that medium is weakly nonlinear. The only nonlinear contribution that we are interested in is $\mathcal{E}^{NL}_1$ which governs the three-wave mixing processes, SPDC included. We are going to consider only non-centrosymmetric crystals, where only the susceptibility tensor of rank $(2)$ is large enough to contribute to the Hamiltonian which is true for weakly nonlinear substances \cite{Garrison2008}. Below calculations follow mostly calculations and reasoning outlined in book \cite{Garrison2008} with an occasional departure in favour of  \cite{Mandel1995}.
We define the Hamiltonian $H_I^{NL}$ which corresponds to the nonlinear part of the total energy of the interaction of radiation field and the crystal as:
\begin{equation}
    H_I^{NL} = \frac{1}{3} \int_{V_C} d^3 r \text{ }  \langle \boldsymbol{P}^{NL}_1(\boldsymbol r, t) \cdot \boldsymbol{E}  (\boldsymbol r, t) \rangle,  \label{eq:interHamiltonian}
\end{equation}
where the integral is performed over the volume of the crystal. A polychromatic electric field $\boldsymbol{E}(\boldsymbol r,t) = \boldsymbol{E}^+(\boldsymbol r, t) + \boldsymbol{E}^-(\boldsymbol r,t) $ can be expressed as \cite{Garrison2008}:
\begin{align}
    \boldsymbol{E}^+(\boldsymbol r,t) &= i\sum_{\beta} \boldsymbol{E}_\beta^+(\boldsymbol r,t) \exp (- i \omega_\beta t) \label{eq:decompositionE+}\\    
    \boldsymbol{E}_\beta^+(\boldsymbol r,t) &=  \sum_{\lambda \in \{1, 2\} }{\sum_{\boldsymbol k}}'\boldsymbol e_{\mathbf{k}, \lambda}  \frac{C_{\mathbf{k},\lambda}}{\sqrt{V}} \alpha_{\mathbf{k}, \lambda} \exp[i (\mathbf{k} \cdot \boldsymbol r - \Delta_{\beta,\mathbf{k}, \lambda} t)], \label{eq:slowlyvarEnv}
\end{align}
where $\beta$ iterates over carrier frequencies - frequencies at which $\boldsymbol{E}^+$ is sharply peaked and the prime symbol in ${\sum_{\mathbf{k}}}'$ signifies that summation is carried over the values of the wavevector such that detunings $\Delta_{\beta, \mathbf{k}, \lambda}$ given by: 
\begin{equation}
    \Delta_{\beta, \mathbf{k}, \lambda} = \omega_{k, \lambda} - \omega_\beta,
\end{equation}
are much smaller than the smallest spacing between the carrier frequencies:
\begin{equation}
    \forall_{\alpha,\beta,\alpha \neq\beta} |\Delta_{\beta, \mathbf{k}, \lambda}| \ll |\omega_\alpha - \omega_\beta|, \label{eq:spacing}
\end{equation}   
Example electromagnetic spectrum corresponding to the above equation is depicted in Fig. \ref{fig:grzebien}.
Although we have already discussed electric field operators in \eref{eq:Eplus}, \eref{eq:Eminus} these relations hold only in free space and need to be reformulated if interaction with a dispersive medium is taken into account. We assume that $\chi$ does not vary significantly across peak frequencies. In (\eref{eq:slowlyvarEnv}), the term $\alpha_{\mathbf{k}, \lambda}$ stands for amplitude and $C_{\mathbf{k}, \lambda}$ (compare with \eref{eq:quantConst}) is normalization constant \cite{Garrison2008}: 
\begin{equation}
    C_{\mathbf{k}, \lambda} = \sqrt{\frac{\hbar \omega_{\mathbf{k}, \lambda} \frac{d \omega_{ks}}{dk}}{2 \epsilon_0 n c}}. \label{eq:quantconst}
\end{equation}
The above constant is introduced to smoothen the transition to the quantum picture. 
Here $k = |\boldsymbol k|$ and angular frequency $\omega_{k, \lambda}$ is a solution of the dispersion relation:
\begin{equation}
k = \frac{n_\lambda (\boldsymbol s, \omega_{k, \lambda}) \omega_{k, \lambda}}{c}, \label{eq:dispersion}
\end{equation}
where $n (\cdot)$ denotes the refractive index of the medium and $\boldsymbol s = \boldsymbol k/k$. In general, $n (\cdot)$ is a function of frequency, polarization and direction of propagation:  
\begin{equation}
n_\lambda (\boldsymbol s, \omega) = n(\boldsymbol s, \boldsymbol e_\lambda, \omega),
\end{equation}
where the wavevector and polarization vectors are set in the medium frame of reference. All the above statements constitute a slowly-varying envelope approximation for polychromatic fields in a dispersive and anisotropic medium. In the slowly varying envelope approximation, Fourier components $\boldsymbol{E} (\boldsymbol r, \omega)$ of polychromatic field $\boldsymbol{E} (\boldsymbol r, t)$ can be expressed as \cite{Garrison2008}:
\begin{equation}
    \boldsymbol{E} (\boldsymbol r, \omega) = \sum_\beta \sum_{\sigma = \pm} \boldsymbol E^{\sigma}_\beta (\boldsymbol r, \omega - \sigma \omega_\beta ),
\end{equation}
where $\boldsymbol E^{\sigma}_\beta (\boldsymbol r, \omega)$ is a continuous function of $\omega$ sharply peaked at $\omega = 0$. Since the only significant amplitudes of the electric field correspond to frequencies $\omega_\beta$, we can approximate the above equation with: 
\begin{equation}
    \boldsymbol{E} (\boldsymbol r, \omega) \approx \sum_\beta \delta_\omega^{ \omega_\beta} \boldsymbol{E} (\boldsymbol r, \omega) = \sum_\beta \sum_{\sigma = \pm} \delta_\omega^{\sigma \omega_\beta} \boldsymbol{E}^{\sigma} (\boldsymbol r, \omega).\label{eq:fourierexpansion}
\end{equation}
Here, we used the Kronecker delta $\delta_a^b$ which is equal to $1$ only if $a=b$ and is $0$ otherwise. The above approximation states that the electric field has significant values only at peaks. Since it corresponds to the plane wave expansion of the field, we can rewrite the interaction Hamiltonian \eref{eq:interHamiltonian}:
\begin{equation}
    H_I^{NL} = \frac{1}{3} \int_{V_C} d^3 r \int d \omega \text{ }  \boldsymbol{P}^{NL}_1(\boldsymbol r, \omega) \cdot \overline{\boldsymbol{E}}  (\boldsymbol r, \omega).  \label{eq:interactionHamiltonian}
\end{equation}
The electric field appears in the above equation in the conjugate form. This is a direct result of the plane wave expansion which allowed us to use the Poynting theorem for time averages of harmonic fields. For a discussion of the Poynting theorem for harmonic fields in the context of derivation of energy density for an electromagnetic wave interacting with a nonlinear medium see \cite{Boyd2003}. In \cite{Garrison2008}, time averaging is done explicitly and the same results follow. In \cite{Mandel1995}, time averaging is omitted and the presented derivation may not be correct in general. 
If we now expand the electric field and polarization terms we get:
\begin{align}
    H_I^{NL} =\sum_{ijk} \frac{\epsilon_0}{3} \int d \omega d \omega' d \omega'' d^3 r\text{ }   \delta(\omega - \omega ' - \omega'') \chi_{ijk}^{(2)}(\omega;\omega', \omega'')\overline{E}_i (\boldsymbol r, \omega) E_j (\boldsymbol r, \omega') E_k (\boldsymbol r, \omega''), \label{eq:interactionHam}     
\end{align}
We can now use equation \eref{eq:fourierexpansion} to expand the Hamiltonian from \eref{eq:interactionHam}:
\begin{align}
    H_I^{NL} =\sum_{ijk} \frac{\epsilon_0}{3} \int d \omega d \omega' d \omega'' d^3 r \sum_{\alpha \beta \gamma} \sum_{\sigma \sigma' \sigma''}   &\delta(\omega - \omega ' - \omega'')\delta_\omega^{\sigma \omega_\alpha} \delta_{\omega'}^{\sigma' \omega_\beta}\delta_{\omega''}^{\sigma'' \omega_\gamma}  \times \nonumber  \\ 
    & \times \chi_{ijk}^{(2)}(\omega;\omega', \omega'')  \overline{E^\sigma_i} (\boldsymbol r, \omega) E^{\sigma'}_j (\boldsymbol r, \omega') E^{\sigma''}_k (\boldsymbol r, \omega'').  
\end{align}
We now perform integrals over frequencies which lead to a Hamiltonian of the form:
\begin{align}
    H_I^{NL} =\sum_{ijk} \frac{\epsilon_0}{3} \int d^3 r \sum_{\alpha \beta \gamma} \sum_{\sigma \sigma' \sigma''}  \delta^{\sigma \omega_\alpha}_{\sigma' \omega_\beta + \sigma'' \omega_\gamma} & \chi_{ijk}^{(2)}(\sigma \omega_\alpha;\sigma' \omega_\beta, \sigma'' \omega_\gamma) \times \nonumber  \\ 
    & \times E^{ - \sigma}_i (\boldsymbol r, \omega_\alpha) E^{\sigma'}_j (\boldsymbol r, \omega_\beta) E^{\sigma''}_k (\boldsymbol r, \omega_\gamma) 
\end{align}
By convention, carrier frequencies are always positive so summation over signs $\sum_{\sigma \sigma' \sigma''} $ gives only two combinations:
\begin{align}
    H_I^{NL} =\sum_{ijk} \epsilon_0 \int d^3 r \sum_{\alpha \beta \gamma}\delta^{\omega_\alpha}_{\omega_\beta + \omega_\gamma} \chi_{ijk}^{(2)}(\omega_\alpha;\omega_\beta, \omega_\gamma)[E^{-}_i (\boldsymbol r, \omega_\alpha) E^{+}_j (\boldsymbol r, \omega_\beta) E^{+}_k (\boldsymbol r,\omega_\gamma) + c.c],
\end{align}
where we have used the fact that nonlinear susceptibility is an even function of its argument for transparent media. By transparent we mean that medium for a given frequency is non-absorbing, but electromagnetic wave might still interact with the medium. The electric field terms present in the Hamiltonian are Fourier components  of \eref{eq:decompositionE+}, which allows us to write Hamiltonian in its final form:
\begin{align}
    H_I^{NL} =\frac{i}{V^{3/2}} \sum_{p,s,i} \delta^{\omega_p}_{\omega_s + \omega_i} g_{p,s,i}(\omega_s, \omega_i) (\alpha_p \overline{\alpha}_s \overline{\alpha}_i  - c.c) \theta(\boldsymbol k_p - \boldsymbol k_s - \boldsymbol k_i), \label{eq:classicalHamiltonian}
\end{align}
where indices $p,s,i$ are short notation for modes: $p \rightarrow \{ \boldsymbol k_0, \lambda_0 \}$,  $s \rightarrow \{ \boldsymbol k_1, \lambda_1 \}$ and $i \rightarrow \{ \boldsymbol k_2, \lambda_2 \}$. In other words sum  i.e $\sum_p$ runs over all possible pumping modes.
Assuming that optical susceptibility does not vary in the spatial domain, function $g_{p,s,i}(\omega_s, \omega_i)$ is:
\begin{equation}
    g_{p,s,i}(\omega_s, \omega_i) = \epsilon_0 C_p C_s C_i \sum_{jkl} [\boldsymbol{e}_p]_j [\boldsymbol{e}_s]_k [\boldsymbol{e}_i]_l \chi^{(2)}_{jkl}(\omega_p;\omega_s, \omega_i), \label{eq:constantHamiltonian}
\end{equation}
and function $\theta(\boldsymbol k)$ stands for:
\begin{equation}
    \theta(\boldsymbol k) = \int_{V_c} d^3r \exp(i \boldsymbol k \cdot \boldsymbol r). \label{eq:phasematch}
\end{equation}
Function $\theta(\boldsymbol k)$ in the limit of a very large crystal is $(2 \pi)^3 \delta^{\boldsymbol k_p}_{\boldsymbol k_s + \boldsymbol k_i}$ and together with term $\delta^{\omega_p}_{\omega_s + \omega_i}$ ensures that the interaction Hamiltonian is non zero only if fields are phase matched:
\begin{align}
\delta^{\omega_p}_{\omega_s + \omega_i} \rightarrow \omega_p &= \omega_s + \omega_i, \label{eq:conservE} \\ 
\delta^{\boldsymbol k_p}_{\boldsymbol k_s + \boldsymbol k_i} \rightarrow \boldsymbol  k_p &= \boldsymbol k_s + \boldsymbol  k_i. \label{eq:conservP}
\end{align}
The above equations constitute the phase-matching conditions. 

The phase matching conditions obtained in this section were derived under certain assumptions and approximations. The approximations we used explicitly are:
\begin{itemize}
    \item the medium is weakly dispersive and weakly nonlinear,
    \item the pumping field can be described in terms of the slowly varying envelope approximation,
    \item the pumping field can be polychromatic but has to be sharply peaked at carrier frequencies,
    \item the medium is transparent at carrier frequencies.
\end{itemize}
For a more extensive list of assumptions see \cite{Garrison2008}.
\section{Phase matching}

Equations \eref{eq:conservE} and \eref{eq:conservP} are very important results describing conditions that have to be met for SPDC to happen. 

\begin{figure}[tbh]
	\centering
	\includegraphics[width=0.9\columnwidth,keepaspectratio]{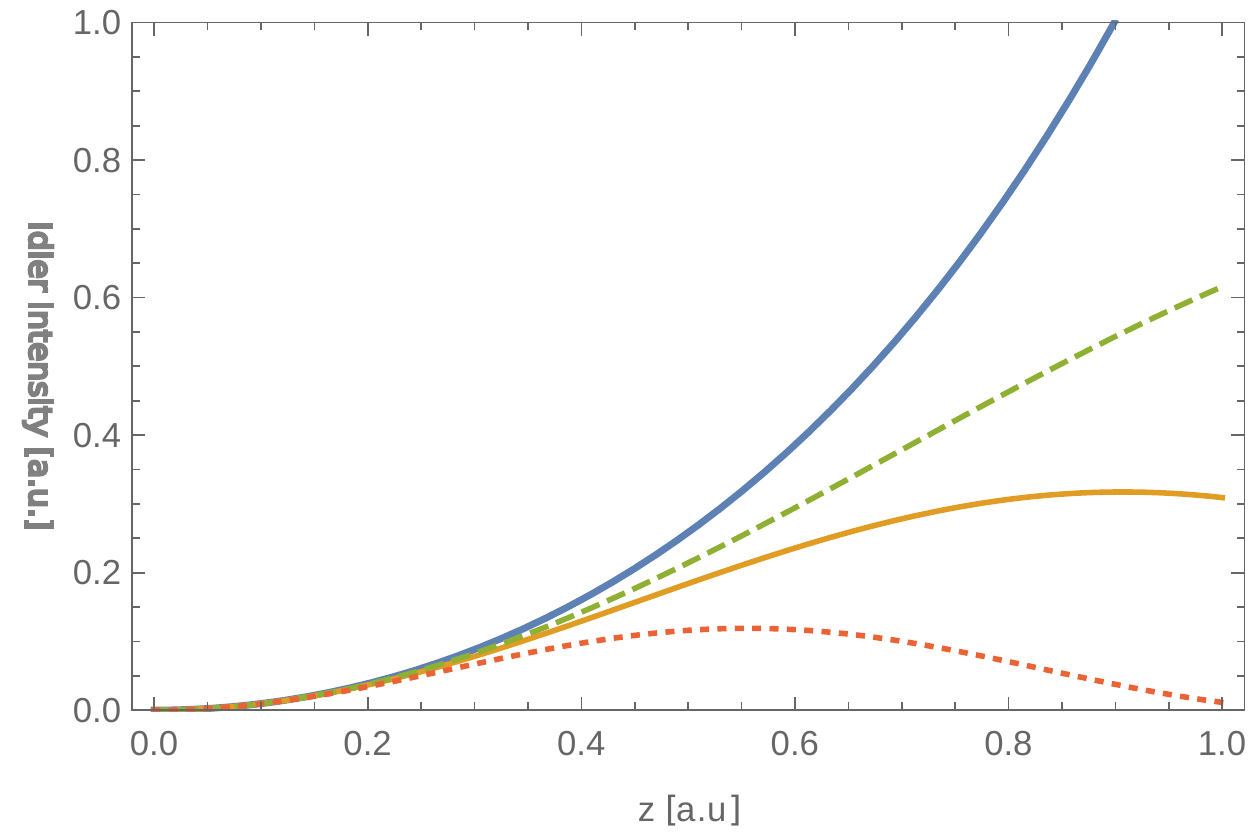}
	\caption{Intensity dependence in the DFG process on crystal length in case of perfect phase matching (blue line) and in case of phase mismatch. The energy transfer is still possible if the mismatch is small in comparison with crystal length (green line). If mismatch or crystal is large enough then oscillatory behaviour inside of the crystal can be observed (orange and red line).}
	\label{fig:DFGphase}
\end{figure}

For the SPDC process to be possible in a given medium there are generally two conditions that need to be met - the medium has to exhibit nonlinearity which allows for three-wave mixing (\eref{eq:interactionHamiltonian}) and the phase-matching conditions (\eref{eq:conservE}, \eref{eq:conservP}) have to be fulfilled \cite{Boyd2003}. One of the most common media used for SPDC is an anisotropic nonlinear crystal \cite{Boyd2003}. Optical anisotropy is a dependence of refractive index on polarization and direction of propagation in a given medium. Medium which exhibits dependence of refractive index on frequency is called a dispersive medium. All solid-state media are dispersive but not all of them are anisotropic, which leads to the phase-matching problem. Crystal anisotropy introduces different refractive index coefficients for different wave polarization and directions of propagation. Choosing a crystal with the right anisotropic characteristic allows for achieving phase matching. An exact match of all three phases leads to efficient transfer of energy from the pumping beam. Even if the phase match is not perfect, the transfer of energy is still possible if the crystal is finite. In Fig.\ref{fig:DFGphase} growth of idler intensity in the DFG process is depicted as a function of crystal length.
Second-order nonlinear susceptibility tensor $\chi_{ijk}^{(2)}$ describes the efficiency of coupling the three beams involved in the process \cite{Boyd2003} as it was shown in \eref{eq:constantHamiltonian}. 
The DFG process is possible when the previously mentioned criteria are met for the waves incident on the crystal and the energy transfer from pumping beam two signal beam or signal and idler beam takes place (see Fig.\ref{fig:downconv} and Fig. \ref{fig:DFG}). If no other beam than the pumping beam is present, then there is no solution to the Maxwell equation that would describe the creation of signal and idler beams (see Fig. \ref{fig:DFGphase}). That kind of solution exists only on the ground of quantum mechanics.
Without the second-order nonlinearity of the medium, SPDC and DFG processes would not be possible, but anisotropy is not required. There are other ways to achieve phase matching e.g modal phase matching \cite{Horn2013}.

\subsection{Anisotropic crystals}

Solving equations \eref{eq:conservE} and \eref{eq:conservP} is a non-trivial problem since the relation between the wavevectors and frequencies has a rather complicated form given in \eref{eq:dispersion} where the refractive index $n(\cdot)$ is in general frequency-dependent. The problem here comes down to choosing a nonlinear medium with a refractive index that would allow phase matching for desired beam frequency.

Usually, the refractive index is an increasing function of frequency. Such dependence is called normal dispersion \cite{Boyd2003}. The refractive index might be a decreasing function of frequency. Such effect is called anomalous dispersion and occurs near absorption frequencies. In principle, it is possible to take advantage of that effect to fulfil the phase-matching condition \cite{Boyd2003}. A simpler and more common approach is to use an optically anisotropic medium. Optical anisotropy is also called birefringence and these terms can be used interchangeably. 

There are two types of anisotropic crystals: uniaxial and biaxial. In anisotropic crystals relative permittivity tensor $ \boldsymbol \epsilon$ is a symmetric second-rank tensor \cite{Saleh2007}. It describes the linear response of a medium to the electric field. In general, the electric displacement field vector does not have to point in the same direction as the electric field vector. Relative permittivity tensor links electric field vector with displacement vector in a medium $\boldsymbol D(\omega)= \boldsymbol \epsilon (\omega) \boldsymbol E (\omega)$. It is always possible to orient a coordinate system in such a way that  off-diagonal elements are equal to $0$:
\begin{equation}
    \boldsymbol \epsilon = \epsilon_0 \left( \begin{array}{ccc} n_x^2 & 0 & 0 \\ 0 & n_y^2 & 0 \\ 0 & 0 & n_z^2 \end{array} \right). \label{eq:relative_permittivity}
\end{equation}
Directions of the axes of such a coordinate system are called principal axes.The electric field polarized along a principal axis will point in the same direction as the displacement vector. Refractive indices $n_x,n_y,n_z$ are called principal refractive indices. 

In uniaxial crystals, two refractive indices are always equal to each other. Without any loss of generality, we can pick $n_x$ and $n_y$ to be equal. We will denote them as $n_o$ where 'o' stands for ordinary. The symbol $n_z$. therefore will be called $n_e$ where 'e' stands for extraordinary. Wave with polarization vector lying on the $xy$-plane is said to be ordinarily polarized. If the polarization is along the $z$-axis then the wave is extraordinarily polarized. Biaxial crystals are more complicated. All three indices are different so there is no 'ordinary' or 'extraordinary' polarization. 

It can be shown that there exist two normal modes for each direction of propagation in the crystal \cite{Saleh2007}. The simplest example is the uniaxial crystal, where a plane wave propagates along $x$. For such waves, there are two normal modes with polarization along the $z$-axis (extraordinary) and the $y$-axis (ordinary). If the wave polarization is a linear combination of these two modes then these two wave components travel with different speeds along the $x$-axis. A phase difference will be acquired between these linearly polarized components, therefore, in general, the wave will no longer be linearly polarized. That is true for other directions of propagation in anisotropic crystal. In uniaxial crystal one of these modes is always ordinarily polarized meaning its phase velocity is $c/n_o$, the other one has a speed of $c/n(\theta)$ where $n(\theta)$ ranges from $n_o$ to $n_e$ depending on $\theta$ angle. Here, the $\theta$ angle is an angle created by the $z$-axis and the wavevector of the electric field (direction of propagation). Note that in uniaxial crystals if a wave propagates along the $z$-axis, then its effective refractive index is polarization independent and is equal to $n_o$. By convention, the wave in the ordinary mode is called an ordinary wave and a wave whose polarization vector has a non-zero component along the extraordinary axis is called an extraordinary wave.
In biaxial crystals where all three refractive indices are different (\eref{eq:relative_permittivity}), the naming convention of polarization is different. The wave component propagating with a higher speed is said to be fast polarized and the other component of the electric field is said to be slow polarized.

\subsection{Phase matching types}
All media, anisotropic crystal included, are dispersive which means that waves with frequencies given by relation \eref{eq:conservE} with the same direction of propagation and the same polarization will not, in most cases, have matching wavevectors as in \eref{eq:conservP}.

Between absorption frequencies of a medium, the refractive index is usually an increasing function. In that case, for the same direction of propagation and the same polarization of all three waves there is no possibility of achieving a phase match.

\begin{figure}[tbh]
	\centering
	\includegraphics[width=0.5\columnwidth,keepaspectratio]{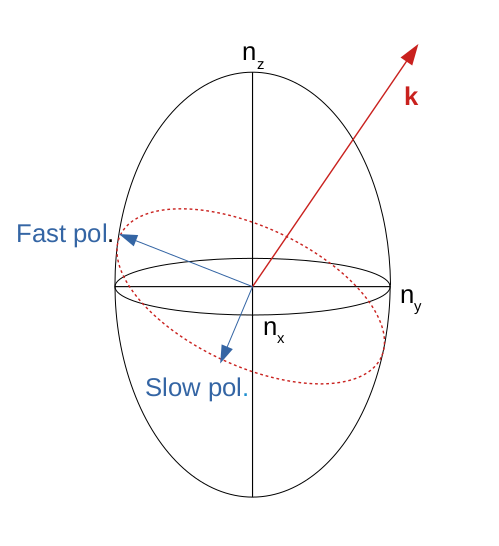}
	\caption{Picture represents polarization indicatrix which is an ellipsoid with a length of semi-axes set to values of the refractive index components of the crystal. The red arrow depicts the wavevector $\boldsymbol k$. Plane to which the wavevector is normal crosses the indicatrix creating an ellipse. Lengths of the semi-axes of the ellipse give values of refractive indices for a fast and slow polarized waves travelling in the direction $\boldsymbol k$.}
	\label{fig:polarizations}
\end{figure}

Anisotropic crystals offer a solution to the above problem in the form:
\begin{equation}
n(\boldsymbol s, \boldsymbol e, \omega_p)\omega_p \boldsymbol s = n(\boldsymbol s', \boldsymbol e', \omega_s) \omega_s \boldsymbol s' + n(\boldsymbol s'', \boldsymbol e'', \omega_i) \omega_i\boldsymbol s''.
\end{equation}
Here, as before, polarization vectors $\boldsymbol e, \boldsymbol e', \boldsymbol e''$ point along polarization directions of normal modes of the crystal and $\boldsymbol s, \boldsymbol s', \boldsymbol s''$ are directions of propagation. Normal modes' polarization directions depend on their propagation direction, but for any given direction of propagation, there are only two basic possibilities: 'fast' polarization and 'slow' polarization as it is shown in Fig. \ref{fig:polarizations}. 
The phase matching condition in an anisotropic crystal can therefore be rewritten as:
\begin{equation}
n(\boldsymbol s,\boldsymbol e_\mathrm{pol_1}, \omega_p)\omega_p\boldsymbol s = n(\boldsymbol s', \boldsymbol e'_\mathrm{pol_2}, \omega_s)\omega_s \boldsymbol s' + n(\boldsymbol s'', \boldsymbol e''_\mathrm{pol_3}, \omega_i) \omega_i \boldsymbol s'',  \thickspace \mathrm{pol_i} \in \{\mathrm{fast},\mathrm{ slow} \}.
\end{equation}
Here, $\mathrm{pol}_1 $ stands for polarization of the pumping beam and $\mathrm{pol}_2$, $\mathrm{pol}_3$ stand for polarizations of the signal and idler beams, respectively. Solution of the form $\mathrm{pol}_1 \neq \mathrm{pol}_2 = \mathrm{pol}_3$ is called type-I phase matching, solution where  $\mathrm{pol}_2 \neq \mathrm{pol}_3$ is called type-II phase matching. There is also a third possibility called type-0 phase matching. In type-0 all polarizations are the same. This type of phase matching is very hard to achieve in regular anisotropic crystals and therefore it requires crystals to be periodically polled. Using periodicity of $\chi$ along the propagating direction allows for compensation of a phase mismatch. This technique is called quasi-phase matching and may be used to achieve phase matching even in isotropic crystals.

\subsection{Quasi-phase matching}

In regular crystals for a particular choice of polarization, frequencies and propagation directions $\chi^{(2)}_{ijk} (\omega_p; \omega_s, \omega_i)$ is a constant. In periodically polled crystals, $\chi^{(2)}$ is a periodic function of one spatial dimension, let us say $z$:
\begin{equation}
    d(z) \equiv \chi^{(2)}_{ijk} (\omega_p; \omega_s, \omega_i, z).
\end{equation}
The usual way of creating periodically polled crystal is to manufacture it in such a way that $d(z)$ changes sign periodically, such that:
\begin{equation}
    d(z) = d(0) \sign [\sin(\frac{2 \pi z}{\Lambda})],
\end{equation}
where $\Lambda$ is a spatial period. Integral over crystal volume can no longer be expressed as in \eref{eq:phasematch}. Instead, the $\theta$ function has to be replaced by
\begin{equation}
    \theta(\boldsymbol k) = \int_{V_c} d^3r \thickspace d'(z) \exp(i \boldsymbol k \cdot \boldsymbol r), \label{eq:quasimismatch}
\end{equation}
where $d'(z) =d(z)/d(0)$. With this change of $\theta(\boldsymbol k)$, the Hamiltonian from equation \eref{eq:classicalHamiltonian} along with the expression from equation \eref{eq:constantHamiltonian} are still valid. The function $d'(z)$ is a periodic function so it can be expanded in a Fourer series with complex coefficients given by:
\begin{equation}
    a_m = \frac{1}{\Lambda} \int_{-\Lambda/2}^{\Lambda/2} \sign(z) \exp(-imKz)dz
\end{equation}
where $m$ is an integer and $K = \frac{2 \pi}{ \Lambda}$. Calculating the above integral leads to the following expression for Fourier coefficients:
\begin{align}
    a_m &= 0, \thickspace \forall |m| \in \{ 0,2,4, \dots \} \\
    a_m &= -\frac{2i}{m \pi}, \thickspace \forall |m| \in \{1,3,5, \dots \}. \label{eq:Fourier}
\end{align}
all coefficients are purely imaginary and $a_{-m} = \overline{a}_m$.
The Fourier series representation of $d'(z)$ is:
\begin{equation}
    d'(z) = \sum_{m = - \infty}^{\infty} a_m \exp(i m K z). 
\end{equation}
We can plug this result into \eref{eq:quasimismatch}:
\begin{equation}
    \theta(\boldsymbol k) = \sum_{m = - \infty}^{\infty} \int_{V_c}  \thickspace a_m \exp[i(k_z + mK)z] \exp[i(k_x x + k_y y)] dxdydz.    
\end{equation}
In the limit of a very large crystal the above integral gives the (quasi) phase matching condition for wavevectors:
\begin{align}
\boldsymbol  k_p &= \boldsymbol k_s + \boldsymbol  k_i + m \boldsymbol K, \label{eq:quasiphasematching}
\end{align}
and $\boldsymbol K$ is the quasi wavevector $\boldsymbol K = K \boldsymbol z$. Equation \eref{eq:quasiphasematching} should be understood not as a condition imposed by the system but as a guideline for designing periodically polled crystals. There is a very small chance that a direction in the crystal that results in the largest effective nonlinear coefficient is the direction that will allow for phase matching for a given set of wave frequencies, no matter their polarization. With the quasi-phase matching technique we can circumvent the resulting phase mismatch with a quasi wavevector $\boldsymbol K$, not only to exploit nonlinearity but also to allow a collinear propagation of all three waves. Periodically polled crystals are therefore designed to allow for an efficient three-wave mixing for a certain combination of polarization directions and frequencies. For that reason, in periodically polled crystals the three-wave mixing process for which the crystal was designed will be phase-matched in the following way:
\begin{equation}
\boldsymbol  k_p = \boldsymbol k_s + \boldsymbol  k_i \pm \boldsymbol K, \label{eq:QPM}
\end{equation}
since the larger the $|m|$ the smaller $a_m$ which results in smaller efficiency. That is why crystals are designed for quasi-phase match with $|m| = 1$, since the next order mismatch ($|m| = 3$) leads to a $3$ times smaller Fourier component (\eref{eq:Fourier}) which in turn means a $9$ times drop in the efficiency of energy transfer.

It is also worth noting that since periodically polled crystals are designed for a given three-wave mixing process they are polled in a direction and in a way that not only allows for phase matching but also allows for collinear phase matching. Therefore the usual form of quasi-phase match is: 
\begin{equation}
k_p =  k_s +   k_i + K,
\end{equation}
where we assumed normal dispersion. 

\subsection{Spontaneous parametric down-conversion}

The largest difference between the classical and quantum theories of three-wave mixing is the following: In SPDC, only the pumping beam is required, while DFG additionally needs an idler beam to generate signal. To describe the SPDC process we need to formulate a Hamiltonian governing dynamics of the system.  Equation \eref{eq:classicalHamiltonian} provides a good starting point. We will replace amplitudes $\alpha_p, \alpha_s, \alpha_i$ by annihilation operators of dressed photonic modes $\hat a_p, \hat a_s, \hat a_i$:
\begin{equation}
    \hat H_I^{NL} =\frac{i}{V^{3/2}} \sum_{p,s,i} \delta^{\omega_p}_{\omega_s + \omega_i} g_{p,s,i}(\omega_s, \omega_i) (\hat a_p \hat a^\dagger_s \hat a^\dagger_i  - h.c) \theta(\boldsymbol k_p - \boldsymbol k_s - \boldsymbol k_i \pm \boldsymbol K), 
\end{equation}
and operator commutation relation is:
\begin{equation}
    [\hat a_{\alpha},\hat a^{\dagger}_{\beta}] = \delta^\beta_{\alpha}, \thickspace \forall \alpha, \beta \in \{p, s, i \}.
\end{equation}
We use the term "dressed mode" since in a material we do not deal with photons but rather quantized excitations of electromagnetic field and material. The quantization constant of a dressed mode was defined in \eref{eq:quantconst} and refers to quantization of photon-dielectric interaction in case of polychromatic field propagating in a linear dielectric. Discussion of the above quantization approach can be found in \cite{Garrison2008}.
The above Hamiltonian describes only the interaction of the nonlinear medium with the incident field. The full form of the Hamiltonian is:
\begin{align}
    \hat H &= \hat H_p + \hat H_s + \hat H_i + \hat H_I^{NL}, \\
    \hat H_{x} &= \hbar \omega_{x}  (\hat n_{x} +\frac{1}{2}), \thickspace x \in \{p, s, i \}.
\end{align}
We will be interested mostly in $\hat H_I^{NL}$. This Hamiltonian can be further simplified by approximation called index matching - we will assume that whole quantization volume is filled with a linear index of refraction matching index of refraction of nonlinear medium . This removes the frequency dependence from the function $g_{p,s,i}(\omega_s, \omega_i)$ \cite{Garrison2008}:
\begin{equation}
    \hat H_I^{NL} =\frac{1}{V^{3/2}} \sum_{p,s,i} g_{p,s,i} \theta(\boldsymbol k_p - \boldsymbol k_s - \boldsymbol k_i \pm \boldsymbol K)\hat a_p \hat a^\dagger_s \hat a^\dagger_i  + h.c. 
    \label{eq:quantumHamiltonian}
\end{equation}
The above Hamiltonian describes well three-photon interaction such as \\ down-conversion and up-conversion. For most applications, this form is not very useful as it does not explicitly take into account the characteristic of the pumping beam. 
In SPDC the pumping beam is usually laser light with intensity much larger than any of the resulting signal and idler beams. Larger intensity allows for undepleted pump approximation - assumption that the number of down-converted photons is much smaller than the number of photons in the pumping beam. Another assumption which we can make is an assumption of continuous wave (CW) pumping. Hence, the pump intensity can be approximated as constant. Continuous laser light can be well described by coherent state $|\alpha_p\rangle$, although it is still an approximation. 
We replace operators $\hat a_p , \hat a^\dagger_p$ with their expectation value:
\begin{align}
    \alpha_p &= \langle \alpha_p | \hat a_p | \alpha_p \rangle \\
    \overline{\alpha_p} &= \langle \alpha_p | \hat a^\dagger_p | \alpha_p \rangle,
\end{align}
which means reducing the size of the Hilbert space. 
CW pumping also results in dropping a sum over $p$ modes since there is now only one pumping frequency. $\alpha_p$ is now a dimensionless amplitude of CW pumping field $E_p$ as in \eref{eq:slowlyvarEnv}. Under the above assumption, the Hamiltonian transforms to:
\begin{align}
    \hat H(t) &= \hbar \omega_p |\alpha_p|^2 + \sum_{x \in \{ s,i \}} \hbar \omega_x  (\hat n_x+ \frac{1}{2}) + \hat H_I^{NL}(t), \\
    \hat H_I^{NL}(t) &= \frac{-i}{V} \sum_{p,s,i} \frac{E_p g_{p,s,i}}{C_p} \theta(\boldsymbol k_p - \boldsymbol k_s - \boldsymbol k_i \pm \boldsymbol K) e^{i( \omega_s + \omega_i - \omega_p )t} \hat a^\dagger_s (0) \hat a^\dagger_i(0)  + h.c,
\end{align}
We switched from the Schrodinger picture to the interaction picture. 
In the interaction picture state produced by the SPDC is given by \cite{Mandel1995}:
\begin{equation}
    | \psi (t) \rangle = \hat \Tau \exp[\frac{1}{i \hbar} \int_0^t \hat H^{NL}_I(t')dt']|0\rangle |0 \rangle . \label{eq:SPDCstate}
\end{equation}
Here, $| \psi (t) \rangle$ is a superposition of squeezed states and $\hat \Tau$ is the Dyson time ordering operator. If we assume for the moment that SPDC is perfectly phase-matched only for one set of carrier frequencies travelling in a well-defined direction and with the same polarization, then $|\psi (t) \rangle$ is given by a single squeezed state:
\begin{equation}
    | \psi (t) \rangle = \exp[Z(t)\hat a^\dagger_s (0) \hat a^\dagger_i(0)  + h.c ] |0\rangle |0 \rangle,
\end{equation}
where the squeezing parameter $Z(t)$ is :
\begin{equation}
 Z(t) = -\frac{g_{p,s,i}}{\hbar V C_p} E_p t.   
\end{equation}
The above exponent is the squeezing operator introduced in \eref{eq:2modesqueezedoperator}. This means that $| \psi (t) \rangle$ is a two-mode squeezed state of the form given by equation \eref{eq:2modesqueezedstate}.
More generally equation \eref{eq:SPDCstate}, for a parallelepiped crystal with side lengths $L_1$, $L_2$ and $L_3$  has a solution of the form which has been derived in analogy to the scheme in \cite{Mandel1995}:
\begin{align}
    | \psi (t) \rangle = &|0\rangle_s |0 \rangle_i - (\hbar V)^{-1}\sum_{s,i} \frac{E_p g_{p,s,i}}{C_p} \Pi^3_{m=1} \sinc [\frac{L_m}{2}(\boldsymbol k_p - \boldsymbol k_s - \boldsymbol k_i \pm \boldsymbol K)_m ] L_m \times \nonumber \\
    & \times e^{i(\omega_s + \omega_i - \omega_p)t/2} \sinc[\frac{t}{2}(\omega_s + \omega_i - \omega_p)]t |1\rangle_s |1 \rangle_i + \dots . \label{eq:BphWaveFun}
\end{align}
This is the general form of a state produced in the SPDC process and emitted (with an assumption of no refraction at crystal interfaces) into free space. Just like in equation \eref{eq:SPDCstate}, the SPDC state is a superposition of states with the same number of photons in the idler and signal modes.

%% file: Chapters/Chapter3.tex
\chapter{Estimating Sellmeier coefficients with single photons}

\label{chap:Sellmeier} 

\label{Chapter3} 

In this chapter, we will present the results of the work summarized in the article \cite{Misiaszek2018}. In that article, a novel method of estimation of Sellmeier coefficients is presented.

Sellmeier coefficients are parameters of the Sellmeier equation which describe the dispersion of light in a given material. The dispersion relation in a medium can be derived from first principles assuming a simple harmonic oscillator model. Although derivation is possible, the final form is very complex and hard to use. Sellmeier equations are uncomplicated phenomenological equations that aim to capture the observed behaviour of light in a material at frequencies that are spectrally far away from absorption frequencies (resonances). The resulting formulas are traditionally called the Sellmeier equations \cite{Marcuse1980a}.
The form of the original equations was modified over time to better conform to experimental observations. Typically, the measurement technique which allows determining the respective coefficients of the equations is based on a technique resorting to the process of second harmonic generation (SHG), sum frequency generation (SFG) or mentioned before difference frequency generation (DFG) \cite{Saleh2007, Jundt1997}. It was used for characterization of nonlinear crystals i.e.: LiInS$_2$ \cite{Kato2014a}, BaGa$_4$Se$_7$ \cite{Kato2017}, GaS$_x$Se$_{1\!^{\_}x}$ \cite{Kato2014, Takaoka1999} , LiGaSe$_2$ \cite{Miyata2017} etc.
These equations are of special interest in nonlinear optics and the SPDC research field since the phase relation between propagating waves in a medium is strongly dependent on the dispersion of the material. Imagine, for example, that a given application requires the generation of a particular set of idler and signal frequencies in the SPDC or DFG process. The crystal which we might choose for this application must not only exhibit strong nonlinearity but must also allow for a phase match which means that its dispersive properties must allow for matching wavevectors.  The accuracy of these dispersive properties is very important, especially if broadband pulses are to be used.

Below, we describe a novel technique that we have proposed to measure the Sellmeier coefficients. The technique is based on the SPDC process and exploits the strong dependence of its efficiency on the dispersion relation of the crystal: even a small phase mismatch may lead to a significant drop in the SPDC performance. Therefore, the approach allows for the determination of the crystal coefficients with high precision. It also provides a possibility of estimation of coefficients governing spectral behaviour of sthe crystal at infrared frequencies that are not directly measured, e.g. due to lack of sufficiently good detectors. At the same time, the concept is simple and experimental setup together with apparatus are typical for quantum optics experiments.

Our method offers an simple solution to a problem that might be encountered in every laboratory doing single-photon experiments with nonlinear crystal - discrepancy between calculated SPDC outcome and the actual outcome. These discrepancies might occur i.e due to the manufacturing process of the crystal, crystal impurities or because of the nature of Sellmeier equations - these equations are phenomenological and are only roughly approximating dispersion of the material. There may be more than one Sellmeier equation describing the dispersion of a given crystal, and every equation has a prescribed range where it reasonably approximates material dispersion.  We show that one can use experimental data to correct a given set of Sellmeier coefficients describing the dispersion of a crystal and as result have a better mathematical model of the SPDC process.
It is worth adding that experimental setup presented in this work is a typical setup used for quantum optical experiments where SPDC is used as source of photons.  

\section{Using SPDC to estimate Sellmeier coefficients}
In biaxial crystals, dispersion of wave propagating in an arbitrary direction in a crystal is in general described by three refractive indices (see Fig. \ref{fig:polarizations}). Each of these refractive indices might be approximated by the phenomenological Sellmeier equation in the form \cite{Saleh2007}:
\begin{equation}
n_j^2(\lambda) = a_{j0} + \frac{a_{j1}}{\lambda^2 - a_{j2}} + \frac{a_{j3}}{\lambda^2 - a_{j4}}, \thickspace j=x,y,z,  \label{eq:Sellmeier}
\end{equation}
where $a_{ji}$ are the Sellmeier coefficients. That form of a Sellemeir equation has an understandable explanation - it describes dispersion between two resonances. Dispersion described by this type of Sellmeier equation is expected to be less accurate near resonances.

For SPDC to be efficient, photons have to strictly fulfil conservation laws (see \eref{eq:conservE} and \eref{eq:conservP}). That is why SPDC photons carry information about the crystal's Sellmeier coefficients. We make use of that property.

\begin{figure}[t]
	\centering
	\includegraphics[width=0.85\columnwidth, clip]{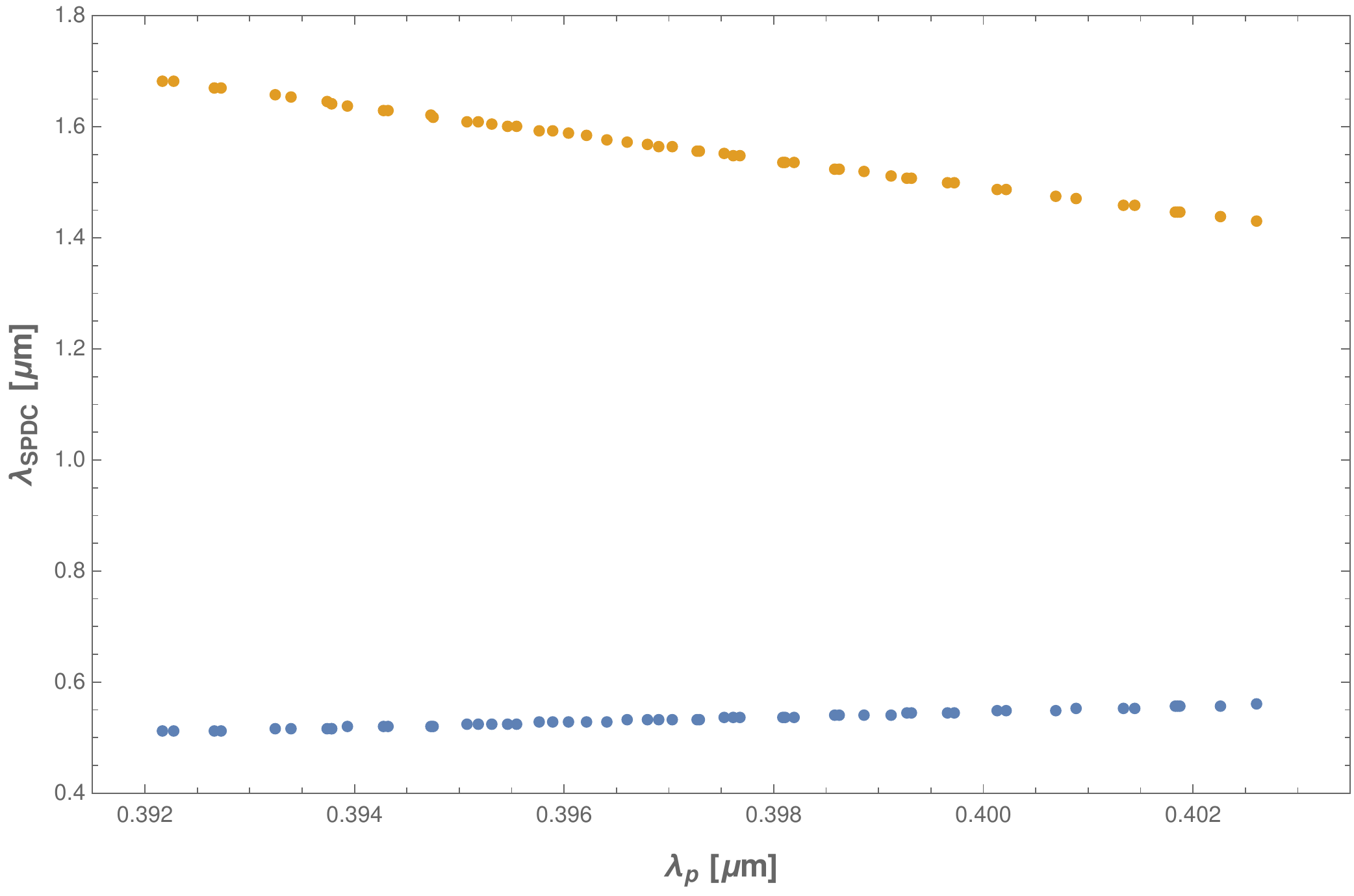}
	\caption{Idler and signal central wavelength as a function of pump central wavelength. Data was generated for SPDC setup described in this chapter with Sellmeier coefficients taken from \cite{Misiaszek2018}.}.
	\label{fig:spdc}
\end{figure}

We propose making a sweep over pump frequencies leading to SPDC. Such sweep allows one to collect data whose example is depicted in Fig. \ref{fig:spdc}. Small changes in the pump frequency may lead to dramatic changes in frequencies of the signal and idler photons especially if SPDC is not degenerated in frequencies. That allows gathering information about dispersion in three spectral ranges: pumping beam sweep range, signal photon spectral range and idler photon spectral range. Note that one range is redundant since given two frequencies it is always possible to calculate the third one. This is the result of energy conservation. That is why measuring the spectrum of one of the down-converted photons is not necessary.

By 'gathering information' we mean changing pumping wavelength and measuring the central wavelength of one of the photons generated in SPDC. Besides central wavelengths, there are other important parameters that have to be known to build a model of the process and, as result, a connection between experimental results and Sellmeier coefficients.

\section{Model}
In this section, I will present a mathematical model which I have built to extract Sellmeier coefficients from experimental data. Theoretical work, along with numerical calculations was performed by the author of this thesis, while experiment and extracting spectra of photons from measurement were performed by my colleague Marta Misiaszek.

Central wavelengths of all three beams have to fulfil phase matching equations such as \eref{eq:conservE} and \eref{eq:QPM}. In this case, our equations have a form:
\begin{align}
    \omega_p &= \omega_{\mathrm{VIS}} + \omega_{\mathrm{IR}}, \\
    \Delta \boldsymbol{k} &= |\boldsymbol{k}_p| \boldsymbol{x} - \boldsymbol{k}_{\mathrm{VIS}} - \boldsymbol{k}_{\mathrm{IR}} - \frac{2 \pi}{ \Lambda (T)} \boldsymbol{x}, \label{eq:phasemismatch}
\end{align}
where $\boldsymbol{x}$ is a unit vector parallel to the $x$-axis. The $\mathrm{VIS}$ and $\mathrm{IR}$ subscripts stand for 'visible' and 'infrared'. Quasi wavevector can have either plus or minus sign. In the case of the SPDC process we are describing here, the quasi wavevector has a minus sign (see \eref{eq:quasiphasematching}). Following the work of \cite{Boeuf2000}, we can split the vector equation into two parts:
\begin{align}
    n_{\mathrm{IR}} \frac{\omega_{\mathrm{IR}}}{\omega_{\mathrm{VIS}}} \sin(\theta_{\mathrm{IR}}) &= n_{\mathrm{VIS}} \sin(\theta_{\mathrm{VIS}}) \label{eq:idlerTheta1} \\
    n_{\mathrm{IR}} \frac{\omega_{\mathrm{IR}}}{\omega_{\mathrm{VIS}}} \cos(\theta_{\mathrm{IR}})  &= n_p \frac{\omega_p}{\omega_{\mathrm{VIS}}} - n_\mathrm{VIS}\cos(\theta_\mathrm{VIS}) - \frac{2 \pi c}{\omega_{\mathrm{VIS}}  \Lambda (T)}, \label{eq:idlerTheta}
\end{align}
where $\theta_\mathrm{VIS}$ stands for an opening angle that the wavevector $\boldsymbol{k}_{\mathrm{VIS}}$ forms with the $x$-axis. Similarly $\theta_\mathrm{IR}$ is an opening angle that $\boldsymbol{k}_{\mathrm{IR}}$ makes with $x$-axis. Effective refractive index coefficients $n_p$, $n_{\mathrm{IR}}$ and $n_{\mathrm{VIS}}$ are in general functions of the direction of propagation, frequency and temperature:
\begin{align}
    n_q = n_q(\varphi_q, \theta_q, \omega_q, T; \boldsymbol S_q), \text{ } q \in \{p, \mathrm{IR}, \mathrm{VIS} \}. 
\end{align}
\begin{figure}[t]
	\centering
	\includegraphics[width=0.95\columnwidth, clip]{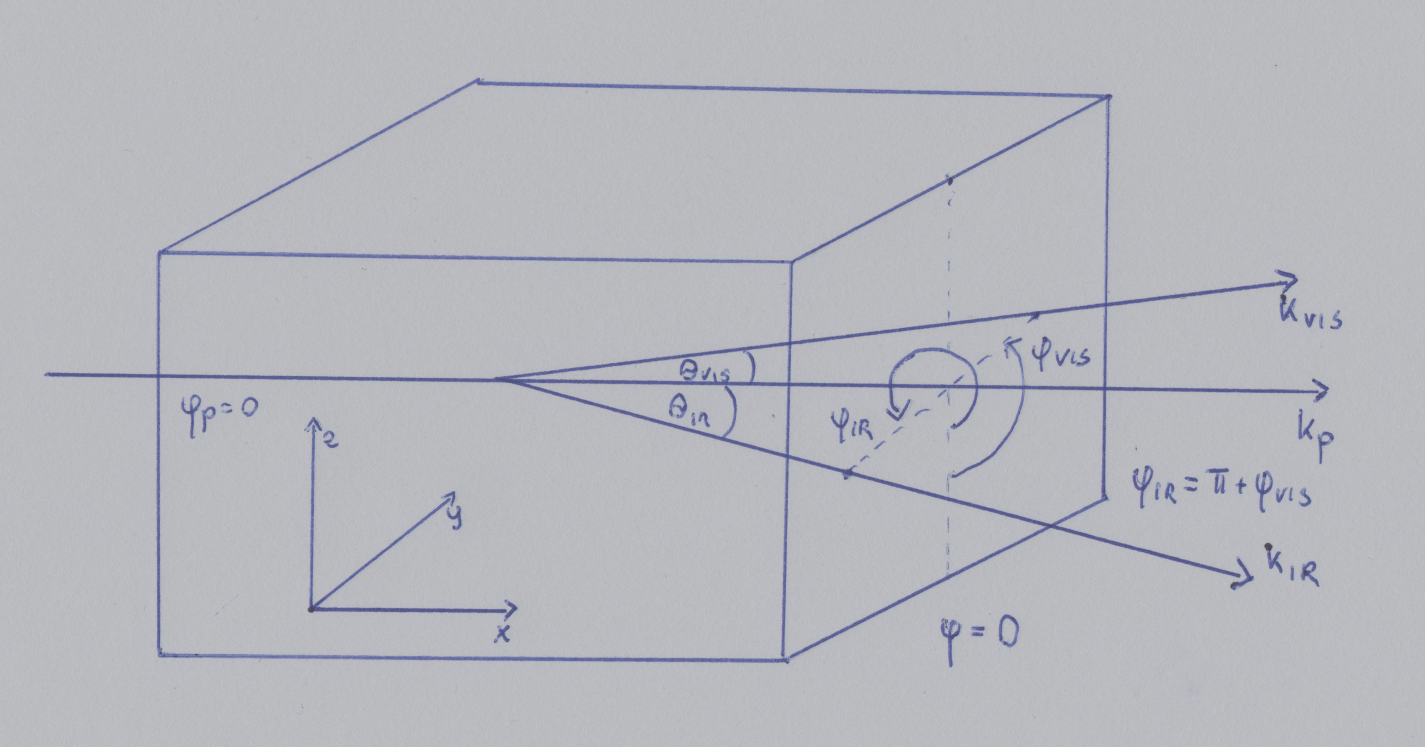}
	\caption{Angles created by SPDC photons wavevectors inside the crystal.}
	\label{fig:crystal}
\end{figure}
Here, $\varphi_q$ stands for rotational angle and we have also explicitly added a vector of the Sellmeier coefficients $\boldsymbol S_q = (a_{q0},a_{q1},...,a_{q4})$ to the function. All angles are defined inside the crystal as depicted in Fig. \ref{fig:crystal}. Their relation to the angles outside the crystal is given by the Snell's law. Due to momentum conservation, rotational angles of the visible and infrared photons are strictly related:
\begin{equation}
    \varphi_\mathrm{IR} = \pi + \varphi_\mathrm{VIS}. 
\end{equation}
We can take squares of equations \eref{eq:idlerTheta1}, \eref{eq:idlerTheta}  and add them up:
\begin{align}
    n_{\mathrm{IR}} \frac{\omega_{\mathrm{IR}}}{\omega_{\mathrm{VIS}}} = \sqrt{n_{\mathrm{VIS}}^2 \sin(\theta_\mathrm{VIS})^2 + (n_p \frac{\omega_p}{\omega_{\mathrm{VIS}}} - n_\mathrm{VIS}\cos(\theta_\mathrm{VIS}) - \frac{2 \pi c}{\omega_{\mathrm{VIS}}  \Lambda (T)})^2} \label{eq:idlerTheta2}. 
\end{align}
With equations \eref{eq:idlerTheta1} and \eref{eq:idlerTheta2} we can express $\theta_\mathrm{IR}$ as:
\begin{equation}
    \theta_\mathrm{IR} = \arcsin(\frac{n_{\mathrm{VIS}} \sin(\theta_{\mathrm{VIS}})}{
    \sqrt{n_{\mathrm{VIS}}^2 \sin(\theta_\mathrm{VIS})^2 + (n_p \frac{\omega_p}{\omega_{\mathrm{VIS}}} - n_\mathrm{VIS}\cos(\theta_\mathrm{VIS}) - \frac{2 \pi c}{\omega_{\mathrm{VIS}}  \Lambda (T)})^2}}).
\end{equation}
This equation can be used only when there is no phase mismatch - $|\Delta \boldsymbol k| \approx 0$. In other words, it can be used to describe phase-matched photons, but it should not be used to estimate the phase mismatch.

In summary, we have shown that the infrared photon wavevector is completely determined by the crystal, pumping and visible photon characteristics. Therefore, visible photon central wavelength might be considered a function of the following parameters:
\begin{equation}
    \lambda_{\mathrm{VIS}} = \lambda_{\mathrm{VIS}} (\omega_p, \theta_p, \varphi_p, \theta_{\mathrm{VIS}},\varphi_{\mathrm{VIS}}, T, \Lambda_0, \boldsymbol S). \label{eq:lambdaVIS}
\end{equation}
In a typical experiment, all four angles do not change - we do not change the direction of propagation of pumping beam nor do we change the position of coupling optics used for coupling visible photons into the waveguide. Although changing pumping beam frequency might change the opening angles of visible and infrared photons, that change is usually small and can be ignored.
We can use equation \eref{eq:lambdaVIS} to calculate visible photon central wavelength which we might expect to detect in our setup. For that, we used Sellmeier coefficients taken from literature \cite{Kato2002}. The result of such investigation is depicted in Fig. \ref{fig:pump}.

\section{PPKTP crystal}

\begin{figure}[t]
	\centering
	\includegraphics[width=0.95\columnwidth, clip]{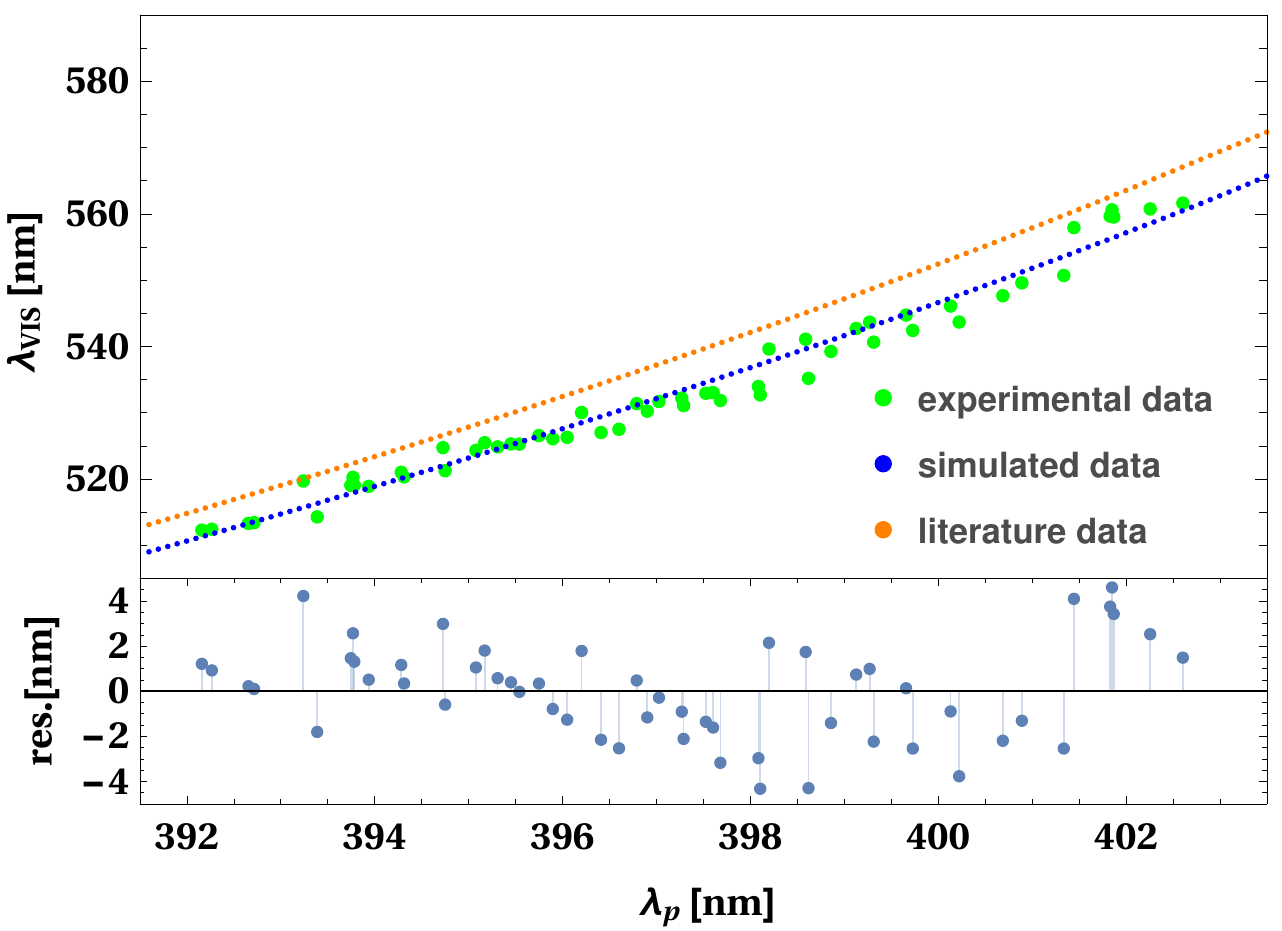}
	\caption{Experimental results of a visible photon central wavelength dependence on the pump central wavelength. Figure and description are taken from \cite{Misiaszek2018}.}
	\label{fig:pump}
\end{figure}

We will demonstrate the application of our method on the periodically poled $\text{KTiOPO}_4$ crystal (PPKTP) crystal. The PPKTP that we used for the experiment was also a source of motivation for this work. If we look at Fig. \ref{fig:pump} and compare the orange line with green dots we can see that the orange line does not cross the experimental data depicted with green dots. The orange line was generated with the mentioned mathematical model. Sellmeier coefficients were taken from \cite{Kato2002}.

Our crystal was fabricated so that it allows for collinear SPDC at room temperature with a pumping laser set to $396$ nm and signal and idler central wavelengths at $532$ nm and $1550$ nm. Periodic polling of our crystal is (in room temperature) $\Lambda(T = 298 K) = \Lambda_0 = 4.01$ $\mu$m. The dependence of periodic polling on temperature is due to the temperature expansion of the material.

PPKTP is a biaxial anisotropic crystal. In our experiment, all three waves are almost collinear and propagate along the $x$-axis. All three waves are 'slow' polarized which means that their dispersion behaviour is dominated by the $n_z$ component of the refractive index tensor.
KTP crystals are extensively used in nonlinear optics so unsurprisingly they were subject to many investigations regarding their thermal and dispersive properties \cite{Kato2002, Manjooran2012, Stoumbou2013, Lee2012, Zhao2010a}. Literature values from \cite{Kato2002} provide a good starting point for our investigation. We will use them as a reference.

\section{Experimental setup}

\begin{figure}[h]
	\centering
	\includegraphics[width=0.75\columnwidth]{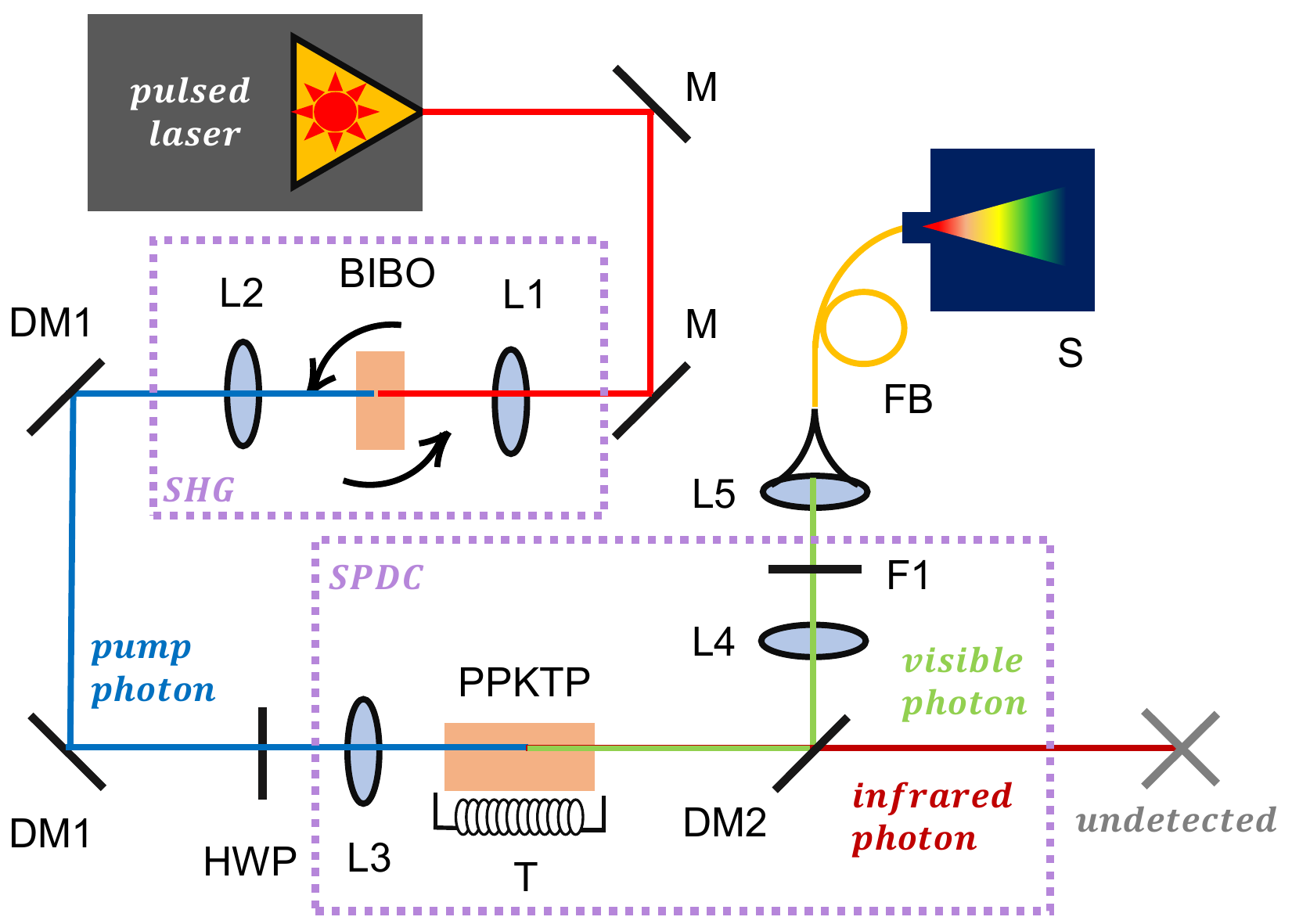}
	\caption{Experimental setup consists of: a pumping pulsed Ti:Sapphire laser, M -- mirror, L1, L2 --  lens (focal lengh f = 7.5 cm), BIBO -- bismuth triborate crystal, DM1 -- dichroic mirror (Semrock T425 LPXR), HWP -- half-wave plate, L3, L4 -- plano-convex lens (f = 10 cm, 12 cm), PPKTP --  periodically poled potassium titanyl phosphate crystal, T --  temperature controller, DM2 -- dichroic mirror (Semrock 76-875 LP), F1 -- set of filters (Chroma ET500, Z532-rdc ), L5 - aspheric lens (f = 1.51 cm), FB -- fiber (Thorlabs SMF460B), S -- spectrometer (Ocean Optics USB2000+). Figure and description are taken from \cite{Misiaszek2018}.}
	\label{fig:setup}
\end{figure}

\begin{figure}[h]
	\centering
	\includegraphics[width=0.5\columnwidth]{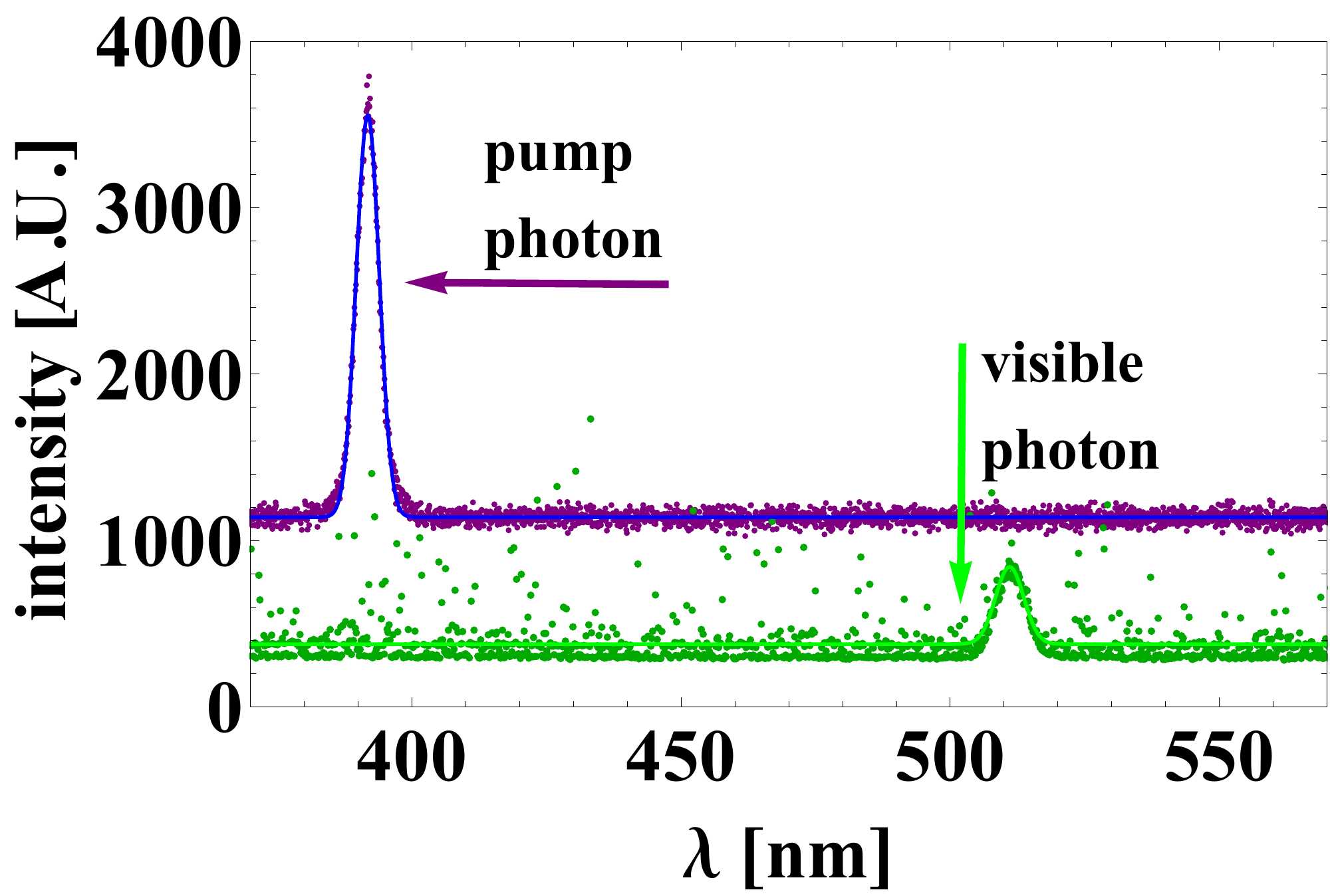}
	\caption{An example of a measurement result of a visible photon and pump photon spectra. Figure is taken from \cite{Misiaszek2018}.}
	\label{fig:inset}
\end{figure}

The measurements were performed by my colleague, Marta Misiaszek. My task was an estimation of the Sellmeier coefficients with the help of gathered data. The results of our investigation were described in \cite{Misiaszek2018}. In the experiment, we used a tunable femtosecond laser to generate central wavelengths between $784$ nm and $806$ nm. The measurement setup is depicted in Fig. \ref{fig:setup}.  Laser beam illuminated bismuth triborate (BiB3O$_6$) nonlinear crystal for generation of the pumping beam in a SHG process. For the separation of the SHG beam from the laser beam, we used a pair of dichroic mirrors. After separation, the pumping beam was manipulated by half-wave plate HWP so that desired polarization was obtained. Then pumping beam was focused on a temperature stabilized PPKTP crystal. The PPKTP crystal was illuminated by a pumping beam with central wavelengths ranging from $392$ nm to $403$ nm.  In the crystal, two additional beams were generated - one at a visible frequency (signal beam) and the other at an infrared frequency (idler beam). The central wavelength of visible photon ranged in the experiment from $500$ nm to $570$. The pumping beam central wavelength and visible beam central wavelength were measured with a spectrometer.  Infrared photon wavelength was not measured but from energy conservation, we know that its central wavelength varied from $1300$ nm to $1900$ nm. Through the use of dichroic mirror and spectral filter infrared beam and pumping beam were separated from the signal beam. The signal beam was coupled to a single-mode fibre and guided to a spectrometer which was used to extract data about the frequency of the signal beam. Infrared photons were not detected. The temperature of PPKTP crystal was set to $T = 298$ K.
An example of a wavelength spectrum of pumping and visible photons is presented in Fig.\ref{fig:inset}. Through fitting, the central wavelength was extracted along with its uncertainty. Gaussian distribution of wavelength was assumed for each spectrum.

Although in our experiment only the frequency of signal photons was measured due to phase match relation we could estimate very well the frequency of idler photon without actually performing its measurement. This is one of the strengths of our proposed technique - we can perform an extensive sweep over infrared frequencies and extract information about the dispersive behaviour of our crystal without actually needing infrared detectors. The result of this sweep is depicted in Fig. \ref{fig:pump} with green dots.

\section{Fitting model to data}

Calculating central wavelength from a set of Sellmeier coefficients is done as follows: based on \cite{Boeuf2000} we constructed the theoretical model described above. This model allows us to express phase mismatch as a function of experimental parameters. The visible photon central wavelength that we are looking for is a value of wavelength that corresponds to phase mismatch being $0$. We noted in  \eref{eq:lambdaVIS} that the resulting function is dependent on experimental and crystal parameters. Finding an explicit solution that would allow for direct calculation of photon wavelength is challenging, so instead I formulated this problem as finding the root of equation \eref{eq:phasemismatch} where $\lambda_{\mathrm{VIS}}$ is a variable. For calculations, I used Mathematica software. From available functions, I decided to use $FindRoot$ function to calculate this value for each of $55$ data points depicted in Fig. \ref{fig:pump}.

Next, we proceed to find a way to use our data to find a correction for literature data of Sellmeier coefficients. First of all, we noted that although there are $15$ coefficients, since all our waves are predominantly polarized along the $z$- axis, only $5$ of them are of interest to us.  Let us write down the equation of refractive index component $n_z$ explicitly: 
\begin{equation}
n^L_z(\lambda) = \sqrt{4.59423 + \frac{0.06206}{\lambda^2 - 0.04763} + \frac{110.807}{\lambda^2 - 86.122}},
\end{equation}
where $\lambda$ is given in micrometres. In the above equation, we have two fractions describing dispersive behaviour of the refractive index between two discontinuity points - one at $0.218$ $\mu$m and the other at $9.28$ $\mu$m. For our wavelength range the second fraction is almost constant: in the range $0.4 - 1.8$ $\mu$m it changes its value by $0.048$ while for the first fraction by $0.533$. This is why we will concentrate on the first three Sellmeier coefficients - our SPDC process is not sensitive enough for changes of Sellmeier coefficients in the far-infrared region.

With the above observation we are ready to calculate corrections for $a_{z0}$, $a_{z1}$ and $a_{z2}$. For that purpose, we used two build-in functions in Mathematica - $NonlinearModelFit$ and mentioned before $FindRoot$. $FindRoot$ function is part of our model which returns visible photon central wavelength based on argument (pumping photon central wavelength) and parameters ($a_{z0}$, $a_{z1}$,$a_{z2}$). $NonLinearModelFit$ takes that model and tries to find parameters that would make our model fit data better. 
To compare results of our computations with literature we calculated the residual sum of squares (RSS):
\begin{equation}
    RSS^{j} = \sum_{i=1}^{N=55} (\lambda_{\mathrm{VIS}}^i - \lambda_{\mathrm{VIS}} (\omega_p^i;a_{z0}^j, a_{z1}^j, a_{z2}^j); \boldsymbol P))^2, \text{ } j=L,C,
\end{equation} 
where $\lambda_{\mathrm{VIS}}^i$ stands for visible photon central wavelength detected in $i$-th measurement. Analogically, $\omega_p^i$ is $i$-th angular frequency used in the experiment. Subscript $L$, as before stand for 'literature' and subscript $C$ for 'computed'. $\lambda_{\mathrm{VIS}}$ is calculated central wavelength for $i$-th measurement by either taking literature or computed Sellmeier coefficients $a_{z0}, a_{z1}, a_{z2}$. Vector $\boldsymbol P$ denotes all other parameters that are constant in experiment and computation - all the rest of Sellmeier coefficients, temperature, crystal polling period and all the angles.

\section{Results}
First, we tested functions provided by Mathematica - $FindRoot$ and $NonLinearModelFit$. Mathematica is not an open source software and it is hard to find out how exactly those functions are implemented. Moreover, they provide the possibility of customization of their behaviour by i.e method selection or setting accuracy for the function. That is why to be sure that our approach is working I tested our model in multiple stages.

\subsection{Model testing}

First, we compared our model (which uses $FindRoot$ function) with results computed by SNLO software for collinear SPDC and achieved perfect agreement. SNLO is popular software developed by Dr. Arlee V. Smith for the simulation of nonlinear optical effects. This software is a freeware made to assist in nonliner optics calculations. A more extensive description of SNLO can be found in the appendix of PhD thesis of Marta Misiaszek.
 
Once I was sure that the mathematical model of phasematching in a nonlinear crystal is working correctly I moved on to testing $NonLinearModelFit$. I wanted to find out if my choice of non-physcial parameters will allow for the accurate estimation of Sellmeier coefficients. These parameters include method, accuracy, precision etc. To do that, I used my mathematical model to produce artificial data. In this step, I used Sellmeier coefficients from \cite{Kato2002} to generate data and then tasked $NonLinearModelFit$ to recover these Sellmeier coefficients from the generated data. This step help me developed general understanding of non-physical parameters influence on computation time, precision and accuracy of estimation.

In the next step, I tested the sensitivity and robustness of our approach by adding Gaussian noise to each of the points. For each point, Gaussian distribution was created where Gaussian peak was centred at the real value of the visible photon central wavelength, and standard deviation was chosen to be some percent of real value. I tested our approach to up to $5\%$. These steps were repeated multiple times and provided information important at the stage of error estimation of the experimentally determined data.

\subsection{Estimation of Sellmeier coefficients}
Once we conducted our test, we were fairly confident in our approach and model. We then proceeded with real experimental data. Results of that part of the investigation are summarized in Tab. \ref{tab:para}. 
\begin{table}
	\centering
	\tabcolsep=0.11cm
	\begin{tabular}{|c|c|c|c|c|c|c|} \hline 
		  & $a_{z0}$   & $a_{z1}$ & $a_{z2}$ & $a_{z3}$ & $a_{z4}$ & $RSS$ \\
	      &           & $[\mu m^2]$ & $[\mu m^2]$ & $[\mu m^2]$ & $[\mu m^2]$ & $[nm^2]$ \\
	    \hline
		liter. & $4.59423$ & $0.06206$ & $0.04763$ & $110.807$ & $86.122$ & $1910$ \\
		comp.   & $4.59423$ & $0.06272$ & $0.04814$ & $-$ & $-$ & $257$ \\
		uncer.& $0.00015$ & $0.0004$ &  $4.5\times 10^{-6}$ & $-$ & $-$ & $-$\\  \hline
		\end{tabular}
	\caption{Comparison of the literature Ref.~\cite{Kato2002} and calculated Sellemeier coefficients as in \eref{eq:Sellmeier}. Table and description are taken from \cite{Misiaszek2018}.}
	\label{tab:para}
\end{table}
We run our code multiple times, changing control parameters of $FindRoot$ and $NonLinearModelFit$. Every execution ended with results exhibiting negligent variance - below $2.1 \times 10^{-3} \%$ for each coefficient. For our $55$ data points, we were able to reduce RSS from $1910$ nm$^2$ for literature data to $257$ nm$^2$ for our computed coefficients. With the RSS we estimated measurement error. For the literature data, the average error was $5.9$ nm and the average error for our computed data - $2.2$ nm (see the bottom part of Fig. \ref{fig:pump}). We used average error as an estimator of measurement error, which in turn allowed us to estimate uncertainties associated with each computed Sellmeier coefficient. Uncertainties values may be found in Tab. \ref{tab:para}.

\begin{figure}[h]
	\centering
	\includegraphics[width=\columnwidth] {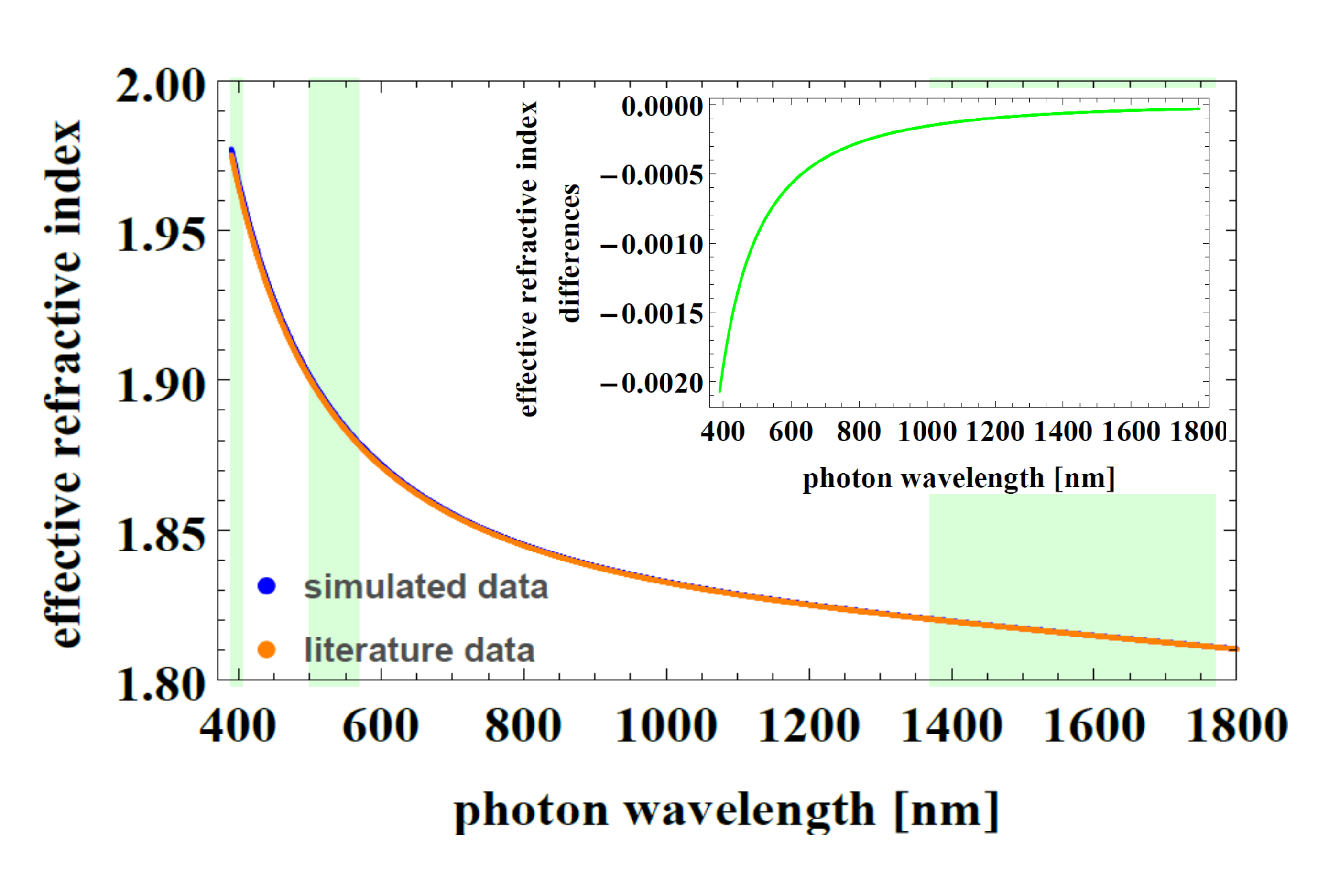}
	\caption{Effective refractive index. The shaded areas depict the wavelength range accessible by our method in our experiment. Inset: differences between literature and simulated data. Figure and description are taken from \cite{Misiaszek2018}.}
	\label{fig:neff}
\end{figure}

It is worth comparing values of effective refractive index computed based on literature values of Sellmeier coefficients with coefficients computed based on computed values of Sellmeier coefficients.  The Fig.\ref{fig:neff} shows how effective refractive index changes in the relevant spectrum. 
New coefficients change the refractive index very slightly, but the corresponding change in photon central wavelengths is significant (see Fig. \ref{fig:pump}). This is proof of the sensitivity of our proposed method of investigating the dispersive properties of a crystal. Although our method allowed for estimation of only three out of fifteen Sellmeier coefficients it is important to note the reason why only three were corrected is that just three coefficients dominate our SPDC process. For that process, our method was capable of significantly correcting the theoretical predictions of SPDC central wavelengths.

\subsection{Final notes}

We presented a very useful method of estimating Sellmeier coefficients based on experimental data. We showed that this simple method allows for accurate estimation of Sellmeier coefficients with an equipment typical to quantum-optical laboratory. We also showed that even small change in Sellmeier coefficients can lead to big diffrences between theoretically predicted wavelengths and the observed ones.

What we did not mention are disadvantages of this method. I expect that there are limitations as to when we can use this method. Here we used periodically polled crystal, which allowed for almost collinear phase matching which in turn allowed for domination of dispersion by one component of refractive index. If we were to consider some regular biaxial crystal, then we would deal with two polarizations, non-collinear phase matching and non-diagonal refractive index tensor. In short, one can expect that all three refractive indices would play important role in the estimation of dispersion. Estimation of all Sellmeier coefficients at once might prove to be challenging if one would decide to use experimental data gathered in only one SPDC setup.

On the other hand, periodically polled crystals which seem to be easier to handle in our method, introduce another source of uncertainty which is periodic polling. The crystal period might not exactly match the value specified by the manufacturer. Also, the crystal has to be temperature stabilised since temperature oscillation would be another source of the noise.

%% file: Chapters/Chapter5.tex
\chapter{Photon pairs states produced by parametric down conversion} 

\label{Chapter5} 

What distinguishes SPDC sources from other single-photon sources is the ability to relatively easily manipulate photon pair characteristics \cite{Gajewski2016}. The work presented in this chapter shows that obtaining an accurate numerical approximation of biphoton wavefunction of photons created via SPDC and coupled into the single-mode fibers is possible. The results of modelling and manipulation of biphoton wavefunction generated in uniaxial BBO crystals were first presented in written form in my master thesis and later in \cite{Gajewski2016}. Those results were based on the work of Piotr Kolenderski who created a mathematical model of SPDC. Necessary measurements were performed to show that this model agrees with experimental results \cite{Lutz2013,Lutz2014}. I improved this approach and used it for the characterisation of BBO based sources in \cite{Gajewski2016}. 

I will start this chapter with a brief introduction to the initial mathematical model. In the following subsections I will show how our numerical prediction compare with measurements performed for two distinct SPDC sources based on periodically polled biaxial crystal PPKTP. For the first comparison we will use the same data sets as were used in the previous chapter \ref{chap:Sellmeier}. We will compare numerical values of predicted angular frequency widths of a photon coupled into a single mode fiber to the widths determined in the experiment. In this setup one of the photons generated in SPDC is coupled to a fiber and the other propagates in free space.

For the second comparison, we will use experimental results presented in \cite{Sedziak2019}. In that work, both photons were coupled into long single-mode fibers. The arrival times of both photons were measured and correlation coefficients were determined. Here, we will compare numerical predictions of temporal widths of both photons and their correlation coefficients with the values obtained in the experiment.

The results of comparisons will demonstrate that our approach is capable of modelling most SPDC sources.

\section{Analytical model of SPDC source}

At the end of chapter \ref{Chapter1}, we introduced the general expression for an SPDC state \eref{eq:BphWaveFun} for the case of both photons propagating in free space.

In most applications, we are not interested in the free space wavefunction of the SPDC state but rather in the wavefunction of the state coupled into single-mode fibres. Fibre coupling provides an easy way of manipulation and transformation of photons. Moreover, some single-photon detectors require that photons are provided with a fibre.

A realistic pumping beam has some distribution of wavevectors and, if a crystal is pumped with pulses, it also has a distribution of frequencies. We will use the notation presented in the work \cite{Kolenderski2009} and shortly outline the calculation presented therein. 
The pumping beam amplitude for a given direction and frequency is assumed to have no spatio-temporal correlations:
\begin{equation}
    A_p (\boldsymbol k_\perp, \omega) = A_p^t (\omega) A_p^{sp} (\boldsymbol k_\perp). \label{eq:factorization}
\end{equation}
Here, the subscript '$\perp$'denotes components of wavevectors perpendicular to the direction of propagation of the pump beam. Superscript $t$ stands for 'temporal' and superscript $sp$ for spatial parts of amplitude function. The temporal part of the amplitude is defined by the duration of the pumping beam pulse. We assume that both factors are Gaussian:
\begin{align}
  A_p^t (\omega) &= \frac{\sqrt \tau_p}{\pi^{1/4}} \exp(-\frac{\tau_p^2}{2} (\omega - 2 \omega_0)^2 )   \\
  A_p^{sp} (\boldsymbol k_\perp) &= \frac{w_p}{ \sqrt{\pi}} \exp(-\frac{w_p^2}{2} \boldsymbol k_\perp^2), 
\end{align}
where $2 \omega_0$ is the central frequency, $\tau_p$ is pulse duration and $w_p$ is spatial width of the pumping beam. Following the work \cite{Kolenderski2009}, we represent the free space state of two photons as:
\begin{align}
    &|\psi \rangle = \int d^2 \boldsymbol k_{s \perp}d^2 \boldsymbol k_{i \perp} d\omega_s d\omega_i \psi(\boldsymbol k_{s \perp}, \omega_s; \boldsymbol k_{i \perp}, \omega_i) \hat a^\dagger(\boldsymbol k_{s \perp}, \omega_s) \hat a^\dagger(\boldsymbol k_{i \perp}, \omega_i) |0\rangle_{s,i}, \\
    &\psi(\boldsymbol k_{s \perp}, \omega_s; \boldsymbol k_{i \perp}, \omega_i) = N \int_{-L/2}^{L/2} dz \thickspace A_p (\boldsymbol k_{s \perp} + \boldsymbol k_{i \perp}, \omega_s + \omega_i) \exp[i \Delta k_z (\boldsymbol k_{s \perp}, \omega_s; \boldsymbol k_{i \perp}, \omega_i)], \label{eq:freeamplitude}
\end{align}
where it was assumed that the crystal is infinitely wide in the $x$ and $y$ directions and the length along the $z$-axis is $L$. We will call $|\psi \rangle $ a biphoton wavefunction in free space. The phase matching function is given by:
\begin{equation}
    \Delta k_z (\boldsymbol k_{s \perp}, \omega_s; \boldsymbol k_{i \perp}, \omega_i) = k_{pz} (\boldsymbol k_{s \perp} + \boldsymbol k_{i \perp},\omega_s + \omega_i) -  k_{sz} (\boldsymbol k_{s \perp} ,\omega_s) - k_{iz} (\boldsymbol k_{i \perp} ,\omega_i) \pm K,
\end{equation}
where we added the quasi wavevector $\boldsymbol{K}$ and assumed that the crystal is polled along the $z$-axis. For non periodically polled crystals $K = 0$. We will refer to these crystals as 'regular'.  Fiber modes are approximated with Gaussian functions:
\begin{equation}
    u_\mu (\boldsymbol k_{\mu \perp}, \omega_\mu)= \frac{w_\mu}{\sqrt{\pi}} \exp[-\frac{w_\mu^2}{2}(\boldsymbol k_{\mu \perp} - \boldsymbol k_{\mu 0 \perp})^2], \mspace{10mu} \mu = s, i 
\end{equation}
The wavefunction of the fiber coupled photons, also called biphoton wavefunction, is given by an overlap of fiber mode functions with free space biphoton wavefunction:
\begin{equation}
    \psi (\omega_s, \omega_i) =  \int d^2 \boldsymbol k_{s \perp}d^2 \boldsymbol k_{i \perp} \overline{u}_s (\boldsymbol k_{s \perp}, \omega_s) \overline{u}_i (\boldsymbol k_{i \perp}, \omega_i) \psi(\boldsymbol k_{s \perp}, \omega_s; \boldsymbol k_{i \perp}, \omega_i).
\end{equation}
Now, if we expand the left side with help of \eref{eq:freeamplitude} and use factorisation from \eref{eq:factorization}, we get:
\begin{align}
    \psi (\omega_s, \omega_i) &= A_p^t(\omega_s + \omega_i) \theta(\omega_s, \omega_i), \label{eq:biphotonwavefunction} \\
    \theta(\omega_s, \omega_i) &= N \int d^2 \boldsymbol k_{s \perp}d^2 \boldsymbol k_{i \perp} d\omega_s d\omega_i \int_{-L/2}^{L/2} dz \thickspace  \overline{u}_s (\boldsymbol k_{s \perp}, \omega_s) \overline{u}_i (\boldsymbol k_{i \perp}, \omega_i) \times \\ \nonumber
    &\times A_p^{sp}(\boldsymbol k_{s \perp} + \boldsymbol k_{i \perp}) \exp(i \Delta k_z (\boldsymbol k_{s \perp}, \omega_s; \boldsymbol k_{i \perp}, \omega_i)). \label{eq:thetaEff}
\end{align}
This result originally presented in \cite{Kolenderski2009} allows for factorisation of biphoton wavefunction into temporal and spatial parts. The temporal part of biphoton wavefunction is dependent only on laser pulse duration. The spatial part, $\theta(\omega_s, \omega_i)$ is determined by the experimental setup - it takes into account lenses, single-mode fibres, crystals and their orientation.

\section{Comparison with experimental data}
The above mathematical model was implemented as a library in the Mathematica software and tested for regular uniaxial crystals. The author of this thesis expanded this model to include periodically polled crystals and helped extending it to biaxial crystals. Below, we prove that with our model we are capable of modelling SPDC sources based on PPKTP crystals. The Sellmeier coefficients needed for mathematical description of this crystal are taken from the previous chapter. We will consider two distinct situations:
\begin{enumerate}
    \item One, where only one photon is coupled to the fiber. Here, we will verify that the computed spectral distribution of a coupled photon is accurately describing  the distribution measured in experiment. Experimental data used for that comparison was the basis of the article \cite{Misiaszek2018}. Descriptions of the experimental setup and the obtained results may be found in chapter \ref{Chapter3}.
    \item The second comparison will done with data coming from the SPDC setup where photons were coupled into $10$ km long single mode fibers. That experimental data along with research on remote temporal wavepacket narrowing was presented in article \cite{Sedziak2019}. I will use that data to verify the accuracy of computed values of temporal width of photons together with their temporal correlation coefficient.
\end{enumerate}

\subsection{SPDC source with one photon coupled to fiber}

Computation of a biphoton wavefunction of two photons where one is coupled into a fiber and second one propagates in free space requires information about the experimental setup. The required parameters are: temperature, crystal length and period, spatial width of the coupling mode, spatial waist of the SHG pumping beam, central angular frequency and angular frequency width of the pumping beam. Spatial width of the coupled mode can be calculated from the specification of optical elements present in the experimental setup. With all that information it is possible to calculate the coupled photon's central angular frequency and angular frequency width.
In this paragraph, we list parameters that are known directly from experiment and were kept constant .
The temperature was set to $31.86$ C. The crystal used was a $L = 10$ mm long PPKTP crystal with $\Lambda_0 = 4.01$ $\mu$m period. Spatial diameter of the coupled photon $2w_{VIS}$ can be estimated to be between $22.25 - 32.58$ $\mu$m. This range results from the range of diameter of the fiber mode which is $2.8 - 4.1$ $\mu$m for Thorlabs SMF460B fiber. Lenses between the crystal and the fiber have focuses of $1.51$ cm and $12$ cm. The spatial diameter of pumping beam  $2w_p$ was estimated to be between $93-99$ $\mu$m.  That data along with central frequencies was available since it was measured for purposes of \cite{Misiaszek2018}. 

\begin{figure}[h!]
     \centering
         \includegraphics[width=0.75\columnwidth]{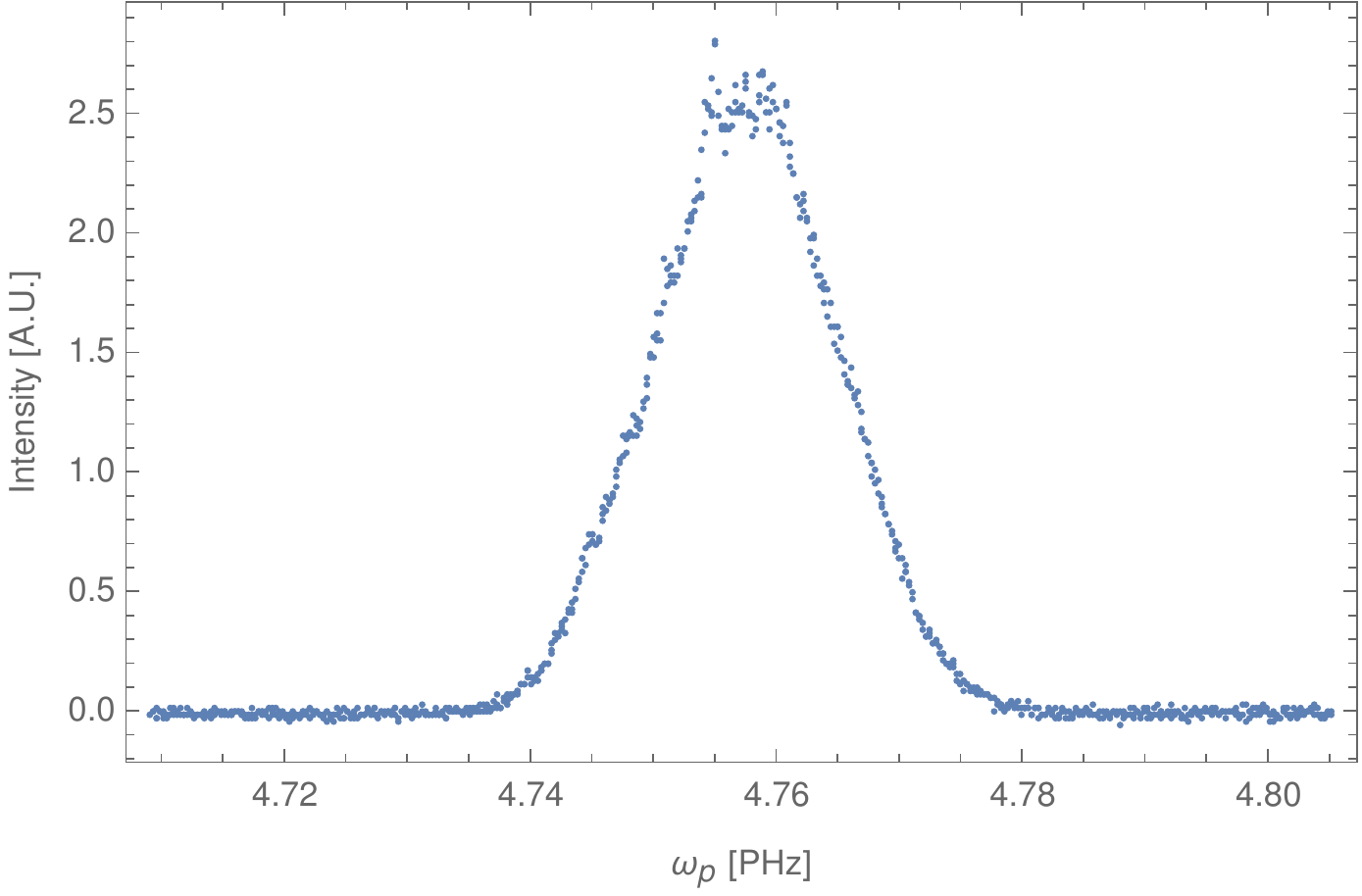}
\caption{Pumping beam spectrum after it passes through a BiBO crystal.}
\label{fig:shg}
\end{figure}

The measurements of coupled photon angular frequency distribution were performed for varied central pump photon frequency. That value of frequency was measured, but the width of angular frequency distribution was not. We are going to assume that frequency width was the same for every measurement. For purposes of this chapter my colleague Kaushik Joarder performed measurements allowing for estimation of the pumping beam angular frequency width. The measurement results are presented in Fig.\ref{fig:shg}.
To extract the necessary parameters from the measurement data we assumed that the frequency data is well approximated with a Gaussian distribution. The same assumption  was used in \cite{Misiaszek2018}. Therefore, to the above and all other measurement data we fitted a general Gaussian function of intensity:
\begin{equation}
    I(\omega) = b + a \exp (- 4 \ln (2) \frac{(\omega -\omega_0)^2}{\mathrm{FWHM}^2}), \label{eq:model}
\end{equation}
where $I(\omega)$ is an intensity of the photon beam at a given frequency,  $b$ stands for the bias, $a$ is amplitude, $\omega_0$ is central angular frequency and $\mathrm{FWHM}$ is full width of distribution measured at half maximum. The relation between $\mathrm{FWHM}$ and width given by standard deviation $\sigma$ is $\mathrm{FWHM} = 2\sqrt{2 \ln 2} \sigma$. 
These are our fit parameters. To obtain values of those parameters from the measurement data we used Mathematica's built-in function $NonLinearModelFit$.

With this approach, we were able to estimate the angular frequency width of the pumping beam $\mathrm{FWHM}$ to be $0.01763$ PHz. This width corresponds to the temporal width of $\tau_p = 314.5$ fs.  All above parameters are summarized in Tab. \ref{tab:spdc_params}.
\begin{table}[h!]
  \begin{center}
    \begin{tabular}{|c|c|}
    \hline
    $T$ &  $31.86$ C \\
    $\Lambda_0$ & $4.01$ $\mu$m \\
    $L$ & $10$ mm \\
    $2w_p$ & $93-99$ $\mu$m \\
    $2w_{VIS}$ & $22.25-32.58$ $\mu$m \\
    $\tau_p$ & $314.5$ fs \\
    \hline
    \end{tabular}
  \end{center}
  \caption{Experimental parameters.}
  \label{tab:spdc_params}
\end{table}

\subsubsection{Measurement data}
\begin{figure}[h!]
     \centering
         \includegraphics[width=0.75\columnwidth]{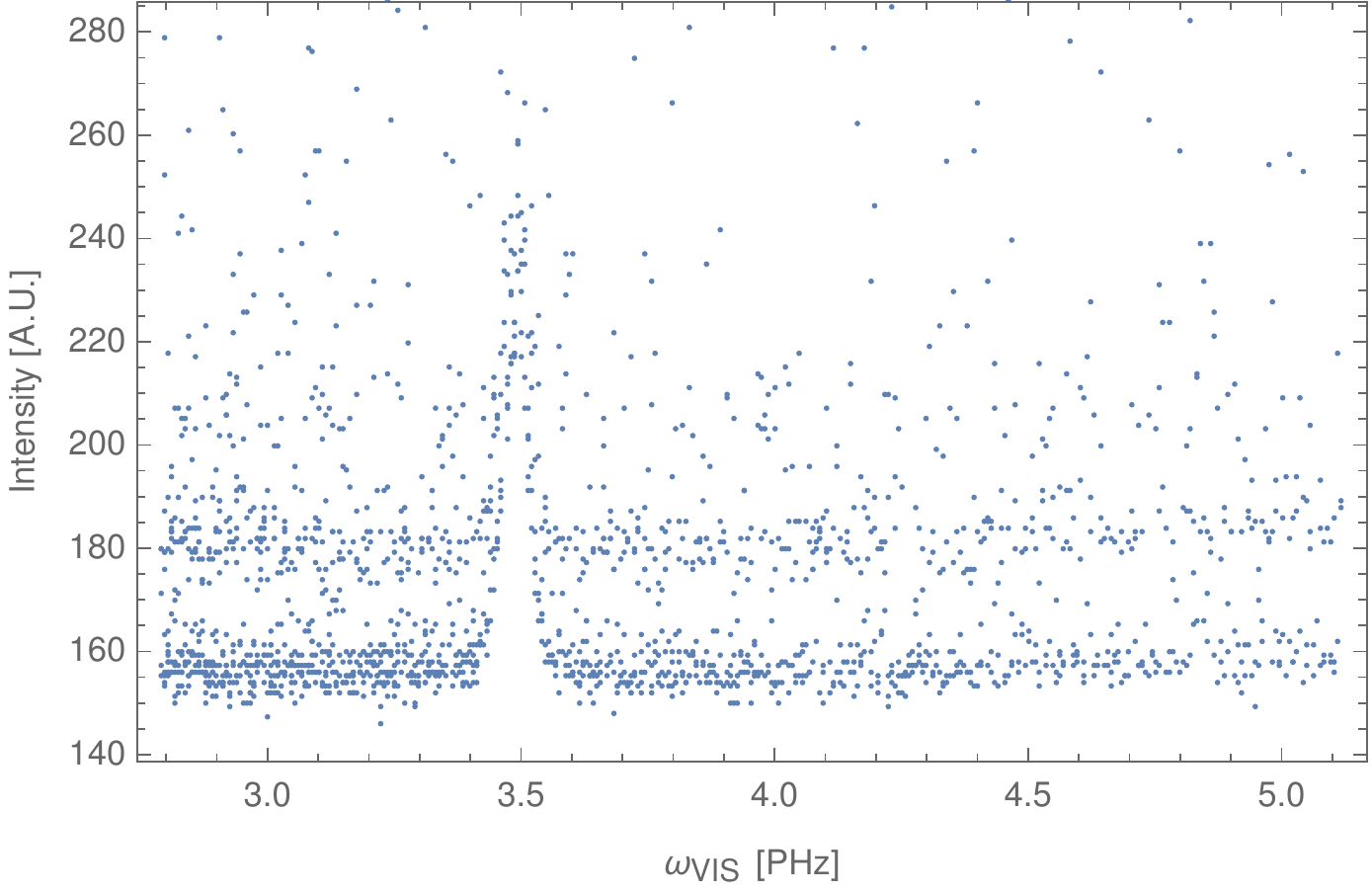}
\caption{One of 55 measured spectral intensity distributions of visible photons.}
\label{fig:measexample}
\end{figure}

To verify the performance of the above model, we extract the central frequency $\omega_p$ of the pumping beam photon and central frequency $\omega_\mathrm{VIS}$ together with frequency width $\mathrm{FWHM}_\mathrm{VIS}$ of the coupled visible photon. An example of measurement data of coupled photon intensity spectrum is depicted in Fig. \ref{fig:measexample}. The values of $\omega_p$ together with parameters from Tab. \ref{tab:spdc_params} are our input data, which we will use to generate predictions for $\omega_{VIS}$ and $\mathrm{FWHM}_{VIS}$of coupled visible photon.

To that data we fitted model \eref{eq:model}. There were $55$ measurements, each consisting of measuring pumping beam central frequency and frequency spectrum of the visible photon. Only $28$ out of $55$ measurements resulted in data allowing an acceptable fit. By acceptable we mean that $p$-value \cite{Bhattacharya2002} corresponding to each fit parameters was below $0.01$. It should be clear that fitting Gaussian distribution from \eref{eq:model} to the data presented in Fig. \ref{fig:measexample} poses a challenge. We expected that some data sets will not allow for an accurate fit. 
All $28$ fit parameters are presented in Tab. \ref{tab:spectralfit}.
\begin{table}[h!]
  \begin{center}
    \begin{tabular}{|c||cc|cc|}
    \hline
    $\omega_p$ [PHz] & $\omega_{VIS}$ [PHz] & std. err. [PHz] & $\mathrm{FWHM}_{VIS}$  [PHz] & std. err. [PHz] \\
    \hline
      4.7650 &  3.5894 & 0.0015 & 0.0518 & 0.0029  \\
      4.7375 &  3.5344 & 0.0034 & 0.0589 & 0.0065 \\
      4.7415 &  3.5400 & 0.0032 & 0.0567 & 0.0061 \\
      4.7472 &  3.5454 & 0.0030 & 0.0523 & 0.0057  \\
      4.7542 &  3.5544 & 0.0032 & 0.0472 & 0.0062  \\
      4.7597 &  3.5778 & 0.0024 & 0.0497 & 0.0046  \\
      4.7666 &  3.5852 & 0.0028 & 0.0504 & 0.0054  \\
      4.7622 &  3.5866 & 0.0019 & 0.0520 & 0.0037  \\
      4.7720 &  3.5902 & 0.0033 & 0.0494 & 0.0064  \\
      4.7774 &  3.6158 & 0.0034 & 0.0510 & 0.0065  \\
      4.7836 &  3.6211 & 0.0049 & 0.0575 & 0.0095  \\
      4.7579 &  3.5811 & 0.0020 & 0.0516 & 0.0039  \\
      4.7770 &  3.6207 & 0.0042 & 0.0575 & 0.0080  \\
      4.7717 &  3.6141 & 0.0040 & 0.0524 & 0.0077  \\
      4.7678 &  3.5931 & 0.0026 & 0.0496 & 0.0050  \\
      4.7633 &  3.5864 & 0.0026 & 0.0507 & 0.0050  \\
      4.7561 &  3.5797 & 0.0027 & 0.0503 & 0.0052  \\
      4.7495 &  3.5714 & 0.0031 & 0.0481 & 0.0061  \\
      4.7444 &  3.5433 & 0.0031 & 0.0561 & 0.0060  \\
      4.7384 &  3.5350 & 0.0034 & 0.0584 & 0.0065  \\
      4.7517 &  3.5747 & 0.0023 & 0.0493 & 0.0043  \\
      4.7317 &  3.5283 & 0.0042 & 0.0614 & 0.0082  \\
      4.7254 &  3.5202 & 0.0069 & 0.0664 & 0.0132  \\
      4.7458 &  3.5529 & 0.0023 & 0.0511 & 0.0045  \\
      4.7412 &  3.5474 & 0.0027 & 0.0542 & 0.0053 \\
      4.7365 &  3.5423 & 0.0034 & 0.0576 & 0.0064 \\
      4.7315 &  3.5367 & 0.0050 & 0.0609 & 0.0095 \\
      4.7226 &  3.4936 & 0.0049 & 0.0583 & 0.0095 \\
      \hline
    \end{tabular}
  \end{center}
  \caption{Table presents $28$ entries corresponding to best fits of the model \eref{eq:model} to experimental data.
  }
  \label{tab:spectralfit}
\end{table}
We will now use the values of $\omega_p$ to compute values of $\omega_{\mathrm{VIS}}$ and $\mathrm{FWHM}_{VIS}$.

\subsubsection{Numerical simulation of biphoton wavefunction}
\begin{figure}[h!]
  \centering
  \includegraphics[width=0.65\columnwidth]{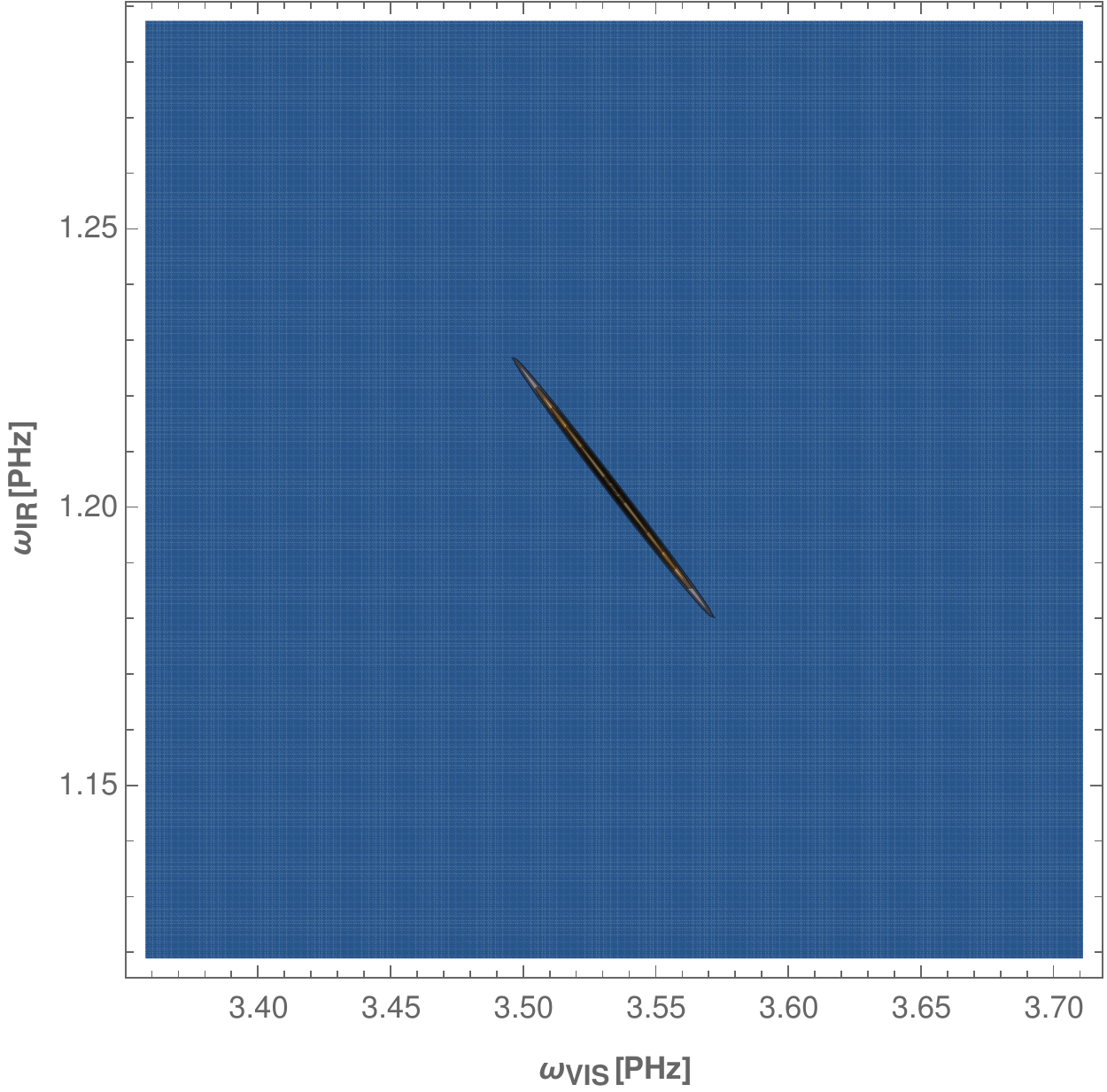}
\caption{Density plot of square modulus of joint probability distribution.}
\label{fig:biphotonRSdensity}
\end{figure}

\begin{figure}[h!]
  \centering
  \includegraphics[width=0.95\columnwidth]{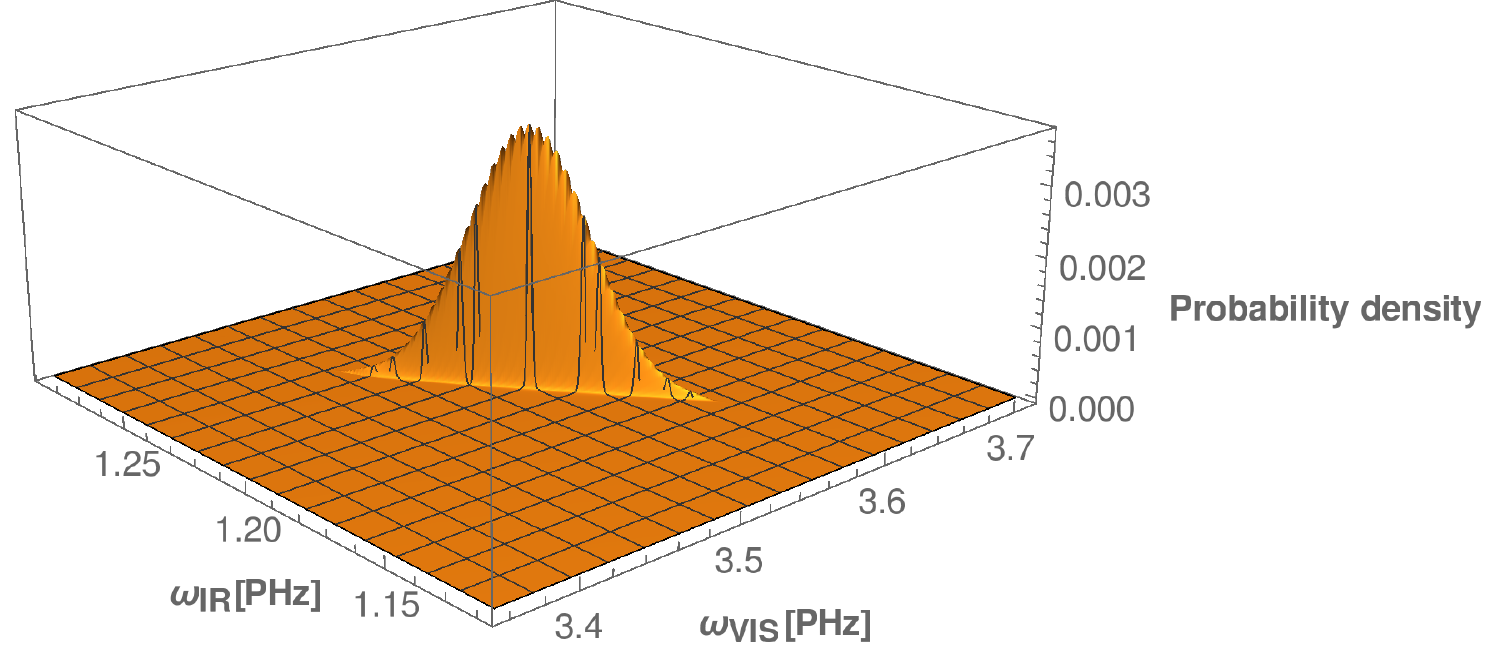}
\caption{3D plot of joint probability distribution.}
\label{fig:biphotonRS3D}
\end{figure}

\begin{figure}[h!]
  \centering
  \includegraphics[width=0.8\columnwidth]{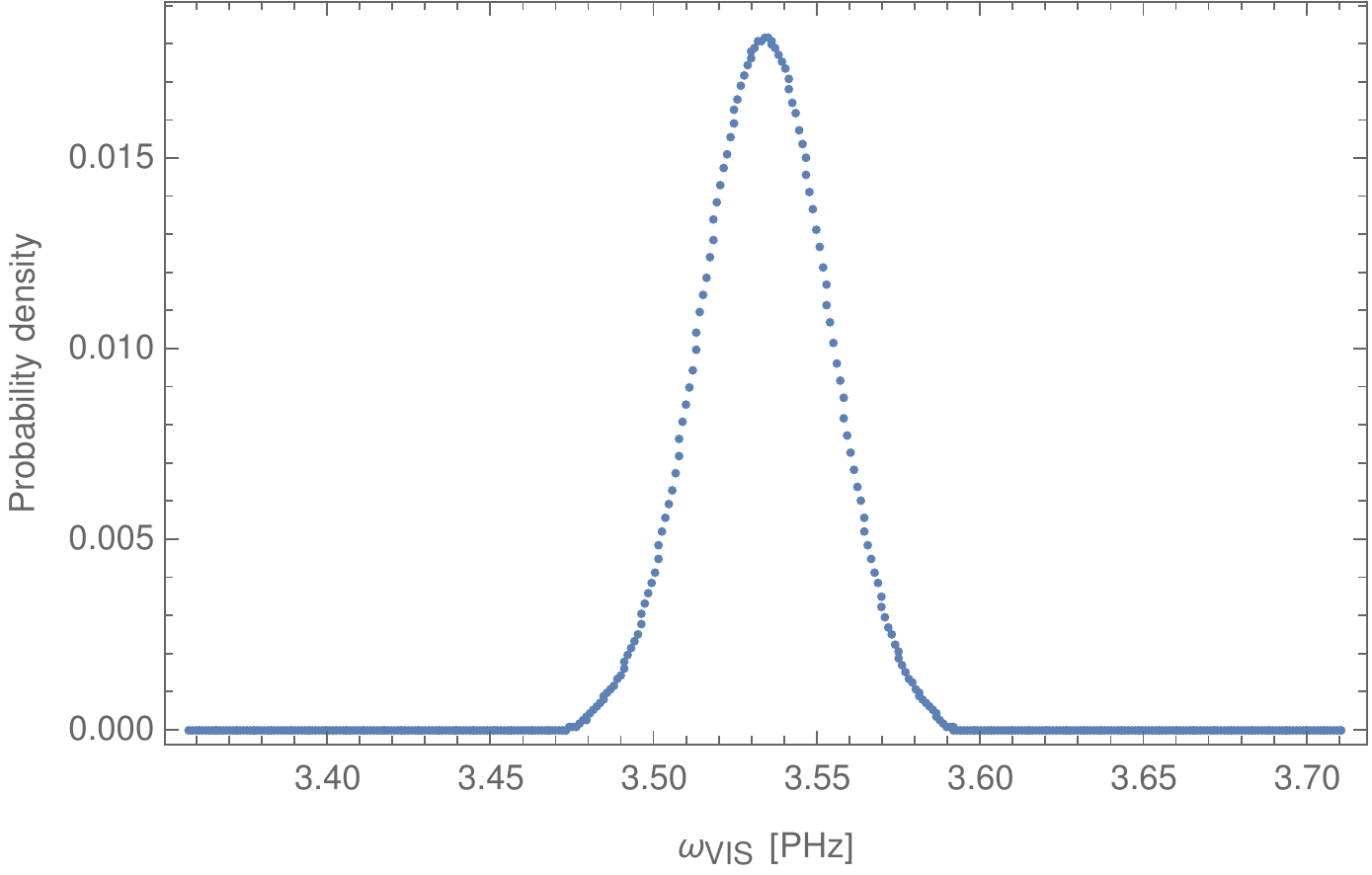}
\caption{Visible photon frequency spectrum. The central frequency is $3.534$ PHz with standard error $\approx 10^{-5}$. The $\mathrm{FWHM}_\mathrm{VIS}$ is estimated to be $0.0461$ PHz with the standard error of $\approx 3\times 10^{-5}$ PHz.}
\label{fig:singleRS}
\end{figure}

After all necessary experimental parameters have been established, we move on to computation of biphoton wavefunction. The SPDC process we are considering is collinear with all the photons having the same "slow" polarization. Each computation result is a tuple consisting of three entries: angular frequency of visible photon, angular frequency of infrared photon and corresponding joint probability desitribution - square modulus of biphoton wavefunction \eref{eq:biphotonwavefunction} for these two values:
\begin{align}
    p(\omega_{VIS}, \omega_{IR}) = |\psi (\omega_{VIS}, \omega_{IR} |^2.
\end{align}
List of such tuples can be plotted as in Fig. \ref{fig:biphotonRSdensity} and Fig. \ref{fig:biphotonRS3D}.
Since we are interested in marginal distributions for visible photon frequencies, we can reduce this list by summing over values of infrared photons frequencies to get a marginal distribution:
\begin{align}
    p(\omega_{VIS}^{(j)}) = \sum_k p(\omega_{VIS}^{(j)}, \omega_{IR}^{(k)}).
\end{align}
Such marginal distribution is depicted in Fig. \ref{fig:singleRS}. In the final step, we fit to that distribution a Gaussian model given by \eref{eq:model} to extract the central frequency $\omega_{\mathrm{VIS}}$ and frequency width $\mathrm{FWHM}_\mathrm{VIS}$ of visible photons along with the standard error for each parameter.

Before we move on, we address two issues. The first one is the range of spatial diameters of the pumping beam and coupling mode presented in Tab. \ref{tab:spdc_params} and the other is the number of frequency pairs $\{\omega_{VIS}, \omega_{IR} \}$ for which we compute joint probability distribution values. 

Spatial diameters of the pumping beam $2w_p$ and the coupling mode $2w_c$ influence the distribution of wavevectors directions. This means they do not influence central frequencies of visible photons, but they may influence its frequency width. That influence is negligibly small  - if we perform computations only for extreme values (minimal and maximal) of $w_p$ and $w_{VIS}$, our parameters of interest will remain almost unchanged. In Tab. \ref{tab:widths}, we present values of computed frequency widths of the visible photon for each situation. Note that values are given in THz instead of PHz.
\begin{table}[h!]
  \begin{center}
    \begin{tabular}{|cc|cc|cc|}
    \hline
    $2w_p$ [$\mu$m] & $2w_{VIS}$ [$\mu$m] &  $\mathrm{FWHM}_{VIS}$ [THz] & std. err. [THz]\\
    \hline
      93 &  22.25 & 46.351 & 0.0308\\
      99 &  22.25 & 46.355 & 0.0308 \\
      93 &  32.58 & 46.082 & 0.0317 \\
      99 &  32.58 & 46.089 & 0.0317  \\
      \hline
    \end{tabular}
  \end{center}
  \caption{Table presents influence of pump beam and coupling mode diameters on frequency width of visible photon.}
  \label{tab:widths}
\end{table}
From this result it is clear that if we pick just one combination of diameter parameters we will introduce negligibly small error. For all following simulations we will assume the average value diameters: $2w_p = 96$ $\mu$m and $2w_{VIS} = 27.4$ $\mu$m.
\begin{figure}[h!]
  \centering
  \includegraphics[width=0.75\columnwidth]{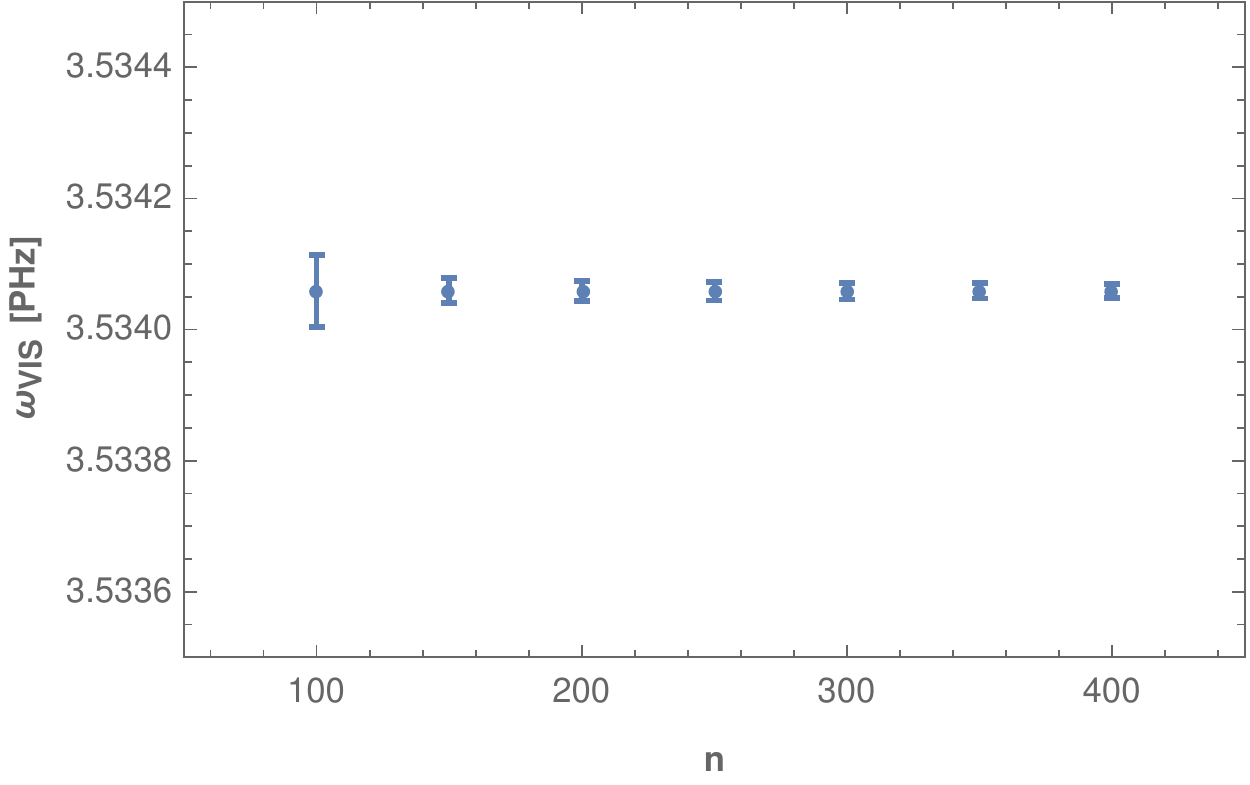}
\caption{Convergence plot for angular frequency $\omega_\mathrm{VIS}$ of the visible photon. On the horizontal axis we have number of visible photon frequencies values $n$. On the vertical axis we have values of central angular frequency $\omega_{VIS}$.}
\label{fig:omega_conv}
\end{figure}

Another parameter of interest is the number of frequencies  $n$ for which we perform computations. 
All presented figures of joint probability distribution (Fig. \ref{fig:biphotonRSdensity} and Fig. \ref{fig:biphotonRS3D}) and marginal distribution \ref{fig:singleRS} were created for $400$ values of $\omega_{IR}$ and $400$ values of $\omega_{VIS}$, which gives $1.6 \times 10^5$ computational points in total. By point we understand a tuple $\{\omega_{VIS}, \omega_{IR}, |\psi (\omega_{VIS}, \omega_{\mathrm{IR}})|^2 \}$. After integration, the marginal distribution of $\omega_{VIS}$ includes $400$ tuples of form $\{\omega_{VIS}, p_{VIS} (\omega_{VIS}) \}$ where $p_{VIS} (\cdot)$ denotes the marginal  distribution of the visible photon frequencies. 

\begin{figure}[h!]
  \centering
  \includegraphics[width=0.75\columnwidth]{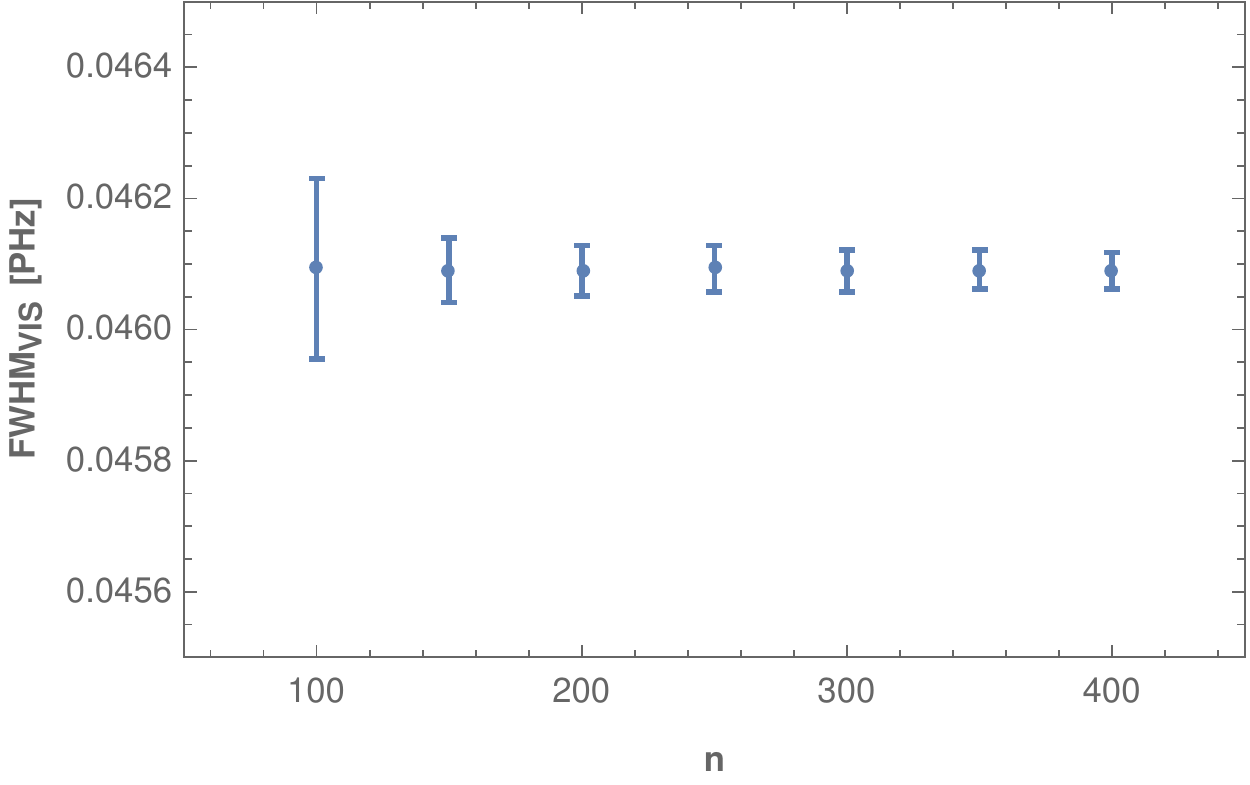}
\caption{Convergence plot for angular frequency width of the visible photon.n the horizontal axis we have number of visible photon frequencies values $n$. On the vertical axis we have a value of angular frequency width $\mathrm{FWHM}_{VIS}$.}
\label{fig:fwhm_conv}
\end{figure}

It is reasonable to ask if that number is sufficient. That question in answered in Fig. \ref{fig:omega_conv} and Fig. \ref{fig:fwhm_conv}. It is clear that going beyond $100$ points does not lead to any significant improvement in estimation of $\omega_\mathrm{VIS}$ or frequency width $\mathrm{FWHM}_\mathrm{VIS}$ in the visible photon marginal distribution. Instead, it results in smaller error of estimation. We will set number of points $n$ to $300$ for all following computations. This number guarantees good estimation and lets us avoid long computations.

\subsection{Results}
At this point we have discussed all the experimental constants (Tab. \ref{tab:spdc_params}) and the numerical parameter $n$. We can now compare numerical predictions of $\omega_{VIS}$ and $\mathrm{FWHM}_\mathrm{VIS}$ with the ones obtained from experimental data. To do that, we are going to use frequencies $\omega_p$ from table Tab. \ref{tab:spectralfit}, as an input data and we will compare the experimentally established central frequencies $\omega_{VIS}$, and frequency widths $\mathrm{FWHM}_{VIS}$ (see Tab. \ref{tab:spectralfit}) with the computed ones.

We start with analysis of predicted values of $\omega_{VIS}$. It is helpful to realized that we have already done something similar in the previous chapter \ref{chap:Sellmeier}. We used a different model to predict visible photon wavelength for a given pumping beam wavelength. If we revisit Fig. \ref{fig:pump} we will note that some experimental values of the visible photon wavelengths do not align with the computed curve depicting predicted visible photon wavelengths. 
That is a problem. If the computed value of the wavevector mismatch for some $\omega_p$ and $\omega_{VIS}$ is too large then, theoretically, measured visible photon intensity at frequency $\omega_{VIS}$ should be close to 0. If we examine \eref{eq:BphWaveFun} we will note that the longer the crystal the smaller wavevector mismatch has to be for SPDC process to be efficient. For $28$ data points presented in Tab. \ref{tab:spdc_params} the computed values of wavevector mismatch for $7$ data points lead to wavevector mismatch $|\Delta k| \geq 10^{-4}$ $\mu m^{-1}$ and the rest $21$ points have a value of mismatch $|\Delta k| \leq 10^{-13}$ $\mu m^{-1}$. If we now multiply these values by the length of the crystal $L = 10^4$ $\mu m$ we can see that for the $7$ points with larger mismatch $L|\Delta k|$ is of the order of $1$ or greater leading to significant drop in SPDC efficiency. 

We do not know why there are no data points leading to wavevector mismatch of values $10^{-4} > |\Delta k| > 10^{-13}$. We can only speculate that data points with large wavevector mismatch are due to an error in measurement. We therefore reject these points and all following computations and verification will be performed for the remaining $21$ data points.
\begin{figure}[h!]
  \centering
  \includegraphics[width=0.75\columnwidth]{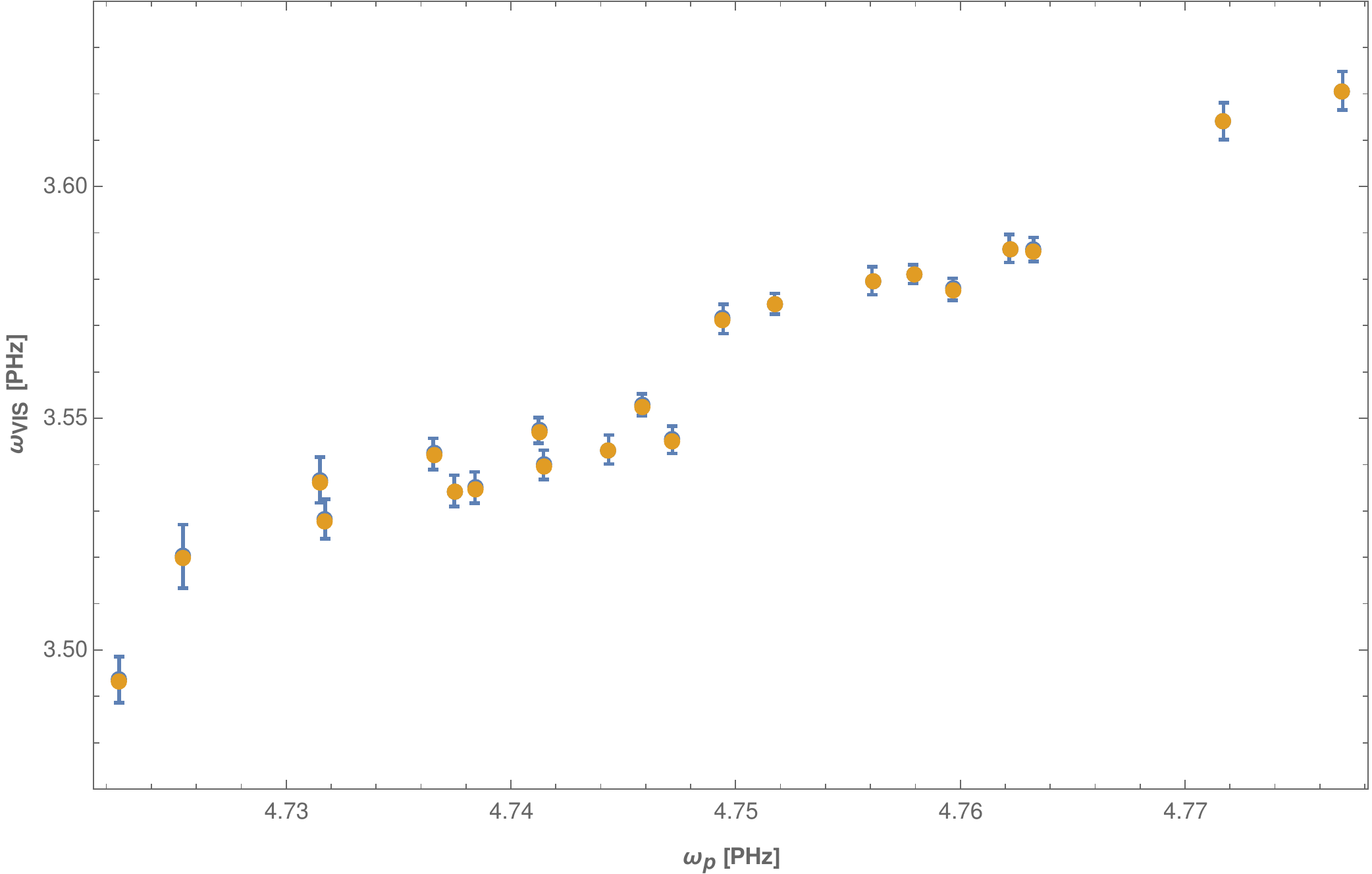}
\caption{Visible photon central frequency dependence on the pump central angular frequency. Blue dots represent experimental results, while orange dots represent simulation results.}
\label{fig:omegas}
\end{figure}
For these 21 points, we compute joint probability distribution and extract central frequencies and frequency widths along with associated errors. The experimental central frequencies match almost exactly the simulated ones. Both frequencies are depicted in Fig. \ref{fig:omegas}. This result means that our model behaves as expected, and for experimental points allowing for a near perfect phasematch it computes $\omega_{VIS}$ with great accuracy. 
\begin{figure}[h!]
  \centering
  \includegraphics[width=0.9\columnwidth]{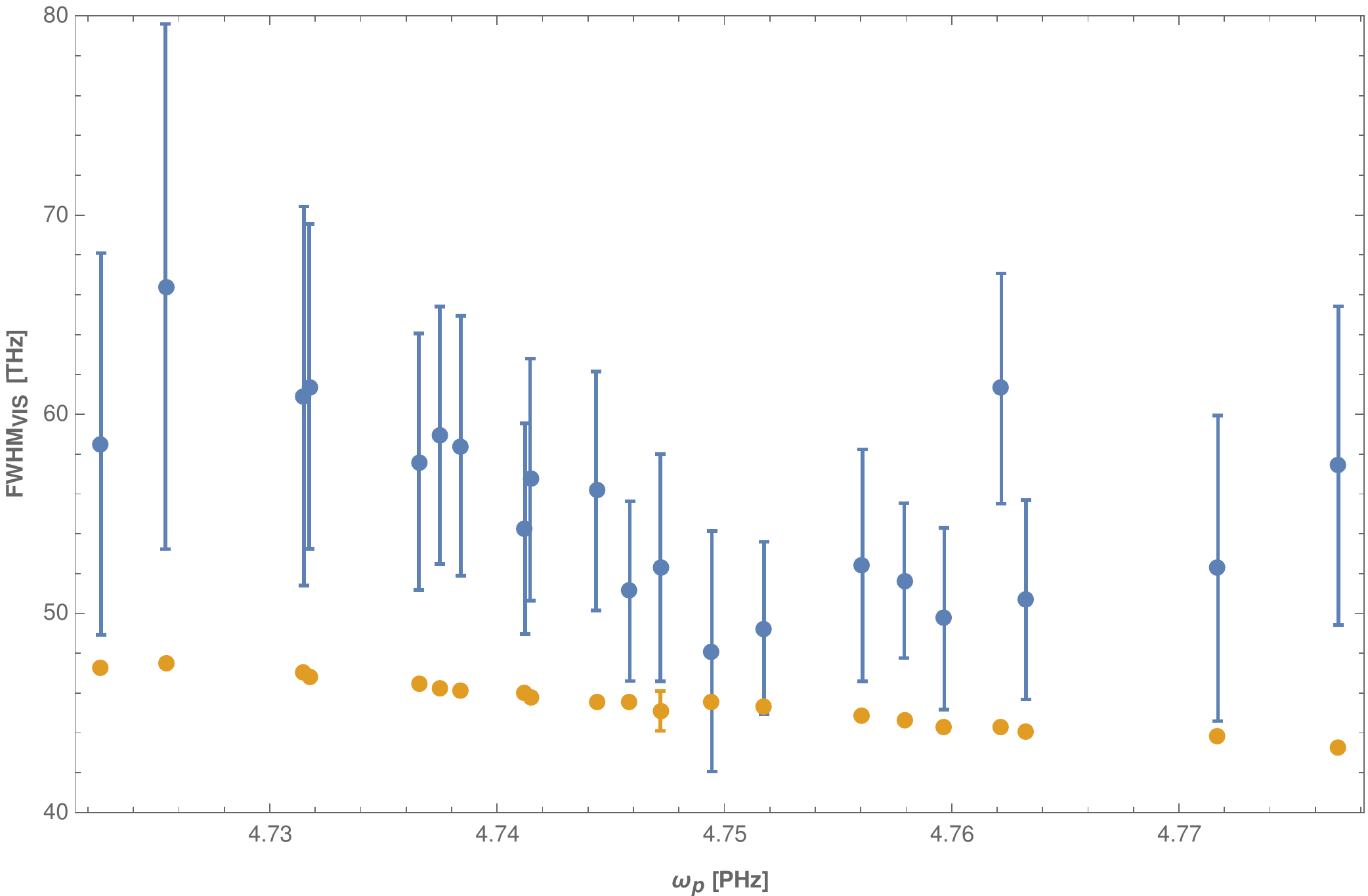}
\caption{Visible photon frequency width  dependence on the pump central angular frequency. Blue dots represent experimental results, while orange ones represent simulation results.}
\label{fig:fwhms}
\end{figure}
The problem with the above comparison is that we are not able to compare our numerical results with the whole set of $55$ data points. We were forced to reject some points based on wavevector mismatch and challenging nature of data sets obtained in experiment.

It could be argued that we are only comparing numerically obtained values of central frequencies with experimental values that lead to such a good agreement. All the values that would lead to inaccurate predictions we had rejected beforehand. 
To show that it is not the case, we first note that neither the Gaussian model we used for fitting nor the measurement technique which was used to obtain the experimental data were chosen for purposes of this comparison. So, there could be no doubt that we did not chose experimental data sets on the grounds of what our model can predict well. Moreover, the presented fitting technique along with the software used were the same as the one used in research presented in article \cite{Misiaszek2018}. Finally, the calculation of the wavevector mismatch which was a basis for the rejection of $7$ points, was performed with the classical model presented in chapter \ref{chap:Sellmeier}. In other words, at no point have we used our model of SPDC to decide if a given data point should be included in verification.
Although the values of central frequencies $\omega_{VIS}$ could be obtained by classical calculations, we believe that the agreement of computed central frequency values with the experimental values shows that our model behaves well. 

The $\mathrm{FWHM}$  parameter provides a better way of verifying our model performance, since we did not reject any of the $28$ data points based on angular frequency width. Comparison of numerical and experimental values of $\mathrm{FWHM}$ is depicted in Fig. \ref{fig:fwhms}. It should be visible that computed data points are correct to an order of magnitude. 

Still, there is a clear bias between the experimental and computed frequency widths. Our computed widths are persistently smaller then the experimental ones. Discrepancies between the computation and simulations are apparent especially at the left and right part of the figure. Our guess is that the measurement were not good enough to capture well not only central frequencies of the visible photon but also its frequency width. If we go back to Fig. \ref{fig:measexample}, we will see that position of the central frequency is reasonably well captured by the experiment, but the value of intensity at the peak has a visible uncertainty. Moreover, the peak itself looks like it is composed out of two peaks. The reason why we are searching for sources of discrepancies in measurement is the final figure Fig. \ref{fig:fwhms}. The computed frequencies almost form a line which is an expected behaviour given that pumping frequency changes only slightly - the maximal pumping frequency is approximately only $1\%$ larger than the minimal pumping frequency. Given that we are far away from any resonances, we expect frequency width to be a smooth and monotonic function of the pumping frequency- which is not true in the case of the measured width.

\subsection{SPDC source with both photons coupled to fiber}
\begin{figure}[h!]
  \centering
  \includegraphics[width=0.75\columnwidth]{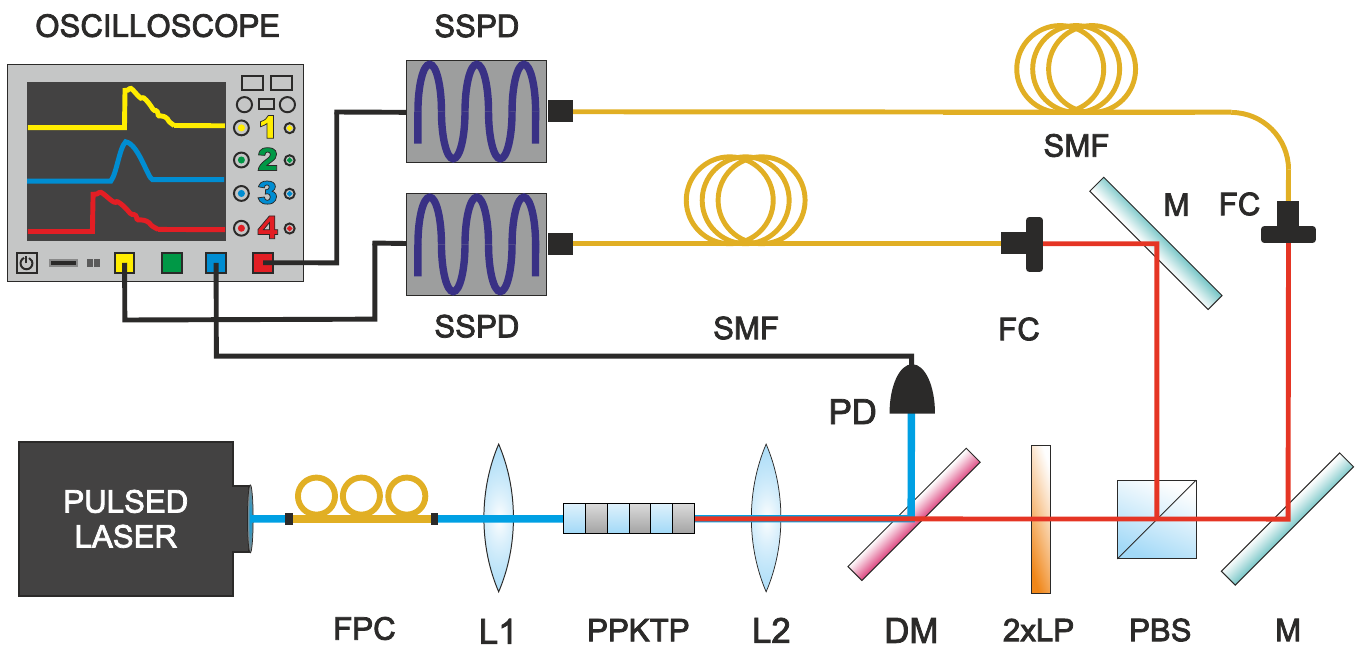}
\caption{The experimental setup. Pulses from the femtosecond Ti:Sapphire laser are coupled to a fiber. Their polarization is set by a fiber polarization controller FPC. The pump is focused using the lens L1 (plano-convex F = $125$ mm) in the type II PPKTP nonlinear crystal. The resulting photon pairs are collimated using the lens L2 (plano-convex, F = $70$ mm). A dichroic mirror DM (Semrock FF875-Di01) separates the unconverted laser beam and the SPDC photons. Two longpass filters LP (Semrock BLP01-1319R-25) remove the remainder of the
pump. Next, a polarizing beam splitter PBS separates the SPDC photons based on their polarization. They are subsequently coupled to single mode fibers SMF (Corning SMF28e+) using fiber collimators FC (F = $8.0$ mm) and mirrors M. The oscilloscope monitors the outputs of superconducting nanowire single-photon detectors SSPD and the fast photodiode PD. Figure and caption were taken from \cite{Sedziak2019}.}
\label{fig:kssetup}
\end{figure}

The second verification will be performed for the joint probability distribution of arrival times of SPDC photons with both photons coupled inside a fiber. This comparison will be based on the work first presented in article \cite{Sedziak2019}. The experimental setup is depicted in Fig. \ref{fig:kssetup}. This time we will be able to compare both frequency widths of photons and we also will be able to compare the Pearson correlation coefficient between the signal and idler photon frequencies. Clearly, this will allow for more robust testing of our numerical approach. On the other hand, in the experiment they were coupled inside two long single-mode fibers and measurements were performed after both photons have propagated through the fibers. Propagation in a dispersive medium, such as fiber, involves an additional transformation of the wavefunction which has to be accounted for. This is especially important in this comparison since we will compare our numerical results to experimental ones indirectly - in the experiment, the arrival times were measured and our model characterises biphoton wavefunction in the frequency domain. The analytical derivation of this transformation can be found in the appendix \ref{AppendixTransform}.

\subsubsection{Experimental setup}
We start with a description of the experimental setup. A \mbox{$L = 10$ mm} long PPKTP crystal with period \mbox{$\Lambda_0 = 46.2$ $\mu$m} is illuminated by laser light with central wavelength set to  \mbox{$\lambda_p = 780.1(1)$ nm}. As a result of collinear SPDC process, two orthogonally polarized photons are generated. The PPKTP crystal was designed to support type-II colinear SPDC where phase matching is achieved for \mbox{$780$ nm} $\rightarrow$ \mbox{$2 \times 1560$ nm.}
After down-conversion, each photon is coupled into a single mode fiber and propagates through distance \mbox{$D = 10$ km}. The group velocity dispersion of each fiber is $2\beta = -2.27 \times 10^{-26} \frac{s^2}{m}$ at \mbox{$1560$ nm} wavelength.
The spatial shape of the laser pumping beam was elliptical: axes lengths were $80$ $\mu$m and $84$ $\mu$m long. For our simulations we are going to assume that the diameter is $2w_p = 82$ $\mu$m. The spatial diameter of the coupled photons $2w_c$ can be estimated to be in the range of $92 - 102$ $\mu$m. This estimation is made based on mode field diameter in the fiber and lenses used to couple SPDC light to that fiber.
The crystal was not temperature stabilized so its temperature $T$ was a laboratory temperature which is in the range $20 - 25$ C. All experimental parameters are summarized in Tab. \ref{tab:spdc_params2}.

\begin{table}[h!]
  \begin{center}
    \begin{tabular}{|c|c|}
    \hline
    $T$ &  $20-25$ C \\
    $\Lambda_0$ & $46.2$ $\mu$m \\
    $L$ & $10$ mm \\
    $\lambda_p$ & $780.1$ nm\\
    $2w_p$ & $82$ $\mu$m \\
    $2w_{c}$ & $92-102$ $\mu$m \\
    $D$ & $10$ km \\
    $2\beta$ & $-2.27 \times 10^{-26} \frac{s^2}{m}$ \\
    \hline
    \end{tabular}
  \end{center}
  \caption{Experimental parameters.}
  \label{tab:spdc_params2}
\end{table}

\subsubsection{Measurement data}
Measurements were performed for three different durations of pump pulses. Standard deviations of duration of those time pulses are $94.58(16)$ fs, $0.7191(14)$ ps and $0.976(2)$ ps. For each duration of pump pulse the duration of the signal and idler pulses was measured along with the Pearson correlation coefficient describing temporal correlation between time of arrival of the signal and idler photons. The Pearson correlation coefficient was established with the assumption that the joint probability distribution of expected time arrival of both photons at times $t_s$ and $t_i$ has a Gaussian shape:
\begin{align}
p(t_s,t_i) &= \frac{1}{2 \pi \tau_s \tau_i \sqrt{1 - \rho_t^2}}\exp (- \frac{1}{2(1 - \rho_t^2)}(\frac{t_s^2}{\tau_i^2} + \frac{t_i^2}{\tau_s^2} - \frac{2 t_s t_i \rho_t}{\tau_s \tau_i})). \label{eq:2dmodel}
\end{align}
The parameters $\tau_s$ and $\tau_i$ stand for standard deviations of arrival times of the signal and idler photons, respectively. The parameter $\rho_t$ is the Pearson correlation coefficient in the time domain.
Measurements outcomes are summarized in Tab. \ref{tab:fitvalues}.
\begin{table}[h]
	\centering
	\begin{tabular}{|c|c|c|c|}
	\hline
	Data set & \#$1$ & \#$2$ & \#$3$ \\ \hline		
		$\Delta\lambda$ & $3.415(6)$ nm & $0.4491(82)$ nm & $0.331(7)$ nm\\ \hline
		$\tau_p$        & $94.58(16)$ fs  & $0.7191(14)$ ps  & $0.976(2)$ ps \\ \hline
		$\rho_t$          & $0.9551(2)$ & $-0.1483(14)$ & $-0.4443(11)$ \\ \hline
		$\tau_s$        & $1.136(2)$ ns & $0.23607(24)$ ns & $0.2146(2)$ ns \\ \hline
		$\tau_i$        & $1.312(2)$ ns & $0.25285(25)$ ns & $0.23130(21)$ ns\\ \hline
	\end{tabular}
	\caption{{\bf Values of the main parameters.} The three pump bandwidths, $\Delta\lambda$, utilized in the experiment (standard deviation), the corresponding pulse duration, $\tau_p$, the best fit parameters $\rho_t$, $\tau_1$, $\tau_2$ for the statistics of arrival times of SPDC photons to the detectors. Table and caption were taken from \cite{Sedziak2019}.}
	\label{tab:fitvalues}
\end{table}
This joint probability distribution can be expressed as (see App. \ref{AppendixTransform}):
\begin{align}
p(t_s,t_i) &= |\hat S(D)  \psi (\omega_s, \omega_i)|^2,
\end{align}
where $\psi (\omega_s, \omega_i)$ is the SPDC biphoton wavefunction and $\hat S(D)$ is operator of the transformation applied by propagation of this function through $10$ km long fibers. 

We assume that the joint probability distribution of photons coupled into single-mode fibers $|\psi (\omega_s, \omega_i)|^2$ is well approximated by following Gaussian distribution: 
\begin{align}
    |\psi (\omega_s, \omega_i)|^2&= \frac{1}{2 \pi \sigma_s \sigma_i \sqrt{1 - \rho_{\omega}^2}}  \times \dots \nonumber \\ 
   \dots \times \exp &(- \frac{1}{2(1 - \rho_{\omega}^2)}(\frac{(\omega_s -\omega_{0s})^2}{\sigma_s^2} + \frac{(\omega_i - \omega_{0i})^2}{\sigma_i^2} - \frac{2 (\omega_s - \omega_{0s}) (\omega_i - \omega_{0i}) \rho_{\omega}}{\sigma_s \sigma_i})),\label{eq:freq2dmodel}
\end{align}
where $\rho_\omega$ is the Pearson correlation coefficient in the frequency domain, $\sigma_s $ and $\sigma_i$ are signal and idler photon angular frequencies' standard deviations, $\omega_{0s}$ and $\omega_{0i}$ are signal and central photon angular frequencies.

In appendix \ref{AppendixTransform} we show that $\rho_t$ is approximately equal to the Pearson correlation coefficient of angular frequencies $\rho_{\omega}$. We will denote both by $\rho$ to avoid confusion. We also show that we can express $\tau_s$ and $\tau_i$ in terms of standard deviations of the signal $\sigma_s$ and idler $\sigma_i$ photons angular frequencies through the equation:
\begin{align}
    \tau_{s(i)} = 2 \beta D \times \sigma_{s(i)}. \label{eq:transformation}
\end{align}
The value of constant term is  $|2 \beta D| = 227$ ns/PHz.

Therefore, we can  use the results of our computation which is a numerical approximation of biphoton wavefunction $\psi (\omega_s, \omega_i)$ to get estimations of parameters $\rho$, $\tau_s$ and $\tau_i$ and compare them with their experimental values. This procedure will provide the second mean of verification of our model. Note that in the work presented in article \cite{Sedziak2019} the mean times of arrivals were not measured and distribution given in \eref{eq:2dmodel} is centered at $t_{s(i)} = 0$. This means that we will not be able to verify accuracy of computed values of $\omega_{0s}$ and $\omega_0i$.

\subsubsection{Numerical simulation of biphoton wavefunction}
The goal of the simulations is to compare experimental values of $\rho$, $\tau_s$ and $\tau_i$ with the computed values. We will do that for all the three pump pulse durations. Before we move on to the direct comparison we first establish influence of uncertainty of temperature $T$ and coupling mode diameter $2w_c$ along with some other numerical parameters - range of frequencies for which we are making calculation which we will denote as $z$ and number of angular frequency values $n$ for which the biphoton wavefunction is computed. The minimal and maximal values of simulated angular frequency are:
\begin{align}
    \omega^{\mathrm{min}}_{s,(i)} &= \omega^{\mathrm{central}}_{s,(i)} (1 - z) \\
    \omega^{\mathrm{max}}_{s,(i)} &= \omega^{\mathrm{central}}_{s,(i)} (1 + z).
\end{align}
These values of $\omega_{s,(i)}$ give us a frequency range. 

\begin{figure}[h!]
     \centering
     \makebox[\textwidth][c]{
        \includegraphics[width=0.45\textwidth]{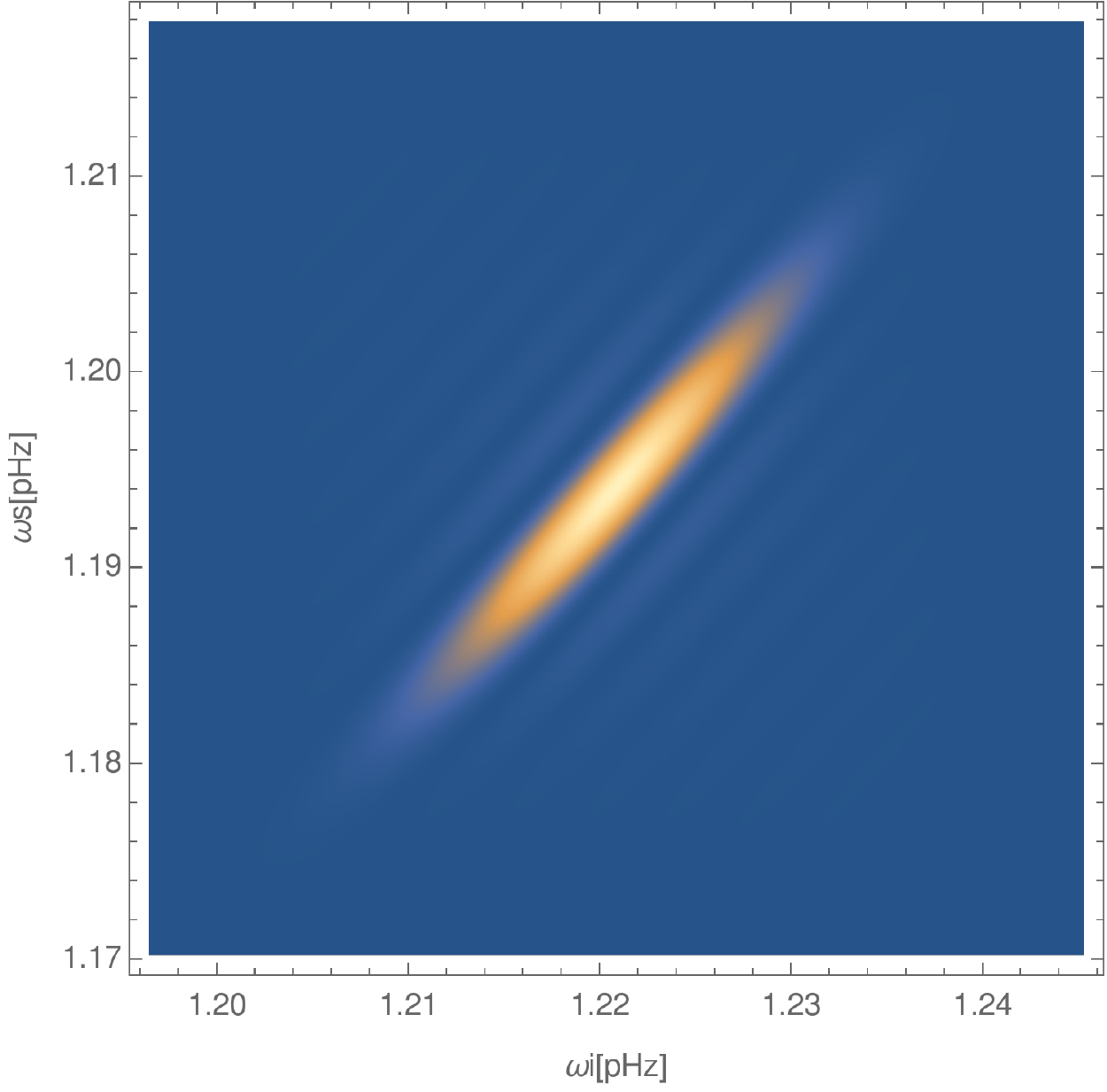}
         \includegraphics[width=0.45\textwidth]{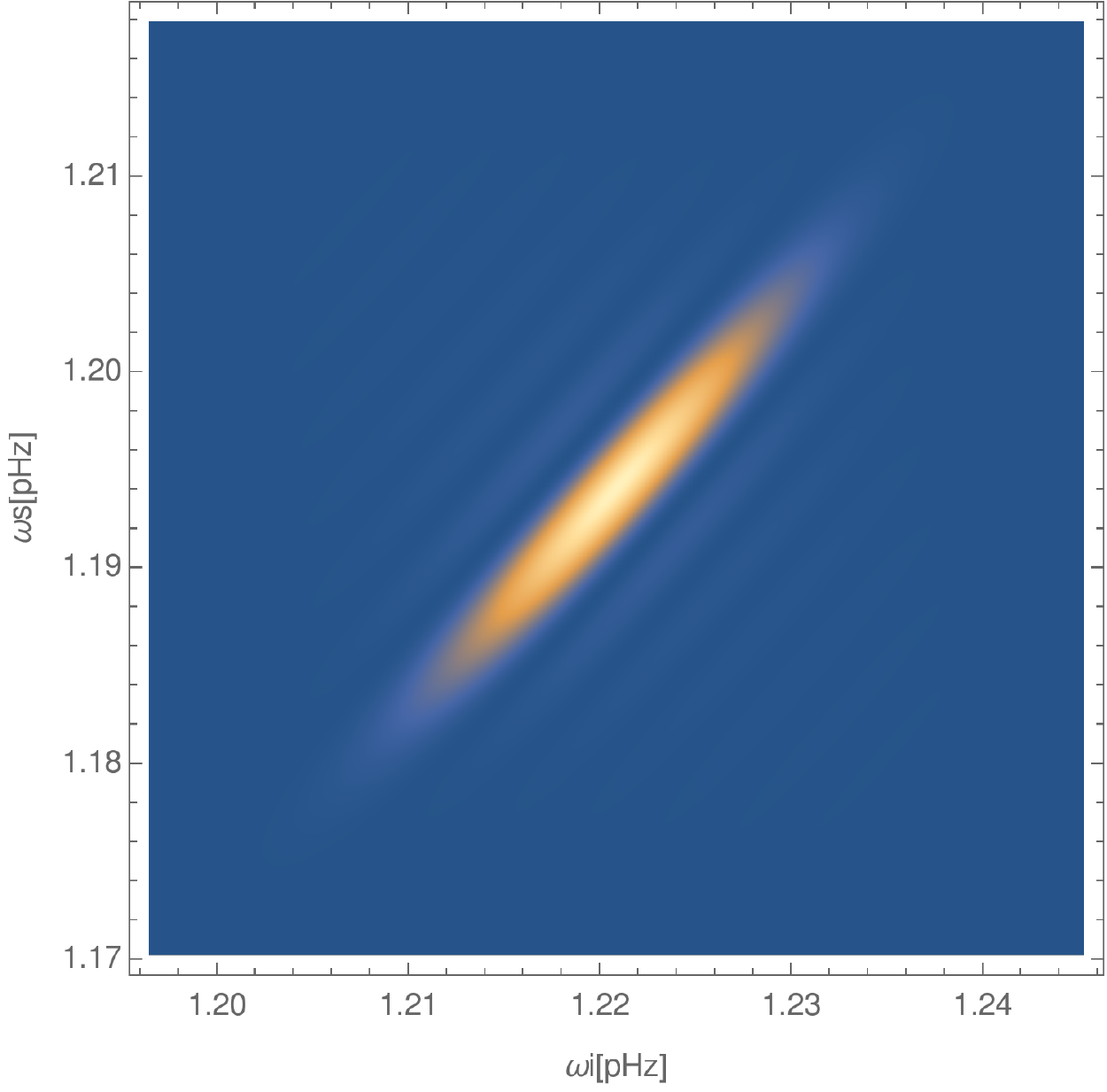}
         }
     \makebox[\textwidth][c]{
         \includegraphics[width=0.45\textwidth]{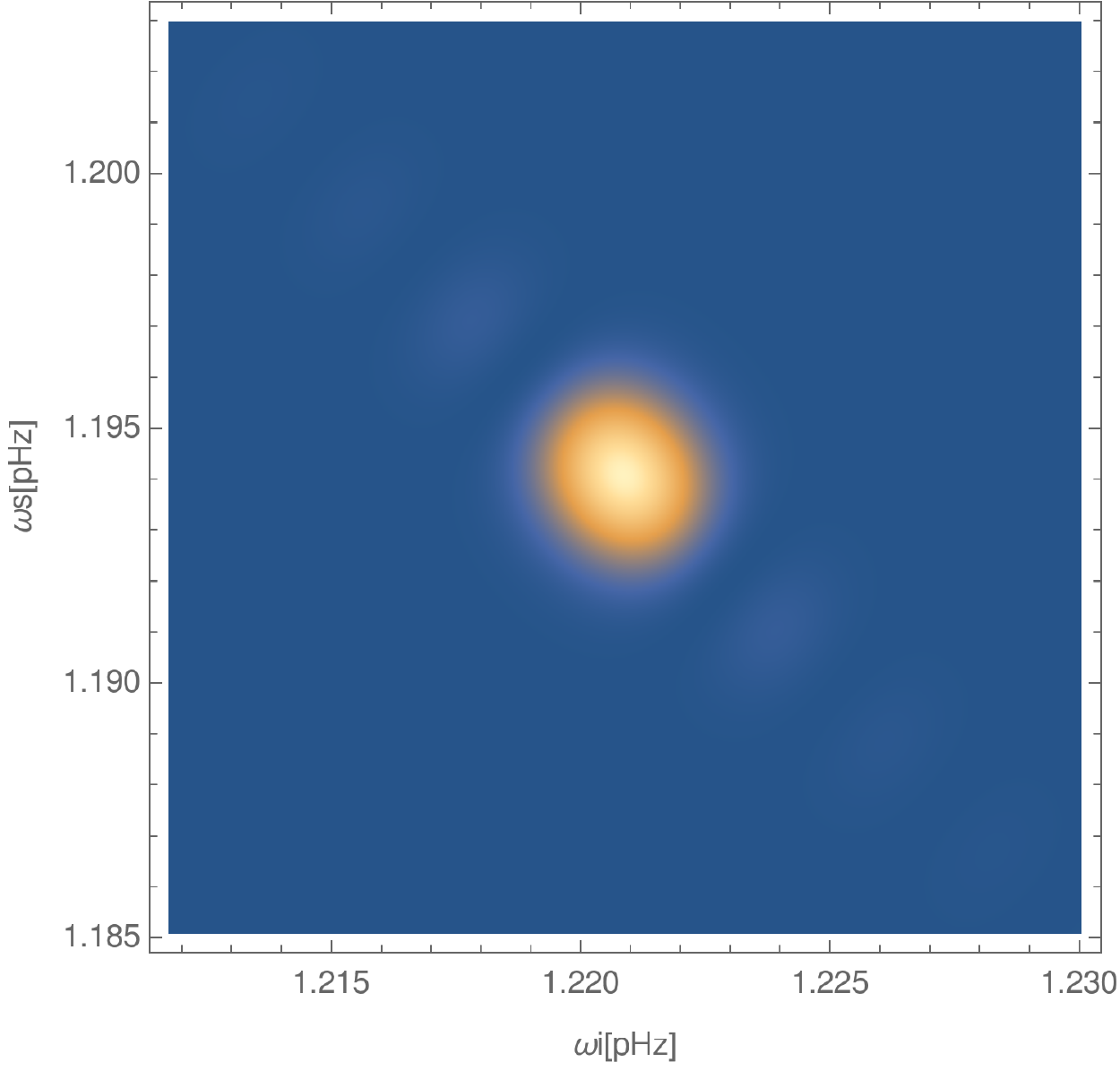}
         \includegraphics[width=0.45\textwidth]{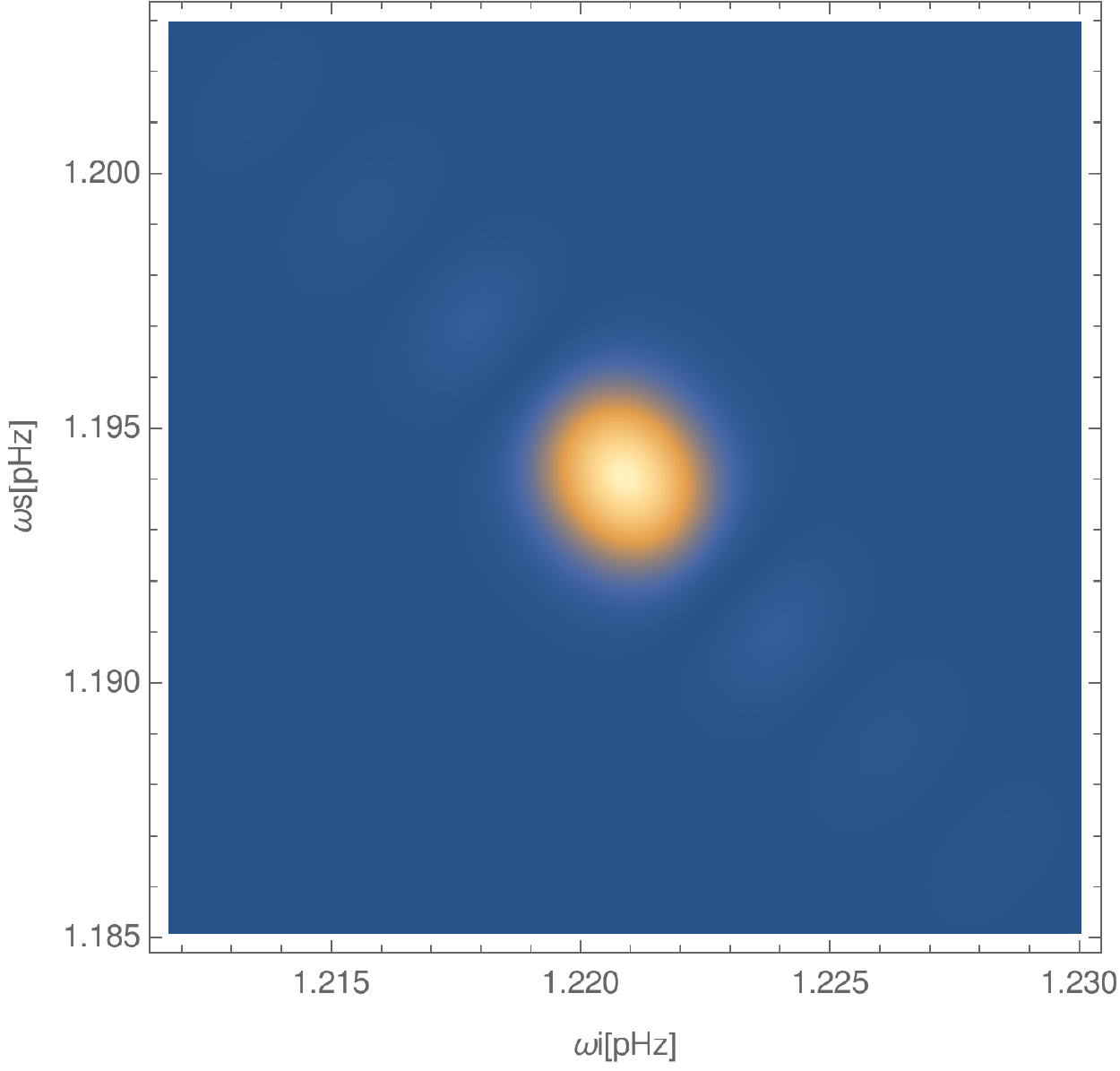}
         }
     \makebox[\textwidth][c]{
         \includegraphics[width=0.45\textwidth]{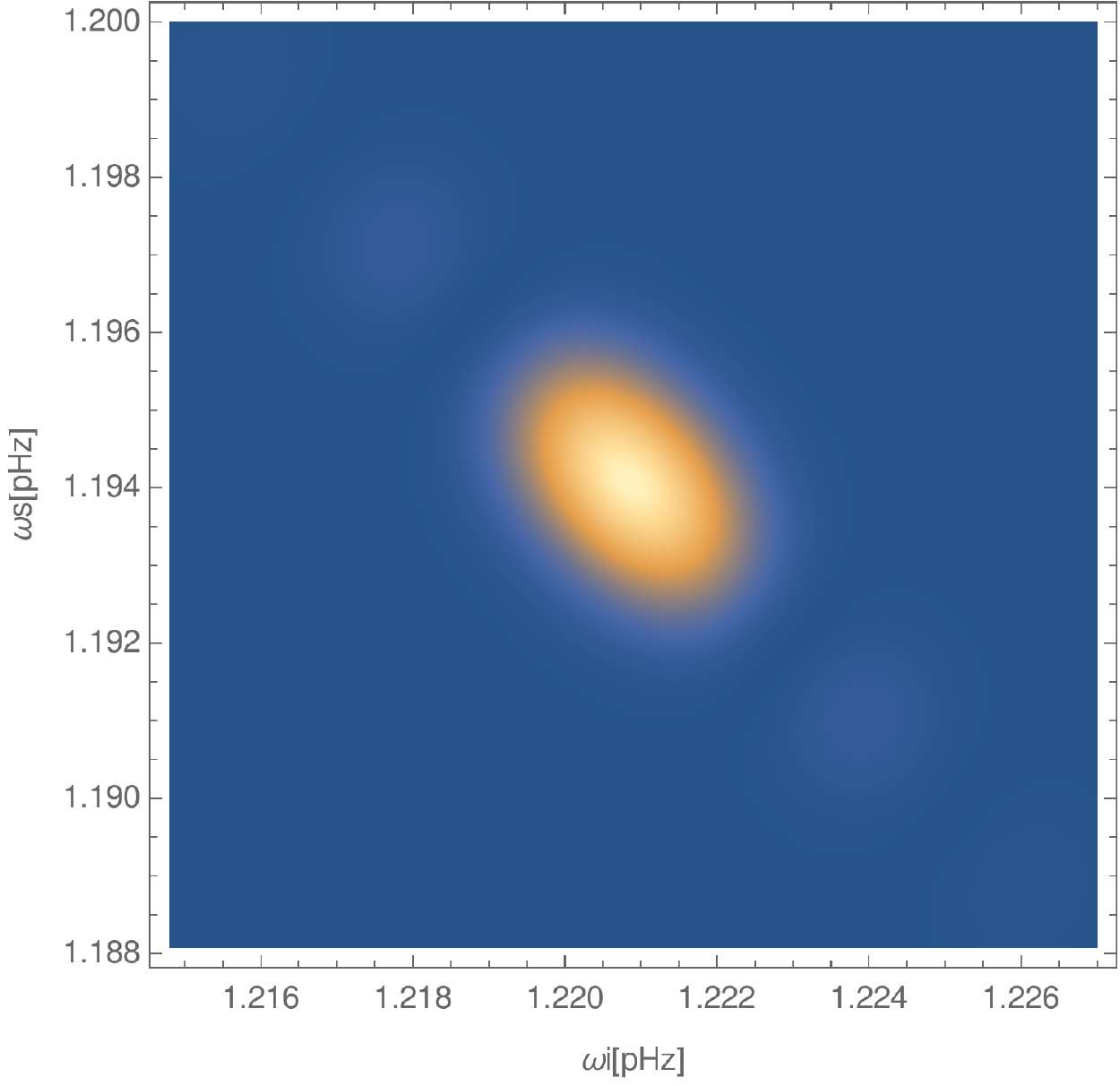}
         \includegraphics[width=0.45\textwidth]{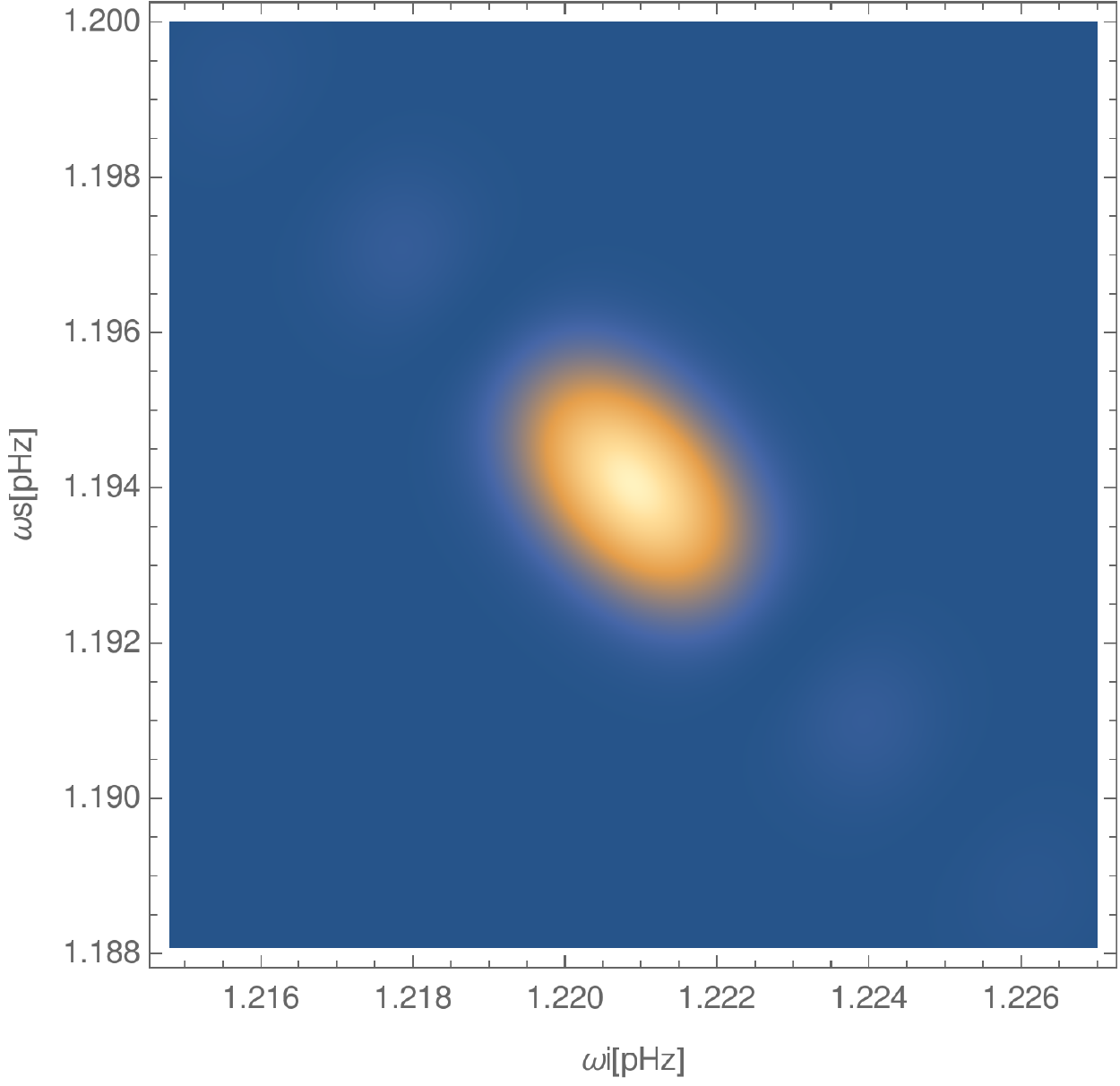}
         }  
        \caption{Plots of joint probability distribution. First column: plots of joint probability distribution calculated for temperature $T = 20$ C. Second column: plots of joint probability distribution for $T = 25$ C. First (top) row depicts calculations for pump pulse duration $\tau_p = 94.58(16)$ fs and range $z = 0.02$. In the middle row we have plots for $\tau_p = 0.7191(14)$ ps  and range $z = 0.0075$. In the last (bottom) row we show plots for $\tau_p = 0.976(2)$ ps  and range $z = 0.005$.}
\label{fig:bphrange}
\end{figure}

We will start with a discussion of the range $z$ and spatial diameter $2w_c$. In the previous comparison, we have noted that small changes of coupling mode diameter do not have a significant influence on the frequency widths of SPDC photons. Similarly here, for all computed temperatures, ranges and numbers of frequencies we do not find any significant change in our numerical predictions. Therefore, for computations we will set the value of diameter $2w_c = 97.5$ $\mu$m.
Frequency ranges defined by the parameter $z$ have to be chosen large enough, so that all frequencies with non zero probabilities are within that range. At the same time, it should be small enough that we do not perform calculations for many frequencies  with negligible contribution. The frequency range can be chosen based on joint probability distribution plot. In Tab. \ref{tab:rangevalues}, we have listed frequency ranges that we picked for each pump pulse duration. The choice of these ranges is based on plots of joint probability distribution for many combinations of parameters. Examples of such plots can be found in Fig. \ref{fig:bphrange}.  We have also checked that these ranges are valid for any temperature in the range given in Tab. \ref{tab:spdc_params2}. In Fig. \ref{fig:bphrange}, we show plots of joint probability distribution made for the smallest ($T= 20$ C) and largest temperature ($T=25$ C) and for three different pump pulse durations. Note that in each row figures are plotted for different $z$ value - the longer the pump duration $\tau_p$ the narrower joint probability distribution gets. The discussion of this effect can be found in \cite{Gajewski2016}.

\begin{table}[h]
	\centering
	\begin{tabular}{|c||c|c|c|c|c|}
	\hline
	$\tau_p$ & range ($z$) & $\omega_s^{min}$ [PHz] & $\omega_s^{max}$ [PHz] & $\omega_i^{min}$ [PHz] & $\omega_i^{max}$ [PHz]\\ \hline		
	$94.58(16)$ fs & $0.02$ & $1.19648$ & $1.24532$ & $1.17016$ & $1.21792$ \\ \hline
	$0.7191(14)$ ps & $0.0075$  & $1.21174$ & $1.23006$ & $1.18508$ & $1.20299$ \\ \hline
	$0.976(2)$ ps  & $0.005$ & $1.21479$ & $1.227$ & $1.18807$ & $1.20001$ \\ \hline
	\end{tabular}
	\caption{Angular frequency ranges for which joint probability distribution were computed. For different pump pulse durations different ranges were used.}
	\label{tab:rangevalues}
\end{table}

\begin{figure}[h!]
     \centering
     \makebox[\textwidth][c]{
        \includegraphics[width=0.45\textwidth]{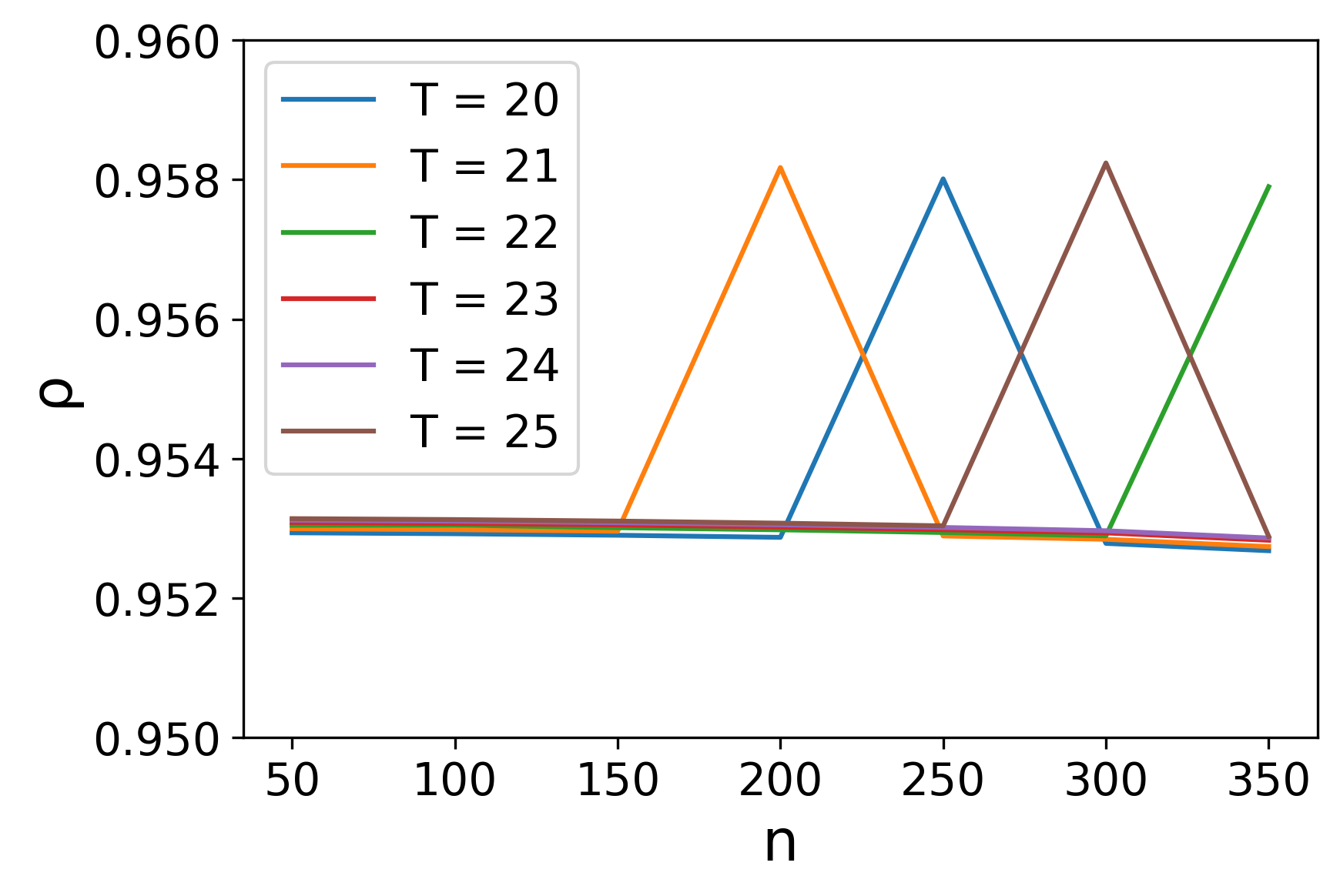}
         \includegraphics[width=0.45\textwidth]{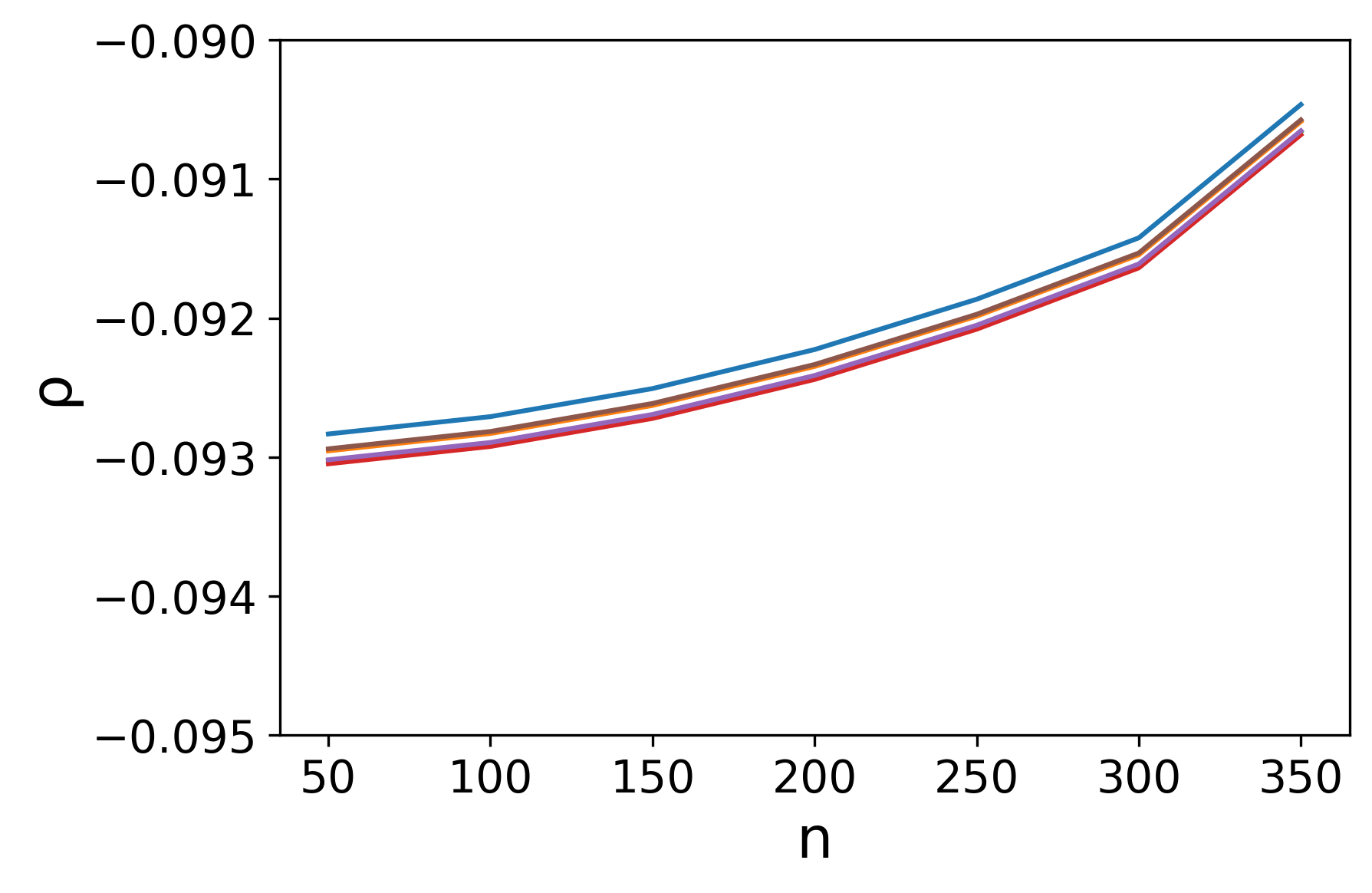}
         \includegraphics[width=0.45\textwidth]{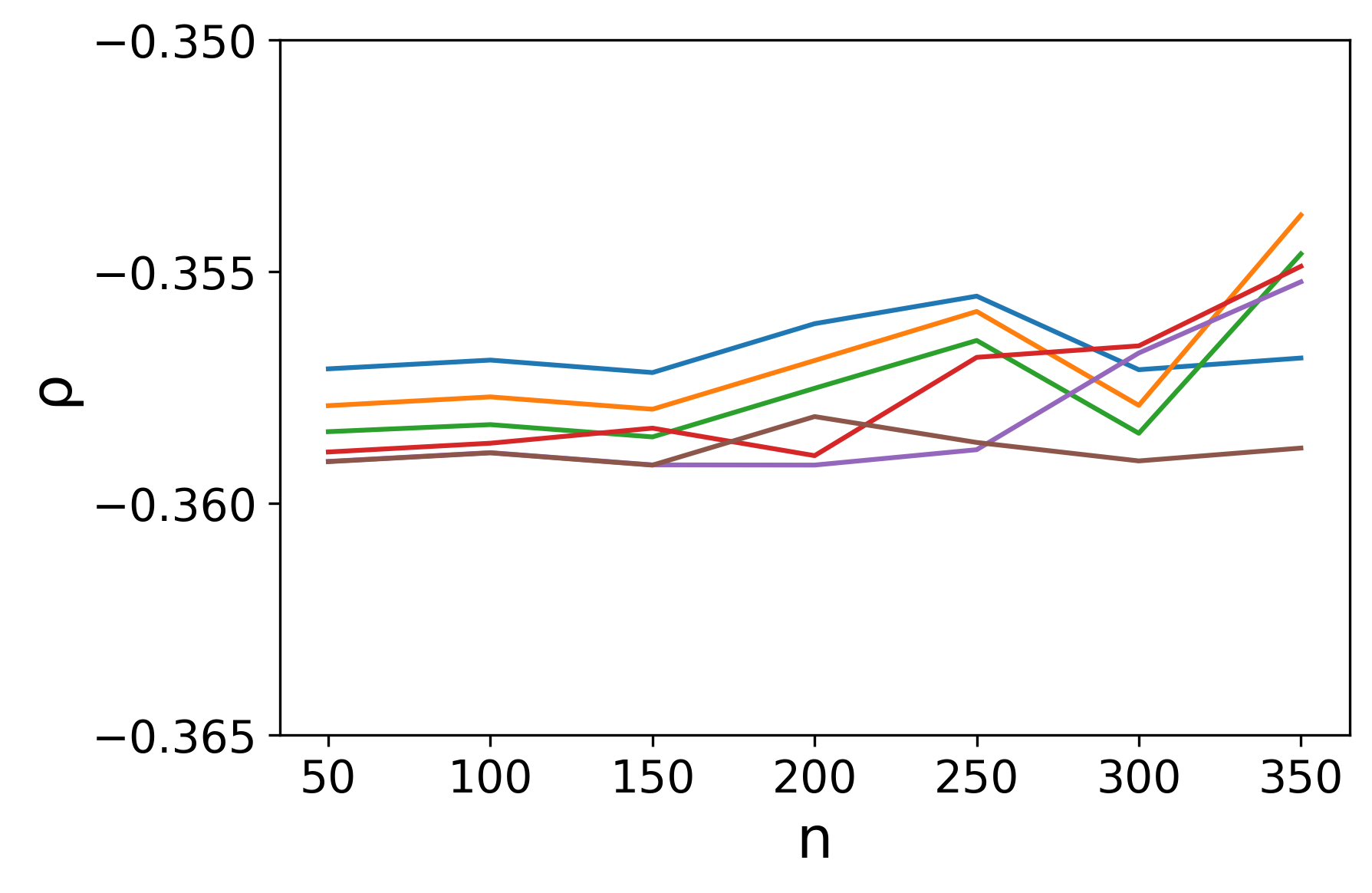}
         }
     \makebox[\textwidth][c]{
         \includegraphics[width=0.45\textwidth]{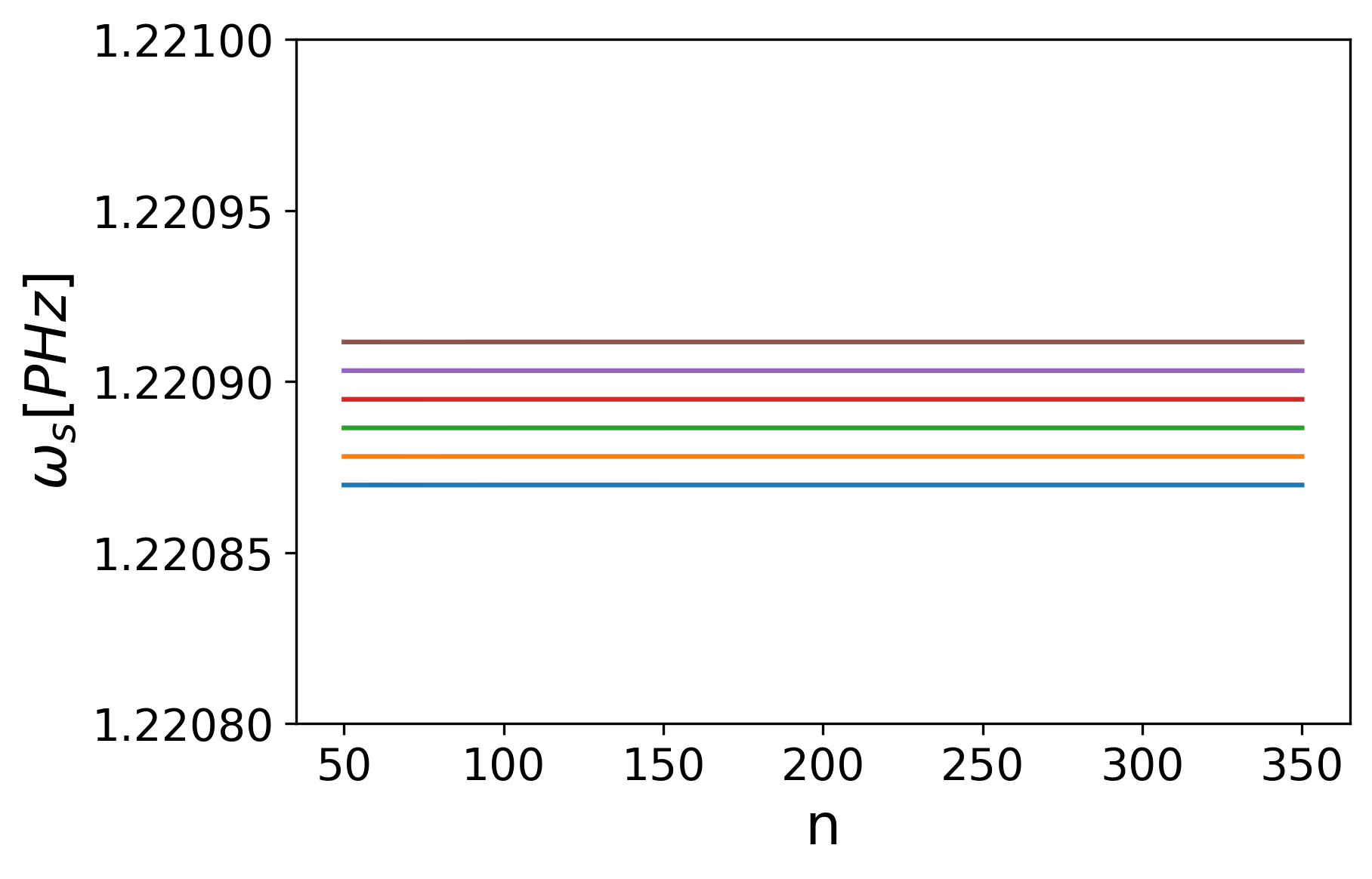}
         \includegraphics[width=0.45\textwidth]{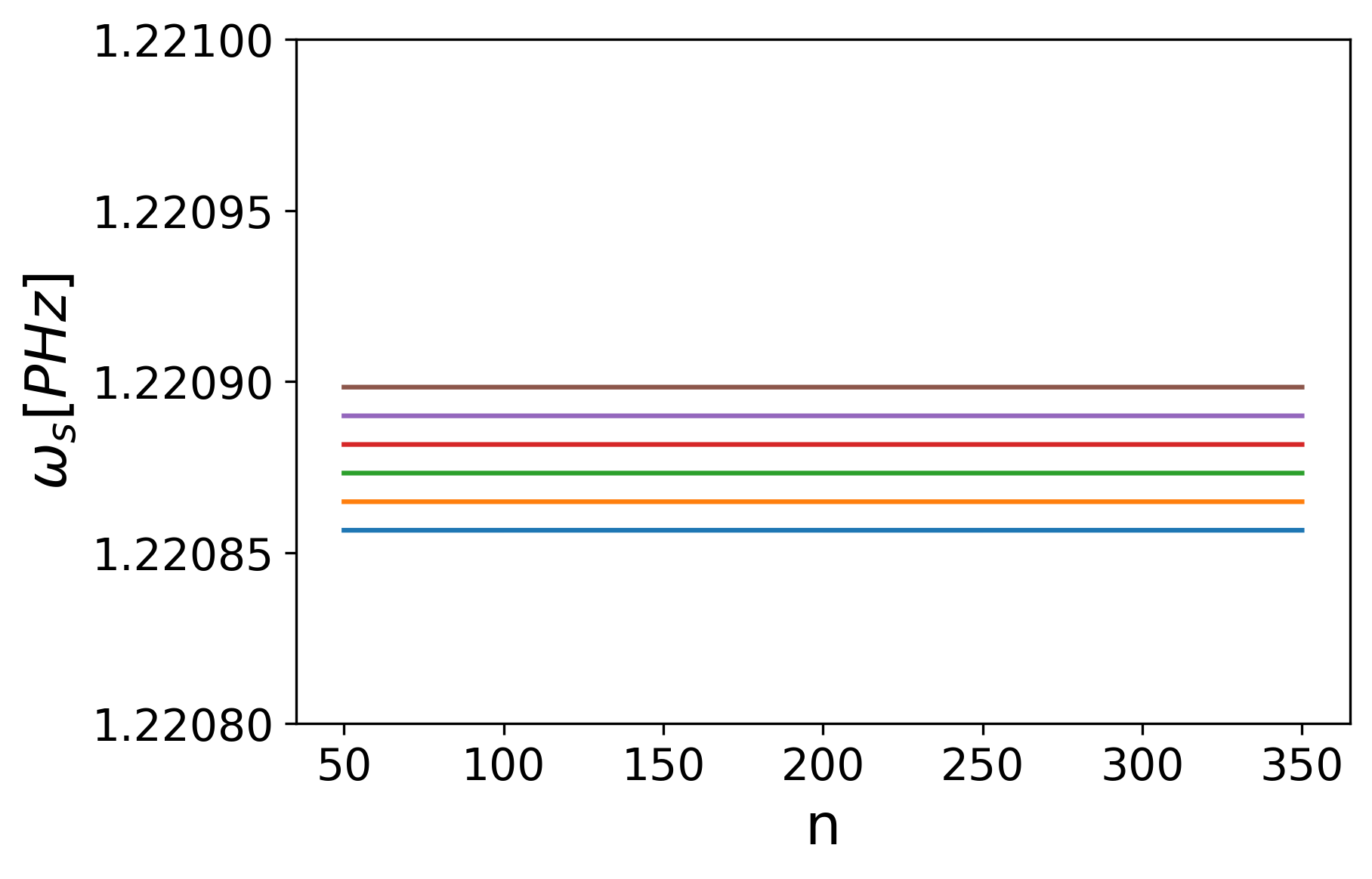}
         \includegraphics[width=0.45\textwidth]{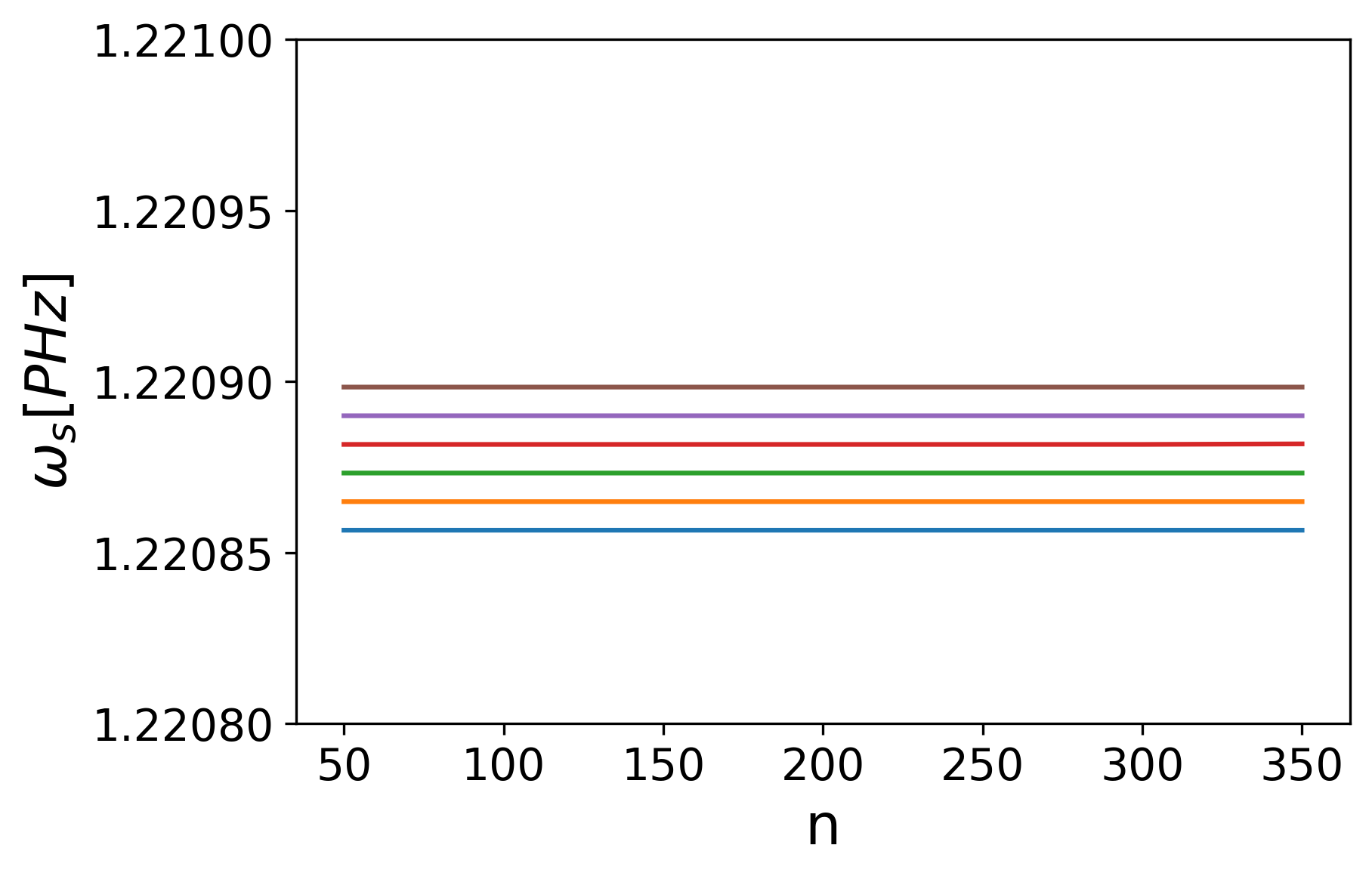}
         }
     \makebox[\textwidth][c]{
         \includegraphics[width=0.45\textwidth]{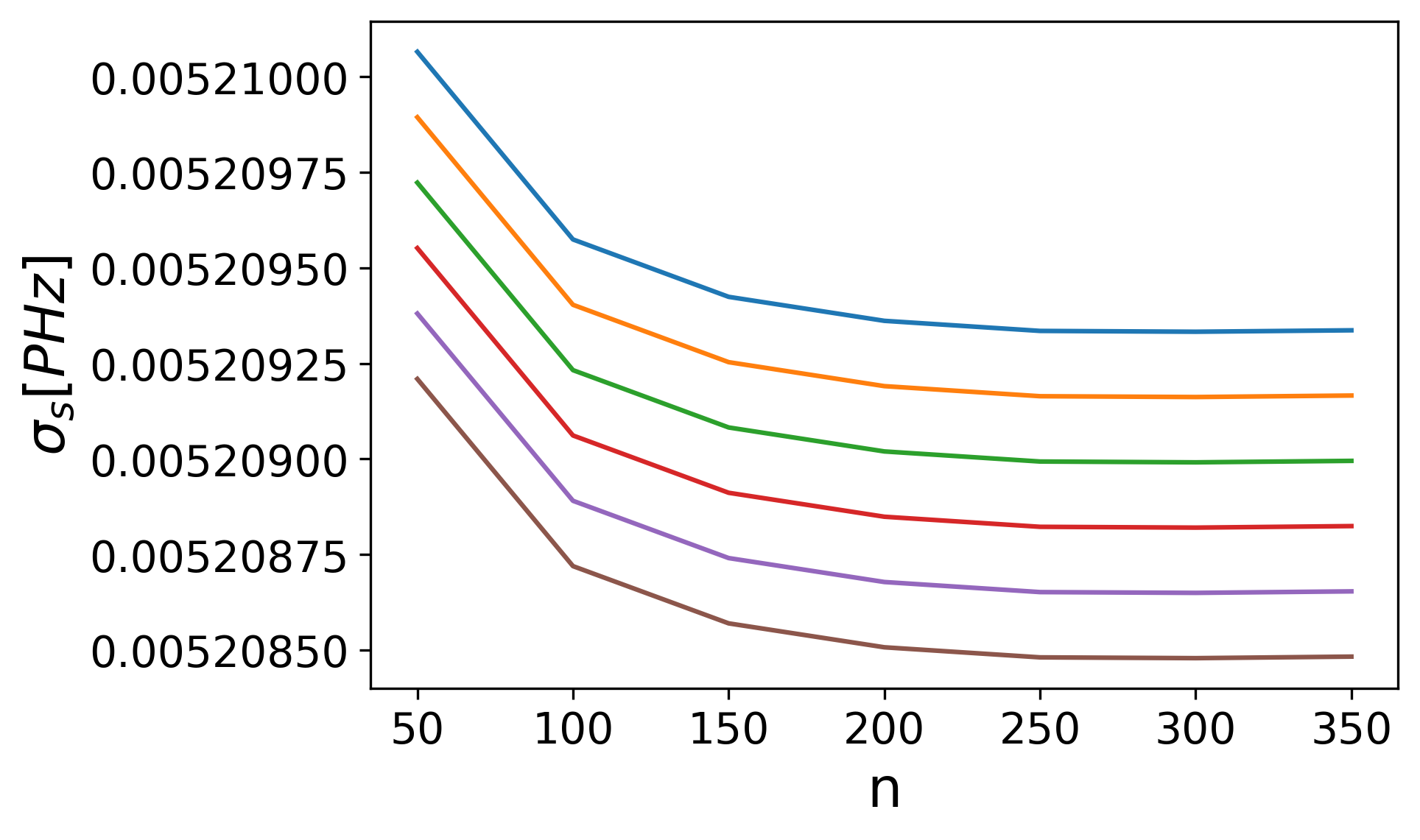}
         \includegraphics[width=0.45\textwidth]{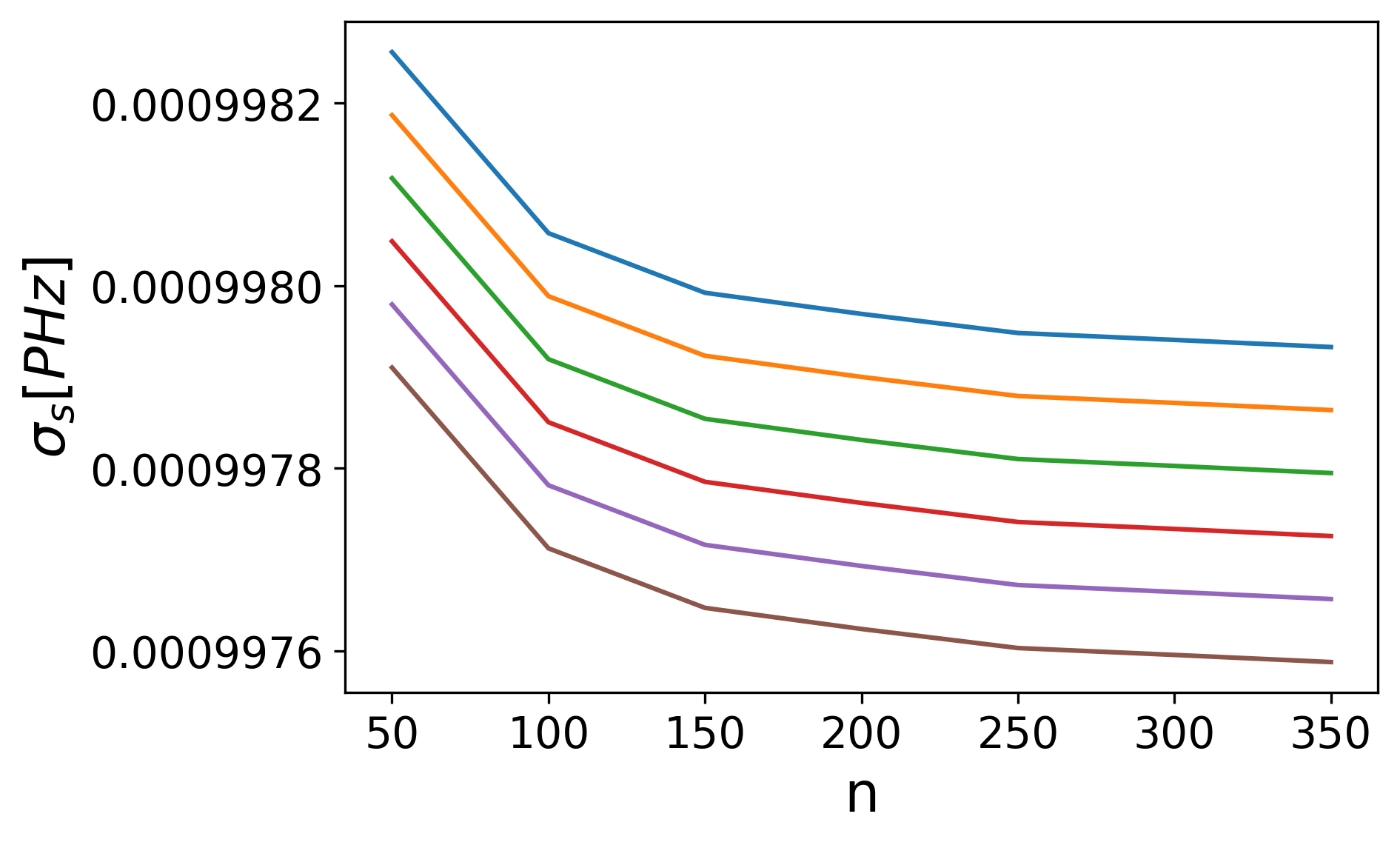}
         \includegraphics[width=0.45\textwidth]{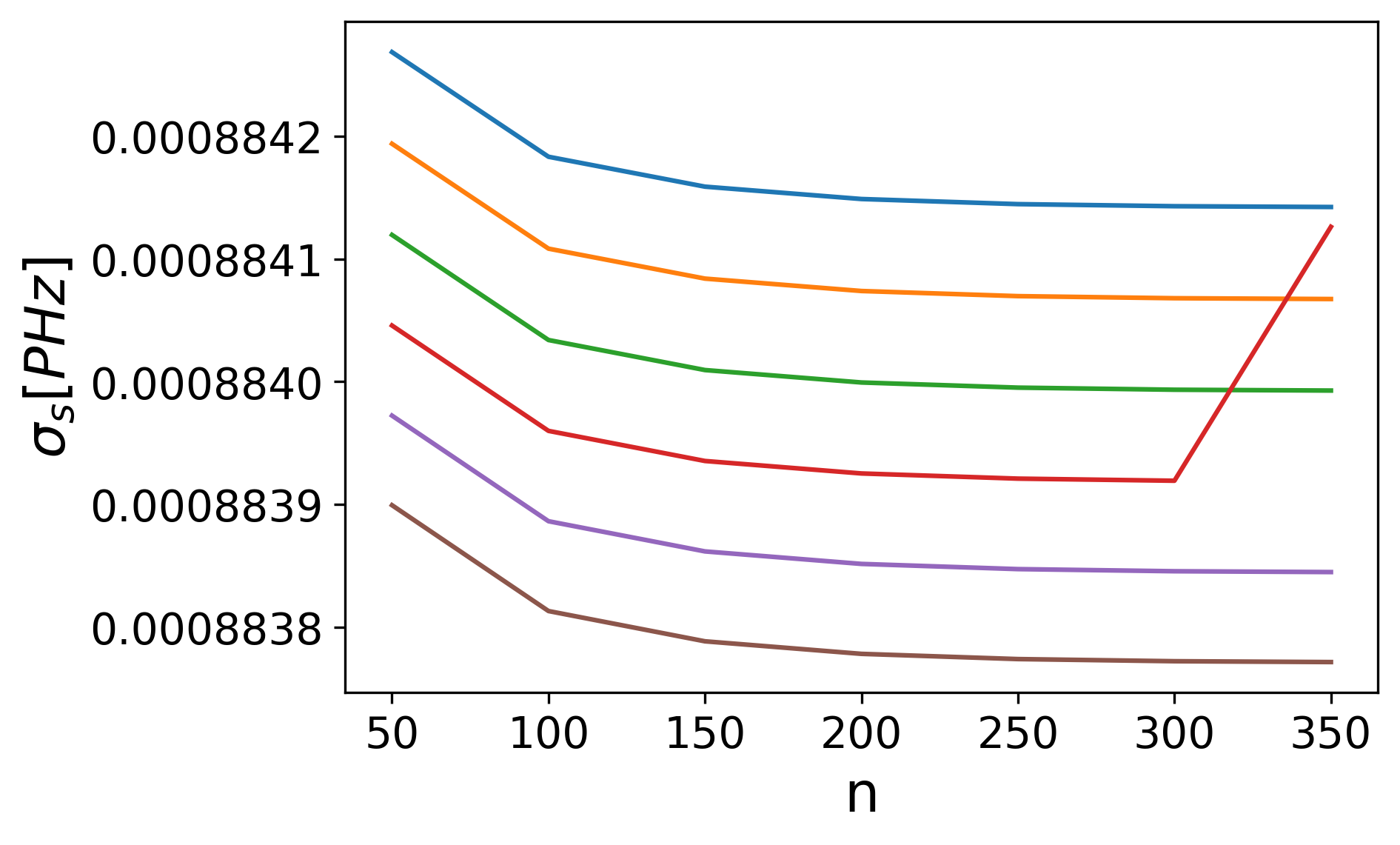}
         }  
        \caption{Convergence plots for three parameters: Pearson correlation coefficient $\rho$, central angular frequency of signal photon $\omega_s$ and angular frequency width (standard deviation) of signal photon $\sigma_s$. Plots in the left column were generated for $\tau_p = 94.58(16)$ fs, in the middle column for $\tau_p = 0.7191(4)$ ps, plots on the right for $\tau_p = 0.976(2)$ ps. In each plot, results of computation for different temperatures are depicted with different colors. The legend from plot in the top left corner denotes colouring which was used for different temperatures and is valid for all the plots.}
\label{fig:ndependence}
\end{figure}

We can move on to the analysis of influence of number of computed angular frequencies $n$ on the predicted values of Pearson correlation coefficient, central angular frequencies and angular frequency widths of signal and idler photons. In Fig. \ref{fig:ndependence}, we can see the convergence plots for three parameters: Pearson correlation coefficient $\rho$, $\omega_{0s}$ and $\sigma_s$. These parameters were extracted from joint probability distribution using the Mathematica software system and its $NonLinearModelFit$ function. For calculation of $\omega_{0s(i)}$ and $\sigma_{s(i)}$ we used the same technique as before- we fitted the single variable Gaussian distribution given in  \eref{eq:model} to marginal distributions of $\omega_s$ and $\omega_i$. For calculation of $\rho$ we fitted the model given in \eref{eq:freq2dmodel} to the joint probability distributions. 

From Fig. \ref{fig:ndependence} it is clear that independent of temperature $T$, number of frequency points $n = 300$ is sufficient to accurately calculate values of the angular frequency $\omega_{0s}$ and frequency standard deviation $\sigma_s$. In case of Pearson correlation coefficient $\rho$, larger numbers of computed frequency points might lead to numerical instabilities. The reason behind this instability is that the Gaussian function only roughly approximates the joint probability distribution - the more points we calculate the more discrepancies from Gaussian distribution become visible.

\subsubsection{Results}
For our verification, we take average of $\rho$, $\omega_s$, $\sigma_s$ over the temperature range for values computed for $n=300$. Averaging over different temperatures is justified since the influence of temperature in this SPDC setup is not significant, which is clearly visible on plots presented in Fig. \ref{fig:ndependence}. Results are summarized in Tab. \ref{tab:paramvalues}. Values of frequencies standard deviations are given in corresponding time units where their relation is given by \eref{eq:transformation}. 
\begin{table}[h]
	\centering
	\begin{tabular}{c|c|c}
	$\tau_p = 94.58(16)$ [fs] & Experimental & Numerical \\ \hline		
	$\rho$  & $0.9551(2)$ & $0.9535$  \\ 
	$\tau_s$ [ns] & $1.136(2)$  & $1.156$ \\ 
	$\tau_i$ [ns] & $1.312$ & $1.182$  \\ \hline
	$\tau_p = 0.7191(4)$ [ps] &  & \\ \hline
	$\rho$  & $-0.1483(14)$ & $-0.0921$  \\ 
	$\tau_s$ [ns] & $0.23607(24)$  & $0.22152$ \\ 
	$\tau_i$ [ns] & $0.25285(25)$ & $0.226509$  \\ \hline
	$\tau_p = 0.976(2)$ [ps] &  & \\ \hline
	$\rho$  & $ -0.4443(11)$ & $-0.35761$  \\ 
	$\tau_s$ [ns] & $0.2146(2)$  & $0.19625$ \\ 
	$\tau_i$ [ns] & $0.23130(21)$ & $0.2007$  
	\end{tabular}
	\caption{Comparison between experimental values and computed values.}
	\label{tab:paramvalues}
\end{table}
From the results presented in Tab. \ref{tab:paramvalues} its is clear that for the shortest pump pulse duration, the computed value of correlation $\rho$ is very similar to the experimental value. Computations for longer pulses yield less accurate predictions of $\rho$. The differences between experimental and computational values are not surprising - the mathematical model of SPDC is itself an approximation and suffers from numerical instability. We also do not have an experimental data for direct comparison between statistical properties of joint probability distribution in the frequency domain - we are not only forced to make the comparison in the time domain, but the relation that we use given by \eref{eq:transformation} is an approximation. That is why numerical values presented in Tab. \ref{tab:paramvalues} are given without estimation of an error - it would be very difficult calculate this compound's error introduced by approximations and final averaging.

Although it is true that our numerical approach has an intrinsic inaccuracy, we think that differences between the estimated and experimental $\rho$ could be of various origin. In both experiment and our computation we extract the Pearson correlation coefficient by fitting the general Gaussian distribution to the corresponding experimental data and compute joint probability distributions. This approach is justified under the assumption that the actual joint probability distribution can be reasonably approximated by a Gaussian distribution. Unfortunately, this model ignores the oscillatory behaviour of joint probability distribution that can be seen in Fig. \ref{fig:bphrange}. The fits are applied to the 'main' part of the distribution depicted in yellow in Fig. \ref{fig:bphrange}. The ignored parts have tendency to lower the values of $\rho$ since they are spaced along the diagonal of the presented plots. Also, they widen the marginal distributions. This widening, which our fitting procedure ignores, might be in part responsible for discrepancies between the predicted temporal widths and the experimental ones.

\section{Summary}
In summary, our model captures well the observed tendencies in the SPDC process. We showed that the central angular frequencies of the signal and idler photons can be well estimated with this model. Moreover, for short pump beam pulse duration, the model was able to predict very well the correlation coefficient in the second experiment. The predicted values of time widths of coupled photons were also well approximated - the estimation difference between the experimental and computed values is below $15\%$ of the experimental values.
For the first experiment, the discrepancies where larger but still accurate to the order of magnitude.

The nature of discrepancies and underestimation of angular frequencies width which can be seen in both experiments is not yet fully understood and requires more investigation. In the research that should follow the presented results, one should go beyond the simple Gaussian model to take into account the oscillatory behaviour present in the numerical results.

%% file: Chapters/Chapter4.tex
\chapter{Bent waveguide modes and their relation to quantum anticentrifugal force}

\label{Chapter4} 

Quantum mechanical descriptions of phenomena arising in low dimensionality systems have been a subject of interest for the past few decades. This interest manifested itself in solid-state physics as growing fields of research in quantum dots (0D), quantum wires (1D), or research regarding surface effects (2D) \cite{Smith1996}. 

One of the effects discussed in 2D geometries with nontrivial topology is the emergence of fictitious quantum anticentrifugal force. The existence of this fictitious force was first described in \cite{Wheeler2001} and since then, it attracted much attention \cite{Cirone2001, Rembielinski2002, Birula2007, Nadareishvili2009, Dandoloff2011, Dandoloff2015}. This anticentrifugal force has no analogue in classical mechanics \cite{Botero2003} and is a consequence of quantum fictitious potential term in the Schrödinger equation of a particle moving in potential with discontinuity, e.g. in the mathematical form of the Dirac's $\delta$ localized in a selected point of space. It was explicitly noted in \cite{Rembielinski2002} that the effect actually requires a 2D space devoid of its origin $R^2 \backslash \{ 0,0 \}$, emphasizing the topological nature of the phenomenon. This area of research belongs to a broader field of inverse square potentials in the Schrödinger equation \cite{Nadareishvili2009}.

Here, we investigate the optical modes in similar semi-circular bent waveguides. We show that we can reformulate mode equations for bent geometry so that they acquire a form analogous to the Schrödinger equation with quantum anticentrifugal potential. We propose a series of approximations to analytically solve these equations for vector components of the electric field and arrive at approximate equations for number of supported waveguide modes. At the end of the chapter, we present a comparison of our analytical solutions with the corresponding results of rigorous simulations of modes in a bent waveguide performed in the commercial COMSOL software. We find that both kinds of solutions are in very good agreement. Besides, these solutions can also be reinterpreted in terms of quantum anticentrifugal force: the electric field distribution in the experimentally feasible geometry of a bent waveguide represents by analogy the wavefunction of a quantum particle subject to such fictitious force.

The outline of this chapter is as follows. We start with well know describition of the modes in hollow waveguide with rectangular cross-section for which analytic solution exists. We move on to the problem of finding modes in dielectric rectangular waveguide for which only approximate solutions are known. We discuss an approximate method of finding modes proposed by Marcatili \cite{Marcatili1969}. In the following section we apply an extension of this method to a bent waveguide, which allows us to arrive at approximate solutions for a bent geometry. We compare our analytical solution with strict numerical simulations. 

\section{Introduction to waveguide modes}

Before we move on to description of light propagation in bent dielectric structures, first we will show how to describe electromagnetic field  in a straight waveguide with rectangular cross-section - both hollow and dielectric. We will present all important calculation and point out the applied approximations.

\subsection{Modes in rectangular waveguide}

We now consider two types of rectangular waveguides: hollow, with perfectly conducting walls, and dielectric. Finding solutions for the first type of waveguide is straightforward and can be done analytically. There exist no exact analytical solution for modes of dielectric rectangular waveguides, therefore, we will outline Marcatili's method \cite{Marcatili1969} which gives a good approximate solution for dielectric waveguides with low contrast of refractive index of the waveguide material and its surroundings.
\begin{figure}[h!]
     \centering
         \includegraphics[width=0.75\columnwidth]{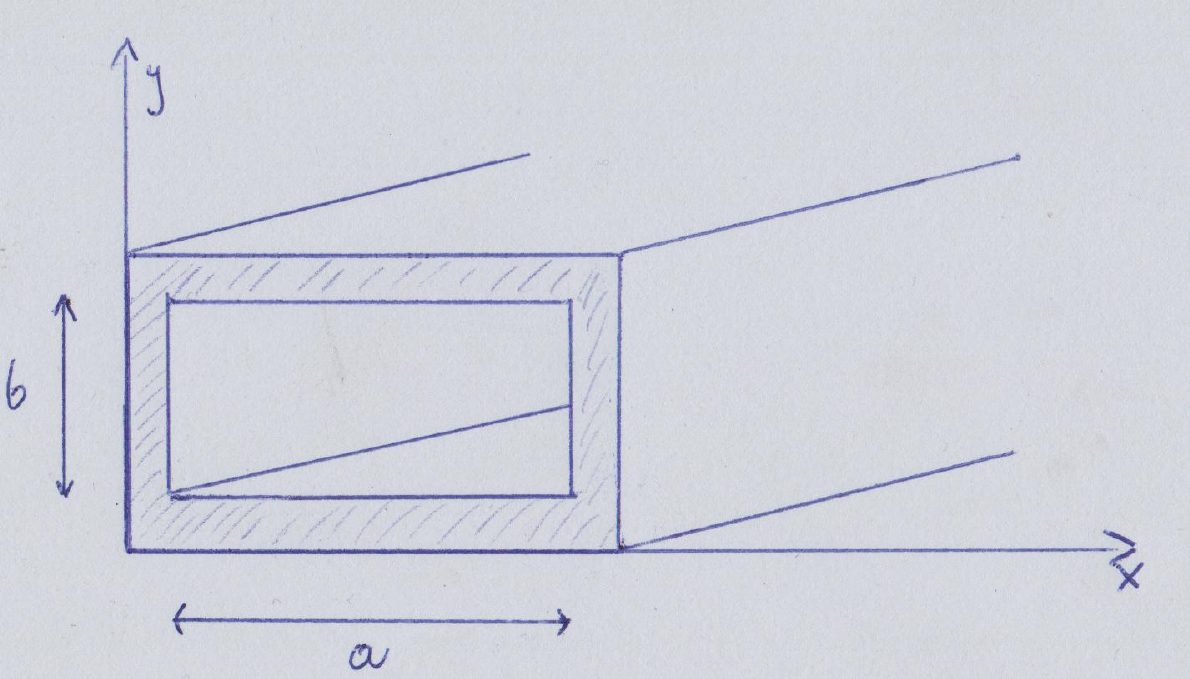}
\caption{Hollow rectangular waveguide. Walls are assumed to be made of a perfectly conducting material.}
\label{fig:HollowWaveguide}
\end{figure}
In both problems, we will assume that the waveguide is infinite in the $z$ direction and its cross-section is rectangular in the $xy$ plane, as it is depicted in Fig.\ref{fig:HollowWaveguide}. The electric and magnetic monochromatic fields can be expressed as \cite{griffithsED}:
\begin{align}
    \boldsymbol{\mathcal{E}}(x,y,z,t) &= \Re \{ \boldsymbol E(x,y) \exp[i (\omega t - k_z z)] \},\label{eq:Emode}\\
    \boldsymbol{\mathcal{B}}(x,y,z,t) &= \Re \{ \boldsymbol B(x,y) \exp[i (\omega t - k_z z)] \}. \label{eq:Bmode}\
\end{align}
The electric and magnetic field components are related through the set of equations \cite{griffithsED}:
\begin{align}
    E_x &= \frac{-i}{k_x^2 + k_y^2} (k_z \frac{\partial E_z}{\partial x} + \omega \frac{\partial B_z}{\partial y}) \\
    E_y &= \frac{-i}{k_x^2 + k_y^2}(k_z \frac{\partial E_z}{\partial y} - \omega \frac{\partial B_z}{\partial x}) \\
    B_x &= \frac{-i}{k_x^2 + k_y^2}(k_z \frac{\partial B_z}{\partial x} - \omega \frac{n^2}{c^2} \frac{\partial E_z}{\partial y}) \\
    B_y &= \frac{-i}{k_x^2 + k_y^2}(k_z \frac{\partial B_z}{\partial y} + \omega  \frac{n^2}{c^2}  \frac{\partial E_z}{\partial x}).
\end{align}
The above set of equation is obtained from Maxwell equations \eref{eq:Erot}, \eref{eq:Brot} when the form of fields given in \eref{eq:Emode}, \eref{eq:Bmode} is assumed. The $E_z$ and $B_z$ components of the field are solutions of the Helmholtz differential equation (see \eref{eq:Helmholtz}):
\begin{align}
    [\frac{\partial^2}{\partial x^2} + \frac{\partial^2}{\partial y^2} + k_x^2 + k_y^2] E_z & = 0, \\
    [\frac{\partial^2}{\partial x^2} + \frac{\partial^2}{\partial y^2} + k_x^2 + k_y^2] B_z & = 0.
\end{align}
These equations have to be solved for the whole space to find the propagating modes in the waveguide.

\subsubsection{Hollow waveguide}

Let us consider a hollow rectangular waveguide with perfectly conducting walls, as in Fig \ref{fig:HollowWaveguide}. That assumption is known as the perfect electric conductor (PEC) boundary condition and it says that the electric field has only components normal to the boundary of PEC and vanishes inside the PEC walls: 
\begin{align}
    \boldsymbol E_{\mathrm{walls}} &= 0, \\ 
    \boldsymbol B_{\mathrm{walls}} &= 0, \\
    \boldsymbol E_{\mathrm{boundary}} \times \boldsymbol e &= 0, \\
    \boldsymbol B_{\mathrm{boundary}} \cdot \boldsymbol e &= 0, 
\end{align}
where $\boldsymbol e$ is the unit vector normal to the boundary. Hollow waveguide means that in the "inside" region (enclosed by PEC walls) $n = 1$ for all frequencies. 

For hollow waveguides, there are two types of mode solutions: transverse electric (TE) and transverse magnetic (TM). For TE modes $E_z =0$, similarly for TM $B_z=0$.  For both mode types, the mathematical expression for fields can be easily obtained through variable separation: 
\begin{align}
    \text{(TE)} \thickspace B_z &= B_0 \cos(k_x x) \cos(k_y y), \\
    \text{(TM)} \thickspace E_z &= E_0 \sin(k_x x) \sin(k_y y),
\end{align}
where wavevectors $k_x$ and $k_y$ are:
\begin{align}
    k_x &= \frac{m \pi}{a}, \thickspace m \in N, \\
    k_y &= \frac{n \pi}{b}, \thickspace n \in N,
\end{align}
and $a$ and $b$ are the height and width of the waveguide cross-section  Fig.\ref{fig:HollowWaveguide}. There are infinitely many modes for a given waveguide geometry but there is a finite number of them that lead to propagating modes. 
The condition for propagating modes is:
\begin{equation}
    k_z^2 = (\frac{\omega}{c})^2 - (\frac{m \pi}{a})^2 - (\frac{n \pi}{b})^2 > 0.
\end{equation}
When this condition is met, $k_z$ is real and the mode can propagate. If $k_z$ is complex then the solution has a form of an evanescent wave along the $z$ direction. A given solution is usually expressed as $\mathrm{TE}_{mn}$ ($\mathrm{TM}_{mn}$) where $m, n$ number the modes. The first index is associated with the oscillation along the wider dimension of the waveguide. For $TM$ modes $n, m$ have to be positive integers. In the case of TE modes, one of the indices can be $0$ but not both. 

Each mode has a cut-off frequency. A cut-off frequency is a frequency below which mode no longer propagates along the waveguide. This frequency is given by the equation:
\begin{equation}
    f_{\text{cut-off}} = \frac{c}{2 \pi} \sqrt{(\frac{m \pi}{a})^2 - (\frac{n \pi}{b})^2}.
\end{equation}
The mode with the lowest value of the cut-off frequency is called the fundamental mode. In case of TE modes this mode is $\text{TE}_{10}$ and for transverse magnetic modes the fundamental mode is $\text{TM}_{11}$.

\begin{figure}[h!]
     \centering
         \includegraphics[width=0.4\columnwidth]{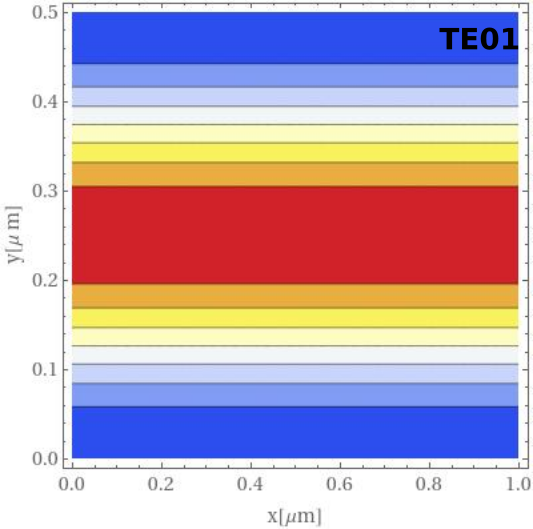}
         \includegraphics[width=0.4\columnwidth]{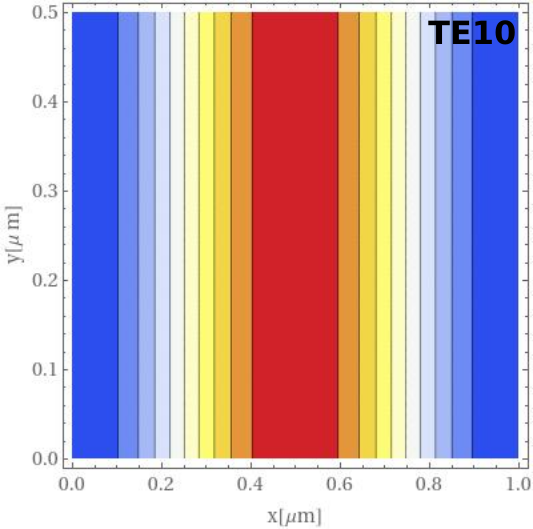}
         \includegraphics[width=0.4\columnwidth]{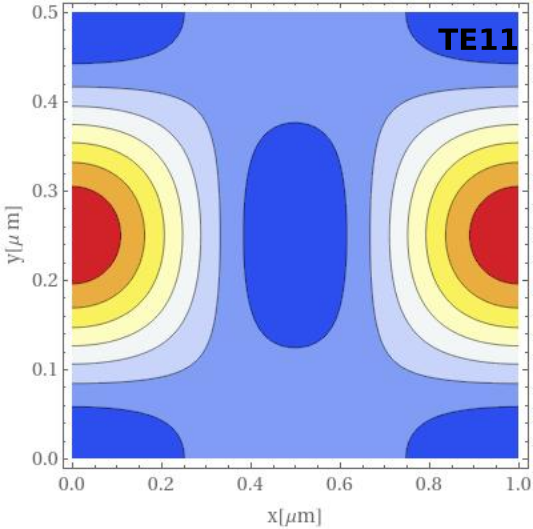}
         \includegraphics[width=0.4\columnwidth]{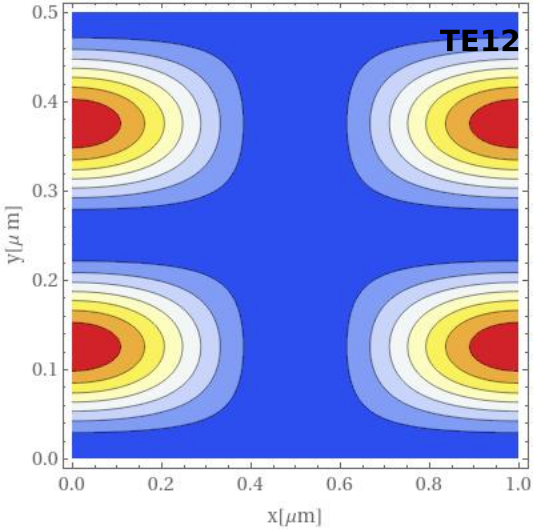}         
        \caption{Electric field intensity of TE modes in waveguide $0.5$ $\mu$m tall and $1$ $\mu$m wide.}
\label{fig:HollowModes}
\end{figure}

\subsection{Dielectric waveguide and Marcatili's method}

There are no known methods for obtaining an exact analytical solution for a rectangular dielectric waveguide. All known solutions are approximated. One of the most used and best known approximate approaches towards solving the problem is the Marcatili's method. In his original article \cite{Marcatili1969}, Marcatili proposed a way to get an approximate solution for a dielectric waveguide embedded in another dielectric with a slightly smaller refractive index. The method has been adapted, so that it can also approximately describe modes in waveguides with high index contrast \cite{Westerveld2012}.

\begin{figure}[tbh]
	\centering
	\includegraphics[width=0.5\columnwidth,keepaspectratio]{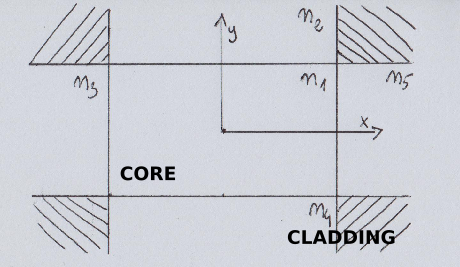}
	\caption{$n_\nu$ stand for refractive indices, and where $n_1$ describes the refractive index of the core of the waveguide and $n_{2-5}$ refractive indices of the cladding. The shaded regions in Marcatili's method are overlooked on purpose.}
	\label{fig:Dielectric}
\end{figure}

The situation considered by Marcatili is depicted in Fig.\ref{fig:Dielectric}. The cross section of the waveguide is divided into 9 regions. We assume that there is no electric and magnetic field at 4 of them. Those regions are depicted by the shaded area int Fig. \ref{fig:Dielectric}. The center region, representing the core, has a refractive index denoted by $n_1$. The rest of the outside regions are part of the cladding and have corresponding refractive indices denoted by $n_j$ where $j=2,..,5$.

Waveguide (core) is surrounded by areas with lower refractive indices (cladding). Waves are guided along the $z$-axis. Marcatili's approximation comes from a few assumptions. The first is of low refractive index contrast $n_1 - n_{2-5} \ll 1$. Low index contrast corresponds to a low number of modes and also the lower the contrast the deeper the penetration of the electromagnetic field into the cladding. Most importantly, low refractive index means that the wavevector of the field propagating inside the waveguide has to be almost parallel to the $z$ axis and that leads to electric and magnetic field components being mostly perpendicular to the wall of the core. Therefore, modes we will be looking for are called
TEM modes, where $E_z, H_z$ are small compared to the dominant field component. We will denote these modes as $E^{x,y}_{pq}$ where the superscripts $x,y$ stand for largest electric field component and the subscripts $p,q$ number the modes.  

It is assumed that the field that penetrates the cladding decays exponentially. This assumption makes it very hard to solve equations for the electromagnetic field in the corners (shaded regions in Fig.\ref{fig:Dielectric}). But the field in the corners should be very weak so omitting these regions should yield only a small error. We will analyse two distinct solutions where the electric field is mostly polarized along the $y$-axis or $x$-axis. Due to symmetry, both solutions have to same form. We will only solve equations for the $y$-polarisation. Solution for the $x$-polarization follows trivially.

\subsubsection{Electric field y-polarized}

Following \cite{Marcatili1969}, we consider solutions in the form:
\begin{align}
    H_{x \nu} &= \exp[i (\omega t - k_z z)] \left\{\begin{array}{cc} 
     A_1 \cos(k_x x+ \alpha) \cos(k_y y+ \beta) & \text{for } \nu = 1 \\ 
     A_2 \cos(k_x x+ \alpha) \exp(-ik_{y2} y) & \text{for } \nu = 2\\
     A_3 \cos(k_y y+ \beta) \exp(-ik_{x2} x) & \text{for } \nu = 3\\
     A_4 \cos(k_x x+ \alpha) \exp(ik_{y4} y) & \text{for } \nu = 4\\
     A_5 \cos(k_y y+ \beta) \exp(ik_{x5} x)  & \text{for } \nu = 5,\\
    \end{array}\right. \\
    H_{y \nu} & = 0, \\
    H_{z \nu} & =  \frac{i}{k_z}(1 - \frac{2 k_{y \nu}^2}{k_0^2 n_{\nu}^2}) \frac{\partial H_{x \nu}}{\partial x}, \\
    E_{x \nu} & = -\frac{1}{\omega \epsilon_0 n_{\nu}^2 k_z} \frac{\partial^2 H_{x \nu}}{\partial x \partial y},\\
    E_{y \nu} & =\frac{k_0^2 n_{\nu}^2 - k_{y \nu}^2}{\omega \epsilon_0 n^2_{\nu} k_z} H_{x \nu},\\
    E_{z \nu} & =\frac{i}{\omega \epsilon_0 n_{\nu}^2} \frac{\partial H_{x \nu}}{ \partial y},
\end{align}
where $k_0$ is the wavevector in free space and $k^2_{x \nu} + k^2_{y \nu} + k_z^2 = n_{\nu}^2 k^2_0$. Index $\nu \in \{1,...,5\}$ indicates the region of space according to Fig.\ref{fig:Dielectric}. Terms $\alpha$ and $\beta$ describe phase shifts. Equations for $H_{x \nu}$ are guessed and other field components are calculated using the Maxwell equations and the fact that $\frac{ \partial^2 H_{x \nu}}{\partial y^2} = - k_{y \nu}^2 H_{x \nu}$. The above set of equations differs from the Marcatili's equations in $H_{z \nu}$, since in his original work he gave an incorrect expression for that field component. 
We will solve the above set of equations following \cite{Marcatili1969} and chapter 5 of  \cite{dudorov2003millimeter}.

We will now consider only the dielectric waveguide in air, $n_{2,3,4,5} = 1$. Due to symmetry $\alpha, \beta \in \{- \pi/2, 0 , \pi/2 \}$. In his work, Marcatili assumed that refractive index difference between cladding and core is small $n_1 - 1 \ll 1$ which leads to $|k_{x,y}/k_z| \ll 1$.  With this assumption, the electromagnetic field component $E_x $ can be disregarded. Now the task is to match the remaining field components $H_x, E_y, H_z, E_z$ at waveguide boundaries. Due to symmetry, we can consider only the part of the waveguide for which $x,y \geq 0$.

The boundary conditions are:
\begin{align}
    \boldsymbol H_1^{||} &= \boldsymbol H_{2,3,4,5}^{||} \thickspace , \\
    \boldsymbol E_1^{||} &= \boldsymbol E_{2,3,4,5}^{||}  \thickspace ,
\end{align}
where the superscript $||$ denotes components tangential to the waveguide boundaries. These boundary conditions mean that we have to match the $H_x, H_z, E_z$ fields at the upper boundary and the $E_y, H_z, E_z$ at the waveguide side.

\begin{figure}[h!]
     \centering
         \includegraphics[width=0.32\columnwidth]{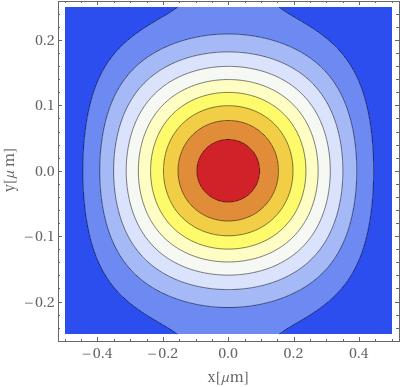}
         \includegraphics[width=0.32\columnwidth]{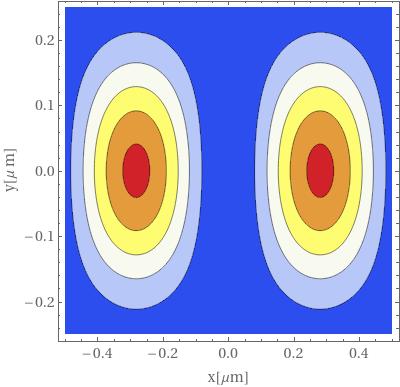}
         \includegraphics[width=0.32\columnwidth]{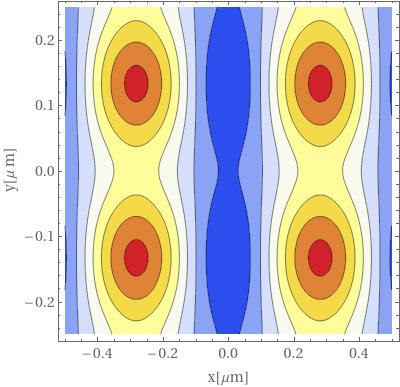}
        \caption{Visualization of electric field intensity of $E^y_{p,q}$ modes for waveguide made of material with refractive index $n_1 =2.3$ and rectangular cross section 1 $\mu m$ wide and $0.5 \mu m$ tall. From the left: $E^y_{1,1}$ (fundamental mode), $E^y_{2,1}$ and $E^y_{2,2}$.}
\label{fig:MarcatiliModes}
\end{figure}

Enforcing the boundary condition on $E_y$ yields the relation:
\begin{equation}
    A_5 = A_1 \cos (k_x \frac{a}{2} + \alpha) \exp(- i k_{x5} \frac{a}{2}).
\end{equation}
The boundary condition for $E_z$ comes down to relation $\frac{1}{n_1^2} = 1$ which in our case is always approximately true since $n_1 \approx 1$.
Finally, matching $H_z$ on the boundary and using the result from $E_y$ matching gives:
\begin{equation}
    k_x \sin(k_x \frac{a}{2} + \alpha) = -i k_{x5} \cos(k_x \frac{a}{2} + \alpha),
\end{equation}
where we ignored the small terms $2k_y^2/k_0^2 n_{\nu}^2$. This equation can be transformed into:
\begin{equation}
    k_x a =\pi m - 2 \arctan (\frac{i k_x}{k_{x5}}), \label{eq:xmodes}
\end{equation}
and $k_{x5} = \sqrt{k_x^2 + k_0^2 (1 - n_1^2)}$. Matching the other set of field components on the other boundary yield a similar result. Enforcing continuity of $H_x$ gives us the relation:
\begin{equation}
    A_2 = A_1 \cos(k_y \frac{b}{2} + \beta) \exp(i k_{y2} \frac{b}{2}).
\end{equation}
Using the above relation for matching $H_z$, we see that $H_z$ is to a good approximation continuous at the boundary. 
Matching the electric field $E_z$ gives us relation:
\begin{equation}
    \frac{k_y}{n_1^2} \sin(k_y \frac{b}{2} + \beta) = i k_{y2} \cos(k_y \frac{b}{2} + \beta),
\end{equation}
which can be reformulated as \cite{Marcatili1969}:
\begin{equation}
    k_y b =\pi q - 2 \arctan (\frac{1}{n_1^2}\frac{-i k_y}{k_{y2}}), \label{eq:ymodes}
\end{equation}
and $k_{y2} = \sqrt{k_y^2 + k_0^2 (1 - n_1^2)}$.
The above results allow us to calculate field components of any mode for a given waveguide geometry and material. In Fig.\ref{fig:MarcatiliModes}, we plotted field intensities for chosen modes and an example of waveguide geometry and material.
Solving both transcendental equations (\eref{eq:xmodes}, \eref{eq:ymodes}) gives us the propagation constant through relation:
\begin{equation}
    k_z = \sqrt{k_0^2 n_1^2 -k_x^2 -k_y^2}.
\end{equation}
Another way of looking at this equation is noting that the propagation constant depends on mode numbers $p,q$ through relations \eref{eq:xmodes}, \eref{eq:ymodes}. This means that for a given waveguide geometry and material it is possible to find all the modes of that waveguide for which the propagation constant is real.
Although the Marcatili method makes use of many approximations, it is still fairly accurate even for a large difference of refractive indices \cite{dudorov2003millimeter}. 

\section{Modelling a bent rectangular waveguide}

\begin{figure}[tbh]
	\centering
	\includegraphics[width=0.5\columnwidth,keepaspectratio]{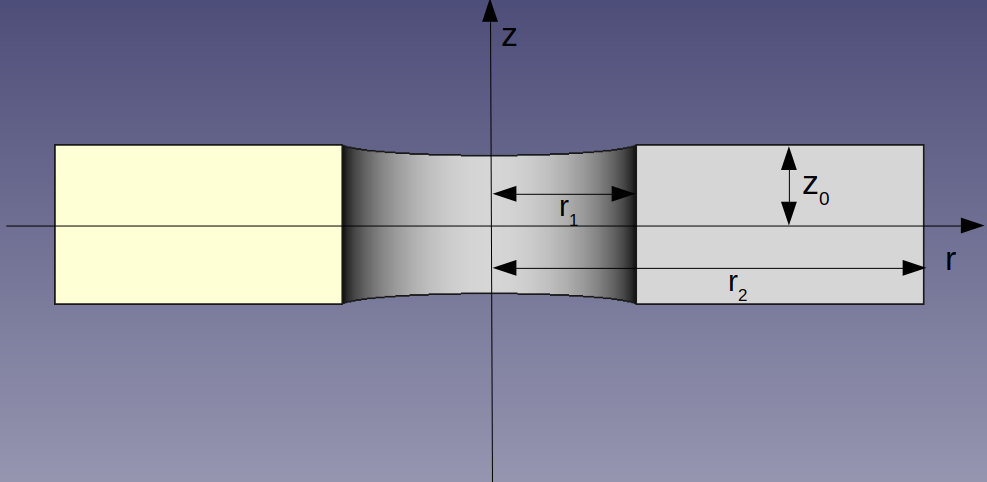}
	\includegraphics[width=0.46\columnwidth,keepaspectratio]{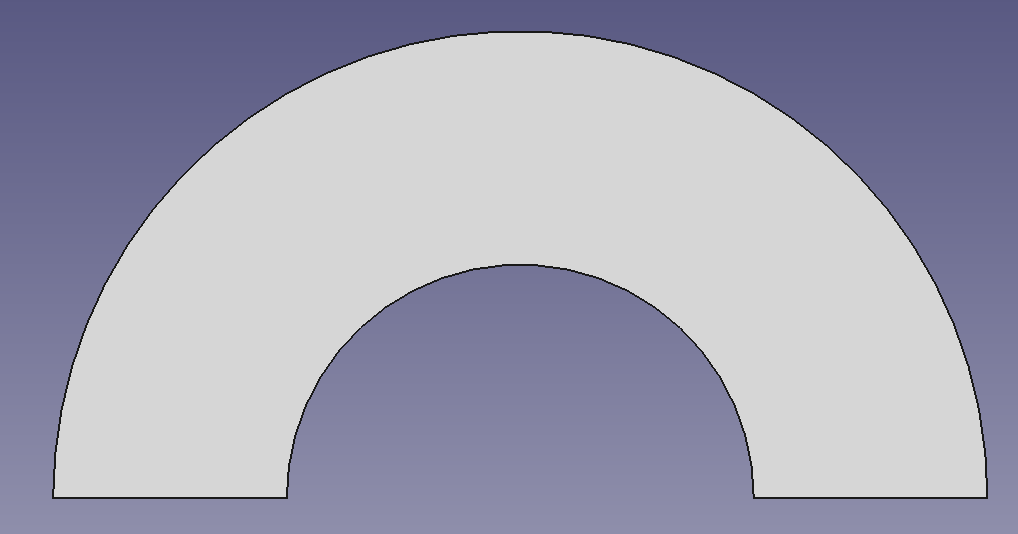}
	\caption{Schematic representation of dielectric bent waveguide. Here, waveguide has a height of $2 z_0$, and inner and outer bending radii of is $r_1$ and $r_2.$}
	\label{fig:Bent}
\end{figure}

To find modes in a bent waveguide geometry like the one depicted in Fig. \ref{fig:Bent} we we solve Maxwell equations for a monochromatic electromagnetic waves. The solutions will require a formulation of the problem in cylindrical coordinates, obtaining an approximate form of the radial component of the electric field, and then solving a set of equations given by the boundary conditions. With the final solutions at hand, we will be able to analyze the behaviour of modes in the waveguide.
\subsubsection{Modes in a bent waveguide}
Analogously to \eref{eq:Emode} and \eref{eq:Bmode}, we postulate the following form of the solutions: 
\begin{align}
    \boldsymbol{\mathcal{E}}(\phi,r,z,t) &= \Re \{ \boldsymbol E(r,z) \exp[ i( m \phi - \omega t)] \}, \label{eq:postulateE}\\
    \boldsymbol{\mathcal{B}}(\phi,r,z,t) &= \Re \{ \boldsymbol B(r,z) \exp[ i( m \phi - \omega t)] \},    
\end{align}
where $(\phi,r,z)$ are cylindrical coordinates and $\omega$ is the mode eigenfrequency. The direction of propagation is indicated by $\hat \phi$. Note that modes in dielectric waveguides do not necessarily have to represent transverse waves - we can expect that electromagnetic field has a component along the direction of propagation.   
With \eref{eq:Erot} we obtain following relation:
\begin{equation}
    \boldsymbol{\mathcal{B}}(\phi,r,z,t) = \frac{1}{\omega}\nabla \times\Re \{i \boldsymbol E(r,z) \exp[ i( m \phi - \omega t)]\}. \label{eq:magneticfield}
\end{equation}
With the above calculations we automatically have $\nabla \cdot  \boldsymbol{\mathcal{B}} = 0$. 
Fields have the form : 
\begin{equation}
    \left[ \begin{array}{c}\boldsymbol B(r,z) \\ \boldsymbol E(r,z) \end{array} \right] = \left[ \begin{array}{c} B_r(r,z) \\  E_r(r,z) \end{array} \right] \boldsymbol r + \left[ \begin{array}{c} B_\phi(r,z) \\  E_\phi(r,z) \end{array} \right] \boldsymbol \phi + \left[ \begin{array}{c} B_z(r,z) \\  E_z(r,z) \end{array} \right]\boldsymbol z,   
\end{equation}
where $(\boldsymbol r, \boldsymbol \phi, \boldsymbol z)$ are unit vectors (see \eref{eq:cylVec}). Relations between the magnetic and electric field components is given by the set of equations:
\begin{align}
   i\omega B_r(r,z) &= \frac{i m}{r} E_z(r,z) - \frac{\partial E_\phi(r,z)}{\partial z}, \label{eq:iomBr} \\
   i\omega  B_\phi(r,z) &= \frac{\partial E_r(r,z)}{\partial z} - \frac{\partial E_z(r,z)}{\partial r}, \label{eq:iomBphi}\\
   i\omega B_z(r,z) &= \frac{1}{r} [\frac{\partial (r E_\phi(r,z))}{\partial r} - im E_r(r,z)] \label{eq:iomBz},
\end{align}
which can be calculated from \eref{eq:cylRot} and \eref{eq:postulateE}.
The Helmholtz equation \eref{eq:Helmholtz} in cylindrical coordinates (\eref{eq:cylVecLap}) has the  form:
\begin{align}
    \nabla_\perp^2 E_r - \frac{m^2 +1}{r^2}E_r - i  \frac{2m}{r^2} E_\phi  &= -\frac{n_j^2\omega^2}{c^2}  E_r,  \label{eq:helmR}\\
    \nabla_\perp^2 E_\phi - \frac{m^2 +1}{r^2} E_\phi+  i \frac{2m}{r^2} E_r &= -\frac{n_j^2\omega^2}{c^2} E_\phi, \label{eq:helmPhi}\\
    \nabla_\perp^2 E_z -\frac{m^2}{r^2} E_z &= -\frac{n_j^2\omega^2}{c^2} E_z, \\
    \nabla_\perp^2 f& \equiv \frac{1}{r} \frac{\partial}{\partial r}(r \frac{\partial f}{\partial r}) + \frac{\partial^2 f}{\partial z^2}.
\end{align}
The symbol $n_j$ stands for the refractive index inside the waveguide for $j = 1$ or refractive index outside the waveguide $j = 2$ .
Any solution of the above set of equations has to have divergence \eref{eq:Ediv} equal to $0$:
\begin{equation}
    \frac{1}{r} \frac{\partial(rE_r)}{\partial r} + \frac{im}{r} E_\phi + \frac{\partial E_z}{\partial z} = 0,
\end{equation}
and the corresponding magnetic field \eref{eq:magneticfield}  also has to be a solution to the Helmholtz equation.
\subsubsection{Components of electric field}
We will assume that bending radius $r_0$ is large and the polarization direction of the electric field is mostly along the $r$ axis during the propagation. 
In other words, this approximation states that:
\begin{equation}
    |E_r|^2 > |E_z|^2 \text{ and } |E_r|^2  \gg |E_\phi|^2 \text { but } |E_r|^2 \sim r_0^4 |\nabla_\perp^2E_\phi|^2.
\end{equation}
Therefore the modes being a solution to above equations can be denotes as $E_{p,g}^r$. Looking for solutions of equation \eref{eq:helmR} we will disregard the $E_\phi$ component. We obtain:
\begin{equation}
\frac{\partial^2 E_r}{\partial r} + \frac{1}{ r}\frac{\partial E_r}{\partial r} + \frac{\partial^2 E_r}{\partial^2 z} + (k_j^2- \frac{m^2 +1}{r^2})  E_r = 0,
\end{equation}
where $k_j \equiv n_j \omega/ c $.
We will solve the above equation by variable separation \mbox{$E_r(r,z) =R(r)Z(z)$:}
\begin{equation}
\frac{1}{R}\frac{\partial^2 R}{\partial^2 r} + \frac{1}{rR}\frac{\partial R}{\partial r} + \frac{1}{Z}\frac{\partial^2 Z}{\partial^2 z} + k_j^2- \frac{m^2 +1}{r^2} = 0.     
\end{equation}
As in Marcatili’s method, we consider only the area near the walls of the waveguide (see Fig. \ref{fig:Dielectric}) and we assume exponential decay outside of the waveguide. Independence of variables leads to the solution of $Z$ in the form:  
\begin{equation}
    \frac{1}{Z} \frac{\partial^2 Z}{\partial z^2} = \begin{cases} -\beta_w^2, \mbox{ if } |z| < z_0, \\ \beta_s^2,  \mbox{ if } |z| > z_0, \end{cases}
\end{equation}
where $2z_0$ is the height of the waveguide as it is depicted in Fig. \ref{fig:Bent}.
The above equation means that we are looking for an oscillatory solution inside the waveguide and an exponentially decaying one outside it.
This finally gives us the equation: 
\begin{align}
& r^2\frac{\partial^2 R}{\partial r^2} + r \frac{\partial R}{\partial r} + (h_j^2r^2 - (m^2+ 1)) R = 0, \label{eq:radialBessel}\\
& h_j = \begin{cases} h_1 = \sqrt{k_1^2-\beta_w^2}, \mbox{ for } |z| < z_0 \\ h_2 =\sqrt{k_2^2-\beta_s^2},  \mbox{ for } |z| > z_0 \end{cases},
\end{align}
which after change of variables $\xi_j = r h_j$ and $\lambda = \sqrt{m^2 + 1}$ gives us the Bessel equation:
\begin{equation}
\xi_j^2\frac{\partial^2 R}{\partial^2 \xi_j} + \xi_j \frac{\partial R}{\partial \xi_j} + (\xi_j^2 - \lambda^2) R = 0.  \label{eq:Bessel}  \end{equation}
In the appendix \ref{apendixQFF} we show how to reformulate equation \eref{eq:radialBessel} so that it is analogous to Schrödinger equation in $R^2$ space in planar coordinates. Discussion of quantum anticentrifugal force can also be found there.

\subsubsection {Boundary conditions}
\eref{eq:Bessel}  has two linearly independent solutions: Bessel function of the first and second kind which we will denote as $J_\lambda (r h_j)$ and $Y_\lambda (r h_j)$, respectively.
Solutions to the above equations for the inside region have the following form:
\begin{align}
    R_{m,j} (r) &= \sin(\gamma_{m,j}) J_\lambda (r h_j) + \cos(\gamma_{m,j}) Y_\lambda (r h_j), \label{eq:Erradial}\\ 
    Z_1(z) &= A_m \sin(\beta_w z) + B_m \cos(\beta_w z), \\
    Z_2(z) &= \begin{cases} C_m \exp(- \beta_s z), \mbox{ if } z > z_0 \\ \pm C_m \exp(\beta_s z),  \mbox{ if } z < - z_0 \end{cases},
\end{align}
where the $\pm$ sign corresponds to two types of possible behaviour of the $Z(z)$ function - at $z=0$ it can either have an extremum or it can be 0. This behaviour is a result of symmetry -  electromagnetic field intensity has to be the same on both boundaries, therefore electric fields on the boundaries can differ only by sign.
At this point, we arrive at a set of simplified equations \eref{eq:iomBphi}, \eref{eq:iomBr} and \eref{eq:iomBz}:
\begin{align}
   i\omega B_r(r,z) &\approx \frac{i m}{r} E_z(r,z), \label{eq:EztoBr}\\
   i\omega  B_\phi(r,z) &= \frac{\partial E_r(r,z)}{\partial z} - \frac{\partial E_z(r,z)}{\partial r}, \label{eq:Bphi}\\
   i\omega B_z(r,z) &\approx- \frac{im}{r}  E_r(r,z) \label{eq:ErtoBz},
\end{align}
a differential equation for $E_z$:
\begin{align}
   \nabla_\perp^2 E_z -\frac{m^2}{r^2} E_z &= -\frac{n_j^2\omega^2}{c^2} E_z \label{eq:helmEz},
\end{align}
which is coupled with $E_r$ through the divergence:
\begin{align}
    \frac{1}{r} \frac{\partial(rE_r)}{\partial r} &+ \frac{\partial E_z}{\partial z} = 0,
\end{align}
and a set of boundary conditions for the electromagnetic field components parallel to the boundaries:
\begin{align}
    E_{r,1}(r, \pm z_0) &= E_{r,2}(r, \pm z_0), \\
    E_{z,1}(r_{1,2}, z) &= E_{z,2}(r_{1,2}, z), \\
    B_{r,1}(r, \pm z_0) &= B_{r,2}(r, \pm z_0), \\
    B_{\phi,1}(r, \pm z_0) &= B_{\phi,2}(r, \pm z_0), \\
    B_{z,1}(r_{1,2}, z) &= B_{z,2}(r_{1,2}, z), \\
    B_{\phi,1}(r_{1,2}, z) &= B_{\phi,2}(r_{1,2}, z).
\end{align}
We are not going to solve the differential equation for $E_z$ but we will make a few notes. Field component $E_z$ has to have similar solutions as $E_r$ because of \eref{eq:helmEz} which means that inside the waveguide the field component $E_z$ has the following form:
\begin{equation}
    E_{z,1}(r,z) = (a_m J_m(r g_1) + b_m Y_m(r g_1))(c_m \sin(\alpha_1 z) + d_m \cos (\alpha_1 z)),
\end{equation}
where parameters $a_m, b_m, g_1, \alpha_1$ are defined analogously to parameters $A_m, B_m, h_1, \beta_1$.
One of the possible solutions is that $E_z$ has nodes on the upper and lower boundaries which means that $E_z(r, \pm z_0) = 0$. From this result and equation \eref{eq:EztoBr}, automatically follows the continuity of $B_r$ which is 0 on the vertical walls. Moreover, the derivative $\frac{\partial E_z(r,z)}{\partial r}|_{z =\pm z_0} = 0$ reduces the problem of matching $B_\phi$ on the horizontal walls to matching the derivative $\frac{\partial E_r(r,z)}{\partial z}|_{z =\pm z_0}$, as follows from \eref{eq:Bphi}.

Discussion of the boundary conditions for $E_r$ is very similar.
We demand that on vertical walls $E_r(r_{1,2},z) = 0$, which leads to $B_z$ being continuous. Similarly as before, the problem of matching $B_\phi$ on horizontal walls is reduced to the matching the derivative of $E_z$.
Making above arguments explicit for $E_r$, the boundary conditions for $E_r$ are:
\begin{align}
    E_{r,1} (r, \pm z_0) &= E_{r,2} (r, \pm z_0), \label{eq:Erboundary1}\\
    E_{r, 1}(r_{1,2}, z) &= 0, \label{eq:Erboundary2}\\
    \frac{\partial E_{r,1} (r,z)}{\partial z} |_{z = \pm z_0} &= \frac{\partial E_{r,2} (r,z)}{\partial z} |_{z = \pm z_0}. \label{eq:Erboundary3}
\end{align}
The resulting boundary conditions do not couple different electric field components. A full solution would require ensuring that in whole space $\nabla \cdot \boldsymbol{E} = 0$. Finding such a solution is very challenging so coupling different components might done i.e by minimizing divergence in some sense. 

We will not try to solve the above equations for both components, we will only consider the larger component $E_r$, simultaneously claiming (without proof) that for every solution of $E_r$ it is possible to find a matching $E_z$ so that the error introduced by our approximation is small. Let us make a last observation that $\oint_S \boldsymbol E \cdot d \boldsymbol \sigma = 0 $ where $S$ is any contour containing the waveguide cross-section. This means that at least an average divergence of the electric field is $0$ inside the waveguide.

\subsubsection{Radial component of electric field}
We will proceed to find a solution for $E_r$.
From the first boundary condition \eref{eq:Erboundary1} it follows that the values of $R_{m,1}$ and $R_{m,2}$ functions have to be the same on the boundary so $h_1 = h_2 \equiv h$ and $\gamma_{1,m} = \gamma_{2,m} \equiv \gamma_m$. This leads to the relation:
\begin{equation}
k_1^2 - \beta_w^2 = k_2^2 +\beta_s^2.    
\end{equation}
If that condition is met then all that is left to do is to match $Z(z)$ and $\frac{\partial Z(z)}{\partial z}$ on the horizontal boundaries and find a function $R_m(r)$ (\eref{eq:Erradial}) such that $R_m(r) = 0$ on the vertical boundaries. 
The explicit form of the boundary conditions reads:
\begin{subequations}
\begin{eqnarray}
    A_m \sin (\beta_w z_0) + B_{m} \cos (\beta_w z_0) &=& C_m e^{ -\beta_s z_0}, \label{eq:BoundEZ1}\\
    \beta_w (A_m \cos (\beta_w z_0) - B_m \sin (\beta_w z_0) ) &=& - \beta_s C_m e^{- \beta_s z_0}, \label{eq:BoundBZ1}\\
    - A_m \sin (\beta_w  z_0) + B_{m} \cos (\beta_w z_0) &=& \pm C_m e^{- \beta_s z_0}, \label{eq:BoundEZ2}\\
    \beta_w (A_m \cos (\beta_w z_0) + B_m \sin (\beta_w z_0) ) &=&  \pm  \beta_s C_m e^{ - \beta_s z_0}, \label{eq:BoundBZ2}\\
    \sin (p) J_{\lambda} (h r_1) +  \cos(p) Y_{\lambda}(h r_1) &=& 0,  \label{eq:R:BoundR1}\\
    \sin (p) J_{\lambda} (h r_2) +  \cos (p) Y_{\lambda}(h r_2) &=& 0.  \label{eq:R:BoundR2}
\end{eqnarray}
\end{subequations}
We will solve the above set of equations separately for both cases: when the field on the horizontal boundaries has the same value and when the field values differ by sign. We will call solutions for the first situation 'even' (subscript 'e') since solutions will be even functions of $z$. The other solutions will be called 'odd' (subscript 'o'). For this analysis we neglect the possible nonlinear response of the waveguide medium. Therefore, we can simplify our equation by assuming $C_m = 1$. After some algebra we obtain a transcendental equation for both types of solutions:
\begin{align}
     \tan (\beta_{w,o} z_0) &=  \frac{\sqrt{k_0^2(n_1^2  - n_2^2) - \beta_{w,o}^2}}{\beta_{w,o}}, \label{eq:oddsolution}\\ 
     \cot (\beta_{w,e} z_0) &= - \frac{\sqrt{k_0^2(n_1^2  - n_2^2) - \beta_{w,e}^2}}{\beta_{w,e}}.
    \label{eq:evensolution}
\end{align} 
Functions $\tan$ and $\cot$ are periodic which means that the above equations can in general have multiple roots. We will estimate the maximal possible number of solutions for $\beta_{w, o (e)}$.

\begin{figure}[h!]
     \centering
         \includegraphics[width=0.48\columnwidth]{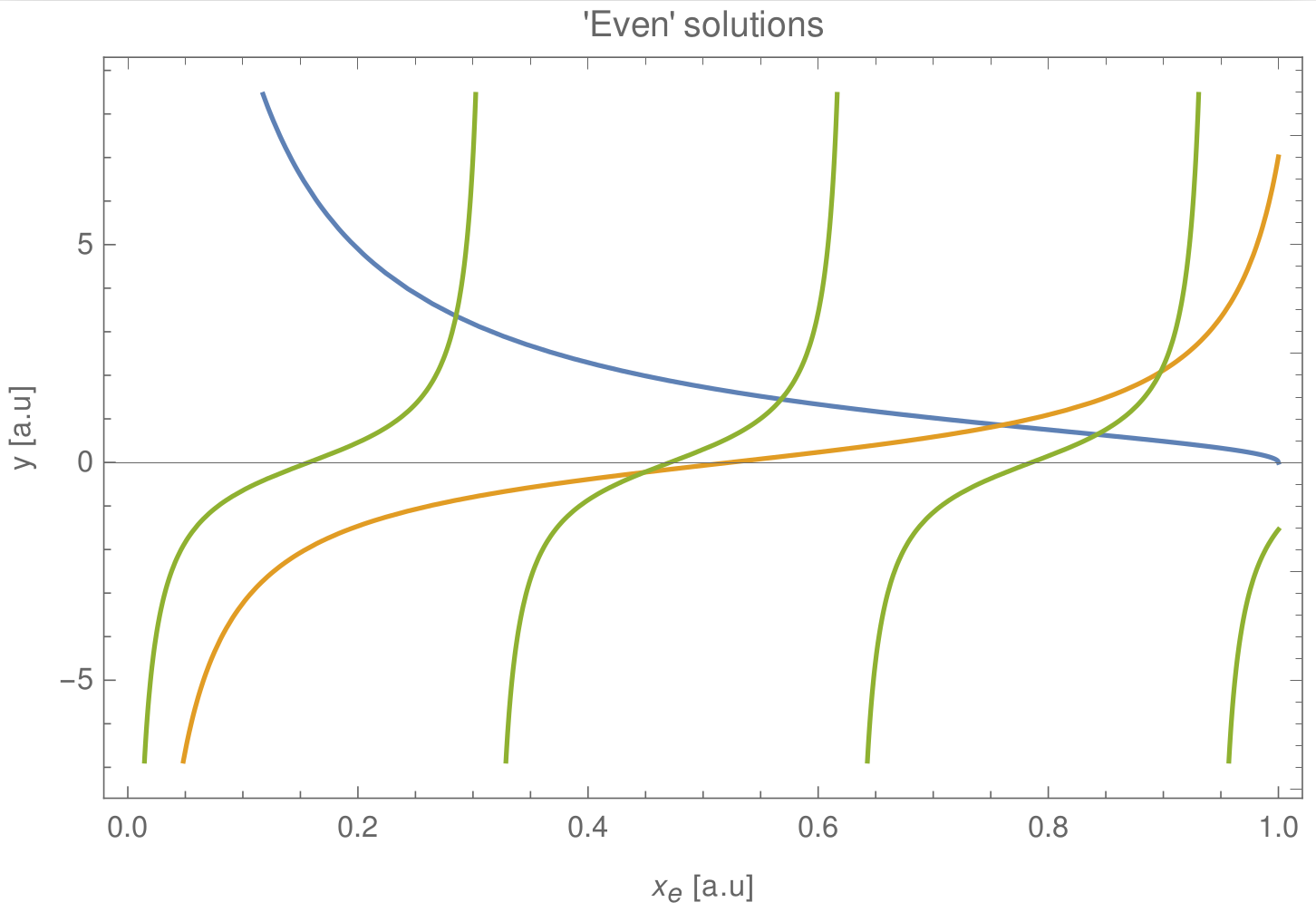}
         \includegraphics[width=0.48\columnwidth]{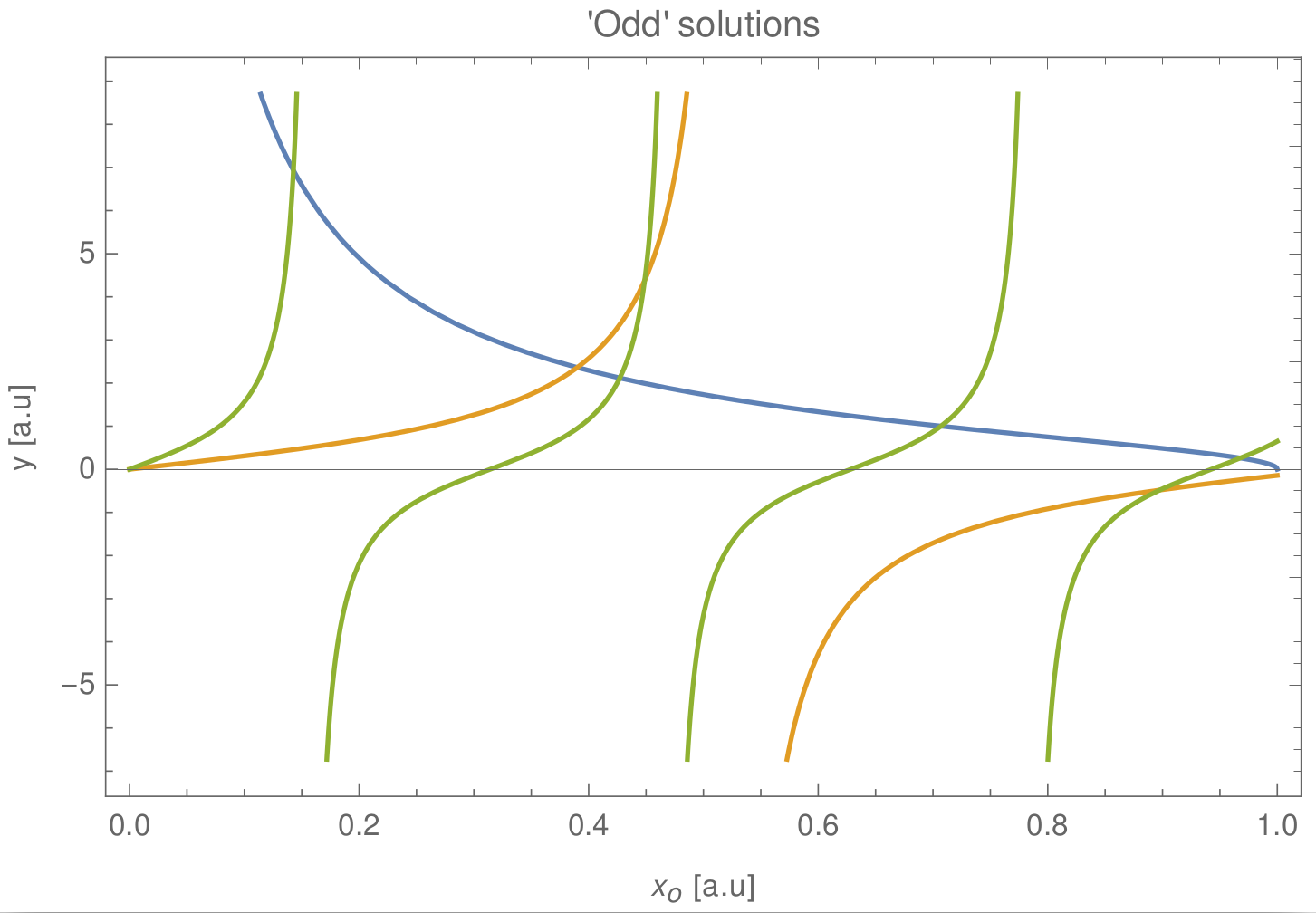}
        \caption{Blue line denotes $\sqrt{1 - x^2}/x$ function. Orange and green lines correspond to $l = 3$ and $l = 10$  respectively. On the left graph ('Even' solution) $\tan(x_e l)$ is depicted, and the right graph ('Odd solution') $-\cot(l x_o)$.}
\label{fig:solutions}
\end{figure}

We change variables so that $x_{o(e)} = \beta_{w,o(e)}/ (k^2\sqrt{n_1^2 - n_2^2})$. Then, the equations take the form:
\begin{align}
     \tan (x_o l) &=  \frac{\sqrt{1 - x_o^2}}{x_o}, \\ 
     - \cot (x_e l) &= \frac{\sqrt{1 - x_e^2}}{x_e},
\end{align} 
where $l = z_0 k^2 \sqrt{n_1^2 -n_2^2}$ and $x_{o,e} \in (0, 1]$. These functions are plotted in Fig. \ref{fig:solutions}. The maximal possible number of roots is given by the number of intersections of two plotted functions.
Each root corresponds to different mode. Moreover, because of relation $h^2 = k_1^2 - \beta_w^2$, it can be seen that $\beta_w$ parameterizes Bessel functions in \eref{eq:Erradial}.
The number of roots in both cases can be estimated by the number of positively valued 'arms' of $\tan$ and $-\cot$ functions. Therefore the number of roots for each equation can be estimated by: 
\begin{equation}
     q_o^\mathrm{max} = \lceil \frac{k z_0}{\pi} \sqrt{n_1^2 - n_2^2}\rceil,\,\, 
     q_e^\mathrm{max} = \lceil \frac{k z_0}{\pi} \sqrt{n_1^2 - n_2^2} -\frac{1}{2}\rceil,
    \label{eq:nmax}
\end{equation} 
where $\lceil$ $\rceil$ is the ceiling function. 

We will now proceed to analyze in a similar way conditions imposed by vertical boundaries. We can rewrite boundary conditions imposed on Bessel functions as:
\begin{equation}
    \left[ \begin{array}{c} 0 \\ 0 \end{array} \right] = \left[ \begin{array}{cc} J_\lambda(h r_1) &  Y_\lambda(h r_1) \\  J_\lambda(h r_2) &  Y_\lambda(h r_2) \end{array} \right]  \left[ \begin{array}{c} \sin(\gamma) \\  \cos(\gamma) \end{array} \right],   
\end{equation}
from which it follows that for solutions to exist the determinant has to be zero:
\begin{equation}
    0= J_\lambda(h r_1)  Y_\lambda(h r_2) - J_\lambda(h r_2)   Y_\lambda(h r_1). \label{eq:determinant}
\end{equation}
The above equation has to be solved together with one of the boundary conditions. These equations are challenging to solve analytically, so to estimate the number of roots we will make approximations. Let us rewrite the Bessel equation in the following form:
\begin{equation}
\frac{\partial^2 R}{\partial r^2} + \frac{1}{r}\frac{\partial R}{\partial r} + \lambda^2 (\frac{1}{r_{av}^2} - \frac{1}{r^2})  R =  (\frac{\lambda^2}{r_{av}^2} - h^2)R, \label{eq:diffequation}
\end{equation}
where $r_{av} = (r_2 + r_1)/2$. We define $\Delta r = r_2 - r_1$ which is waveguide width. For waveguides where $\frac{\Delta r}{r_{av}} \ll 1$ quantity $1 / r_{av}^2 - 1/ r^2$ is small and we can discard this term. The resulting differential equation has as solutions Bessel functions: $J_0$ and $Y_0$. These Bessel functions do not bring us closer to obtaining a simple form for the number of solutions for  \eref{eq:Bessel}. We are therefore making another approximation and replace $1/r$ by $1/r_{av}$:
\begin{equation}
\frac{\partial^2 R}{\partial r^2} + \frac{1}{r_{av}}\frac{\partial R}{\partial r}  + (h^2 -\frac{\lambda^2}{ r_{av}^2})R =  0.
\end{equation}
This is a second-order differential equation and therefore it has two solutions:
\begin{equation}
  R(r) = C_1 \exp[- \frac{r}{2}(\sqrt{\frac{1}{r_{av}^2} -4 (h^2 - \frac{\lambda^2}{r_{av}^2})} + \frac{1}{r_{av}})] + C_2 \exp[\frac{r}{2}(\sqrt{\frac{1}{r_{av}^2} -4 (h^2 - \frac{\lambda^2}{r_{av}^2})}- \frac{1}{r_{av}})].  
\end{equation}
We simplify this form to:
\begin{equation}
    R(r) = \exp[-\frac{r(1 + r_{av} \sqrt{\frac{1 + 4 \lambda^2 }{r_{av}^2} - 4h^2})}{2 r_{av}}](C_1 + C_2 \exp[r \sqrt{\frac{1 + 4 \lambda^2 }{r_{av}^2} - 4h^2}]).
\end{equation}
Now, we apply boundary conditions, that is $R(r_{av} \pm \Delta r /2) = 0$ :
\begin{align}
    C + \exp[(r_{av} \pm \frac{\Delta r}{2}) \sqrt{\frac{1 + 4 \lambda^2 }{r_{av}^2} - 4h^2}] &= 0,
\end{align}
where $C = C_1/C_2$. The boundary condition for $-\Delta r/2$ gives us $C$. The second boundary condition gives us an equation for $\lambda$:
\begin{align}
    - \exp[- \frac{\Delta r}{2} \sqrt{\frac{1 + 4 \lambda^2 }{r_{av}^2} - 4h^2}] + \exp[\frac{\Delta r}{2} \sqrt{\frac{1 + 4 \lambda^2 }{r_{av}^2} - 4h^2}] &= 0,
\end{align}
Roots of this equations are:
\begin{equation}
    \lambda^2 =  h^2 r_{av}^2 - \frac{1}{4} - \frac{\pi^2 r_{av}^2 p^2}{\Delta r^2}, \thickspace p \in \mathcal{N},
\end{equation}
which finally brings us to equation for $m$:
\begin{equation}
    m = r_{av} \sqrt{h^2 - \frac{\pi^2 p^2}{\Delta r^2}  - \frac{5}{4r_{av}^2}}, \thickspace p\in \mathcal{N} . \label{eq:mvalues}
\end{equation}
We use these solutions to get the final expression for $R(r)$:
\begin{equation}
    R(r) = \exp(-\frac{r}{2 r_{av}} - i \frac{\pi p r}{\Delta r})[\exp(i\frac{2 \pi p r}{\Delta r}) - \exp(i\frac{\pi p (2 r_{av} + \Delta r)}{\Delta r})].
\end{equation}
These solutions are almost orthogonal for different $p$ values in a sense that:
\begin{equation}
    \forall_{p_1 \neq p_2} \int_{r - \Delta r/2}^{r + \Delta r/2} R(r, p_1) \overline R(r, p_2) \approx 0.
\end{equation}
This property can be used to further improve our solutions - differential equation \eref{eq:diffequation} might be replaced by matrix eigenvalue equation with perturbation. Solving that equation would yield corrections to functions and eigenvalues but would not change the number of roots.    

The number of roots can be estimated from \eref{eq:mvalues}. Since $m$ must be a real parameter, $h^2 - \frac{\pi^2 p^2}{\Delta r^2}  - \frac{5}{4r_{av}^2} > 0$. This leads us to the expression for the number of solutions:
\begin{equation}
    p_{max} \approx \lfloor \frac{h \Delta r}{\pi} \rfloor, \label{eq:radialsolnum}
\end{equation}
where $\lfloor \rfloor$ is a floor function. This expression agrees with the one that was presented in the article \cite{Gajewski2020} of which I was a co-author.
It is worth noting that the presented result is asymptotically correct. If we use for calculations the asymptotic forms of the Bessel functions that is \cite{WolframBesselFirst, WolframBesselSecond}:
\begin{align}
    J_\lambda (r) & \approx \sqrt{\frac{2}{\pi r}} \cos(r- \frac{\lambda \pi}{2} - \frac{\pi}{4}), \\
    Y_\lambda(r) & \approx  \sqrt{\frac{2}{\pi r}}  \sin(r - \frac{\lambda \pi}{2} - \frac{\pi}{4}), 
\end{align}
and we plug them into \eref{eq:determinant} we get an expression:
\begin{equation}
    \sin (h \Delta r) = 0
\end{equation}
which has the same solution for maximal number of roots as \eref{eq:radialsolnum}.
\subsection{Characterisation of modes}
Mode frequency and refractive index contrast give us the largest possible value of parameter $\beta_w$ which describes the number of oscillation in the $z$ direction (\eref{eq:oddsolution}, \eref{eq:evensolution}). The height of the waveguide together with $\beta_w$ indicates the number of nodes in the $z$- direction and hence it also indicates one of the mode characteristics. The bigger the values of $\beta_w$ the smaller the value of $h$. Value of $h$ specifies the oscillatory behaviour of the electric field along the $r$-axis. Just as in the case of dielectric waveguides, we can denote our modes as $E^r_{pq}$, where $r$ indicates that mode has the largest electric field component along the $r$-axis. The largest field component is used for describing the mode number with indices $p, q$. Mode numbers $p,q$ are associated with oscillations along the $r$-axis and $z$-axis, respectively. 

Let us now consider a waveguide with the inner bending radius $r_1 = 0.5 \,\mu \mathrm{m}$, outer radius $r_2 = 1.5 \mu m$ and height $2 z_0 = 0.5 \,\mu \mathrm{m}$. Refractive indices are $n_w = 2.3, \thickspace n_s = 1$. We want to couple to such waveguide electromagnetic field with a wavelength of $800$ nm.

\begin{figure}[h!]
     \centering
         \includegraphics[width=0.48\columnwidth]{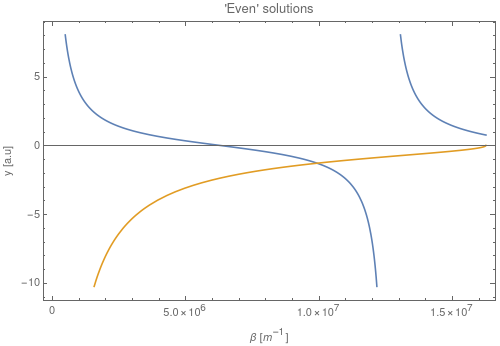}
         \includegraphics[width=0.48\columnwidth]{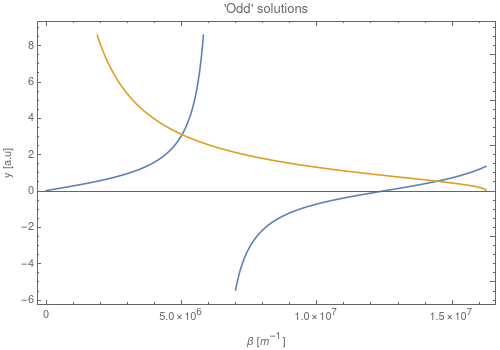}
        \caption{Visualisation of graphical solutions for equations \eref{eq:evensolution} (graph on the left) and \eref{eq:oddsolution} (graph on right) and waveguide described in text. }
\label{fig:solutions_example}
\end{figure}

From equation \eref{eq:nmax}, it follows immediately that $q_{max} = 3$ since $q_o^{max} = 2$ and $q_e^{max} = 1$. We can check if those numbers are correct by looking at graphs in Fig. \ref{fig:solutions_example} - values of $\beta$ at the intersection points, are the solutions $\beta_w$ we are looking for.  These values of $\beta_w$ can be found in Tab. \ref{tab:out}.

For each value of $q$ we have corresponding $p_{max}$ which can be estimated with \eref{eq:radialsolnum}. The number of solutions together with calculated values of $m$ for each mode can be found in Tab. \ref{tab:out}. 
\begin{table}[h!]

  \begin{center}
    \begin{tabular}{c|ccc}
    \hline
   $n$ & $1$ & $2$ & $3$\\
    \hline
    $\beta_w$  [ $\mu m^{-1}$ ] & 5.03 & 9.94 & 14.46\\
    $h$  [ $\mu m^{-1}$ ] & 17.35 & 15.09  & 10.8\\
    \hline
    $(m, n_\mathrm{eff})$ & $q=1$ & $q=2$ & $q=3$ \\
    \hline
    $p=1$ & (20.54, 2.03) &  (17.391, 1.75) & (11.53, 1.22) \\
    $p=2$ & (16.50, 1.86) & (13.54, 1.58) & (8.08, \textcolor{red}{1.04}) \\
    $p=3$ & (13.23, 1.69) & (10.41, 1.40) & (4.56, \textcolor{red}{0.6}) \\
    $p=4$ & (10.26, 1.46) &  (7.15, \textcolor{red}{1.02}) & --- \\
    $p=5$ & (6.03, \textcolor{red}{0.9}) & --- & --- \\ 
    \end{tabular}
  \end{center}
  \caption{Table is adapted from \cite{Gajewski2020}.The parameters of the spatial modes for the waveguide geometry specified in the main text. The value of $m$ is computed based on \eref{eq:determinant}. The values of $n_\mathrm{eff} \lesssim 1$ marked in red correspond to non-physical solutions.
  }
  \label{tab:out}
\end{table}
In the table, there is also another parameter called $n_{eff}$ which stand for effective refractive index. The effective refractive index is a very handy parameter used i.e by the COMSOL software to search for modes in waveguide structures. It can be also used to sort the modes. Its value can be approximated by:
\begin{equation}
    n_{eff} \approx \frac{m}{k \langle r \rangle_{p,q}},
\end{equation}
where $\langle r \rangle_{pq}$ is defined as :
\begin{align}
    \langle r \rangle_{pq} = \frac{\iint |E^r_{pq}(r, 0,z)|^2  r \text{d}r \text{d}z}{\iint |E^r_{pq}(r, 0,z)|^2 \text{d}r \text{d}z}. 
\end{align}
We have calculated the values of $ \langle r \rangle_{pq}$. The comparison of $ \langle r \rangle_{pq}$ for our analytical modes and numerical modes is summarized in Tab. \ref{tab:my_label}.
Another useful property of an effective refractive index is to indicate if a given mode is guided or not. For a mode to be guided $n_{eff} > n_2$. In the tables Tab. \ref{tab:out} and Tab. \ref{tab:my_label} we marked in red all the $n_{eff}$ values which might correspond to non-physical (not guided) solutions. 
\begin{table}[]
    \centering
    \scalebox{0.9}{
    \begin{tabular}{ccccccc}
    \hline
    Mode  & \multicolumn{6}{c}{Average Radial Position $\langle r \rangle_{pq}$ ($\mu m$)}\\
    \cline{2-7}
    Number & \multicolumn{3}{c}{Analytical} & \multicolumn{3}{c}{Numerical} \\
    \hline
   &$p=1$&$q=2$&$q=3$&$q=1$&$q=2$&$q=3$ \\
    $p=1$ & 1.29 & 1.27 & 1.21 & 1.21 & 1.17 & 1.05 \\
    $p=2$ & 1.13 & 1.09 & \textcolor{red}{0.99} & 1.04 & 0.98 &  -  \\
    $p=3$ & 1.00 & 0.95 & \textcolor{red}{0.9} & 0.95 &  -   &  - \\
    $p=4$ & 0.90 & \textcolor{red}{0.89} & - & 0.95 &  -   &  - 
          \end{tabular}
          }
    \caption{Table is adapted from \cite{Gajewski2020}. Average radial position of waveguide modes computed analytically and numerically. The values marked in red correspond to non-physical solutions (See Tab.~\ref{tab:out}).}
    \label{tab:my_label}
\end{table}

\subsection{Numerical calculations}

\begin{figure}[h!]
     \centering
         \includegraphics[width=0.7\columnwidth]{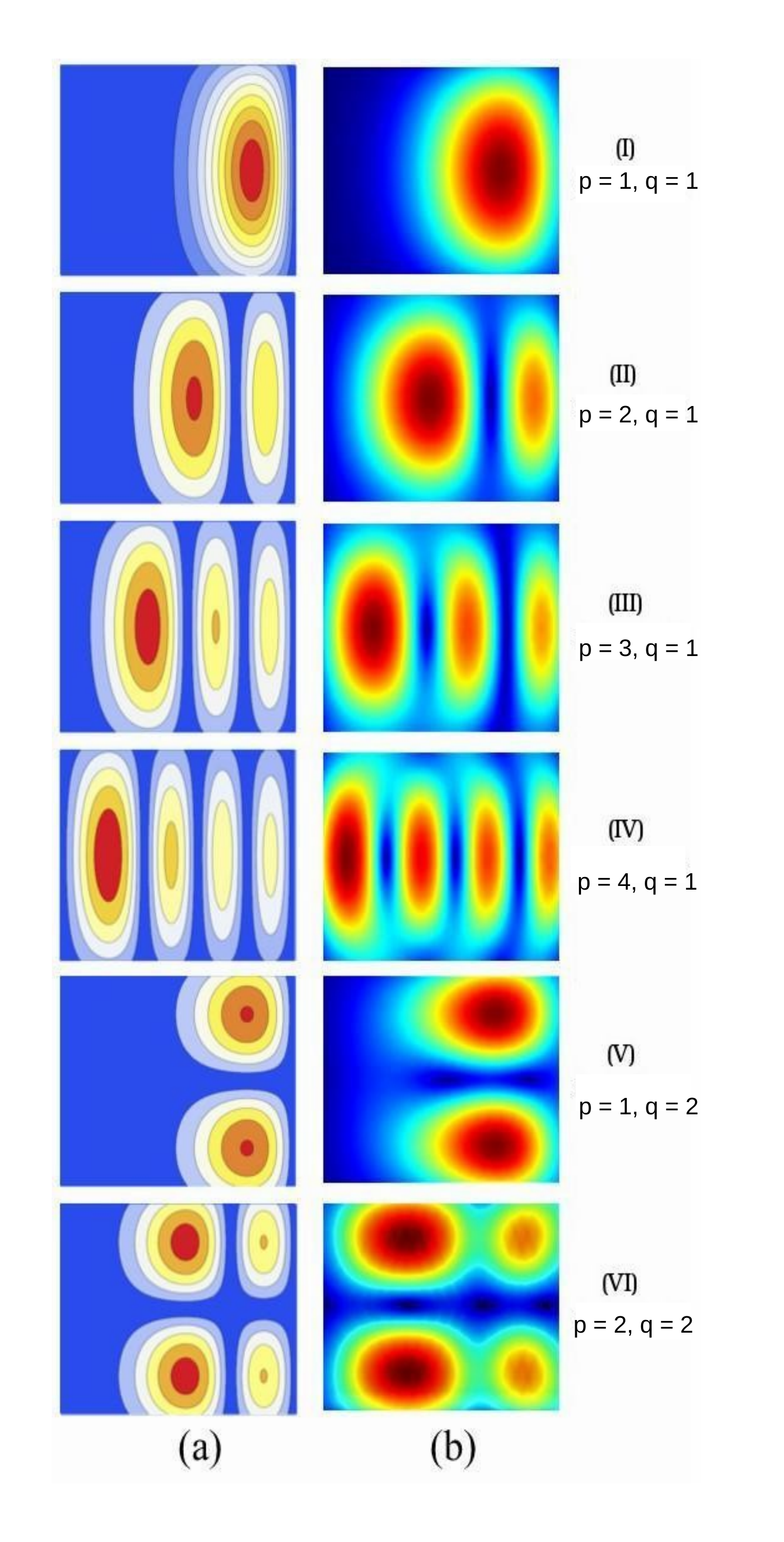}
         \caption{Graphs are adapted from \cite{Gajewski2020} and show spatial modes in bent waveguides. Graphs from the left column visualize analytical results presented in this chapter. Graphs on the right were generated with COMSOL software. Numbers in each row indicate mode numbers.}
\label{fig:modesCOMSOL}
\end{figure}

My colleague Nor Roshidah Yusof performed numerical calculations in the COMSOL Multiphysics software. COMSOL is a commercial software design for assistance in modelling and simulation of problems arising in many areas of physics and engineering. It uses the finite element method \cite{jin2015} for solving differential equations and has methods dedicated to eigenfrequency problems arising in waveguide simulations. Roshidah used that module to find the modes in a bent waveguide under the assumption of full axial symmetry. In COMSOL, we built a fully axisymmetric model which corresponds to a full ring instead of a bent waveguide.  Such structure allows for a great simplification of calculations, but it also means that $m$ parameter, which in our analysis could be any positive real number, in COMSOL simulation is assumed to be a positive integer. In Fig. \ref{fig:modesCOMSOL}, we present a comparison between our analytical method and the numerical result obtained with COMSOL software.
In that figure, we have plotted a $|E^r|$ component of the electric field for 6 different modes. Clearly, our analytical model captures well the general behaviour observed in numerical simulations. In numerical and analytical results we can see that spacing between nodes and their position (Tab. \ref{tab:my_label}) is similar. Numerically obtained modes are approximately $0$ at the sidewalls of the waveguide and are shifted toward the upper and lower wall. This shifting can be observed also in our analytical model and is a result of assumed exponential decay outside the waveguide.

It can also be observed that our analytical approximation works better for modes with a high $n_{eff}$ value (see. Tab. \ref{tab:out}). When that value approaches the value of the refractive index of the surrounding medium, modes are no longer well contained. When that happens, modes lose their symmetry which we attribute to the influence of the ignored corners. Just like in Marcatili’s approach (see Fig. \ref{fig:Dielectric}), we assumed that field in the corners (shaded areas in \ref{fig:Dielectric})is always equal to $0$.  
This observation is confirmed by the analysis of the average radial position $\langle r \rangle_{pq}$. If we compare the values presented in the Tab. \ref{tab:my_label} we will note that for the low mode numbers, our analytically computed $\langle r \rangle_{pq}$ agrees very well with the one obtained from COMSOL simulation. 

It can be seen from the Tab. \ref{tab:my_label} that our analytical approach led to more solutions than we were able to obtain through numerical simulations. But with the additional analysis of $n_{eff}$ we were able to further improve our analytical model and reject solutions which lead to non-physical modes. There is only one mode that our combined approach predicted to exist but which did not appear in our simulations - $E^r_{32}$. It is not however clear if that is because of our approximate method or because of the challenging nature of COMSOL simulations. 

\subsection{Summary}

We have formulated an eigenvalue problem for a waveguide with a bent structure and show that it has an analogous form to the radial Schrödinger equation. We presented approximate solutions to this eigenvalue problem given the boundary condition enforced by the waveguide geometry. The mathematical analysis presented in this chapter is more detailed than the one presented in the article ref. \cite{Gajewski2020} and also more precise. 

We have also formulated an expression for the maximal number of modes. We showed that this expression together with solutions to the eigenvalue problem are similar to the results obtained through numerical simulation in the COMSOL software. We managed to find approximate equations for the maximal mode numbers of modes (\eref{eq:radialsolnum} and \eref{eq:nmax}) for a given waveguide and wavelength. 

The research regarding modes in the bent waveguide was motivated by our interest in quantum fictitious forces, especially a quantum anticentrifugal force. The equations for the radial component of the electric field \eref{eq:radialBessel} arise naturally in the context of Schrödinger equation in axial symmetry. Here we presented a solution to that equation in a geometry of a bent waveguide. Since the equations are the same we fully expect that our analysis can be directly translated to an analysis of particle trapped between two semi-circular walls as proposed in ref. \cite{Dandoloff2011}.

Our research offers not only insight into the behaviour of a particle in axially symmetrical two-dimensional space, but we have also shown that it offers a means for designing an experimental investigation where the theory can be tested. The research might include an investigation for i.e a "pizza slice" part of a disk,  for cases where $m$ is not a positive integer but has a complex value (by introducing absorption) or for a series of coaxial bent waveguides where tunnelling effect occurs. All these experiments are physically realizable in contrast to two-dimensional space with a Dirac delta potential. 

%% file: Chapters/summary.tex
\chapter{Summary} 

\label{summ}

In this dissertation, the author presented the results within three research areas that the author dealt with during his doctoral studies. The presentation of the results was preceded by a theoretical introduction, which included the concepts and issues related to the presented results. In this introduction, the author tried to briefly, but  mathematically strictly present the key physical issues for his work.

In the first chapter concerning research, the author presented a method for determining Sellmeier coefficients in nonlinear crystals under phase-matching conditions. This method is based on a mathematical model of the phase-matching phenomenon. This model was developed by the author. At the end of this chapter, the author presents the Sellmeier coefficients corrected against the reported once in the literature, showing that the Sellmeier equation with the newly determined coefficients better corresponds to the experimental results.

The fourth chapter deals with the modelling of the SPDC effect in nonlinear crystals. These studies are a continuation of the author's research that started during the author's master's studies. In this chapter, author have extended the mathematical formalism to include quasi phase-matching. The author have used this formalism to model the SPDC effect in biaxial periodic nonlinear crystals. The model makes it possible to obtain an approximate form of the wave function of two photons in the frequency domain, where one or both photons from the SPDC are coupled to the optical fibre. Author used the wave function obtained in this way to calculate the statistical parameters, which was compared directly with the results from two separate experiments. Author showed that the model allows us to obtain a very good approximation of the experimental results. The sources of differences between  the numerical predictions and the experimental results were also indicated. 

In the fifth chapter, the author presented the results of his theoretical research on the propagation of light in a bent waveguide. Author solved an approximate form of the eigenmode equation. The analytical modes were compared with modes obtained in numerical simulation. It was shown that the solutions are in agreement with numerical prediction not only for the mode numbers but also for their spatial form. Author presented the equations which allow for the calculation of maximal mode numbers for a given geometry of the bent waveguide and given light wavelength. The analogy of obtained equation to the radial Schrödinger equation was shown and it is discussed in detail in the appendix which concludes this dissertation. 

%% file: Appendices/AppendixA.tex

\chapter*{Appendix} 


\section{Nabla operator in cylindrical coordinates \label{AppendixNabla}}
 
In cylindrical coordinates, any vector can be expressed as:
\begin{align}
    \boldsymbol A &= A_r \boldsymbol r + A_\phi \boldsymbol \phi +  A_z \boldsymbol z \label{eq:cylVec},
\end{align}
where $\boldsymbol r$, $\boldsymbol \phi$ and $\boldsymbol z$ stand for unit vectors. In the following work we are going to use explicit forms of divergence and curl of a vector field in cylindrical coordinates:
\begin{align}    
    \nabla \cdot \boldsymbol A &= \frac{1}{r} \frac{\partial(r A_r)}{\partial r} + \frac{1}{r}\frac{\partial A_\phi}{\partial \phi} + \frac{\partial A_z}{\partial z} \label{eq:cylDiv} \\
    \nabla \times \boldsymbol A &=(\frac{1}{r} \frac{\partial A_z}{\partial \phi} - \frac{\partial A_\phi}{\partial z})\boldsymbol r  + (\frac{\partial A_r}{\partial z} - \frac{\partial A_z}{\partial r})\boldsymbol \phi + \frac{1}{r}(\frac{\partial(r A_\phi)}{\partial r} - \frac{\partial A_r}{\partial \phi})\boldsymbol z \label{eq:cylRot},
\end{align}
and also an explicit form of Laplace operator ($\nabla^2$) of both vector and scalar fields:
\begin{align}
    \nabla^2 \boldsymbol A &= (\nabla^2 A_r - \frac{A_r}{r^2} - \frac{2}{r^2}\frac{\partial A_\phi}{\partial \phi}) \boldsymbol r + (\nabla^2 A_\phi - \frac{A_\phi}{r^2} - \frac{2}{r^2}\frac{\partial A_r}{\partial \phi}) \boldsymbol \phi + (\nabla^2 A_z) \boldsymbol z \label{eq:cylVecLap} \\
    \nabla^2 f &= \frac{1}{r} \frac{\partial}{\partial r}(r \frac{\partial f}{\partial r}) + \frac{1}{r^2}\frac{\partial^2 f}{\partial \phi^2} + \frac{\partial^2 f}{\partial z^2} \label{eq:cylScalLap}
\end{align}
These equations will be used later in the mode analysis of the bent waveguide.

\section{Transformation of biphoton wavefunction by a fiber \label{AppendixTransform}}

The wavevector of the photon coupled to a single-mode fiber undergoes transformation while it propagates through the fiber. We can denote that transformation as $\hat S_D$:
\begin{align}
    | \psi_{out} \rangle = \hat S_D | \psi_{in} \rangle, \label{eq:Soperator}
\end{align}
where $\psi_{in}$ is a wavevector of photon coupled to the fiber at one end and $\psi_{out}$ is its wavevector after propagating through distance $D$ in the fiber.
Photon wavevector can be described in terms of its temporal part and spatial parts: 
\begin{align}
\langle r, t| \psi \rangle = \psi_{\text{temporal}}(t) \psi_{\text{spatial}}(\vec r).
\end{align}
The spatial part $\psi_{\text{spatial}}(\vec r)$ is completely  defined by the fiber mode. We are interested in the transformation of the temporal part $\psi_{\text{temporal}}(t)$. We will call that part a wavefunction.
We can rewrite equation \eref{eq:Soperator} as:
\begin{align}
    \psi_{out} (t) &= \int \psi_{in}(\tau)S_D(t - \tau) d\tau,\\
    S_D(t) &= \frac{1}{\sqrt{i \pi 4 \beta D}} \exp{(i \frac{t^2}{4 \beta D})},
\end{align}
where $\beta$ is group velocity dispersion of the fiber. The above equation is taken from \cite{Sedziak2017} and can be found in master and doctoral theses of Karolina Sedziak. We can generalize the above formulation to include a description of the transformation of the biphoton wavefunction, where the two photons are coupled to two fibers \cite{Sedziak2017}:
\begin{align}
\psi_{out} (t_1,t_2) = \iint \psi_{in} (\tau_1, \tau_2) S_D(t_1 - \tau_1)&S_D(t_2 - \tau_2) d\tau_1 d\tau_2. \label{eq:generaltrans}
\end{align}
In the scenario considered here, both fibers have the same length and the same group velocity dispersion. The temporal biphoton wavefunction is a Fourier transform of the biphoton wavefunction in the frequency domain:
\begin{align}
 \psi (t_1,t_2) = \frac{1}{2 \pi} \iint \tilde{\psi}(\omega_1,\omega_2) & \exp(-i (t_1 \omega_1 + t_2 \omega_2)) d\omega_1 d\omega_2 \label{eq:psifourier}.    
\end{align}
We replace the temporal functions in \eref{eq:generaltrans} with their Fourier transforms:
\begin{align}
\iint\tilde \psi_{out}(\omega_1,\omega_2) e^{-i (t_1 \omega_1 + t_2 \omega_2)} d\omega_1 d\omega_2 &= \iiiint \tilde \psi_{in} (\omega_1',\omega_2') e^{-i (\tau_1 \omega_1' + \tau_2 \omega_2')} \times \dots  \\ & \dots \times S_D(t_1 -\tau_1)S_D(t_2 -\tau_2) d\tau_1 d\tau_2 d\omega_1' d\omega_2'.
\end{align}
The above equation is true if integrated functions are equal:
\begin{align}
\tilde \psi_{out}(\omega_1,\omega_2) e^{-i (t_1 \omega_1 + t_2 \omega_2)} &= \iint \tilde \psi_{in} (\omega_1,\omega_2) e^{-i (\tau_1 \omega_1 + \tau_2 \omega_2)}  S_D(t_1,\tau_1)S_D(t_2,\tau_2) d\tau_1 d\tau_2.
\end{align}
We can transform the above equation into the following:
\begin{align}
\tilde \psi_{out}(\omega_1,\omega_2) &= \frac{1}{i \pi 4 \beta D}  \tilde  \psi_{in} (\omega_1,\omega_2) \iint e^{i\phi(\omega_1,\omega_2;\tau_1, \tau_2)} d\tau_1 d\tau_2, \\
\phi(\omega_1,\omega_2;\tau_1, \tau_2) &= \frac{1}{4 \beta D} [(t_1-\tau_1)^2 + (t_2-\tau_2)^2]+ (t_1-\tau_1)\omega_1 + (t_2-\tau_2)\omega_2.
\end{align}
Finally, after performing integration we obtain an analytical equation describing the transformation which the biphoton wavefunction undergoes during the propagation:
\begin{align}
\tilde\psi_{out}(\omega_1,\omega_2) &=  \tilde \psi_{in} (\omega_1,\omega_2) \exp(-i\beta D (\omega_1^2 + \omega_2^2)).
\end{align}
If we now use the above result in \eref{eq:psifourier} we obtain:
\begin{align}
\psi_{out}(t_1,t_2) &= \frac{1}{2 \pi} \iint \tilde \psi_{in}(\omega_1,\omega_2) \exp[-i(\beta D (\omega_1^2 + \omega_2^2) +  \omega_1 t_1 + \omega_2 t_2)] d \omega_1 d \omega_2 \label{eq:psiout}
\end{align}
Using above formula might be challenging - for every pair of $t_1$ and $t_2$ we have to perform a double integral over whole frequency domain. Although it is very difficult to obtain an exact analytical relation which would circumvent that problem, we can obtain an approximate relation. First, we note that the phase term in \eref{eq:psiout} has a stationary point at $\{ \omega_1, \omega_2 \} = \{- \frac{t_1}{2 \beta D},- \frac{t_2}{2 \beta D} \}$. In the vicinity of this point first derivatives are approximately equal to $0$. This observation allows us to obtain an approximate expression of \eref{eq:psiout} with the stationary approximation method \cite{Arfken2013}. First, we rewrite \eref{eq:psiout}:
\begin{align}
 \psi_{out}(t_1,t_2) &= \frac{1}{2 \pi} \iint \tilde \psi_{in}(\omega_1,\omega_2) \exp[-i x f(\omega_1, \omega_2)] d \omega_1 d \omega_2, \\
f(\omega_1, \omega_2) &= (\omega_1^2 + \omega_2^2) +  \frac{\omega_1 t_1 + \omega_2 t_2}{x}, \thickspace x = \beta D.
\end{align}
We expand in Taylor series $f(\omega_1,\omega_2)$ around stationary point:
\begin{align}
    f(\omega_1,\omega_2) \approx f(\Omega_1 , \Omega_2) +& \frac{\partial^2 f}{\partial \omega_1^2}(\Omega_1 , \Omega_2) \frac{(\omega_1 - \Omega_1)^2}{2} + \frac{\partial^2 f }{\partial \omega_2^2}(\Omega_1 , \Omega_2) \frac{(\omega_2 - \Omega_2)^2}{2} \\
    &\Omega_1 = - \frac{t_1}{2 x} , \Omega_2 = - \frac{t_2}{2 x} 
\end{align}
Both second derivatives are equal to $2$. With the above Taylor expansion we can approximate our integral:
\begin{align}
\psi_{out}(t_1,t_2) & \approx \frac{e^{-ixf(\Omega_1 , \Omega_2)}}{2 \pi} \iint \tilde \psi_{in}(\omega_1,\omega_2) e^{-i x [(\omega_1 - \Omega_1)^2 + (\omega_2 - \Omega_2)^2]} d \omega_1 d \omega_2.
\end{align}
For $x$ large enough, function $\tilde \psi_{in}(\omega_1,\omega_2)$ does not change in the vicinity of the stationary point:
\begin{align}
\psi_{out}(t_1,t_2) & \approx \frac{e^{-ixf(\Omega_1 , \Omega_2)}}{2 \pi} \tilde \psi_{in}(\Omega_1 , \Omega_2) \iint e^{-i x [(\omega_1 - \Omega_1)^2 + (\omega_2 - \Omega_2)^2]} d \omega_1 d \omega_2.
\end{align}
Calculation of the integral gives us our approximation of the fiber transformation:
\begin{align}
\psi_{out}(t_1,t_2) & \approx \frac{-i e^{-ixf(- \frac{t_1}{2 \beta D},- \frac{t_2}{2 \beta D})}}{2 \beta D} \tilde \psi_{in}(- \frac{t_1}{2 \beta D},- \frac{t_2}{2 \beta D}) .
\end{align}
Probability density of detecting photons at times $t_1$ and $t_2$ at the ends of two fibers of length $D$ is:
\begin{align}
p(t_1,t_2) = |\psi_{out}(t_1,t_2)|^2 \approx \frac{|\tilde \psi_{in}(- \frac{t_1}{2 \beta D},- \frac{t_2}{2 \beta D})|^2}{(2 \beta D)^2}  .
\end{align}
Let us change variables so that $\langle t_{1,2}\rangle = 0$. In this coordinate system we have:
\begin{align}
p(t_1,t_2)  \approx \frac{|\tilde \psi_{in}'(\langle \omega_1 \rangle - \frac{t_1}{2 \beta D}, \langle \omega_2 \rangle - \frac{t_2}{2 \beta D})|^2}{(2 \beta D)^2}.
\end{align}
We can now relate statistical properties of time variables $t_1$ and $t_2$ given by distribution $p(t_1,t_2)$ with statistical properties of $\omega_1$ and $\omega_2$ given by distribution: 
\begin{align}
p(\omega_1,\omega_2) = |\psi_{in}'(\langle \omega_1 \rangle - \frac{t_1}{2 \beta D}, \langle \omega_2 \rangle - \frac{t_2}{2 \beta D})|^2 d\omega_1 d\omega_2.    
\end{align}
The expectation values $\langle t_{1,(2)} \rangle$ are $0$, therefore the expression for the time variances $\sigma_{t_{1,(2)}}^2$ is:
\begin{align}
    \sigma_{t_{1,(2)}}^2 &= \iint t_{1,(2)}^2 p(t_1,t_2) dt_1 dt_2 \approx \iint t_{1,(2)}^2 \frac{|\tilde \psi_{in}'(\langle \omega_1 \rangle - \frac{t_1}{2 \beta D}, \langle \omega_2 \rangle  - \frac{t_2}{2 \beta D})|^2}{(2 \beta D)^2} dt_1 dt_2.
\end{align}
We calculate above integral by substitution $\omega_{1,(2)}' = \frac{t_{1,(2)}}{2 \beta D}$:
\begin{align}
    \sigma_{t_{1,(2)}}^2 & \approx (2 \beta D)^2 \iint (\omega'_{1,(2)})^2 |\tilde \psi_{in}'(\langle \omega_1 \rangle - \omega'_1, \langle \omega_2 \rangle  - \omega'_2)|^2 d\omega'_1 d\omega'_2 = (2 \beta D \sigma_{\omega_{1,(2)}})^2.
\end{align}
As we can see, the variance of photon arrival time at the end of the fiber is equal to the variance of the angular frequency of the coupled photon multiplied by a constant $(2 \beta D)^2$. The same can be shown for the covariance function $Cov(t_1, t_2) = (2 \beta D)^2 Cov (\omega_1, \omega_2)$ which in turn means that the Pearson correlation coefficient given by \eref{eq:rho} of the arrival time of both photons $\rho(t_1, t_2)$ is the same as the Pearson correlation coefficient of frequencies of coupled photons $\rho(\omega_1, \omega_2)$. We use that result in chapter \ref{Chapter5}.

\section {Quantum anticentrifugal force analogy in a bent waveguide} \label{apendixQFF}
We start with the time independent Schrödinger equation for a particle in two-dimensional space \cite{shankar2012principles}:
\begin{align}
    \hat H \psi (x, y) = E \psi(x, y), \label{eq:indepSch}
\end{align}
where $\hat H$ is a Hamiltonian and $E$ stands for the energy of the particle. The Hamiltonian of a free particle in two-dimensional space is:
\begin{align}
    \hat H = - \frac{\hbar^2}{2 M} \Delta,
\end{align}
where $\hbar$ is a reduced Planck constant and $\Delta$ is a Laplacian. The Laplacian in Cartesian coordinates can be expressed as:
\begin{align}
    \Delta = \frac{\partial^2}{\partial x^2} + \frac{\partial^2}{\partial y^2}.
\end{align}
The choice of Cartesian coordinates is understandable in case of $R^2$ space, but it is not very practical for $R^2 \backslash \{ 0,0 \}$ or (equivalently) for $R^2$ space with Dirac delta potential. The problem arises due to normalization - for either pointed plane or Dirac delta potential normalization equation has a form:
\begin{align}
   \lim_{\epsilon \rightarrow 0^+} \iint_{R^2 \backslash  C(\epsilon)}  |\psi (x, y)|^2 dxdy \equiv 1,
\end{align}
where $C(\epsilon)$ denotes a circular area with radius $\epsilon$ centered at the origin. The same equation in polar coordinates has a form:
\begin{align}
   \lim_{\epsilon \rightarrow 0^+} \int_0^{2 \pi} d\phi \int_{\epsilon}^{\infty}  |\psi (r, \phi)|^2 r dr  \equiv 1.
\end{align}
This form is easier to work with and justifies our choice of planar coordinates. The Laplacian in planar coordinates is:
\begin{align}
    \Delta = \frac{\partial^2}{\partial r^2} + \frac{1}{r}\frac{\partial}{\partial r} +\frac{1}{r^2} \frac{\partial^2}{\partial \phi^2}.
\end{align}
We rewrite \eref{eq:indepSch}:
\begin{align}
    - \frac{\hbar^2}{2 M} (\frac{\partial^2}{\partial r^2} + \frac{1}{r}\frac{\partial}{\partial r} +\frac{1}{r^2}\frac{\partial^2}{\partial \phi^2}) \psi (r, \phi) = E \psi(r, \phi).
\end{align}
The space is axially symmetrical which allows us to make a following \textit{ansatz} \cite{Cirone2001}:
\begin{align}
    \psi(r, \phi) = e^{im\phi}\frac{u_m(r)}{\sqrt{r}}.
\end{align}
In that notation probability density of finding particle at distance $r$ from the origin is $|u_m(r)|^2$ \cite{Rembielinski2002}. We use the above formulation to arrive at the radial Schrödinger equation \cite{Wheeler2001, Cirone2001}:
\begin{align}
    [\frac{d^2 }{d r^2}  + \frac{2M}{\hbar^2}(E -  V_m(r))] u_m(r) = 0,
\end{align}
where $V_m(r)$ an effective potential \cite{Wheeler2001}:
\begin{align}
    V_m(r) = \frac{\hbar^2}{2M}\frac{m^2 - 1/4}{r^2}.
\end{align}
This potential is a sum of two potentials each with separate physical interpretation. The first term $\frac{\hbar^2}{2M}\frac{m^2}{r^2}$ is a result of the angular motion of the particle and manifest itself as a centrifugal force. The second term $- \frac{\hbar^2}{2M}\frac{1}{4r^2}$ is present even if particle has angular momentum equal to $0$. This term manifest itself as a quantum anticetrifugal force \cite{Wheeler2001}. Let us denote this term by $V^{(2)}$:
\begin{align}
    V_{QFF}^{(2)} = - \frac{\hbar^2}{2M}\frac{1/4}{r^2}. \label{eq:qffradial}
\end{align}
The quantum anticentrifugal force is a result of a radial motion and has been linked to interference effect \cite{Wheeler2001}. In $R^2 \backslash \{ 0,0 \}$ space it acts as an attractive force toward the origin of coordinate system. The analysis of equation of motion in $D$-dimensional space for particle with vanishing angular momentum allowed for a formulation of general equation for such potential \cite{Botero2003}:
\begin{align}
    V_{QFF}^{(D)} = \frac{\hbar^2}{2M}\frac{(D-1)(D-3)}{4r_D^2},
\end{align}
where $D$ stands for number of dimensions and $r_D = \sqrt{x_1^2 + x_2^2 + \dots + x_D^2}$. For $D = 2$,  the fictitious potential  $V_{QFF}$ is  negative and gives rise to quantum anticetrifugal force. The fictitious potential arises in 2D geometries but is equal to $0$ in one- and three-dimensional systems. For systems with higher dimensionality it manifests itself as a repulsive force \cite{Birula2007}, but only in two-space dimensional system it acts as an attractive force \cite{Cirone2001, Rembielinski2002, Birula2007, Dandoloff2011, Dandoloff2015}. 

\subsubsection{Bent waveguide analogy}
Let us rewrite \eref{eq:radialBessel} into following form:
\begin{align}
\frac{\partial^2 R}{\partial r^2} + \frac{1}{r} \frac{\partial R}{\partial r} - \frac{\lambda^2 - h_j^2r^2}{r^2} R = 0. 
\end{align}
In the above equation we substitute function $R(r)$ with function $\frac{u(r)}{\sqrt{r}}$. After simple calculations we arrive at the equation:
\begin{align}
[\frac{\partial^2 }{\partial r^2}  - \frac{\lambda^2 - \frac{1}{4}}{r^2}] u(r) = - h_j^2 u(r).  \label{eq:qffBessel}
\end{align}
The solutions of \eref{eq:qffBessel} must be the same as solutions of \eref{eq:qffradial} for $\lambda^2 = m^2$ and $E = -  \frac{\hbar^2 h_j^2}{2M}$, given the same boundary conditions. It is therefore reasonable to treat a bent waveguide as a platform for researching quantum anticentrifugal force. 

%% file: main.bbl
\begin{thebibliography}{10}

\bibitem{Harris1967}
S.~E. Harris, M.~K. Oshman, and R.~L. Byer.
\newblock Observation of tunable optical parametric fluorescence.
\newblock {\em Phys. Rev. Lett.}, 18:732--734, May 1967.

\bibitem{Shalm2015}
L.~K. Shalm, E.~Meyer-Scott, B.~G. Christensen, P.~Bierhorst, M.~A. Wayne,
  M.~J. Stevens, T.~Gerrits, S.~Glancy, D.~R. Hamel, M.~S. Allman, K.~J.
  Coakley, S.~D. Dyer, C.~Hodge, A.~E. Lita, V.~B. Verma, C.~Lambrocco,
  E.~Tortorici, A.~L. Migdall, Y.~Zhang, D.~R. Kumor, W.~H. Farr, F.~Marsili,
  M.~D. Shaw, J.~A. Stern, C.~Abell{\'a}n, W.~Amaya, V.~Pruneri, T.~Jennewein,
  M.~W. Mitchell, P.~G. Kwiat, J.~C. Bienfang, R.~P. Mirin, E.~Knill, and S.~W.
  Nam.
\newblock Strong loophole-free test of local realism$^{*}$.
\newblock {\em Phys. Rev. Lett.}, 115(25):250402, December 2015.

\bibitem{Giustina2015}
Marissa Giustina, Marijn A.~M. Versteegh, SŲren Wengerowsky, Johannes
  Handsteiner, Armin Hochrainer, Kevin Phelan, Fabian Steinlechner, Johannes
  Kofler, Jan Ňke Larsson, Carlos AbellŠn, Waldimar Amaya, Valerio Pruneri,
  Morgan~W. Mitchell, JŲrn Beyer, Thomas Gerrits, Adriana~E. Lita, Lynden~K.
  Shalm, Sae~Woo Nam, Thomas Scheidl, Rupert Ursin, Bernhard Wittmann, and
  Anton Zeilinger.
\newblock Significant-loophole-free test of bell's theorem with entangled
  photons.
\newblock {\em Phys. Rev. Lett.}, 115:250401, 2015.

\bibitem{Aharon2008}
N.~Aharon and L.~Vaidman.
\newblock Quantum advantages in classically defined tasks.
\newblock {\em Phys. Rev. A}, 77(5):052310, May 2008.

\bibitem{Kolenderski2011}
Piotr Kolenderski, Urbasi Sinha, Li~Youning, Tong Zhao, Matthew Volpini, Adan
  Cabello, Raymond Laflamme, and Thomas Jennewein.
\newblock Playing the aharon-vaidman quantum game with a young type photonic
  qutrit.
\newblock {\em Phys. Rev. A}, 86:012321, July 2012.

\bibitem{Khan2018}
Imran Khan, Bettina Heim, Andreas Neuzner, and Christoph Marquardt.
\newblock Satellite-based qkd.
\newblock {\em Optics and Photonics News}, 29(2):26--33, 2018.

\bibitem{Liao2017}
Sheng-Kai Liao, Wen-Qi Cai, Wei-Yue Liu, Liang Zhang, Yang Li, Ji-Gang Ren,
  Juan Yin, Qi~Shen, Yuan Cao, Zheng-Ping Li, et~al.
\newblock Satellite-to-ground quantum key distribution.
\newblock {\em Nature}, 549(7670):43--47, 2017.

\bibitem{Pugh2017}
Christopher~J Pugh, Sarah Kaiser, Jean-Philippe Bourgoin, Jeongwan Jin, Nigar
  Sultana, Sascha Agne, Elena Anisimova, Vadim Makarov, Eric Choi, Brendon~L
  Higgins, and Thomas Jennewein.
\newblock Airborne demonstration of a quantum key distribution receiver
  payload.
\newblock {\em Quantum Sci. Technol.}, 2(2):024009, 2017.

\bibitem{kako2006gallium}
Satoshi Kako, Charles Santori, Katsuyuki Hoshino, Stephan G{\"o}tzinger,
  Yoshihisa Yamamoto, and Yasuhiko Arakawa.
\newblock A gallium nitride single-photon source operating at 200 k.
\newblock {\em Nature materials}, 5(11):887--892, 2006.

\bibitem{shields2007semiconductor}
Andrew~J Shields.
\newblock Semiconductor quantum light sources nat.
\newblock In {\em Photonics}, volume~1, pages 215--223, 2007.

\bibitem{yonezu2017efficient}
Yuya Yonezu, Kentaro Wakui, Kentaro Furusawa, Masahiro Takeoka, Kouichi Semba,
  and Takao Aoki.
\newblock Efficient single-photon coupling from a nitrogen-vacancy center
  embedded in a diamond nanowire utilizing an optical nanofiber.
\newblock {\em Scientific reports}, 7(1):1--9, 2017.

\bibitem{Marcuse1980a}
Dietrich Marcuse.
\newblock {\em Principles of Quantum Electronics}.
\newblock Academic Press, May 28 1980.

\bibitem{Saleh2007}
B.~E.~A. Saleh and M.~C. Teich.
\newblock {\em Fundamentals of Photonics}.
\newblock John Wiley \& Sons, INC, 2019.

\bibitem{Kato2017}
Kiyoshi Kato, Kentaro Miyata, and Valentin Petrov.
\newblock Phase-matching properties of baga4se7 for shg and sfg in the
  0.901--10.5910\&\#x2009;\&\#x2009;\&\#x03bc;m range.
\newblock {\em Appl. Opt.}, 56(11):2978--2981, Apr 2017.

\bibitem{Miyata2017}
Kentaro Miyata, Valentin Petrov, and Kiyoshi Kato.
\newblock Phase-matching properties of ligase 2 for shg and sfg in the
  1.026--10.5910 $\mu$m range.
\newblock {\em Appl. Opt.}, 56(22):6126--6129, 2017.

\bibitem{Yin2017}
Juan Yin, Yuan Cao, Yu-Huai Li, Sheng-Kai Liao, Liang Zhang, Ji-Gang Ren,
  Wen-Qi Cai, Wei-Yue Liu, Bo~Li, Hui Dai, Guang-Bing Li, Qi-Ming Lu, Yun-Hong
  Gong, Yu~Xu, Shuang-Lin Li, Feng-Zhi Li, Ya-Yun Yin, Zi-Qing Jiang, Ming Li,
  Jian-Jun Jia, Ge~Ren, Dong He, Yi-Lin Zhou, Xiao-Xiang Zhang, Na~Wang, Xiang
  Chang, Zhen-Cai Zhu, Nai-Le Liu, Yu-Ao Chen, Chao-Yang Lu, Rong Shu,
  Cheng-Zhi Peng, Jian-Yu Wang, and Jian-Wei Pan.
\newblock Satellite-based entanglement distribution over 1200 kilometers.
\newblock {\em Science}, 356(6343):1140--1144, 2017.

\bibitem{Lutz2013}
Thomas Lutz, Piotr Kolenderski, and Thomas Jennewein.
\newblock Toward a downconversion source of positively spectrally correlated
  and decorrelated telecom photon pairs.
\newblock {\em Opt. Lett.}, 38(5):697--699, Mar 2013.

\bibitem{Lutz2014}
Thomas Lutz, Piotr Kolenderski, and Thomas Jennewein.
\newblock Demonstration of spectral correlation control in a source of
  polarization entangled photon pairs at telecom wavelength.
\newblock {\em Opt. Lett.}, 39(6):1481, September 2014.

\bibitem{Gajewski2016}
Andrzej Gajewski and Piotr Kolenderski.
\newblock Spectral correlation control in down-converted photon pairs.
\newblock {\em Phys. Rev. A}, 94(1):013838, Jul 2016.

\bibitem{Wheeler2001}
M.~A. Cirone, K.~Rza\ifmmode \mbox{\c{}}\else \c{}\fi{}\ifmmode~\dot{z}\else
  \.{z}\fi{}ewski, W.~P. Schleich, F.~Straub, and J.~A. Wheeler.
\newblock Quantum anticentrifugal force.
\newblock {\em Phys. Rev. A}, 65:022101, Dec 2001.

\bibitem{Cirone2001}
M.~A. Cirone, K.~{Rza{\.z}ewski}, W.~P. {Schleich}, F.~{Straub}, and J.~A.
  {Wheeler}.
\newblock Quantum anticentrifugal force.
\newblock {\em Phys. Rev. A}, 65(2):022101, February 2001.

\bibitem{Rembielinski2002}
Krzysztof Kowalski, Krzysztof Podlaski, and Jakub Rembielinski.
\newblock Quantum mechanics of a free particle on a pointed plane revisited.
\newblock 06 2002.

\bibitem{Birula2007}
Iwo Bialynicki-Birula, M.~Cirone, J.~Dahl, T.~Seligman, F.~Straub, and
  W.~Schleich.
\newblock {\em Quantum Fictitious Forces}, pages 170 -- 178.
\newblock 11 2007.

\bibitem{Dandoloff2011}
R.~Dandoloff and V.~Atanasov.
\newblock Quantum anticentrifugal potential in a bent waveguide.
\newblock {\em Annalen der Physik}, 523(11):925--930, Sep 2011.

\bibitem{Dandoloff2015}
Victor Atanasov and Rossen Dandoloff.
\newblock The curvature of the rotating disk and its quantum manifestation.
\newblock {\em Physica Scripta}, 90:065001, 06 2015.

\bibitem{leighton1965feynman}
Robert~B Leighton and Matthew Sands.
\newblock {\em The Feynman lectures on physics}.
\newblock Addison-Wesley Boston, MA, USA, 1965.

\bibitem{Jackson1999}
J.~D. Jackson.
\newblock {\em Classical electrodynamics}.
\newblock John Wiley \& Sons, INC, 3 edition, 1999.

\bibitem{gallager2013stochastic}
R.G. Gallager.
\newblock {\em Stochastic Processes: Theory for Applications}.
\newblock Stochastic Processes: Theory for Applications. Cambridge University
  Press, 2013.

\bibitem{loudon1973}
R.~Loudon.
\newblock {\em The Quantum Theory of Light}.
\newblock OUP Oxford, 1973.

\bibitem{Brown1956}
R.~H. Brown and R.~Q. Twiss.
\newblock Correlation between photons in two coherent beams of light.
\newblock 177:27--29, January 1956.

\bibitem{Knight2005}
Christopher~C. Gerry and Peter~L. Knight.
\newblock {\em Introductory Quantum Optics}.
\newblock Cambridge University Press, 1 edition, 2008.

\bibitem{Glauber1963a}
Roy~J Glauber.
\newblock The quantum theory of optical coherence.
\newblock {\em Phys. Rev.}, 130(6):2529, 1963.

\bibitem{Mandel1995}
Leonard Mandel and Emil Wolf.
\newblock {\em Optical coherence and quantum optics}.
\newblock Cambridge, 1995.

\bibitem{beveratos2002room}
Alexios Beveratos, Sergei K{\"u}hn, Rosa Brouri, Thierry Gacoin, J-P Poizat,
  and Philippe Grangier.
\newblock Room temperature stable single-photon source.
\newblock {\em The European Physical Journal D-Atomic, Molecular, Optical and
  Plasma Physics}, 18(2):191--196, 2002.

\bibitem{Garrison2008}
J.~C. Garrison and R.~Y. Chiao.
\newblock {\em Quantum Optics}.
\newblock Oxford University Press, 2008.

\bibitem{Boyd2003}
Robert~W. Boyd.
\newblock {\em Nonlinear Optics}.
\newblock Academic Press, 2 edition, 2003.

\bibitem{Horn2013}
Rolf~T. Horn, Piotr Kolenderski, Dongpeng Kang, Carmelo Scarcella,
  Adriano~Della Frera, Alberto Tosi, Lukas~G. Helt, Sergei~V. Zhukovsky,
  John~E. Sipe, Gregor Weihs, Amr~S. Helmy, and Thomas Jennewein.
\newblock Inherent polarization entanglement generated from a monolithic
  semiconductor chip.
\newblock {\em Sci. Rep.}, 3:2314, April 2013.

\bibitem{Misiaszek2018}
Piotr~Kolenderski Marta~Misiaszek, Andrzej~Gajewski.
\newblock Dispersion measurement method with down conversion process.
\newblock {\em J. Phys. Commun.}, 2(6):065014, 2018.

\bibitem{Jundt1997}
Dieter~H Jundt.
\newblock Temperature-dependent sellmeier equation for the index of refraction,
  n e, in congruent lithium niobate.
\newblock {\em Opt. Lett.}, 22(20):1553--1555, 1997.

\bibitem{Kato2014a}
Kiyoshi Kato and Nobuhiro Umemura.
\newblock Sellmeier and thermo-optic dispersion formulas for liins 2.
\newblock {\em Appl Opt}, 53(33):7998--8001, 2014.

\bibitem{Kato2014}
Kiyoshi Kato and Takuya Mikami.
\newblock Set of sellmeier equations for gas and gase and its applications to
  the nonlinear optics in gas x se 1- x.
\newblock {\em Appl. Opt.}, 53(10):2177--2179, 2014.

\bibitem{Takaoka1999}
E.~Takaoka and K.~Kato.
\newblock Temperature phase-matching properties for harmonic generation in
  gase.
\newblock {\em Jpn. J. Appl. Phys.}, 38:2755--+, May 1999.

\bibitem{Boeuf2000}
N.~Boeuf, D.~{Branning}, I.~{Chaperot}, E.~{Dauler}, S.~{Guerin}, G.~{Jaeger},
  A.~{Muller}, and A.~L. {Migdall}.
\newblock Calculating characteristics of noncollinear phase matching in
  uniaxial and biaxial crystals.
\newblock {\em Opt. Eng.}, 39:1016--1024, April 2000.

\bibitem{Kato2002}
Kiyoshi Kato and Eiko Takaoka.
\newblock Sellmeier and thermo-optic dispersion formulas for ktp.
\newblock {\em Appl. Opt.}, 41(24):5040, Aug 2002.

\bibitem{Manjooran2012}
S.~Manjooran, H.~Zhao, I.~T. Lima, and A.~Major.
\newblock Phase-matching properties of ppktp, mgo:ppslt and mgo:ppcln for
  ultrafast optical parametric oscillation in the visible and near-infrared
  ranges with green pump.
\newblock {\em Laser Phys.}, 22(8):1325--1330, 2012.

\bibitem{Stoumbou2013}
Eleni Stoumbou, Ilias Stavrakas, George Hloupis, Alex Alexandridis, Dimos
  Triantis, and Konstantinos Moutzouris.
\newblock A comparative study on the use of the extended-cauchy dispersion
  equation for fitting refractive index data in crystals.
\newblock {\em Opt. Quantum. Electron.}, 45(8):837--859, 2013.

\bibitem{Lee2012}
H.~J. Lee, H.~Kim, M.~Cha, and H.~S. Moon.
\newblock Simultaneous type-0 and type-ii spontaneous parametric
  down-conversions in a single periodically poled ktiopo4 crystal.
\newblock {\em Appl. Phys. B}, 108(3):585--589, Jun 2012.

\bibitem{Zhao2010a}
H.~Zhao, I.~T. Lima, Jr., and A.~Major.
\newblock Peculiarities of temperature-dependent sellmeier equations for
  periodically poled ktiopo4 crystal in the near-infrared and visible ranges.
\newblock {\em Proc. SPIE}, 7750:77501D--77501D--8, 2010.

\bibitem{Sedziak2019}
Karolina Sedziak-Kacprowicz, Miko{\l}aj Lasota, and Piotr Kolenderski.
\newblock Remote temporal wavepacket narrowing.
\newblock {\em Scientific Reports}, 9(1):1--9, 2019.

\bibitem{Kolenderski2009}
Piotr Kolenderski, Wojciech Wasilewski, and Konrad Banaszek.
\newblock Modelling and optimization of photon pair sources based on
  spontaneous parametric down-conversion.
\newblock {\em Phys. Rev. A}, 80:013811, May 2009.

\bibitem{Bhattacharya2002}
Bhaskar Bhattacharya and Desale Habtzghi.
\newblock Median of the p value under the alternative hypothesis.
\newblock {\em The American Statistician}, 56(3):202--206, 2002.

\bibitem{Smith1996}
C~G Smith.
\newblock Low-dimensional quantum devices.
\newblock {\em Reports on Progress in Physics}, 59(2):235--282, feb 1996.

\bibitem{Nadareishvili2009}
T.~Nadareishvili and Anzor Khelashvili.
\newblock Some problems of self-adjoint extension in the schrodinger equation.
\newblock 03 2009.

\bibitem{Botero2003}
J.~Botero, M.A. Cirone, J.P. Dahl, F.~Straub, and W.P. Schleich.
\newblock Geometry, commutation relations and the quantum fictitious force.
\newblock {\em Appl. Phys. B: Lasers Opt.}, 76(2):129, Feb 2003.

\bibitem{Marcatili1969}
E.~A.~J. Marcatili.
\newblock Dielectric rectangular waveguide and directional coupler for
  integrated optics.
\newblock {\em The Bell System Technical Journal}, 48(7):2071--2102, 1969.

\bibitem{griffithsED}
David~J. Griffiths.
\newblock {\em INTRODUCTION TO ELECTRODYNAMICS}.
\newblock Pearson Education, Inc, 4 edition, 1942.

\bibitem{Westerveld2012}
Wouter~J. Westerveld, Suzanne~M. Leinders, Koen W.~A. van Dongen, H.~Paul
  Urbach, and Mirvais Yousefi.
\newblock Extension of marcatili's analytical approach for rectangular silicon
  optical waveguides.
\newblock {\em Journal of Lightwave Technology}, 30(14):2388--2401, 2012.

\bibitem{dudorov2003millimeter}
D.L.S.T.S. Dudorov, B.~D{\"u}nweg, D.~Lioubtchenko, D.P. Landau, S.~Tretyakov,
  A.I. Milchev, S.~Dudorov, North Atlantic Treaty Organization.
  Scientific~Affairs Division, and Scientific Affairs Division~Staff North
  Atlantic Treaty~Organization.
\newblock {\em Millimeter-Wave Waveguides}.
\newblock NATO Science Series. Springer US, 2003.

\bibitem{Gajewski2020}
Andrzej Gajewski, Daniel Gustaw, Nor~Roshidah Yusof, Norshamsuri Ali, Karolina
  S{\l}owik, and Piotr Kolenderski.
\newblock Waveguide platform for quantum anticentrifugal force.
\newblock {\em Opt. Lett.}, 45(13):3373, jun 2020.

\bibitem{WolframBesselFirst}
Eric~W. Weisstein.
\newblock Bessel function of the first kind.
\newblock {\em From MathWorld--A Wolfram Web Resource.}

\bibitem{WolframBesselSecond}
Eric~W. Weisstein.
\newblock Bessel function of the second kind.
\newblock {\em From MathWorld--A Wolfram Web Resource.}

\bibitem{jin2015}
Jian-Ming Jin.
\newblock {\em The finite element method in electromagnetics}.
\newblock John Wiley \& Sons, 2015.

\bibitem{Sedziak2017}
Karolina Sedziak, Mikolaj Lasota, and Piotr Kolenderski.
\newblock Reducing detection noise of a photon pair in a dispersive medium by
  controlling its spectral entanglement.
\newblock {\em Optica}, 4(1):84, jan 2017.

\bibitem{Arfken2013}
Frank~Harris G.B.~Arfken, H.J~Weber.
\newblock {\em Mathematical Methods for Physicists}.
\newblock 2013.

\bibitem{shankar2012principles}
Ramamurti Shankar.
\newblock {\em Principles of quantum mechanics}.
\newblock Springer Science \& Business Media, 2012.

\end{thebibliography}
